Doctoral Dissertation

# Study of local conserved quantities in the one-dimensional Hubbard model

A Dissertation Submitted for the Degree of Doctor of Philosophy
December 2023

Department of Physics, Graduate School of Science,
The University of Tokyo

Kohei Fukai

# Abstract

In this thesis, I study local charges in the one-dimensional Hubbard model, derive their explicit expressions, and show their completeness. Quantum integrable lattice models can be exactly solvable via the Bethe Ansatz method, and the characteristic feature is that they possess an extensive number of local charges. Their existence has been well understood through the quantum inverse scattering method; local charges are generated from the transfer matrix. Alternatively, one can obtain local charges by using the Boost operator; the local charges are recursively generated from the charges with a smaller support.

Although the existence of local charges and the formalism to obtain them are well known, it is practically difficult to derive general expressions of the local charges. There is no universal understanding of the structure of the local charges applicable to all integrable models, and there are few integrable models whose local charges are fully understood. Understanding the structure of local charges has notably advanced for some integrable spin chains. The structures of the local charges are now well understood for some integrable spin chains, such as the Heisenberg chain, the spin-$1/2$ XYZ chain, and the models represented by the Temperley-Lieb algebra.

The situation is quite different for integrable electron systems such as the one-dimensional Hubbard model. There have been no cases in which the structure of the local charges is completely understood for integrable electron systems. One of the difficulties in the electron systems is the absence of the Boost operator, which comes from the violation of the Lorentz invariance due to the spin-charge separation. This makes the study of their local charges even more challenging, and their investigation is highly desirable.

In this thesis, I study the local charges $\{Q_k\}$ in the one-dimensional Hubbard model, which is an integrable system used for theoretical studies of non-perturbative effects in strongly correlated electron systems. I obtained their explicit expressions in a closed form. An expression of the $k$-local charge $Q_k$ for $k > 5$, where $k$ denotes the support, had been unknown in previous studies. I find that $Q_k$ is a linear combination of a particular kind of products of the hopping terms. I introduce a diagrammatic notation to represent these products efficiently, which enables the prediction of the general formula of the local charges. These linear combinations have nontrivial coefficients for $k > 5$, which were not known in previous studies. Some of the coefficients were proved to be identical to the generalized Catalan numbers, which also appear in the local charges of the Heisenberg chain. I derive the recursion equation for these coefficients.

I also prove that the obtained local charges are exhaustive. Thus, any local charge is written as a linear combination of $\{Q_k\}$. Although the completeness of the local charges is a natural conjecture in the transfer-matrix formulation, its proof has not been presented for almost all cases. I show that the local charges generated by the transfer matrix are represented as linear combinations of my results $\{Q_k\}$.

Lastly, I emphasize that this is the first study demonstrating general explicit expressions of the local charges for an electron system that lacks the boost operator. One can use my result to construct the generalized current operators, which have an essential role in the study of the factorization of correlation functions, and generalized hydrodynamics, a novel way to study the nonequilibrium dynamics in integrable models.

# Contents

# List of Figures

# List of Publications

The bulk of the original research content of this thesis is contained in Chapters 5 and 6. The result of Chapter 5, the explicit expressions for the local charges in the one-dimensional Hubbard model, follows the paper [1]:

- Kohei Fukai, “All Local Conserved Quantities of the One-Dimensional Hubbard Model”, Phys. Rev. Lett. **131**, 256704 (2023).

The content in Section 5.5 is a new result and not included in Ref. [1].

The result of Chapter 6, the proof of the completeness of the local charges in the one-dimensional Hubbard model, follows the paper [2]:

- Kohei Fukai, "Proof of completeness of the local charges in the one-dimensional Hubbard model," arXiv:2309.09354.

The following four publications of mine are not directly related to this thesis.

- Y. Nozawa and K. Fukai, “Explicit Construction of Local Conserved Quantities in the XYZ Spin-1/2 Chain”, Phys. Rev. Lett. **125**, 90602 (2020).
- K. Fukai, Y. Nozawa, K. Kawahara, and T. N. Ikeda, “Noncommutative generalized Gibbs ensemble in isolated integrable quantum systems,” Phys. Rev. Res. **2**, 33403 (2020).
- K. Yamada and K. Fukai, “Matrix product operator representations for the local conserved quantities of the Heisenberg chain,” SciPost Phys. Core **6**, 069 (2023).
- K. Fukai, R. Kleinemühl, B. Pozsgay, and E. Vernier, “On correlation functions in models related to the temperley-lieb algebra,” SciPost Phys. **16**, 003 (2024).

# Chapter 1

# Introduction

Quantum integrability and local conservation laws are two sides of the same coin. Quantum many-body systems are called "integrable" when their dynamics or thermodynamics can be computed exactly [3, 4]. Although the rigorous criterion of quantum integrability has yet to be formulated, it is widely accepted that quantum integrability is characterized by the existence of an extensive number of local charges, which bears an analogy to the Liouville-Arnold definition of classical integrability [5, 6]. Quantum integrable systems are usually exactly solvable by the Bethe Ansatz [7], and the existence of an extensive number of local charges $\{Q_k\}_{k\geq 2}$ can be shown from the quantum inverse scattering method [8]. An infinite set of the local charges is generated from the expansion of the transfer matrix $T(\lambda)$ made from the R-matrix satisfying the Yang-Baxter equation [8]. They are local in the sense that they are a linear combination of operators acting on finite-range lattice sites.

Recently, quantum integrable systems are becoming an arena for studying nonequilibrium quantum dynamics, inspired by their experimental realization with ultracold atoms [9–12], for which $Q_k$ play a crucial role; the existence of an extensive number of local charges leads to the absence of thermalization [13–15] and the conjectured long-time steady-state is the generalized Gibbs ensemble [16–18], involving all local and also quasi-local charges as well as the Hamiltonian [19–23] with associated generalized temperatures. The large-scale nonequilibrium behavior in quantum integrable lattice models is described by the generalized hydrodynamics [24, 25], which is based on the local continuity equations for the local charges and its current.

Although the existence of local conservation laws in quantum integrable systems itself has been well known, deriving their general expressions has still been a challenging problem. This difficulty lies not only in the computational expense of calculating the higher-order charges but also in their complex structure, for which one cannot easily find any patterns or regularities. The mysterious structure of local charges has been investigated in the spin-$1/2$ XXX chain and its $SU(N)$ generalization [26–29], the spin-$1/2$ XYZ chain [30], and the Temperley-Lieb models [31] including the spin-$1/2$ XXZ chain. For these models, the boost operator $B$ exists, with which one can efficiently calculate the local charges recursively by $[B, Q_k] = Q_{k+1}$. As outlined above, there has been significant recent progress in understanding the structure of local charges in integrable spin chains. However, in integrable electron systems, such as the one-dimensional Hubbard model, which feature both spin and charge degrees of freedom, there has yet to be an instance at which the general structure of local conservation laws is revealed, and a comprehensive understanding of their general structure is highly anticipated. A significant distinction of integrable electron systems from integrable spin chains is that the boost operator does not function in a recursive way unlike local charges in an integrable spin system. In this thesis, I study the local charges in the one-dimensional Hubbard model, a famous and basic integrable electron system.

The one-dimensional Hubbard model is the most fundamental model for strongly correlated electron systems and has been of considerable interest not only in solid-state physics but also in mathematical physics and high-energy physics [32]. The exact solvability of the one-dimensional Hubbard model is firstly revealed by Lieb and Wu [33], using the so-called nested Bethe ansatz [34]. Several thermodynamic quantities of the one-dimensional Hubbard model were also calculated with the string hypothesis and the thermodynamic Bethe ansatz [35, 36]. From the point of view of the quantum inverse scattering method [8], the R-matrix of the one-dimensional Hubbard model was first found by Shastry [37], and the algebraic Bethe-ansatz framework was constructed in Refs. [38, 39]. Despite the progress in understanding the integrability of the one-dimensional Hubbard model, its local conservation laws are not well understood; the local charges are found up to $Q_5$ [28], and its general structure has not been found until today.

In this thesis, I reveal the general structure of the local charges in the one-dimensional Hubbard model [1]. I found the operator basis that constructs the linear combination of the local charges. The operator basis can be viewed as a product of hopping terms and also represented by a product of the densities of local charges in the XX chain when expressed in the coupled spin chain representation via the Jordan-Wigner transformation [40]. The coefficients appearing in the one-dimensional Hubbard model are conjectured as trivial in [28] from the observation up to $Q_5$. Contrary to the conjecture of Ref. [28], I found non-trivial coefficients appear in the linear combination of the operator basis of the local charges beyond $Q_6$, which has not been accessed before. Some of the coefficients are the generalized Catalan numbers, which also appear in the local charges of the spin-$1/2$ Heisenberg chain and its $\mathrm{SU}(N)$ generalization [27, 28]. Considering the strong-coupling limit of my charges in the one-dimensional Hubbard model, I can derive another family of mutually commuting charges that is closely related to the $t$-0 model [32]. This is the first study that revealed the explicit expressions for the local charges in the integrable electron systems, for which there is no boost operator to calculate recursively local charges.

I also prove that there are no other local charges independent of what I have obtained [2]. The proof is based on the same spirit as the proof of the non-integrability of the spin-$1/2$ XYZ spin chain with a non-zero magnetic field [41] and the mixed-field Ising model [42]. Their strategy is straightforward; first, write down the linear combination, which consists of a candidate of local charges, using all local operators acting on finite-range sites, and demonstrate that if it is conserved, then all of the coefficients in this linear combination must be zero. I apply this straightforward method to the proof of the completeness of my local charges. This proof does not need detailed knowledge of the structure of the local charges that I have obtained.

This thesis is organized as follows. Chapter 2 is devoted to a review of the quantum inverse scattering method, and I will see that the existence of local charges in quantum integrable lattice models is assured abstractly from the Yang-Baxter algebra. Chapter 3 is devoted to a brief review of the integrability of the one-dimensional Hubbard model. Chapter 4 is devoted to a review of the progress in the understanding of the local conservation law in quantum integrable lattice models. In Chapter 5, I explain the main result of this thesis; explicit expressions for the local charges in the one-dimensional Hubbard model, and derive another mutually commuting family from their strong-coupling limit. In Chapter 6, I prove that there are no other local charges independent of those that I obtained in Chapter 5. Chapter 7 summarizes this thesis and the future outlook of my work.

# Chapter 2

# Review of quantum inverse scattering method

In this chapter, we briefly review the quantum inverse scattering method [8]. We overview that the quantum integrable lattice models are made from the solution of the Yang-Baxter equation, and they have infinitely many local conservation laws as a direct consequence of the Yang-Baxter algebra. We also explain that when the R-matrix is of a difference form, we can recursively derive the local charges utilizing the boost operator [43].

## 2.1 Yang-Baxter algebra

We first abstractly describe quantum integrability. We introduce an associative quadratic algebra $\mathcal{Y}$, called the *Yang-Baxter algebra*, defined in terms of the generators $T_\alpha^\beta(\lambda)$ ($\alpha, \beta = 1, \ldots, d$, $\lambda \in \mathbb{C}$). We define the *monodromy matrix* $T(\lambda)$ with its element being the generators of $\mathcal{Y}$ by

$$T(\lambda) \coloneqq \sum_{\alpha,\beta} T_\alpha^\beta(\lambda) e_\beta^\alpha \doteq \begin{pmatrix} T_1^1(\lambda) & \ldots & T_d^1(\lambda) \\ \vdots & \ddots & \vdots \\ T_1^d(\lambda) & \ldots & T_d^d(\lambda) \end{pmatrix}, \tag{2.1}$$

where $e_\beta^\alpha \in \mathrm{End}(\mathbb{C}^d)$ is the $d \times d$ matrix with a single non-zero element at $(\alpha, \beta)$ acting on the auxiliary space $V_a = \mathbb{C}^d$: $e_\beta^\alpha \left|e_\gamma\right\rangle = \delta_\gamma^\alpha \left|e_\beta\right\rangle$ and $\{\left|\gamma\right\rangle \in \mathbb{C}^d | \gamma = 1, \ldots, d\}$ is the basis of $\mathbb{C}^d$, and the rightmost matrix in Eq. (2.1) is the matrix representation of $T(\lambda)$ in terms of the basis of the auxiliary space. Here, $d \in \mathbb{N}$ is the dimension of the auxiliary space.

The generators of the Yang-Baxter algebra $\mathcal{Y}$ satisfy the quadratic relation, which is the so-called RTT relation:

$$R(\lambda, \mu) T_1(\lambda) T_2(\mu) = T_2(\mu) T_1(\lambda) R(\lambda, \mu), \tag{2.2}$$

where $\forall\lambda, \mu \in \mathbb{C}$, and $R(\lambda,\mu) \in \mathrm{End}(\mathbb{C}^d \otimes \mathbb{C}^d)$ is a c-number $d^2 \times d^2$ matrix, which is called the ***R-matrix***, and $T_1(\lambda)$ and $T_2(\mu)$ are defined by

$$T_1(\lambda) := T_\alpha^\beta(\lambda) e_\beta^\alpha \otimes I_d = T(\lambda) \otimes I_d, \tag{2.3}$$

$$T_2(\lambda) := T_\alpha^\beta(\lambda) I_d \otimes e_\beta^\alpha = I_d \otimes T(\lambda), \tag{2.4}$$

where $I_d$ is the $d \times d$ identity matrix, and the summations over $\alpha$ and $\beta$ are omitted and implicitly contracted. Here and in the following, repeated indices denote implicit summation. We assume that $R(\lambda,\mu)$ is invertible for almost all $\lambda, \mu \in \mathbb{C}$. We write explicitly the relation (2.2) as

$$R^{\alpha_1\alpha_2}_{\beta_1\beta_2}(\lambda,\mu) T^{\beta_1}_{\gamma_1}(\lambda) T^{\beta_2}_{\gamma_2}(\mu) = T^{\alpha_2}_{\beta_2}(\mu) T^{\alpha_1}_{\beta_1}(\lambda) R^{\beta_1\beta_2}_{\gamma_1\gamma_2}(\lambda,\mu), \tag{2.5}$$

where $R^{\alpha_1\alpha_2}_{\beta_1\beta_2}(\lambda,\mu)$ is the matrix element of the R-matrix: $R(\lambda,\mu)\left|e_{\beta_1}\right\rangle \otimes \left|e_{\beta_2}\right\rangle = R^{\alpha_1\alpha_2}_{\beta_1\beta_2}(\lambda,\mu)\left|e_{\alpha_1}\right\rangle \otimes \left|e_{\alpha_2}\right\rangle$.

The Yang-Baxter algebra includes rich commutative subalgebra. We define the following element of $\mathcal{Y}$:

$$t(\lambda) := \mathrm{tr}_{\mathrm{aux}}\, T(\lambda) = T^\alpha_\alpha(\lambda). \tag{2.6}$$

In the following subsection, we will see that $t(\lambda)$ becomes the transfer matrix, which is the source of the local charges. The most important fact is that $t(\lambda)$'s mutually commute:

$$[t(\lambda), t(\mu)] = 0. \tag{2.7}$$

Equation (2.7) indicates that $t(\lambda)$ serves as the generating function for a commutative subalgebra. When we expand $t(\lambda)$ with respect to $\lambda$ as in

$$t(\lambda) = I_0 + \lambda I_1 + \lambda^2 I_2 + \cdots, \tag{2.8}$$

then from Eq. (2.7), we have

$$[I_k, I_l] = 0 \qquad (\forall k, l \geq 0). \tag{2.9}$$

The commutativity (2.7) can be proved from the relation (2.5) as follows:

$$\begin{aligned} t(\mu)t(\lambda) &= T^{\alpha_1}_{\alpha_1}(\mu) T^{\alpha_2}_{\alpha_2}(\lambda) \\ &= R^{\alpha_1\alpha_2}_{\beta_1\beta_2}(\lambda,\mu) T^{\beta_1}_{\gamma_1}(\lambda) T^{\beta_2}_{\gamma_2}(\mu) R^{-1}{}^{\gamma_1\gamma_2}_{\alpha_1\alpha_2}(\lambda,\mu) \\ &= (R(\lambda,\mu)R^{-1}(\lambda,\mu))^{\gamma_1\gamma_2}_{\beta_1\beta_2} T^{\beta_1}_{\gamma_1}(\lambda) T^{\beta_2}_{\gamma_2}(\mu) \\ &= \delta^{\gamma_1}_{\beta_1}\delta^{\gamma_2}_{\beta_2} T^{\beta_1}_{\gamma_1}(\lambda) T^{\beta_2}_{\gamma_2}(\mu) \\ &= T^{\gamma_1}_{\gamma_1}(\lambda) T^{\gamma_2}_{\gamma_2}(\mu) \\ &= t(\lambda)t(\mu). \end{aligned} \tag{2.10}$$

Quantum integrable lattice models can be derived from the representation of the Yang-Baxter algebra, as seen in the following subsections.

## 2.2 Yang-Baxter equation

The defining relation of the Yang-Baxter algebra (2.2) imposes some restrictions on the R-matrix. In the following, every auxiliary space $V_a$ ($a = 1, 2, 3$) is assumed to be $V_a = \mathbb{C}^d$. The R-matrix acting nontrivially on the auxiliary spaces $V_a$ and $V_b$ is denoted by $R_{ab}(\lambda, \mu)$. Then, the R-matrix in (2.2) is written as $R_{12}(\lambda, \mu)$. In the same way, the monodromy matrix $T(\lambda)$ acting nontrivially on the auxiliary space $V_a$ is denoted by $T_a(\lambda)$. To write explicitly,

$$R_{ab}(\lambda, \mu) \coloneqq R^{\beta_1\beta_2}_{\alpha_1\alpha_2}(\lambda, \mu) e_{a\beta_1}^{\ \alpha_1} e_{b\beta_2}^{\ \alpha_2}, \tag{2.11}$$

$$T_a(\lambda) \coloneqq T^{\beta}_{\alpha}(\lambda) e_{a\beta}^{\ \alpha}, \tag{2.12}$$

where

$$e_{1\beta}^{\ \alpha} = e^{\alpha}_{\beta} \otimes I_d \otimes I_d, , \tag{2.13}$$

$$e_{2\beta}^{\ \alpha} = I_d \otimes e^{\alpha}_{\beta} \otimes I_d, \tag{2.14}$$

$$e_{3\beta}^{\ \alpha} = I_d \otimes I_d \otimes e^{\alpha}_{\beta}. \tag{2.15}$$

The RTT relation (2.2) is rewritten as

$$R_{ab}(\lambda, \mu) T_a(\lambda) T_b(\mu) = T_b(\mu) T_a(\lambda) R_{ab}(\lambda, \mu). \tag{2.16}$$

From the relation (2.16) and the associativity of $\mathcal{Y}$, we can see that the action of the R-matrix on the product of the monodromy matrix induces their permutation.

Let us consider the product of the monodromy matrix $T_1(\lambda_1)T_2(\lambda_2)T_3(\lambda_3)$ and its reverse order $T_3(\lambda_3)T_2(\lambda_2)T_1(\lambda_1)$. There are two ways of permutation to connect these two products:

$$T_1T_2T_3 \to T_2T_1T_3 \to T_2T_3T_1 \to T_3T_2T_1, \tag{2.17}$$

and

$$T_1T_2T_3 \to T_1T_3T_2 \to T_3T_1T_2 \to T_3T_2T_1. \tag{2.18}$$

We write explicitly these two ways of permutation. From the first path (2.17), we have

$$\begin{aligned}
T_1(\lambda_1)T_2(\lambda_2)T_3(\lambda_3) &= (T_1(\lambda_1)T_2(\lambda_2))T_3(\lambda_3) \\
&= R_{12}^{-1}(\lambda_1, \lambda_2)(T_2(\lambda_2)T_1(\lambda_1))T_3(\lambda_3)R_{12}(\lambda_1, \lambda_2) \\
&= R_{12}^{-1}(\lambda_1, \lambda_2)T_2(\lambda_2)(T_1(\lambda_1)T_3(\lambda_3))R_{12}(\lambda_1, \lambda_2) \\
&= (R_{13}(\lambda_1, \lambda_3)R_{12}(\lambda_1, \lambda_2))^{-1} T_2(\lambda_2)(T_3(\lambda_3)T_1(\lambda_1))R_{13}(\lambda_1, \lambda_3)R_{12}(\lambda_1, \lambda_2) \\
&= (R_{13}(\lambda_1, \lambda_3)R_{12}(\lambda_1, \lambda_2))^{-1} (T_2(\lambda_2)T_3(\lambda_3))T_1(\lambda_1)R_{13}(\lambda_1, \lambda_3)R_{12}(\lambda_1, \lambda_2) \\
&= (R_{23}(\lambda_2, \lambda_3)R_{13}(\lambda_1, \lambda_3)R_{12}(\lambda_1, \lambda_2))^{-1} T_3(\lambda_3)T_2(\lambda_2)T_1(\lambda_1) \\
&\qquad \times R_{23}(\lambda_2, \lambda_3)R_{13}(\lambda_1, \lambda_3)R_{12}(\lambda_1, \lambda_2).
\end{aligned} \tag{2.19}$$

In a similar manner, from the path (2.18), we have

$$T_1(\lambda_1)T_2(\lambda_2)T_3(\lambda_3) = (R_{12}(\lambda_1,\lambda_2)R_{13}(\lambda_1,\lambda_3)R_{23}(\lambda_2,\lambda_3))^{-1}T_3(\lambda_3)T_2(\lambda_2)T_1(\lambda_1) \times R_{12}(\lambda_1,\lambda_2)R_{13}(\lambda_1,\lambda_3)R_{23}(\lambda_2,\lambda_3). \tag{2.20}$$

From Eqs. (2.19) and (2.20), we have

$$[T_3(\lambda_3)T_2(\lambda_2)T_1(\lambda_1), M(\lambda_1,\lambda_2,\lambda_3)] = 0, \tag{2.21}$$

where

$$M(\lambda_1,\lambda_2,\lambda_3) \equiv R_{12}(\lambda_1,\lambda_2)R_{13}(\lambda_1,\lambda_3)R_{23}(\lambda_2,\lambda_3) \times (R_{23}(\lambda_2,\lambda_3)R_{13}(\lambda_1,\lambda_3)R_{12}(\lambda_1,\lambda_2))^{-1}. \tag{2.22}$$

When the Yang-Baxter algebra $\mathcal{Y}$ is well defined as an associative algebra, i.e., Eqs. (2.19) and (2.20) is compatible, Eq. (2.21) must holds true for any $\lambda_1, \lambda_2, \lambda_3 \in \mathbb{C}$. We can see that Eq. (2.21) hold if $M(\lambda_1,\lambda_2,\lambda_3)$ is just an identity operator:

$$M(\lambda_1,\lambda_2,\lambda_3) = I, \tag{2.23}$$

and rewriting this, we have the so-called *Yang-Baxter equation*:

$$R_{12}(\lambda_1,\lambda_2)R_{13}(\lambda_1,\lambda_3)R_{23}(\lambda_2,\lambda_3) = R_{23}(\lambda_2,\lambda_3)R_{13}(\lambda_1,\lambda_3)R_{12}(\lambda_1,\lambda_2). \tag{2.24}$$

We can see that the Yang-Baxter equation (2.24) is a sufficient condition for Eqs. (2.19) and (2.20) to be compatible. Within the algebra (2.2), the Yang-Baxter equation emerges as a natural way to impose the associativity of the Yang-Baxter algebra $\mathcal{Y}$.

We introduce the permutation operator $P \in \mathrm{End}(\mathbb{C}^d \otimes \mathbb{C}^d)$ by

$$P := e_\alpha^\beta \otimes e_\beta^\alpha, \tag{2.25}$$

and when the permutation operator non-trivially acts on the space $V_a$ and $V_b$, we use the notation $P_{a,b} = e_{a\alpha}^{\ \beta} e_{b\beta}^{\ \alpha}$. Then we define an alternative form of the R-matrix by $\check{R}_{a,b}(\lambda,\mu) = P_{a,b}R_{a,b}(\lambda,\mu)$. The Yang-Baxter equation is rewritten as

$$\check{R}_{23}(\lambda,\mu)\check{R}_{12}(\lambda,\nu)\check{R}_{23}(\mu,\nu) = \check{R}_{12}(\mu,\nu)\check{R}_{23}(\lambda,\nu)\check{R}_{12}(\lambda,\mu)\,. \tag{2.26}$$

The R-matrix is invertible, and we assume that the R-matrix satisfies the unitarity condition

$$R_{a,b}(\lambda,\mu)R_{b,a}(\mu,\lambda) = I, \tag{2.27}$$

where $I$ is the identity operator, and this equation is also rewritten as

$$\check{R}(\lambda,\mu)\check{R}(\mu,\lambda) = I_d \otimes I_d. \tag{2.28}$$

We call the R-matrix *regular* if there exist spectral parameters $\lambda_0$ and $\mu_0$ that satisfy

$$R(\lambda_0, \mu_0) = P, \tag{2.29}$$

and we call such $(\lambda_0, \mu_0)$ the *shift point*.

We introduce a graphical representation of the R-matrix and the Yang-Baxter equation. The R-matrix $R(\lambda, \mu)$ can be represented by the two crossing arrows as in

$$R^{\alpha_2\beta_2}_{\alpha_1\beta_1}(\lambda, \mu) =$$

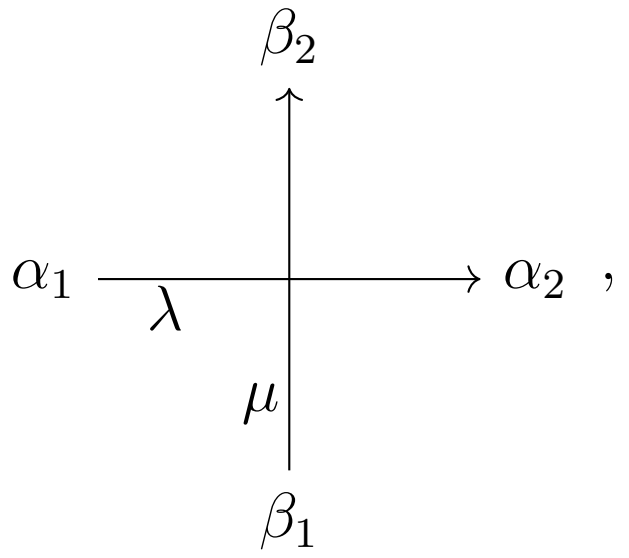

(2.30)

where the horizontal arrow 'carries' the spectral parameter $\lambda$ and the vertical arrow 'carries' the spectral parameter $\mu$, and hence we attached the corresponding indices to the arrows in Eq. (2.30). The R-matrix can be considered as the "scattering" of two particles with the rapidity (spectral parameter) $\lambda$ and $\mu$. The R-matrix can also be understood as a (directed) vertex and a Boltzmann weight of a two-dimensional vertex model.

Using this graphical representation of the R-matrix, we can represent the Yang-Baxter equation (2.24) graphically as

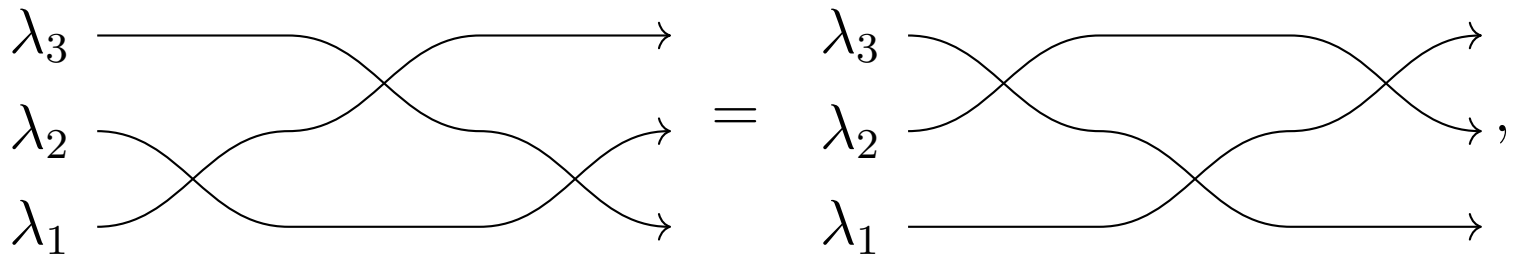

(2.31)

where we omit the indices of the R-matrices and only show the spectral parameters. We assume the contraction of indices by the connection of lines.

We represent the identity as straight line:

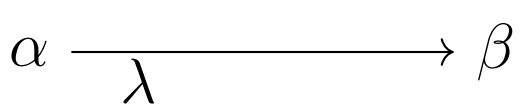

$$= \delta^{\beta}_{\alpha}. \tag{2.32}$$

Then, we represent the unitarity conditions (2.27) and (2.28) as

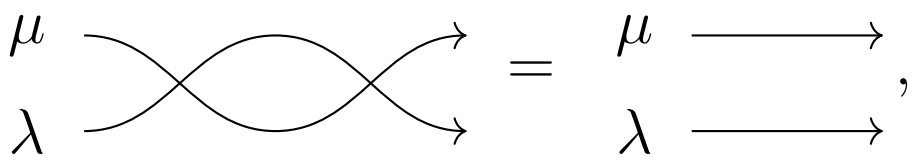

(2.33)

and a regular R-matrix (2.29) with a shift point $(\lambda_0, \mu_0)$ is represented as

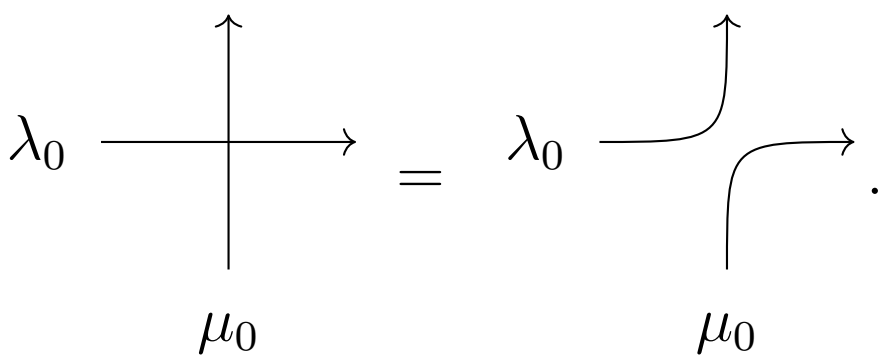

(2.34)

Below, we will see that the realization of the Yang-Baxter algebra is made from the solution of the Yang-Baxter equation (2.24).

## 2.3 Representation of Yang-Baxter algebra

Let us see the physical representation of the Yang-Baxter algebra, i.e., the representation of the generators $T_\alpha^\beta(\lambda)$ in terms of the operator on the physical Hilbert space. In the following, we let the physical Hilbert space denoted by $\mathcal{H} = \bigotimes_{j=1}^L V_j = (\mathbb{C}^d)^{\otimes L}$, where $V_j = \mathbb{C}^d$ is the local physical Hilbert space on the $j$ th site and $L$ is the system size. The Hilbert space $\mathcal{H}$ also denotes the quantum space and $V_j$ also denotes the local quantum space. In the following, the indices at the subscript of the operator, such as $R_{a,i}$, indicate the local space that the operator non-trivially acts on. The indices $a$ and $b$ indicate the auxiliary spaces $V_a$ and $V_b$, respectively, and the indices $j\ (=1, 2, \dots L)$ indicate the local quantum space $V_j$.

We first introduce the *Lax operator*, which non-trivially acts on the auxiliary space $V_a$ and the quantum space $V_j$, and is required to satisfy the following intertwining relation, which is so-called the RLL equation:

$$R_{ab}(\lambda,\mu)\mathcal{L}_{a,j}(\lambda)\mathcal{L}_{b,j}(\mu) = \mathcal{L}_{b,j}(\mu)\mathcal{L}_{a,j}(\lambda)R_{ab}(\lambda,\mu). \tag{2.35}$$

Here, $R_{ab}(\lambda,\mu)$ nontrivially acts on the two auxiliary spaces $V_a$ and $V_b$. We can construct the representation of the Yang-Baxter algebra from the Lax operator:

$$T_a(\lambda) = \mathcal{L}_{a,L}(\lambda)\cdots\mathcal{L}_{a,1}(\lambda). \tag{2.36}$$

This is an operator non-trivially acting on $V_a \otimes \mathcal{H}$. It can be shown that the RLL relation (2.35) leads to the RTT relation (2.2):

$$\begin{aligned}
&R_{ab}(\lambda,\mu)T_a(\lambda)T_b(\mu)\\
=&R_{ab}(\lambda,\mu)\mathcal{L}_{a,L}(\lambda)\mathcal{L}_{b,L}(\mu)\mathcal{L}_{a,L-1}(\lambda)\mathcal{L}_{b,L-1}(\mu)\cdots\mathcal{L}_{a,1}(\lambda)\mathcal{L}_{b,1}(\mu)\\
=&\mathcal{L}_{b,L}(\mu)\mathcal{L}_{a,L}(\lambda)R_{ab}(\lambda,\mu)\mathcal{L}_{a,L-1}(\lambda)\mathcal{L}_{b,L-1}(\mu)\cdots\mathcal{L}_{a,1}(\lambda)\mathcal{L}_{b,1}(\mu)\\
=&\mathcal{L}_{b,L}(\mu)\mathcal{L}_{a,L}(\lambda)\mathcal{L}_{b,L-1}(\mu)\mathcal{L}_{a,L-1}(\lambda)R_{ab}(\lambda,\mu)\cdots\mathcal{L}_{a,1}(\lambda)\mathcal{L}_{b,1}(\mu)\\
&\quad\vdots\\
=&\mathcal{L}_{b,L}(\mu)\mathcal{L}_{a,L}(\lambda)\mathcal{L}_{b,L-1}(\mu)\mathcal{L}_{a,L-1}(\lambda)\cdots\mathcal{L}_{b,1}(\mu)\mathcal{L}_{a,1}(\lambda)R_{ab}(\lambda,\mu) &(2.37)\\
=&T_b(\mu)T_a(\lambda)R_{ab}(\lambda,\mu). &(2.38)
\end{aligned}$$

We can make the Lax operator satisfying the RLL relation (2.35) from the solution of the Yang-Baxter equation (2.24):

$$\mathcal{L}_{a,j}(\lambda) = R_{a,j}(\lambda,\xi_j), \tag{2.39}$$

where $\xi_j$ is a fixed parameter. We can immediately prove from the Yang-Baxter equation (2.24) that the Lax operator defined by Eq. (2.39) satisfies the RLL equation (2.35). The representation of $\mathcal{Y}$ in Eq. (2.39) is called a *fundamental representation*. The transfer matrix (2.6) is now an operator on $\mathcal{H}$ in the representation of Eq. (2.39):

$$t(\lambda) = \mathrm{tr}_a\left[R_{a,L}(\lambda,\xi_L)R_{a,L-1}(\lambda,\xi_{L-1})\cdots R_{a,1}(\lambda,\xi_1)\right], \tag{2.40}$$

where the trace is taken over the auxiliary space $V_a$. The transfer matrix (2.40) is written graphically as

$t(\lambda) =$

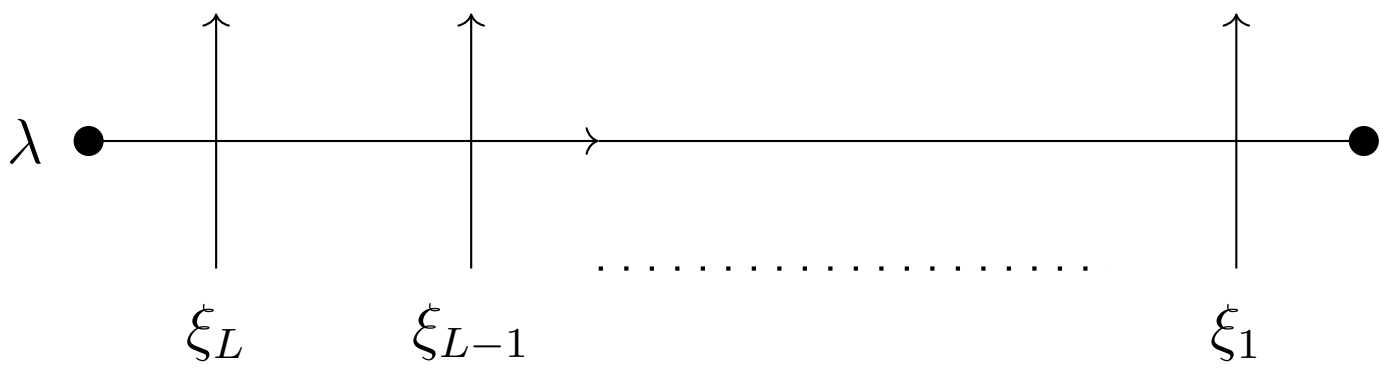

, (2.41)

where the filled circles at the right and left ends denote the contraction of the indices. While we already know the mutual commutativity of the transfer matrix on the algebra level (2.7), we give the proof of the mutual commutativity again using the graphical notation:

$t(\lambda)t(\mu) =$

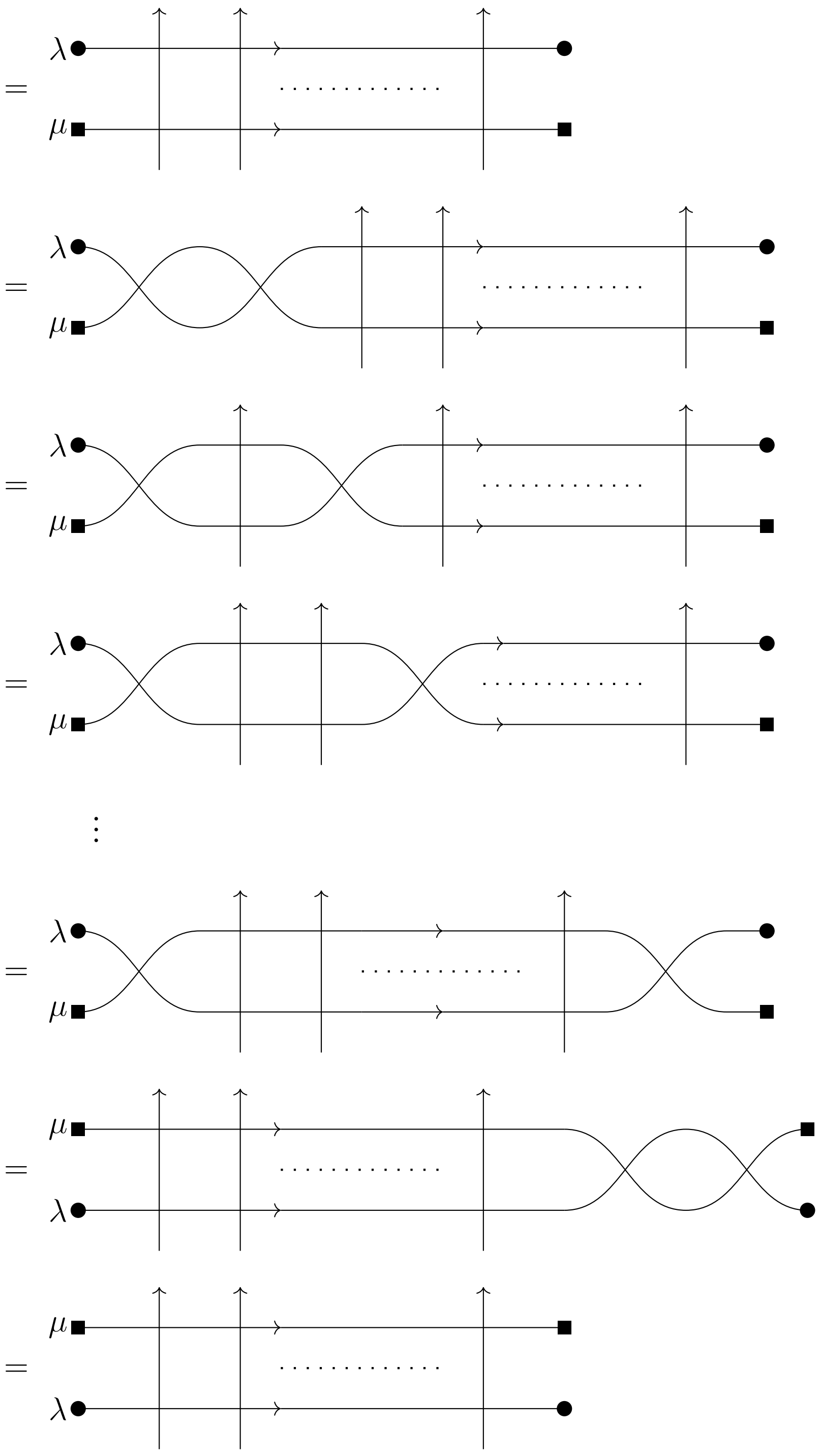

$$= t(\mu)t(\lambda)\,, \tag{2.42}$$

where we have used the Yang-Baxter equation (2.31) repeatedly as well as the unitarity condition (2.33), and we omit the spectral parameter of the vertical arrows. Note that the transfer matrix takes the form of the matrix-product operator.

Quantum integrable systems for which the Yang-Baxter algebra exists attract much interest because they offer a robust method of solving the eigenspectrum problem. The algebraic Bethe ansatz is one such method, employing the direct use of quadratic commutation relations of the Yang-Baxter algebra (2.2). Remember the algebraic solution of the harmonic oscillator; the algebraic Bethe ansatz has spirits similar to the algebraic solution of the harmonic oscillator in some sense. Within the algebraic Bethe ansatz, it is crucial to identify elements of the monodromy matrix as "quasi-particle" creation and annihilation operators. An important point is that there must be a pseudo-vacuum annihilated by all annihilation operators, which often require the $\mathrm{U}(1)$ symmetry to the system. Therefore, the XXZ chain or the one-dimensional Hubbard model can be successfully solvable within the framework of the algebraic Bethe ansatz, for which the $\mathrm{U}(1)$ symmetry is present; however, the algebraic Bethe ansatz for the XYZ chain, which does not possess the $\mathrm{U}(1)$ symmetry, has not been known. In every instance to which the algebraic Bethe ansatz has been successfully applied to date, the elements of the monodromy matrix have been organized in such a manner that the monodromy matrix is arranged as an upper triangular matrix on the pseudo-vacuum. Additionally, the pseudo vacuum serves as an eigenstate of diagonal elements of the monodromy matrix. We leave the detail of the algebraic Bethe ansatz for the book [8].

## 2.4 Local conservation law and transfer matrix

In this subsection, we explain that the transfer matrix generates the Hamiltonian of a quantum integrable lattice model and its local charges. We already know that there exists an infinite number of mutually commuting quantities on the level of the algebra (2.9). In the following, we see that they become the local charges in the fundamental representation. We set the parameter $\xi_j$ to the shift point: $\xi_j = \mu_0$ in (2.39).

Local charges are generated from the expansion of the logarithm of the transfer matrix with respect to the spectral parameter at the shift point $\lambda = \lambda_0$:

$$\ln T(\lambda) = \sum_{k=0}^{\infty} \frac{(\lambda - \lambda_0)^k}{k!} Q_{k+1} \, ; \tag{2.43}$$

in other words,

$$Q_k = \frac{\partial^{k-1}}{\partial \lambda^{k-1}} \ln T(\lambda) \bigg|_{\lambda=\lambda_0} \, , \tag{2.44}$$

where $\widetilde{Q}_1 = iP$ is the total momentum operator, which is related to the translational one-site shift operator $\hat{U}$ as

$$\hat{U} = e^{iP} = P_{1,2} P_{2,3} \cdots P_{L-1,L} . \tag{2.45}$$

From the mutual commutativity (2.7) of the transfer matrix, the local charges also mutually commute:

$$[Q_k, Q_l] = 0 \quad (k, l \geq 0). \tag{2.46}$$

In many cases, $Q_2 = H$ often corresponds to the Hamiltonian of a quantum integrable lattice model:

$$H = \hat{U}^{-1}\partial_\lambda t(\lambda)\Big|_{\lambda=\lambda_0} = \sum_{j=1}^{L} h_{j,j+1}, \tag{2.47}$$

where the periodic boundary condition $h_{L,L+1} \equiv h_{L,1}$ is imposed, and the Hamiltonian density is given by the first derivative of the R-matrix:

$$h_{j,j+1} = \partial_\lambda \check{R}_{j,j+1}(\lambda, \mu_0)\big|_{\lambda=\lambda_0}. \tag{2.48}$$

The quantities $\{Q_k\}_{k\geq 3}$ are the higher-order local charges, where $Q_k$ is written as a linear combination of $k$-local operators, i.e., operators acting on adjacent $k$ sites at most. In this way, the existence of an extensive number of the local charges in quantum integrable lattice models itself is guaranteed from the underlying Yang-Baxter algebra.

## 2.5 Local conservation law and Boost operator

The other way to construct $Q_k$ is the usage of the boost operator [43], denoted by $B$ if it exists. We can construct the boost operator if the R-matrix is of a difference form, i.e., the R-matrix depends on the difference of the two spectral parameters, as in

$$R(\lambda, \mu) = R(\lambda - \mu), \tag{2.49}$$

where we set the shift point as $R(0) = P$ in this case and the Hamiltonian density $h_{j,j+1} = \check{R}'_{j,j+1}(0)$. When the R-matrix is of a difference form (2.49), the boost operator is given by

$$B = \sum_{j=1}^{L} j h_{j,j+1}. \tag{2.50}$$

With the boost operator, $Q_k$ can be calculated recursively by

$$[Q_k, B] = Q_{k+1}. \tag{2.51}$$

We present the proof of Eq. (2.50) [43]. We assume the space reflection symmetry of the Hamiltonian density of $h_{a,b} = h_{b,a}$ and the thermodynamic limit $L \to \infty$ here for simplicity. When the R-matrix is of a difference form (2.49), the Yang-Baxter equation (2.24) is rewritten as

$$R_{12}(\theta)R_{13}(\theta+\mu)R_{23}(\mu) = R_{23}(\mu)R_{13}(\theta+\mu)R_{12}(\theta). \tag{2.52}$$

Differentiating the Yang-Baxter equation (2.52) with respect to $\mu$ and then substitute $\mu = 0$, we have

$$R'_{a,j}(\theta)R_{a,j+1}(\theta) - R_{a,j}(\theta)R'_{a,j+1}(\theta) = [R_{a,j}(\theta)R_{a,j+1}(\theta), h_{j,j+1}]\,, \tag{2.53}$$

where we redesignate the space in Eq. (2.52) in the following manner: $(1,2,3) \to (a, j, j+1)$. Multiplying Eq. (2.53) from the left by the product $\prod_{i<j} R_{a,i}(\theta)$ and from the right by $\prod_{j+1<i} R_{a,i}(\theta)$, we have

$$\left(\prod_{i<j} R_{a,i}(\theta)\right) R'_{a,j}(\theta) \left(\prod_{j<i} R_{a,i}(\theta)\right) - \left(\prod_{i<j+1} R_{a,i}(\theta)\right) R'_{a,j+1}(\theta) \left(\prod_{j+1<i} R_{a,i}(\theta)\right)$$
$$= \left[\prod_{i=-\infty}^{\infty} R_{a,i}(\theta), h_{j,j+1}\right]. \tag{2.54}$$

Multiplying Eq. (2.54) by $j$, and then taking the summation over $j$, we have

$$\frac{\partial T(\theta)}{\partial \theta} = [T(\theta), B]\,. \tag{2.55}$$

Substituting Eq. (2.55) in Eq. (2.43), we have the boost recursion relation (2.50).

We next explain that the existence of the boost operator, i.e., the case in which the boost recursion relation (2.51) holds, immediately leads to the self-conserving currents [44]. We let the local charges denoted by

$$Q_k = \sum_x q_k(x), \tag{2.56}$$

where $q_k(x)$ is the density of $Q_k$ that satisfies $Uq_k(x)U^\dagger = q_k(x+1)$. In the following, we assume that the operator with a site index such as $a(x)$ satisfies the site translational relation: $Ua(x)U^\dagger = a(x+1)$. The current operator $J_\alpha = \sum_x J_\alpha(x)$ is the quantities that satisfy the continuity equation:

$$[H, q_\alpha(x)] = J_\alpha(x) - J_\alpha(x+1). \tag{2.57}$$

The meaning of the current operator $J_\alpha$ is the flow of $Q_\alpha$ under the time evolution dictated by $H$. When we set $\alpha = 2$, multiply $x$ for both sides, and then take the summation over $x$, we have

$$Q_3 = [H, B] = J_2, \tag{2.58}$$

where the first equality holds from the boost recursion relation (2.51). From Eq. (2.58), we can see that the energy current $J_2$ corresponds to the 3-local charge $Q_3$ and $J_2$ is a self-conserved current. In this case, the density of the first non-trivial local charge can be written as

$$q_3(x) = [q_2(x), q_2(x+1)]\,. \tag{2.59}$$

In the same way, the generalized current operator $J_\alpha^\beta = \sum_x J_\alpha^\beta(x)$ [45, 46], which describes the flow of $q_\alpha(x)$ under the time evolution generated $Q_\beta$, is defined through the generalized continuity equation

$$[Q_\beta, q_\alpha(x)] = J_\alpha^\beta(x) - J_\alpha^\beta(x+1), \tag{2.60}$$

where $J_\alpha^\beta$ is the current of $Q_\alpha$ under the time evolution dictated by $Q_\beta$. When we set $\beta = 2$, the generalized continuity equation (2.60) reduces to the usual continuity equation (2.57). It has been observed that $J_2^\beta$ is also a self-conserved current and satisfies $J_2^\beta = Q_{\beta+1}$ [44].

There are many quantum integrable lattice models for which the R-matrix is of difference form (2.49) and the local charges can be constructed by the boost recursion relation (2.51): the spin-$1/2$ XYZ chain, the Temperley-Lieb models, which include the spin-$1/2$ XXZ chain, and the $SU(N)$ invariant chain. In these models, the energy current is conserved thanks to the boost recursion equation. For the one-dimensional Hubbard model, the R-matrix is not of a difference form. Thus, there is no boost recursion relation, i.e., no recursive way to obtain the local charges, and the energy current is not self-conserved.

When the R-matrix is not of a difference form (2.49), the boost operator (2.50) does not satisfy the boost recursive relation (2.51). With some modification, we can formally make a "generalized" boost operator which includes the derivative terms with respect to the spectral parameter [47]. However, the connection to the currents (2.58) is then lost, and the calculation to obtain the local charges becomes much more complicated, and it may be impractical to obtain the local charges using the generalized boost operator.

# Chapter 3

# Review of the integrability of the one-dimensional Hubbard model

In this chapter, we briefly review the exact solvability of the one-dimensional Hubbard model. The Hubbard model is an electron system on a lattice, incorporating the Coulomb interaction by assuming a repulsive force only when two electrons occupy the same site [48]. In a related topic, C. N. Yang and Gaudin derived the exact solution of a Hamiltonian for a continuous-space electron system with delta function-type interactions by using a generalization of the Bethe ansatz method, known as the so-called nested Bethe-ansatz [34, 49]. Lieb and Wu applied the nested Bethe ansatz for the exact solution of the one-dimensional (1D) Hubbard model and derived the so-called Lieb-Wu equation [33]. In 1986, B. S. Shastry pioneered a novel approach to analyzing the Hubbard model by incorporating it into the framework of the quantum inverse scattering method. Shastry found the R-matrix for the one-dimensional Hubbard model [37], and Shiroishi and Wadati showed that Shastry's R-matrix actually satisfies the Yang-Baxter equation [50]. In the following, we briefly overview the above progress.

## 3.1 Hamiltonian and fundamental symmetry of the 1D Hubbard model

In this subsection, we introduce the Hamiltonian of the one-dimensional Hubbard model and briefly review its fundamental symmetry.

The Hamiltonian of the one-dimensional Hubbard model is

$$H = -t\sum_{j=1}^{L}\sum_{\sigma=\uparrow,\downarrow}\left(c_{j,\sigma}^{\dagger}c_{j+1,\sigma}+\text{h.c.}\right)+4U\sum_{j=1}^{L}\left(n_{j,\uparrow}-\frac{1}{2}\right)\left(n_{j,\downarrow}-\frac{1}{2}\right), \tag{3.1}$$

where $t$ is the hopping amplitude, $U$ is the coupling constant of the on-site Coulomb interaction, and $c_{j,\sigma}$ is the fermion annihilation operator of flavor $\sigma=\uparrow,\downarrow$ at the $j$ th site. Here we impose the periodic boundary condition: $c_{L+1,\sigma}=c_{1,\sigma}$. We set $t=1$ in the following.

The one-dimensional Hubbard model enjoys the $\mathrm{SU}(2)$ symmetry. The generators are

$$S^{+}=\sum_{i=1}^{L} c_{j,\uparrow}^{\dagger} c_{j,\downarrow}, \qquad S^{-}=\sum_{i=1}^{L} c_{j,\downarrow}^{\dagger} c_{j,\uparrow}, \qquad S^{z}=\frac{1}{2} \sum_{i=1}^{L}\left(n_{j,\uparrow}-n_{j,\downarrow}\right), \tag{3.2}$$

and we define

$$S^{x}=\frac{1}{2}\left(S^{+}+S^{-}\right), \qquad S^{y}=-\frac{\mathrm{i}}{2}\left(S^{+}-S^{-}\right). \tag{3.3}$$

We can see that they obey the commutation relations of $su(2)$:

$$\left[S^{\alpha}, S^{\beta}\right]=\epsilon^{\alpha \beta \gamma} S^{\gamma}, \tag{3.4}$$

where $\epsilon^{\alpha \beta \gamma}$ is the totally anti-symmetric tensor. These operators commute with the Hamiltonian:

$$\left[H, S^{\alpha}\right]=0. \tag{3.5}$$

When the system size $L$ is even, the one-dimensional Hubbard model enjoys another $\mathrm{SU}(2)$ symmetry, so called $\eta$-pairing symmetry [51, 52]. The generators for the $\eta$-pairing charges are:

$$\eta^{+}=\sum_{i=1}^{L}(-1)^{i+1} c_{j,\uparrow}^{\dagger} c_{j,\downarrow}^{\dagger}, \qquad \eta^{-}=\sum_{i=1}^{L}(-1)^{i+1} c_{j,\downarrow} c_{j,\uparrow}, \qquad \eta^{z}=\frac{1}{2}(N-L), \tag{3.6}$$

where $N \equiv \sum_{i=1}^{L}(n_{i,\uparrow}+n_{i,\downarrow})$ is the particle number, and we define

$$\eta^{x}=\frac{1}{2}\left(\eta^{+}+\eta^{-}\right), \qquad \eta^{y}=-\frac{\mathrm{i}}{2}\left(\eta^{+}-\eta^{-}\right), \tag{3.7}$$

where $N \equiv \sum_{i=1}^{L}(n_{i,\uparrow}+n_{i,\downarrow})$ is the particle number, and they obey the commutation relations of $su(2)$:

$$\left[\eta^{\alpha}, \eta^{\beta}\right]=\epsilon^{\alpha \beta \gamma} \eta^{\gamma}. \tag{3.8}$$

These operators commute with the Hamiltonian when the system size $L$ is even:

$$\left[H, \eta^{\alpha}\right]=0. \tag{3.9}$$

We note the $\mathrm{U}(1)$ charge $\eta^{z}$ is conserved for both even and odd $L$. We also note the mutual commutativity of the two $su(2)$ generators:

$$\left[S^{\alpha}, \eta^{\beta}\right]=0. \tag{3.10}$$

Not all representations of $\mathrm{SU}(2) \times \mathrm{SU}(2)$ are present for even $L$. This is because

$$S^{z}+\eta^{z}=N_{\uparrow}-\frac{L}{2}=\text{integer}, \tag{3.11}$$

where $N_\uparrow = \sum_{i=1}^{L} n_{i,\uparrow}$. We can see $(S^z, \eta^z) = (\text{even}, \text{even}), (\text{odd}, \text{odd})$. This means that the true symmetry is $\mathrm{SO}(4) \simeq (\mathrm{SU}(2) \times \mathrm{SU}(2))/\mathbb{Z}_2$ [52]. In the rest of this chapter, we assume that $L$ is even.

All those $\mathrm{SU}(2)$ charges above are linear combinations of one-local operators. In Appendix C, we demonstrate that there are no one-local charges other than those mentioned above.

The full $\mathrm{SO}(4)$ symmetry is only realized for the Hubbard Hamiltonian of the form (3.1). Adding a magnetic field term $-2BS_z$ breaks the rotational invariance, while the $\eta$-pairing invariance is preserved. Adding a chemical potential term $-\mu\hat{N}$, on the other hand, breaks the $\eta$-pairing symmetry but preserves the invariance under rotations. The chemical potential $\mu$ plays the same role of a symmetry-breaking field for the $\eta$-spin as the magnetic field $B$ for the spin.

## 3.2 Exact solution of the one-dimensional Hubbard model

We briefly explain the Lieb-Wu equations derived by Lieb and Wu [33]. By solving these equations, one can obtain eigenstates and the corresponding eigenenergy for the one-dimensional Hubbard model. We refer the interested reader who wants to understand the details of this chapter to a comprehensive review [32]. Consider the eigenstates of the one-dimensional Hubbard model for which the number of electrons is $N$ and the number of down-spin electrons is $M$, subject to the conditions $0 \leq 2M \leq N \leq L$. It is understood that the eigenstates with $N_\downarrow = M$ can be transformed into eigenstates with $N_\downarrow = N - M$ under the reversal of all spins, and eigenstates with the particle number $N$ can be transformed into eigenstates with the particle number $2L - N$ under the particle-hole transformation. Hence, the above conditions do not lose generality.

We denote the charge momenta by $k_j$ for $j = 1, \ldots, N$ and spin rapidities by $\lambda_l$ for $l = 1, \ldots, M$. They are the solution of the Lieb-Wu equation [33]:

$$\mathrm{e}^{\mathrm{i}k_j L} = \prod_{\ell=1}^{M} \frac{\lambda_\ell - \sin k_j - \mathrm{i}U}{\lambda_\ell - \sin k_j + \mathrm{i}U}, \quad j = 1, \ldots, N, \tag{3.12}$$

$$\prod_{j=1}^{N} \frac{\lambda_\ell - \sin k_j - \mathrm{i}U}{\lambda_\ell - \sin k_j + \mathrm{i}U} = \prod_{\substack{m=1 \\ m \neq \ell}}^{M} \frac{\lambda_\ell - \lambda_m - 2\mathrm{i}U}{\lambda_\ell - \lambda_m + 2\mathrm{i}U}, \quad \ell = 1, \ldots, M. \tag{3.13}$$

From the solution of the Lieb-Wu equation (3.12) and (3.13), we can construct the eigenstate of the Hamiltonian as in

$$|\psi_{\mathbf{k},\boldsymbol{\lambda}}\rangle = \sum_{\mathbf{x},\boldsymbol{\sigma}} \psi(\mathbf{x}; \boldsymbol{\sigma}|\mathbf{k}; \boldsymbol{\lambda}) c^\dagger_{x_N,\sigma_N} \cdots c^\dagger_{x_2,\sigma_2} c^\dagger_{x_1,\sigma_1} |0\rangle \,, \tag{3.14}$$

where $\mathbf{k} = \{k_1, \ldots, k_N\}$ and $\boldsymbol{\lambda} = \{\lambda_1, \ldots, \lambda_M\}$ denote the rapidities, while $\mathbf{x} = \{x_1, \ldots, x_N\}$ denotes the locations of the electrons and $\boldsymbol{\sigma} = \{\sigma_1, \ldots, \sigma_N\}$ denotes the corresponding spins.

The amplitude of the wave function (3.14) is given by

$$\psi(\mathbf{x};\boldsymbol{\sigma}|\mathbf{k};\boldsymbol{\lambda}) = \sum_{P\in\mathfrak{S}^N} \mathrm{sign}(PQ)\,\langle\boldsymbol{\sigma}Q|\mathbf{k}P,\boldsymbol{\lambda}\rangle \exp\left[\mathrm{i}\sum_{j=1}^{L} k_{P(j)}x_{Q(j)}\right], \tag{3.15}$$

where $\mathfrak{S}^N$ denotes the symmetric group of order $N$. The permutation $Q \in \mathfrak{S}^N$ is defined by

$$1 \le x_{Q(1)} \le x_{Q(2)} \le \cdots \le x_{Q(N)} \le L, \tag{3.16}$$

and the amplitude $\langle\boldsymbol{\sigma}|\mathbf{k},\boldsymbol{\lambda}\rangle$ is defined by

$$\langle\boldsymbol{\sigma}|\mathbf{k},\boldsymbol{\lambda}\rangle = \sum_{R\in\mathfrak{S}^M} A(\boldsymbol{\lambda}R)\prod_{\ell=1}^{M} F_{\mathbf{k}}\left(\lambda_{R(\ell)};y_\ell\right), \tag{3.17}$$

$$F_{\mathbf{k}}(\lambda;y) = \frac{2\mathrm{i}U}{\lambda-\sin k_y+\mathrm{i}U}\prod_{j=1}^{y-1}\frac{\lambda-\sin k_j-\mathrm{i}U}{\lambda-\sin k_j+\mathrm{i}U}, \tag{3.18}$$

$$A(\boldsymbol{\lambda}) = \prod_{1\le m<n\le M}\frac{\lambda_m-\lambda_n-2\mathrm{i}U}{\lambda_m-\lambda_n}, \tag{3.19}$$

where $y_l$ denotes the $l$ th index of the down spin in the sequence $\sigma_1,\ldots,\sigma_N$. If the number of down spins in $\sigma_1,\ldots,\sigma_N$ is different from $M$, we define $\langle\boldsymbol{\sigma}|\mathbf{k},\boldsymbol{\lambda}\rangle = 0$.

Not all eigenstates are obtained from the solution of the Lieb-Wu equation; the Bethe eigenstate (3.14) corresponds to the highest-weight state of the $\mathrm{SO}(4)$ multiplets. The action of the ladder operator of both spin and $\eta$-spin on a Bethe eigenstate generates degenerate multiplets of eigenstates. Using the string hypothesis [35], it was proved that the Bethe eigenstates and accompanied multiplets form a complete set of the Hilbert space for even $L$ [53].

The eigenenergy corresponding to the Bethe roots $\mathbf{k} = \{k_1,\ldots,k_N\}$ and $\boldsymbol{\lambda} = \{\lambda_1,\ldots,\lambda_M\}$, which are the solution of the Lieb-Wu equation (3.12) and (3.13), are

$$H\,|\psi_{\mathbf{k},\boldsymbol{\lambda}}\rangle = E_{\mathbf{k}}\,|\psi_{\mathbf{k},\boldsymbol{\lambda}}\rangle\ , \tag{3.20}$$

where

$$E_{\mathbf{k}} = -2\sum_{j=1}^{N}\cos k_j + U(L-2N) \tag{3.21}$$

and the corresponding momentum is

$$P_{\mathbf{k}} = \sum_{j=1}^{N} k_j \qquad (\mathrm{mod}\ 2\pi). \tag{3.22}$$

The Bethe state is also the eigenstate of the higher-order charge $Q_l$:

$$Q_l\,|\psi_{\mathbf{k},\boldsymbol{\lambda}}\rangle = Q_l(\mathbf{k})\,|\psi_{\mathbf{k},\boldsymbol{\lambda}}\rangle\ , \tag{3.23}$$

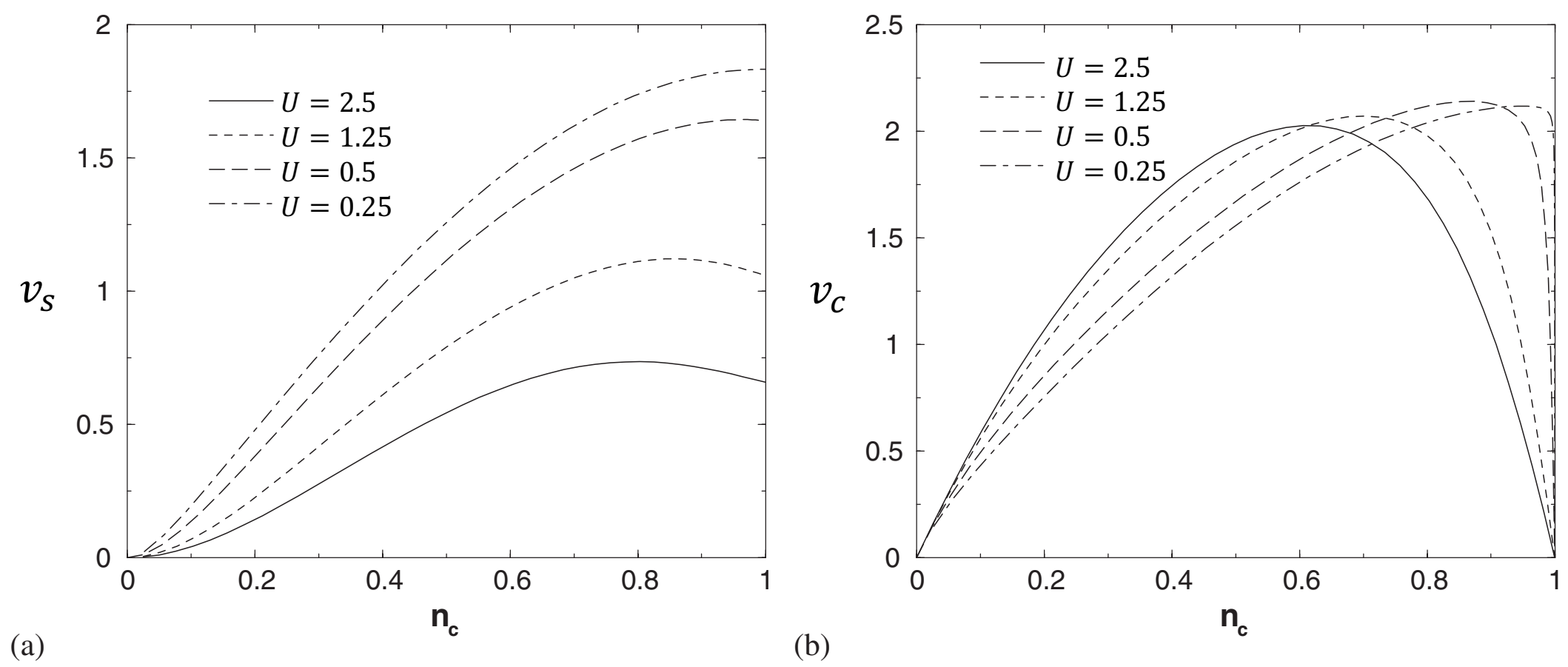

Figure 3.1: (a) Spin velocity and (b) charge velocity as functions of the particle density $n_c$ for several values of $U$ in the zero magnetic field. This figure is quoted from [32].

where $Q_l(\mathbf{k})$ is a function of the Bethe roots $\mathbf{k} = \{k_1, \ldots, k_N\}$ and $Q_2(\mathbf{k}) = E_{\mathbf{k}}$.

The solutions of Eqs. (3.12) and (3.13) completely determine the spectrum of the Hubbard Hamiltonian (3.1). They are the foundation for the investigation into the physical properties of the Hubbard model, such as the ground-state properties and elementary excitations in the thermodynamic limit [36, 54] and finite-temperature calculation of the thermodynamic quantities with the thermodynamic Bethe ansatz [35]. Moreover, from the finite-size corrections to the thermodynamic limit, we can calculate the asymptotic of correlation functions within the framework of the conformal field theory [54–59].

From the analysis of the ground-state properties with the Bethe Ansatz, we can see that the low-energy physics of the system is governed by spin-charge separation. This is a direct demonstration of the non-trivial effects of strong correlations in one-dimensional systems, which lead to the breakdown of the conventional Fermi liquid picture. In Figure 3.1, we show the spin and charge velocities $v_s$ and $v_c$, which can be calculated from the Bethe ansatz, as functions of the particle density $n_c$ for several values of $U$ in the zero magnetic field. The point $n_c = 1$ corresponds to half-filling, which produces the Mott insulating phase, where the charge excitation becomes gapped. When $n_c$ goes to $1$, the charge velocity $v_c$ goes to zero, whereas the spin velocity $v_s$ is finite, which indicates the occurrence of a quantum phase transition to the Mott insulating phase [54]. For the detail of the calculation of $v_s$ and $v_c$, we refer to Ref. [32].

## 3.3 R-matrix of the 1D Hubbard model and algebraic Bethe Ansatz

In this section, we briefly overview the history of the studies on the integrability of the one-dimensional Hubbard model from the perspective of the quantum inverse scattering method.

We explain the R-matrix of the one-dimensional Hubbard model. The R-matrix of the one-dimensional Hubbard model is first obtained by Shastry [37, 60]. He mapped the Hubbard model from the usual fermion notation to the spin variable notation via the Jordan-Wigner transformation [40] and obtained the R-matrix for the spin model. The Hamiltonian in the spin variable representation takes the form of two coupled XX chains:

$$H = t\sum_{j=1}^{L}\sum_{s=\uparrow,\downarrow}\sum_{a=x,y}\left(\sigma^{+}_{j,s}\sigma^{-}_{j+1,s} + \text{H.c.}\right) + U\sum_{j=1}^{L}\sigma^{z}_{j,\uparrow}\sigma^{z}_{j,\downarrow}, \tag{3.24}$$

and the transformation from the fermion variable to the spin variable is

$$c_{j,\uparrow} = \exp\left(\mathrm{i}\pi\sum_{l=1}^{j-1}\sigma^{+}_{l,\uparrow}\sigma^{-}_{l,\uparrow}\right)\sigma^{-}_{j,\uparrow}, \tag{3.25}$$

$$c_{j,\downarrow} = \exp\left(\mathrm{i}\pi\sum_{l=1}^{L}\sigma^{+}_{l,\uparrow}\sigma^{-}_{l,\uparrow}\right)\exp\left(\mathrm{i}\pi\sum_{l=1}^{j-1}\sigma^{+}_{l,\downarrow}\sigma^{-}_{l,\downarrow}\right)\sigma^{-}_{j,\downarrow}, \tag{3.26}$$

where $\sigma^{a}_{j,s}$ ($a \in \{x,y,z\}$ , $s \in \{\uparrow,\downarrow\}$) is the Pauli matrix on the $j$ th site of flavor $s$, which satisfies the commutation relation

$$\left[\sigma^{a}_{j,s}, \sigma^{b}_{j',s'}\right] = \delta_{s,s'}\delta_{j,j'}\epsilon^{a,b,c}\sigma^{c}_{j,s} \tag{3.27}$$

with $\sigma^{\pm}_{j,s} = \frac{1}{2}\left(\sigma^{x}_{j,s} \pm \mathrm{i}\sigma^{y}_{j,s}\right)$. Then, the relation $\sigma^{z}_{j,s} = 2n_{j,s} - 1$ and $\sigma^{+}_{j,s}\sigma^{-}_{j,s} = \frac{1}{2}\left(1 + \sigma^{z}_{j,s}\right) = n_{j,s}$ holds. When the periodic boundary condition is imposed for the fermion variable notation, the boundary condition for the spin variable notation is twisted as

$$\sigma^{\pm}_{L+1,\uparrow} = (-1)^{N_{\uparrow}}\sigma^{\pm}_{1,\uparrow} \tag{3.28}$$

$$\sigma^{\pm}_{L+1,\downarrow} = (-1)^{N}\sigma^{\pm}_{1,\downarrow}\,, \tag{3.29}$$

where the twist depends on the parity of the particle numbers. Nevertheless, we assume that the periodic boundary condition is assumed for the Hamiltonian in the spin variable notation (3.24).

Shastry extensively investigated the integrability of the one-dimensional Hubbard model in the spin variable notation (3.24) in the late 1980s [37, 60, 61]. In Ref. [37], he constructed a one-parameter family of transfer matrices of a two-dimensional classical vertex model that commutes with the Hamiltonian of the 1D Hubbard model (3.24), finding a Lax operator. Therefore, he showed that the one-dimensional Hubbard model has infinitely many local charges. In Ref. [60], he showed that any two transfer matrices of the family commute mutually, deriving the R-matrix of the one-dimensional Hubbard model, so that the R-matrix satisfies the RLL relation, which relies on brute-force calculations involving extensive computer algebra. Shastry's R-matrix $\check{R}(\lambda,\mu) \in \mathrm{end}(\mathbb{C}^4 \otimes \mathbb{C}^4)$ is given by

$$\begin{aligned}\check{R}(\lambda,\mu) &= \cos(\lambda+\mu)\cosh(h(\lambda,U) - \cosh(\mu,U))\check{r}(\lambda-\mu)\\ &\quad + \cos(\lambda-\mu)\sinh(h(\lambda,U) - \cosh(\mu,U))\check{r}(\lambda+\mu)\sigma_{1,\uparrow}\sigma_{1,\downarrow}\,,\end{aligned} \tag{3.30}$$

where $h(\lambda, U)$ is defined by

$$h(\lambda, U) \equiv \frac{1}{2} \sinh^{-1} (U \sin(2\lambda)) , \tag{3.31}$$

and $r_a(\lambda)$ is the R-matrix for the XX chain:

$$r_a(\lambda) = \frac{\cos(\lambda) + \sin(\lambda)}{2} + \frac{\cos(\lambda) - \sin(\lambda)}{2} \sigma^z_{1,a} \sigma^z_{2,a} + \left( \sigma^+_{1,a} \sigma_{2,a} + \sigma^-_{1,a} \sigma^+_{2,a} \right) , \tag{3.32}$$

and we defined $r(\lambda) \equiv r_\uparrow(\lambda) r_\downarrow(\lambda)$ and $\check{r}(\lambda) \equiv Pr(\lambda)$. In Ref. [61], Shastry developed a more elegant algebraic method, the so-called decorated star-triangle relation, to derive the R-matrix and a peculiar representation of the Yang-Baxter algebra simultaneously. Alternative derivations, which also include the formulation in terms of the fermion notation, were developed in Refs. [62–64]. It was first shown that Shastry's R-matrix actually satisfies the Yang-Baxter equation in Ref. [50]. The algebraic Bethe ansatz for the Hubbard model was established by Ramos and Martins [38, 39]. We can rederive the Lieb-Wu equation (3.12) and (3.13) in the algebraic Bethe ansatz. For a detailed explanation of the algebraic Bethe ansatz applied to the Hubbard model, we direct readers to the book [32].

This result was crucial for the quantum transfer matrix (QTM) method of calculating the thermodynamics quantities in the Hubbard model [65]. The QTM method drastically simplified the treatment of thermodynamics because we only have to solve a finite number of sets of nonlinear integral equations. In contrast, in the thermodynamic Bethe ansatz (TBA) approach [35], we have to consider an infinite number of sets of integral equations. While the calculations from QTM and those from TBA are believed to coincide with each other, the rigorous proof of their equivalence has not been established. For the demonstration of the equivalence in the case of the Heisenberg model, refer to Refs. [66, 67]. Within the QTM approach, thermodynamic quantities can be calculated numerically with a very high precision. The QTM method can also be applied to calculating correlation lengths at finite temperatures in the Hubbard model at half-filling [68–70]. For the case of less than half-filling, finite temperature-correlation lengths have not yet been calculated.

## 3.4 The 1D Hubbard model and experiments

The one-dimensional Hubbard model has been highly instrumental in interpreting experiments on quasi-one-dimensional materials. While not an exact representation of any specific materials, many of its qualitative aspects appear to be reflected in the real world. Currently, there is a significant number of materials whose electronic properties are thought to be captured by 'Hubbard-like' Hamiltonians. Notable examples include the chain cuprates $Sr_2CuO_3$ [71, 72], $SrCuO_2$ [73, 74], and organic conductors such as TTF-TCNQ [75].

However, the corresponding electronic Hamiltonians quantitatively differ from a simple one-band Hubbard model in all these examples. For example, in the angle-resolved photoemission (ARPES) experiment for $SrCuO_2$ [73], it was suggested that the spectra observed in $SrCuO_2$ can be understood quantitatively by the 1D $t$-$J$ model, rather than 1D Hubbard model. From the ARPES experiment for the organic conductor TTF-TCNQ [75], it was observed that the unusual behavior of the ARPES spectra of TTF-TCNQ is consistent with the 1D Hubbard model near half-filling, which shows the spin-charge separation over an energy scale of the conduction bandwidth. However, the nearly linear spectral onset at low energy cannot be reproduced by the simple 1D Hubbard model. Near the fermi energy, the density of states of an interacting 1D conductor is expected to behave as the power law $|E-E_F|^\alpha$, and it is known that for the only on-site Coulomb interaction, the Hubbard model yields $\alpha < 1/8$, conflicting with the ARPES data observation. When considering extended Hubbard models, including the next-nearest-neighbor interactions, the exponents can be up to $\alpha < 1$ [76]. Thus, replacing the Coulomb repulsion with a basic on-site Hubbard interaction is generally not a sufficient approximation.

# Chapter 4

# Review of progress on the structure of local charges

This chapter reviews previous studies on the local conservation laws in quantum integrable lattice models. The existence of local charges became widely known after the invention of the quantum scattering method [8]. Formally, it is widely known that local charges can be obtained through the expansion of the transfer matrix.

However, cases in which their general explicit formula is explicitly derived are quite rare. The general expression of local charges is first revealed for the spin-$1/2$ isotropic Heisenberg (XXX) chain independently by Anshelevich [26] and Grabowski and Mathieu [27]. They are subsequently obtained for the $\mathrm{SU}(N)$ fundamental spin chain, which is the natural generalization of the XXX chain, by Grabowski and Mathieu [28]. Recently, it has been reported that the expression for the local charges can be simplified in the matrix-product operator [29]. The explicit expressions for the local charges in the spin-$1/2$ XYZ chain, which is the anisotropic version of the above Heisenberg chain, are obtained by Nozawa and Fukai [30] using the doubling product [41], and for the XXZ case, independently obtained by Nienhuis and Huijgen [31] utilizing the Temperley-Lieb algebra [77]. The result in Ref. [77] can be applied to all the integrable systems that are represented by the Temperley-Lieb algebra, such as the Potts chain [78] and the Golden chain [79], and used to derive the factorization formula for the correlation function for the Temperley-Lieb models [80].

As mentioned above, for some integrable spin chains, the structure of local charges has been clarified. However, there are no cases for integrable electron systems, such as the one-dimensional (1D) Hubbard model, where the general expression for local charges is obtained. The local charges for the 1D Hubbard model have been obtained up to $Q_5$ [28, 60, 81, 82], and the general expression is given in Chapter 5. Grabowski and Mathieu also found the explicit expressions for the local charges up to the first order of the coupling constant [28].

Below, we explain the general expression of the local charges in the above spin systems and the previous study of local charges in the 1D Hubbard model. Moreover, we explain how to construct the current and generalized current from the explicit expression of the local charges [80, 83]. The periodic boundary condition is imposed for the integrable Hamiltonians in this chapter.

## 4.1 Spin-$1/2$ isotropic Heisenberg chain

The Hamiltonian of the spin-$1/2$ Heisenberg chain is given by

$$H = \sum_{i=1}^{L} \boldsymbol{\sigma}_i \cdot \boldsymbol{\sigma}_{i+1}, \tag{4.1}$$

where $\boldsymbol{\sigma}_i = (X_i, Y_i, Z_i)$ represents the vector of the standard Pauli matrices acting non-trivially on the $i$ th site, and $L$ denotes the system size. We assume the periodic boundary condition: $\boldsymbol{\sigma}_{i+L} = \boldsymbol{\sigma}_i$. The Hamiltonian (4.1) is integrable [7] and possesses an extensive number of local charges $\{Q_k\}_{k=2,3,4,\ldots}$.

To find the expression of $Q_k$, we introduce some notations. A sequence of $n$ sites, denoted as $\mathcal{C} = \{i_1, i_2, \ldots, i_n\}$, where $i_1 < i_2 < \ldots < i_n$, is referred to as a *cluster* of order $n$. A cluster $\mathcal{C}$ may be further classified by its *hole*, defined as $i_n - i_1 + 1 - n$, representing the count of sites between $i_1$ and $i_n$ that are not encompassed in $\mathcal{C}$. For example, the cluster $\mathcal{C} = \{1, 3, 4, 7\}$ is the cluster of order $4$ and hole $3$.

For a cluster $\mathcal{C} = \{i_1, i_2, \ldots, i_n\}$ of order $n$, we define the nested products of Pauli matrices $f_n(\mathcal{C})$ as in

$$f_n(\mathcal{C}) \coloneqq \boldsymbol{\sigma}_{i_1} \cdot \left(\boldsymbol{\sigma}_{i_2} \times \left(\boldsymbol{\sigma}_{i_3} \times \left(\cdots \times \left(\boldsymbol{\sigma}_{i_{n-1}} \times \boldsymbol{\sigma}_{i_n}\right) \cdots\right)\right)\right) \tag{4.2}$$

and we introduce a component of the local charges

$$F_{n,m} = \sum_{\mathcal{C} \in \mathcal{C}^{(n,m)}} f_n(\mathcal{C}), \tag{4.3}$$

where $\mathcal{C}^{(n,m)}$ denotes the set of clusters of order $n$ and $m$ holes and $1 \leq i_1 \leq L$.

The general formula for $Q_k$ is given by [26, 27]

$$Q_k = F_{k,0} + \sum_{\substack{1 \leq n+m < \lfloor k/2 \rfloor \\ 0 \leq n, 1 \leq m}} C_{n+m-1,n} F_{k-2(n+m),m}, \tag{4.4}$$

where

$$C_{k,n} \equiv \binom{k+n}{n} - \binom{k+n}{n-1} = \frac{k+1-n}{k+1} \binom{k+n}{n} \tag{4.5}$$

is the generalized Catalan number, [1] which also appears in the local charges in the 1D Hubbard model, as will be explained in Chapter 5.

[1]The coefficients of local charges of Eq. (4.2) in Ref. [28] are complicated. Then, we used a simpler notation here. The relation between our $C_{k,n}$ and $\alpha_{k,l}$ of Eq. (4.2) in Ref. [28] is $C_{k,n} = \alpha_{k+1,k+1-n}$. This can be easily seen by using Eq. (4.5) in Ref. [28]: $\alpha_{k+1,k+1-n} = \binom{k+n-1}{n} - \binom{k+n-1}{n-2} = \left(\binom{k+n}{n} - \binom{k+n-1}{n-1}\right) - \left(\binom{k+n}{n-1} - \binom{k+n-1}{n-1}\right) = \binom{k+n}{n} - \binom{k+n}{n-1} = C_{k,n}$. Note that the definition of the symbol $C_{n,m}$ employed in Ref. [28] is distinct from the generalized Catalan number in this work.

## 4.2 $\mathrm{SU}(N)$ generalization

The structure of the local charges of the $\mathrm{SU}(N)$ generalization of the spin-$1/2$ Heisenberg chain [84] is the same as that of the spin-$1/2$ Heisenberg chain [28]. The Hamiltonian of $\mathrm{SU}(N)$ fundamental chain is

$$H=\sum_{i=1}^{L}\sum_{a=1}^{N^2-1} t_i^a t_{i+1}^a, \tag{4.6}$$

where $t_i^a, a=1,\dots,N^2-1$ are the generators of $su(N)$ in the fundamental representation. In the case $N=2$, Eq. (4.6) reduces to the Hamiltonian of the spin-$1/2$ Heisenberg chain. We choose the convention that the generators $t_a$ are the $su(N)$ Gell-Mann matrix, satisfying the following relations:

$$\left[t^a, t^b\right]=2if^{abc}t^c, \tag{4.7}$$

$$t^at^b+t^bt^a=\frac{4}{N}\delta_{ab}+2d^{abc}t^c, \tag{4.8}$$

where $f^{abc}$ is the structure constant of $su(N)$ and $d^{abc}$ is a completely symmetric tensor, which is zero in the case $N=2$, which is the case of the usual spin-$1/2$ isotropic Heisenberg chain.

For the $\mathrm{SU}(N)$ case, the nested product $f_n(\mathcal{C})$ for a cluster $\mathcal{C}=\{i_1,i_2,\dots,i_n\}$ is defined by

$$f_n(\mathcal{C}) \coloneqq \mathbf{t}_{i_1}\cdot\left(\mathbf{t}_{i_2}\times\left(\mathbf{t}_{i_3}\times\left(\cdots\times\left(\mathbf{t}_{i_{n-1}}\times\mathbf{t}_{i_n}\right)\cdots\right)\right)\right), \tag{4.9}$$

where $\mathbf{t}_i=\left(t_i^1,\dots,t_i^{N^2-1}\right)$ represents the vector of the $su(N)$ Gell-Mann matrices acting non-trivially on the $i$ th site. The outer product of the vectors $\mathbf{A}$ and $\mathbf{B}$ with $N^2-1$ elements is defined by $(\mathbf{A}\times\mathbf{B})^c \equiv f^{abc}A^aB^b$.

The local charges in the $\mathrm{SU}(N)$ invariant chain (4.6) are expressed in the same form as (4.4), replacing $f_n(\mathcal{C})$ with that defined in Eq. (4.9).

Recently, it has been shown that the local charges in the spin-$1/2$ XXX chain and its $\mathrm{SU}(N)$ generalization can be simply written in terms of the matrix-product operator representation [29].

## 4.3 Spin-$1/2$ XYZ chain

We explain the explicit expressions of the local charges of the spin-$1/2$ XYZ chain, firstly obtained in Ref. [30]. The Hamiltonian of the spin-$1/2$ XYZ chain is

$$H=\sum_{i=1}^{L}\left[J_XX_iX_{i+1}+J_YY_iY_{i+1}+J_ZZ_iZ_{i+1}\right], \tag{4.10}$$

where $X_i, Y_i, Z_i$ are the usual Pauli matrices acting non-trivially on the $i$ th site, $L$ is the system size, and $J_X$, $J_Y$ and $J_Z$ are the coupling constants. We assume the periodic boundary condition: $X_{i+L}=X_i$, etc. The Hamiltonian (4.10) is integrable [3] and has a macroscopic number of local charges $\{Q_k\}_{k=2,3,4,\dots}$, where $Q_2=H$ is the Hamiltonian itself.

### 4.3.1 Doubling product notation

To represent the expression of $Q_k$, we define the doubling product notation [30, 41]:

$$\overline{A_1 A_2 \cdots A_n} := \sum_{i=1}^{L} (A_1)_i (A_1 A_2)_{i+1} (A_2 A_3)_{i+2} \cdots (A_{n-1} A_n)_{i+n-1} (A_n)_{i+n} \, , \tag{4.11}$$

where $A_i \in \{X, Y, Z\}$, $A_l A_{l+1}$ is the product of $A_l$ and $A_{l+1}$, and $(\cdot)_i$ denotes the operator acting on the $i$ th site. The hole is equal to the number of $j$ that satisfies $A_j = A_{j+1}$. We define the support of an operator as the range of sites on which it acts. The support of the doubling product of Eq. (4.11) is $n+1$. To represent the number of holes, we introduce the abbreviated notation for the doubling product:

$$\overline{A_1^{m_1+1} A_2^{m_2+1} \cdots A_t^{m_t+1}} := \overline{\underbrace{A_1 \ldots A_1}_{m_1+1} \underbrace{A_2 \ldots A_2}_{m_2+1} \ldots \underbrace{A_t \ldots A_t}_{m_t+1}} \times (-i)^{t-1} \, , \tag{4.12}$$

where the letter $A_i$ repeats $m_i + 1$ times, and $A_i \neq A_{i+1}$, and $m_i \geq 0$. The hole of the doubling product of Eq. (4.12) is $m = \sum_i^t m_i$. In the following, we use the abbreviated notation (4.12) for the doubling product. With the doubling product notation, the Hamiltonian (4.10) is rewritten as

$$H = J_X \overline{X} + J_Y \overline{Y} + J_Z \overline{Z} \, . \tag{4.13}$$

### 4.3.2 Local charges and simplification of the coefficients

The general expression of $Q_k$ was first obtained in Ref. [30]. Here, we give the simplified coefficients appearing in the local charges from those introduced in Ref. [30]. The general formula for the local charges in the spin-1/2 XYZ chain is

$$Q_k = \sum_{\substack{0 \leq n+m < \lfloor k/2 \rfloor, \\ n,m \geq 0}} \sum_{\overline{\boldsymbol{A}} \in \mathcal{S}_k^{n,m}} R^n_{\left(N_X^{\boldsymbol{A}}+m,\, N_Y^{\boldsymbol{A}}+m,\, N_Z^{\boldsymbol{A}}+m\right)} J_{\boldsymbol{A}} \, \overline{\boldsymbol{A}} \, , \tag{4.14}$$

where $\mathcal{S}_k^{n,m}$ is the set of all the doubling product with $k - 2n - m$ support and $m$ holes, $\overline{\boldsymbol{A}} \equiv \overline{A_1^{m_1+1} A_2^{m_2+1} \cdots A_t^{m_t+1}}$, $J_{\boldsymbol{A}} \equiv (J_X J_Y J_Z)^m \prod_{i=1}^t J_{A_i}^{1-m_i}$, $t \equiv k - 2(n+m) - 1$, and $m \equiv \sum_{i=1}^t m_i$. The symbols $N_X$, $N_Y$ and $N_Z$ denote the numbers of $X$, $Y$ and $Z$ in the doubling-letter $A_1 A_2 \cdots A_t$, respectively. The coefficient $R^n_{(N_X, N_Y, N_Z)}$ is given by

$$R^n_{(N_X,N_Y,N_Z)} := 4T^n_{(N_X,N_Y,N_Z)} - T^n_{(N_X+1,N_Y,N_Z)} - T^n_{(N_X,N_Y+1,N_Z)} - T^n_{(N_X,N_Y,N_Z+1)} \, , \tag{4.15}$$

where $T^n_{(N_X,N_Y,N_Z)}$ is defined by

$$T^n_{(N_X,N_Y,N_Z)} := S^n_{(n+N_X, n+N_Y, n+N_Z)} \, , \tag{4.16}$$

and $S^n_{(N_X,N_Y,N_Z)}$ is defined by

$$S^n_{(N_X,N_Y,N_Z)} := \sum_{\substack{n_X+n_Y+n_Z=n \\ n_X,n_Y,n_Z\geq 0}} \binom{n_X+N_X-1}{N_X-1}\binom{n_Y+N_Y-1}{N_Y-1}\binom{n_Z+N_Z-1}{N_Z-1} J_X^{2n_X} J_Y^{2n_Y} J_Z^{2n_Z} \,. \tag{4.17}$$

Here $\binom{n}{m}$ is the binomial coefficient, and for the special case of the binomial coefficient, we define $\binom{n}{-1} = \delta_{n,-1}$.

The coefficient for the isotropic case is

$$R^n_{(N_X,N_Y,N_Z)}\big|_{J_X=J_Y=J_Z=1} = \frac{m}{4n+m}\binom{4n+m}{n}\,, \tag{4.18}$$

where $m = N_X + N_Y + N_Z$. The reason why Eq. (4.18) does not coincide with the generalized Catalan number (4.5) is that the linear combinations of the lower-order charges are different from Eqs. (4.14) and (4.4).

The local charges in the spin-$1/2$ XYZ chain can be also written simply in terms of a matrix-product operator representation [85].

### 4.3.3 Structure of local charges in the XYZ chain

In this subsection, we explain the structure of the local charges in the XYZ chain. The cancellation of the operators in $[Q_k, H]$ is very complicated and we cannot easily see how the doubling products cancel each other. Here, we define the *structure* as the classification of the operator components that construct the local charges and with which we can easily see how the components in local charges cancel each other.

The structure of the local charges in the XYZ chain is understood from the general expression (4.14). The doubling products in them are classified by the characteristic quantities, support and hole [28, 30]. We denote the doubling product with support $s$ and hole $m$ as $(s,m)$-doubling product. The commutator of $(s,m)$-doubling product with the Hamiltonian can generate $(s\pm 1, m)$-doubling product and $(s, m\pm 1)$-doubling product. Thus, the contributions of the cancellation of $(s,m)$-doubling product in $[Q_k, H]$ come from $(s\pm 1, m)$-doubling products and $(s, m\pm 1)$-doubling products in $Q_k$.

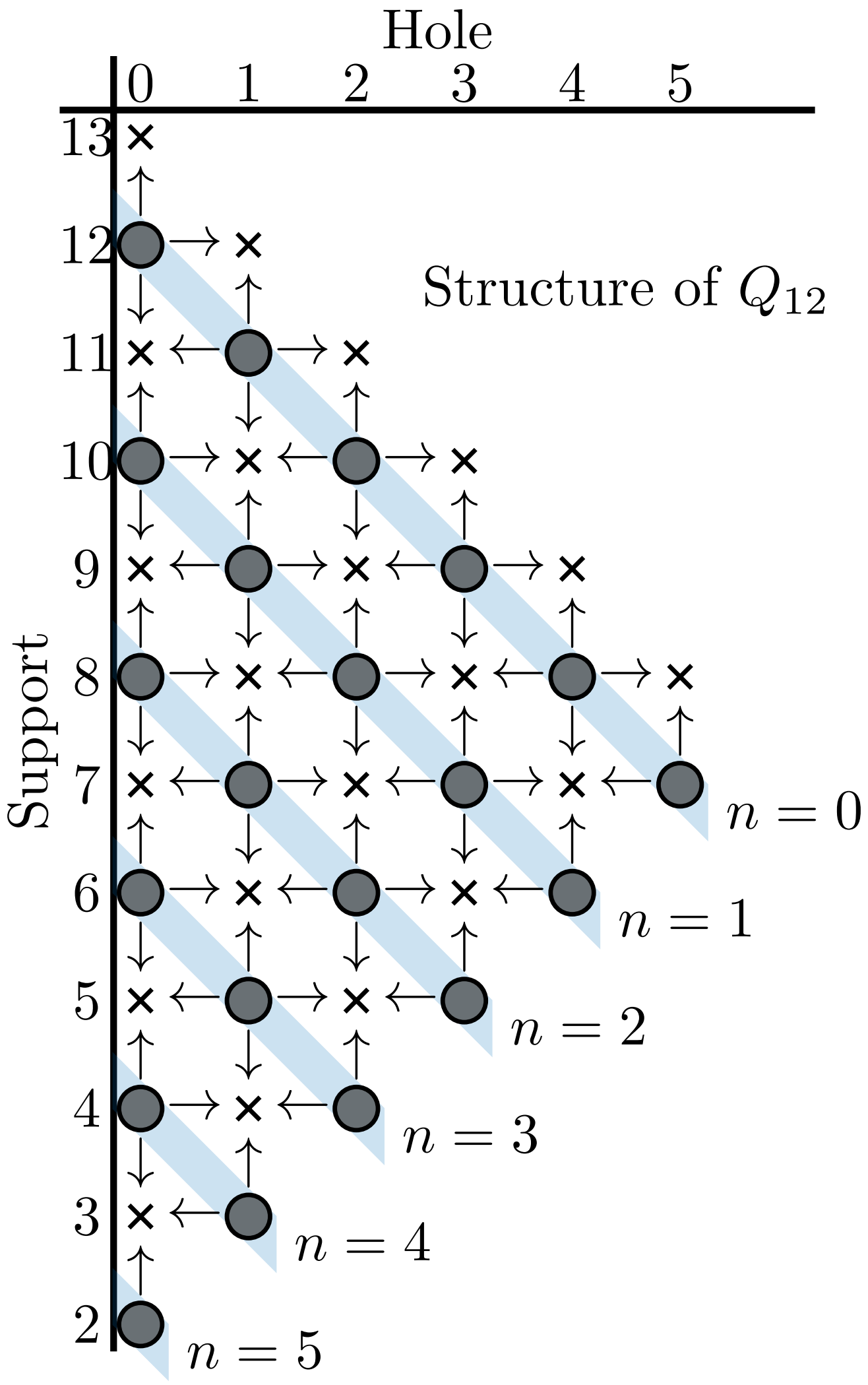

Figure 4.1: Structure of the local charge $Q_k$ in the XYZ chain for $k = 12$. Circles represent $(s, m)$-doubling products in $Q_{12}$ and $s + m = k - 2n$. Crosses represent doubling products generated from the commutator of doubling products represented by the circles and the Hamiltonian. The generated doubling products at the crosses are to be canceled due to the conservation law.

In Fig. 4.1, we show the structure of the local charge $Q_k$ in the XYZ chain for $k = 12$. The circles at $(s, m)$ represent the $(s, m)$-doubling products in $Q_k$. More precisely, the circle at

$(k-2n-m, m)$ represents the following operator $Q_k(k-2n-m, m)$:

$$Q_k(k-2n-m,m) = \sum_{\overline{\boldsymbol{A}} \in \mathcal{S}_k^{n,m}} R^n_{\left(N_X^{\boldsymbol{A}}+m,\, N_Y^{\boldsymbol{A}}+m,\, N_Z^{\boldsymbol{A}}+m\right)} J_{\boldsymbol{A}} \, \overline{\boldsymbol{A}} \,, \tag{4.19}$$

and the local charge $Q_k$ is written as the summation of these operators:

$$Q_k = \sum_{\substack{0 \leq n+m < \lfloor k/2 \rfloor, \\ n,m \geq 0}} Q_k(k-2n-m,m) \,. \tag{4.20}$$

The doubling products in the circle at $(s, m)$ generate the doubling products in the crosses at $(s \pm 1, m)$ and $(s, m \pm 1)$. The generated doubling products at the crosses are to be canceled for the conservation law $[Q_k, H] = 0$.

It is crucial to elucidate the structure of the local charges to derive their general expressions.

## 4.4 Temperley-Lieb models

In this section, we briefly explain the local charges in integrable lattice models which can be represented by the Temperley-Lieb algebra (TL algebra) [77]. Here, we collectively refer to them as the Temperley-Lieb models (TL models). The TL models include various quantum spin systems: the spin-$1/2$ XXZ chain [86–88], the quantum Potts chain [78], the golden (anyon) chain [79], and trace representation [89, 90]. The explicit expressions of the local charges in the TL models are derived in Ref. [31]. The factorization formula of short-range correlation functions in the TL models based on the mean value of current operators is studied in Ref. [80]. In the following, we briefly review the TL algebra and its several representations, and the lower order charges. For comprehensive details on the general expression of the local charges, refer to Ref. [31].

### 4.4.1 Temperley-Lieb algebra

The TL algebra is defined as follows. There are generators $e_j$ with index $j = 1, \ldots, L-1$, which satisfy the relations [77]

$$e_j^2 = d e_j \,, \qquad e_j e_{j\pm 1} e_j = e_j \,, \qquad [e_j, e_k] = 0 \quad \text{for } |j-k| > 1 \,. \tag{4.21}$$

Here, $d \in \mathbb{C}$ is a free parameter of the TL algebra.

In the case of the periodic TL algebra, another generator $e_L$ is introduced, satisfying [91]

$$e_L^2 = d e_L, \quad e_L e_b e_L = e_L, \quad e_b e_L e_b = e_b, \quad [e_L, e_j] = 0, \tag{4.22}$$

where $b = 1, L-1$ and $j \neq 1, L-1$. The TL Hamiltonian with periodic boundary conditions is abstractly defined on the level of the algebra as

$$H = \sum_{j=1}^{L} e_j \,. \tag{4.23}$$

### 4.4.2 Representations of the Temperley-Lieb algebra

We next review several representations of the TL algebra that satisfy the relations (4.21). Each representation gives rise to the corresponding TL model.

#### XXZ representation

In the XXZ representation, the generators act on the Hilbert space of a spin-1/2 chain and are defined by

$$e_j = h^{\mathrm{XXZ}}_{j,j+1}\,, \tag{4.24}$$

where $h^{\mathrm{XXZ}}_{j,j+1}$ is the Hamiltonian density of the XXZ chain [92]

$$h^{\mathrm{XXZ}}_{j,j+1} = -\frac{1}{2}\left[2e^{i\phi/L}\sigma^+_j\sigma^-_{j+1} + 2e^{-i\phi/L}\sigma^-_j\sigma^+_{j+1} + \cos\gamma\big(\sigma^z_j\sigma^z_{j+1} - 1\big) + i\sin\gamma\big(\sigma^z_j - \sigma^z_{j+1}\big)\right]\,, \tag{4.25}$$

where $\gamma$ is the anisotropic parameter related to the parameter of the TL algebra by $d = 2\cos\gamma$, and in practice, restricted to be real or pure imaginary. The parameter $\phi \in \mathbb{C}$ is the twist angle. In our representation, we choose to apply a homogeneous distribution of the twist and focus on the translational invariant case. This definition of the TL generator actually satisfies the relation of the TL algebra (4.21).

The XXZ chain is solved by the Bethe Ansatz. The Bethe eigenstates are made of interacting spin waves and characterized by a set of rapidities $\boldsymbol{\lambda} = \{\lambda_1, \dots, \lambda_N\}$, which satisfy the Bethe equations [86–88]

$$e^{i\phi}\left(\frac{\sinh(\lambda_j + i\gamma/2)}{\sinh(\lambda_j - i\gamma/2)}\right)^L \prod_{k\neq j}\frac{\sinh(\lambda_j - \lambda_k - i\gamma)}{\sinh(\lambda_j - \lambda_k + i\gamma)} = 1. \tag{4.26}$$

Here, $N$ denotes the number of down spins for the corresponding eigenstates. If $\gamma \in \mathbb{R}$, the configuration of the rapidities corresponding to the ground state consists of real roots.

The eigenenergy of the Bethe state characterized by the Bethe root $\boldsymbol{\lambda}$ are written as

$$E = \sum_{j=1}^{N}\epsilon(\lambda_j) \tag{4.27}$$

with

$$\epsilon(\lambda) = \frac{\sin^2\gamma}{\sinh(\lambda + i\gamma/2)\sinh(\lambda - i\gamma/2)}\,. \tag{4.28}$$

#### Potts representation

The $Q$-states quantum Potts chain is defined on a chain of $L/2$ sites with a $Q$-dimensional local Hilbert space. Here, we assume that $L$ is even. On each sites we define matrices $X$ and $Z$, which are the generalizations of the Pauli matrices $\sigma^x$ and $\sigma^z$ that satisfy the $\mathbb{Z}_Q$ clock algebra

$$X^\dagger = X^{Q-1}, \quad Z^\dagger = Z^{Q-1}, \quad X^Q = Z^Q = 1\,, \quad XZ = \omega ZX\,, \tag{4.29}$$

where $\omega = e^{i\frac{2\pi}{Q}}$ and the concrete representation of $X$ and $Z$ are given as

$$Z = \begin{pmatrix} 1 & & & \\ & \omega & & \\ & & \ddots & \\ & & & \omega^{Q-1} \end{pmatrix}, \qquad X = \begin{pmatrix} 0 & 1 & & \\ & \ddots & \ddots & \\ & & \ddots & 1 \\ 1 & & & 0 \end{pmatrix}. \tag{4.30}$$

It follows that the generators $e_1, \dots e_L$ are defined as [93]

$$e_{2j} = \frac{1}{\sqrt{Q}} \sum_{a=0}^{Q-1} \left( X_j^\dagger X_{j+1} \right)^a, \qquad e_{2j+1} = \frac{1}{\sqrt{Q}} \sum_{a=0}^{Q-1} Z_j^a, \tag{4.31}$$

and we can confirm that they satisfy the relation of the TL algebra with $d = \sqrt{Q}$ (4.21), and periodic boundary conditions. Note that they are Hermitian operators.

We note that the quantum Potts chain is critical for $Q \leq 4$, corresponding to the critical temperature in the square-lattice Potts cases, and non-critical for $Q \geq 5$.

### Golden chain representation

The golden chain was first introduced in Ref. [79]. It is a special case of the family of Hamiltonians related to the restricted solid-on-solid (RSOS) models [94, 95]. The Hilbert space of the golden chain is constrained: it is spanned by the states of the computational spin-$1/2$ basis, which do not have two neighboring down spins.

Let us define the local projection operators

$$P_j = (1 + \sigma_j^z)/2\,, \qquad N_j = (1 - \sigma_j^z)/2\,. \tag{4.32}$$

Then, the constraint on the Hilbert space can be written as

$$N_j N_{j+1} = 0\,. \tag{4.33}$$

The golden chain representation of the TL algebra is given by

$$e_j = h_{j,j+1,j+2}\,, \tag{4.34}$$

where $h_{j,j+1,j+2}$ is a three-site operator acting on the constrained Hilbert space, defined by

$$h_{j,j+1,j+2} = -\varphi \left[ (P_j + P_{j+2} - 1) - P_j P_{j+2} \big( \varphi^{-3/2} \sigma_{j+1}^x + \varphi^{-3} P_{j+1} + \varphi^{-2} + 1 \big) \right], \tag{4.35}$$

where $\varphi = (1 + \sqrt{5})/2 = 2\cos(\pi/5)$ is the golden ratio. The TL parameter in the golden chain representation is $d = \varphi$. we can confirm that this definition of $e_j$ satisfies the relation of the TL algebra (4.21).

#### Trace representation

In the trace representation, the local Hilbert spaces is $\mathbb{C}^d$ with $d \geq 2$ and the TL generators are given by

$$e_j = K_{j,j+1}\,, \tag{4.36}$$

where $K$ is the so-called trace operator defined by

$$K = \sum_{a,b=1}^{d} |aa\rangle \langle bb|\ . \tag{4.37}$$

In this case, the parameter of the TL algebra is equal to the local dimension $d$. We note that the trace representation in which $d$ is a non-zero integer can be constructed from graded vector spaces [89]. These models appeared recently in the study of Hilbert-space fragmentation [90].

### 4.4.3 Temperley-Lieb equivalence

It is known that eigenenergies of the TL models can be found in the eigenspectrum of the XXZ chain of the same TL parameter $d$ and an appropriate twist. The eigenspectrum of the TL models is decomposed into the sector of the eigenspectrum of the XXZ representation $\mathcal{W}_{j,z}$ with the following correspondence:

$$j = |S_z|\,, \qquad z = e^{\mathrm{i}\phi}\,, \tag{4.38}$$

where $j$ indicates the magnetization sector and $z$ indicates the twist. The dimension of the sector is

$$\mathrm{dim}\mathcal{W}_{j,z} = \binom{L}{L/2 - j}\,. \tag{4.39}$$

These correspondences of the eigenspectrum in the TL models are called the Temperley-Lieb equivalence [3, 77, 78, 91].

The equivalence of these TL models and the XXZ representation can also be understood from the fact that the corresponding two-dimensional classical model is mapped to the six-vertex models [3]. For example, the square lattice Potts model at the critical temperature can be mapped to the six-vertex model [96]. This follows from the equivalence of the quantum Potts chain and the XXZ chain explained above.

For example, the eigenspectrum of the trace representation for $d = 3$ and $L = 6$ is decomposed as [80]

$$\mathcal{H}^{\mathrm{trace}} = \mathcal{W}_{0,\mathfrak{q}^2} \oplus 7\mathcal{W}_{1,1} \oplus 27\mathcal{W}_{2,1} \oplus 20\mathcal{W}_{2,-1} \oplus 322\mathcal{W}_{3,1}\,, \tag{4.40}$$

where $\mathfrak{q} = e^{i \arccos d/2}$ and the integer coefficients denote the degeneracy of the sector. We can confirm that the dimensions are the same for both sides: the dimension of the Hilbert space of the trace representation for $d = 3$ and $L = 6$ is $\dim\mathcal{H}^{\text{trace}} = d^L = 3^6$, and we can see

$$3^6 = \binom{6}{3} + 7\binom{6}{2} + 27\binom{6}{1} + 20\binom{6}{1} + 322\binom{6}{0}, \tag{4.41}$$

where $\binom{n}{m}$ is the binomial coefficient.

### 4.4.4 Integrability of the Temperley-Lieb models

The integrability of the TL models can be assured abstractly on the level of the TL algebra, and there are an extensive number of local charges in the TL models defined by Eq. (4.23).

We can confirm the integrability by confirming the Yang-Baxter equation abstractly. In this subsection, we consider the infinite-chain case $L \to \infty$ for simplicity. We define the R-matrix for the TL models by

$$\check{R}_j(u) = 1 + ue_j\,, \tag{4.42}$$

where $u \in \mathbb{C}$ is the spectral parameter. With this definition of the R-matrices, the Yang-Baxter equation holds [97]:

$$\check{R}_j(u)\check{R}_{j+1}(u+v)\check{R}_j(v) = \check{R}_{j+1}(v)\check{R}_j(u+v)\check{R}_{j+1}(u)\,. \tag{4.43}$$

Then, the transfer matrices can be given by

$$t(u) = \prod_j \check{R}_j(u) \tag{4.44}$$

and the mutual commutativity is guaranteed from the Yang-Baxter equation (4.43):

$$[t(u), t(v)] = 0\,. \tag{4.45}$$

### 4.4.5 Local charges in the Temperley-Lieb models

The local charges in the TL modules are universally represented in terms of the TL generators. The expansion of the transfer matrix gives the local charges:

$$\log t(u) = \sum_{k=2} \frac{u^{k-1}}{(k-1)!} Q_k\,, \tag{4.46}$$

where $Q_2 = H$ is the Hamiltonian of the TL models.

Because the R-matrix for the TL models (4.42) is of a difference form, we can define the boost operator

$$B = \sum_j j e_j\,, \tag{4.47}$$

and the local charges are recursively obtained as

$$Q_{k+1} = [B, Q_k] \ . \tag{4.48}$$

The local charges are extensive and are expressed as

$$Q_k = \sum_j q_k(j) \ , \tag{4.49}$$

where $q_k(j)$ is a short-range density operator of local charges, which can be expressed using the generators of the TL algebra. With this notation, the densities of the lower-order local charges are given by

$$q_2(j) = e_j \ , \tag{4.50}$$
$$q_3(j) = e_j e_{j+1} - e_{j+1} e_j \ , \tag{4.51}$$
$$q_4(j) = 2(e_{j+2}e_{j+1}e_j - e_{j+1}e_je_{j+2} - e_je_{j+2}e_{j+1} + e_je_{j+1}e_{j+2}) + de_{j+1}e_j + de_je_{j+1} \ , \tag{4.52}$$
$$\begin{aligned} q_5(j) = &6(e_je_{j+1}e_{j+2}e_{j+3} - e_{j+3}e_{j+2}e_{j+1}e_j + e_{j+2}e_{j+1}e_je_{j+3} + e_{j+1}e_je_{j+3}e_{j+2} \\ &+ e_je_{j+3}e_{j+2}e_{j+1} - e_{j+1}e_je_{j+2}e_{j+3} - e_je_{j+2}e_{j+1}e_{j+3} - e_je_{j+1}e_{j+3}e_{j+2} \\ &- de_{j+2}e_{j+1}e_j + de_je_{j+1}e_{j+2}) + (2+d^2)\left(e_je_{j+1} - e_{j+1}e_j\right) \ , \end{aligned} \tag{4.53}$$
$$\begin{aligned} q_6(j) = &24(e_{j+4}e_{j+3}e_{j+2}e_{j+1}e_j - e_{j+3}e_{j+2}e_{j+1}e_je_{j+4} - e_{j+2}e_{j+1}e_je_{j+4}e_{j+3} - e_{j+1}e_je_{j+4}e_{j+3}e_{j+2} \\ &- e_je_{j+4}e_{j+3}e_{j+2}e_{j+1} + e_{j+2}e_{j+1}e_je_{j+3}e_{j+4} + e_{j+1}e_je_{j+3}e_{j+2}e_{j+4} + e_{j+1}e_je_{j+2}e_{j+4}e_{j+3} \\ &+ e_je_{j+3}e_{j+2}e_{j+1}e_{j+4} + e_je_{j+2}e_{j+1}e_{j+4}e_{j+3} + e_je_{j+1}e_{j+4}e_{j+3}e_{j+2} - e_{j+1}e_je_{j+2}e_{j+3}e_{j+4} \\ &- e_je_{j+2}e_{j+1}e_{j+3}e_{j+4} - e_je_{j+1}e_{j+3}e_{j+2}e_{j+4} - e_je_{j+1}e_{j+2}e_{j+4}e_{j+3} + e_je_{j+1}e_{j+2}e_{j+3}e_{j+4}) \\ &+ 36d(e_{j+3}e_{j+2}e_{j+1}e_j + e_je_{j+1}e_{j+2}e_{j+3}) - 12d(e_{j+2}e_{j+1}e_je_{j+3} + e_{j+1}e_je_{j+3}e_{j+2} \\ &+ e_je_{j+3}e_{j+2}e_{j+1} + e_{j+1}e_je_{j+2}e_{j+3} + e_je_{j+2}e_{j+1}e_{j+3} + e_je_{j+1}e_{j+3}e_{j+2}) \\ &+ (16+14d^2)(e_{j+2}e_{j+1}e_j + e_je_{j+1}e_{j+2}) + (-16-2d^2)(e_{j+1}e_je_{j+2} + e_je_{j+2}e_{j+1}) \\ &+ (8d+d^3)(e_{j+1}e_j + e_je_{j+1}) - 24de_je_{j+2} \ . \end{aligned} \tag{4.54}$$

The general expression of $Q_k$ is obtained by Nienhuis and Huijgen [31]. The local charge $Q_k$ can be written as a linear combination of products of the TL generator with simple binomial coefficients. For the details of the general expression, we refer to Ref. [31]. We note that the convention of the local charges obtained in Ref. [31] is different from those obtained from the expansion of the transfer matrix: the local charges in Ref. [31] are some linear combinations of the transfer matrix charges.

It is profoundly intriguing to derive local charges at the level of abstract algebra, as it implies simultaneously deriving the local charges for all quantum integrable systems belonging to that class of abstract algebra. In the case of the TL algebra, local charges in the XXZ chain, the Potts chain, the golden chain, and trace representations have all been determined simultaneously.

## 4.5 Construction of current and generalized current operators

The knowledge of the explicit expression for the local charges is helpful for constructing current and generalized current. The construction of the current operator is established in Ref. [83]. Later, this result has been extended to the generalized current operator [80]. These constructions of the current and generalized current operators are useful in finding the factorization formula for the correlation functions in integrable systems [80]. In this section, we review these methods.

### 4.5.1 Current operators

We start by reviewing the procedure for constructing the current operators [83]. Henceforth, we will describe an operator that acts on the $x$ th site and locally on sites beyond the $x$ th site as "starting from the $x$ th site." We consider a general integrable Hamiltonian written by a two-site operator and denote its higher-order charges by

$$Q_\alpha = \sum_x q_\alpha(x) \,, \tag{4.55}$$

where we assume that $q_\alpha(x)$ is starting from $x$ th site. We assume the thermodynamic limit $L \to \infty$ in this section for the simplicity of the following proof, and the summation over $x$ runs over $-\infty < x < \infty$. Nevertheless, the same argument holds for finite periodic systems. The current operators $J_\alpha(x)$ are defined by the continuity equation:

$$[H, q_\alpha(x)] = J_\alpha(x) - J_\alpha(x+1) \,. \tag{4.56}$$

Since the Hamiltonian is a two-site operator, $[H, q_\alpha(x)]$ is constructed from the operator starting from the $(x-1)$ th, $x$ th, and $(x+1)$ th sites, which we denote by $F_\alpha^{-1}(x-1)$, $F_\alpha^{0}(x)$, and $F_\alpha^{1}(x+1)$, respectively:

$$[H, q_\alpha(x)] = F_\alpha^{-1}(x-1) + F_\alpha^{0}(x) + F_\alpha^{1}(x+1) \,. \tag{4.57}$$

Then, the current operator is given by [83]

$$J_\alpha(x) = F_\alpha^{-1}(x-1) - F_\alpha^{1}(x) \,. \tag{4.58}$$

The proof of Eq. (4.58) is as follows: by substituting Eq. (4.58) into the right-hand side of Eq. (4.56), we obtain

$$J_\alpha(x) - J_\alpha(x+1) = [H, q_\alpha(x)] - \delta_\alpha(x) \,, \tag{4.59}$$

where $\delta_\alpha(x)$ is defined as $\delta_\alpha(x) \equiv F_\alpha^{-1}(x) + F_\alpha^{0}(x) + F_\alpha^{1}(x)$. Summing Eq. (4.57) over $x$, we have $0 = \sum_{x=1}^{L} \delta_\alpha(x)$. Since $\{\delta_\alpha(x)\}_{x=1,2,\ldots,L}$ are mutually linear independent, each $\delta_\alpha(x)$ should be zero itself:

$$\delta_\alpha(x) = 0 \,. \tag{4.60}$$

Thus, we have proved Eq. (4.58) satisfies the continuity equation (4.56).

### 4.5.2 Generalized current operators

In Ref. [80], the construction of the current operator [83] has been extended to the generalized current operator. The generalized continuity equation that the generalized currents $J_{\alpha,\beta}(x)$ satisfies is

$$[Q_\beta, q_\alpha(x)] = J_{\alpha,\beta}(x) - J_{\alpha,\beta}(x+1)\,. \tag{4.61}$$

We note that $J_{\alpha,2} = J_\alpha$. In our convention here, $Q_\beta$ is a $\beta$-site operator, and we can express the left-hand side of Eq. (4.61) as

$$[Q_\beta, q_\alpha(x)] = \sum_{y=-(\beta-1)}^{\beta-1} F^y_{\alpha,\beta}(x+y)\,, \tag{4.62}$$

where $F^y_{\alpha,\beta}(x+y)$ is an operator starting from the $(x+y)$ th site. Summing up Eq. (4.62) over $x$, we have

$$\begin{aligned} 0 = [Q_\beta, Q_\alpha] &= \sum_x \sum_{y=-(\beta-1)}^{\beta-1} F^y_{\alpha,\beta}(x+y) \\ &= \sum_x \sum_{y=-(\beta-1)}^{\beta-1} F^y_{\alpha,\beta}(x)\,. \end{aligned} \tag{4.63}$$

Because $\sum_{y=-(\beta-1)}^{\beta-1} F^y_{\alpha,\beta}(x)$ is an operator starting from the $x$ th site, we have

$$\sum_{y=-(\beta-1)}^{\beta-1} F^y_{\alpha,\beta}(x) = 0\,. \tag{4.64}$$

The generalized current operator is given by

$$J_{\alpha,\beta}(x) = \sum_{b=1}^{\beta-1} \sum_{y=1}^{b} \left[F^{-b}_{\alpha,\beta}(x-y) - F^b_{\alpha,\beta}(x+y-1)\right]\,. \tag{4.65}$$

The case $\beta = 2$ of Eq. (4.65) recovers (4.58).

We prove that Eq. (4.65) in the following. We define

$$E^y_\pm(x) \equiv \sum_{b=y}^{\beta-1} F^{\pm b}_{\alpha,\beta}(x)\,, \tag{4.66}$$

and we can rewrite the generalized current as $J_{\alpha,\beta}(x) = \sum_{y=1}^{\beta-1} [E^y_-(x-y) - E^y_+(x+y-1)]$.

Substituting Eq. (4.65) in the right-hand side of Eq. (4.61), we have

$$\begin{aligned}
J_{\alpha,\beta}(x) - J_{\alpha,\beta}(x+1) &= \sum_{y=1}^{\beta-1} [E_-^y(x-y) + E_+^y(x+y)] - \sum_{y=1}^{\beta-1} [E_-^y(x-y+1) + E_+^y(x+y-1)] \\
&= \sum_{\beta-1 \geq |y| > 0} F^y(x+y) + \sum_{y=1}^{\beta-2} \left[E_-^{y+1}(x-y) + E_+^{y+1}(x+y)\right] \\
&\qquad - \sum_{y=1}^{\beta-1} [E_-^y(x-y+1) + E_+^y(x+y-1)] \\
&= [Q_\beta, q_\alpha(x)] - F_{\alpha,\beta}^0(x) - E_+^1(x) - E_-^1(x) \\
&= [Q_\beta, q_\alpha(x)] - \sum_{y=-\beta+1}^{\beta-1} F_{\alpha,\beta}^y(x) \\
&= [Q_\beta, q_\alpha(x)] \ ,
\end{aligned} \tag{4.67}$$

where we have used $E_\pm^y(x) = E_\pm^{y+1}(x) + F_{\alpha,\beta}^y(x)$ and $E_\pm^\beta(x) = 0$ in the second equality, and we have used Eq. (4.64) in the last equality. Thus, we have proved that Eq. (4.65) satisfies Eq. (4.61).

One can derive the explicit expressions of the generalized currents by first calculating the commutator in the continuity equation, then determining $F_{\alpha,\beta}^y(x)$, and finally utilizing Eq. (4.65), if we know the explicit expression of the local charges $Q_\alpha$.

It is important to note that the aforementioned index $x$ does not necessarily have to indicate a physical site index; the sole necessity is that the operators corresponding to different indices are linearly independent.

We give examples of the generalized current operators in the Temperley-Lieb models, which are calculated from Eqs. (4.58) and (4.65) below.

$$J_{3,2}(j) = e_{j+1}e_je_{j-1} - e_je_{j-1}e_{j+1} - e_{j-1}e_{j+1}e_j + e_{j-1}e_je_{j+1} - 2e_j\,, \tag{4.68}$$

$$\begin{aligned}J_{3,3}(j) = &- e_{j+2}e_{j+1}e_je_{j-1} - e_{j+1}e_je_{j-1}e_{j-2} + e_{j+1}e_je_{j-1}e_{j+2} + e_je_{j-1}e_{j+2}e_{j+1} + e_je_{j-1}e_{j-2}e_{j+1}\\ &+ e_{j-1}e_{j+2}e_{j+1}e_j + e_{j-1}e_{j-2}e_{j+1}e_j + e_{j-2}e_{j+1}e_je_{j-1} - e_je_{j-1}e_{j+1}e_{j+2} - e_{j-1}e_{j+1}e_je_{j+2}\\ &- e_{j-1}e_je_{j+2}e_{j+1} - e_{j-1}e_{j-2}e_je_{j+1} - e_{j-2}e_je_{j-1}e_{j+1} - e_{j-2}e_{j-1}e_{j+1}e_j + e_{j-1}e_je_{j+1}e_{j+2}\\ &+ e_{j-2}e_{j-1}e_je_{j+1} - de_{j+1}e_je_{j-1} + de_{j-1}e_je_{j+1} + e_{j+1}e_j + e_je_{j-1} - e_je_{j+1} - e_{j-1}e_j\,,\end{aligned} \tag{4.69}$$

$$\begin{aligned}J_{4,2}(j) = &- 2e_{j+2}e_{j+1}e_je_{j-1} + 2e_{j+1}e_je_{j-1}e_{j+2} + 2e_je_{j-1}e_{j+2}e_{j+1} + 2e_{j-1}e_{j+2}e_{j+1}e_j\\ &- 2e_je_{j-1}e_{j+1}e_{j+2} - 2e_{j-1}e_{j+1}e_je_{j+2} - 2e_{j-1}e_je_{j+2}e_{j+1} + 2e_{j-1}e_je_{j+1}e_{j+2}\\ &- de_{j+1}e_je_{j-1} - de_je_{j-1}e_{j+1} + de_{j-1}e_{j+1}e_j + de_{j-1}e_je_{j+1} + 2e_{j+1}e_j - 2e_je_{j+1}\,,\end{aligned} \tag{4.70}$$

$$\begin{aligned}J_{4,3}(j) = 2(&e_{j+3}e_{j+2}e_{j+1}e_je_{j-1} + e_{j+2}e_{j+1}e_je_{j-1}e_{j-2} - e_{j+2}e_{j+1}e_je_{j-1}e_{j+3} - e_{j+1}e_je_{j-1}e_{j+3}e_{j+2}\\ &- e_{j+1}e_je_{j-1}e_{j-2}e_{j+2} - e_je_{j-1}e_{j+3}e_{j+2}e_{j+1} - e_je_{j-1}e_{j-2}e_{j+2}e_{j+1} - e_{j-1}e_{j+3}e_{j+2}e_{j+1}e_j\\ &- e_{j-1}e_{j-2}e_{j+2}e_{j+1}e_j - e_{j-2}e_{j+2}e_{j+1}e_je_{j-1} + e_{j+1}e_je_{j-1}e_{j+2}e_{j+3} + e_je_{j-1}e_{j+2}e_{j+1}e_{j+3}\\ &+ e_je_{j-1}e_{j+1}e_{j+3}e_{j+2} + e_je_{j-1}e_{j-2}e_{j+1}e_{j+2} + e_{j-1}e_{j+2}e_{j+1}e_je_{j+3} + e_{j-1}e_{j+1}e_je_{j+3}e_{j+2}\\ &+ e_{j-1}e_je_{j+3}e_{j+2}e_{j+1} + e_{j-1}e_{j-2}e_{j+1}e_je_{j+2} + e_{j-1}e_{j-2}e_je_{j+2}e_{j+1} + e_{j-2}e_{j+1}e_je_{j-1}e_{j+2}\\ &+ e_{j-2}e_je_{j-1}e_{j+2}e_{j+1} + e_{j-2}e_{j-1}e_{j+2}e_{j+1}e_j - e_je_{j-1}e_{j+1}e_{j+2}e_{j+3} - e_{j-1}e_{j+1}e_je_{j+2}e_{j+3}\\ &- e_{j-1}e_je_{j+2}e_{j+1}e_{j+3} - e_{j-1}e_je_{j+1}e_{j+3}e_{j+2} - e_{j-1}e_{j-2}e_je_{j+1}e_{j+2} - e_{j-2}e_je_{j-1}e_{j+1}e_{j+2}\\ &- e_{j-2}e_{j-1}e_{j+1}e_je_{j+2} - e_{j-2}e_{j-1}e_je_{j+2}e_{j+1} + e_{j-1}e_je_{j+1}e_{j+2}e_{j+3} + e_{j-2}e_{j-1}e_je_{j+1}e_{j+2})\\ &+ d(+3e_{j+2}e_{j+1}e_je_{j-1} + e_{j+1}e_je_{j-1}e_{j-2} - e_{j+1}e_je_{j-1}e_{j+2} - e_je_{j-1}e_{j+2}e_{j+1}\\ &+ e_je_{j-1}e_{j-2}e_{j+1} - e_{j-1}e_{j+2}e_{j+1}e_j - e_{j-1}e_{j-2}e_{j+1}e_j - e_{j-2}e_{j+1}e_je_{j-1} - e_je_{j-1}e_{j+1}e_{j+2}\\ &- e_{j-1}e_{j+1}e_je_{j+2} - e_{j-1}e_je_{j+2}e_{j+1} - e_{j-1}e_{j-2}e_je_{j+1} - e_{j-2}e_je_{j-1}e_{j+1}\\ &+ e_{j-2}e_{j-1}e_{j+1}e_j + 3e_{j-1}e_je_{j+1}e_{j+2} + e_{j-2}e_{j-1}e_je_{j+1} - 2e_je_{j-1}e_{j+1}e_j + 3e_{j+1}e_j + e_je_{j-1}\\ &+ 3e_je_{j+1} + e_{j-1}e_j) + d^2(e_{j-1}e_je_{j+1} + e_{j+1}e_je_{j-1} + 2e_j) + 4(e_j + e_{j+1})\,.\end{aligned} \tag{4.71}$$

The higher-order generalized currents have a more complicated structure.

## 4.6 Factorization of correlation function

Local charges and associated currents are important in the study of correlation functions in integrable systems. In this subsection, we explain how short-range correlation functions are expressed in terms of the mean values of current operators, which can be calculated from the explicit form of local charges, as exemplified in the Temperley-Lieb models.

Even if the system is exactly solvable via the Bethe ansatz, calculating correlation functions is still a challenging problem. We can calculate the eigenspectrum from the solution of the Bethe equations, but calculating the matrix elements of observables with respect to the Bethe eigenstates is a difficult task, and thus, the calculation of the correction function is also very difficult.

The theory of factorized correlation functions was initiated for the XXZ chain [98]. The theory states that the mean values of short-range correlation functions can be expressed as the sum of the products of the Taylor coefficients of a function of one or two variables, which determines the whole information of correlation functions in the system. The factorization of the correlation functions has been extensively studied, especially in the XXZ chain for ground-state correlation functions and finite-temperature correlation functions [99–108]. Later, it was reported that the Taylor coefficient used in the factorization is equal to the mean value of the current operators [45, 46, 109]. The factorization formula, which may be applicable to all the Temperley-Lieb models, was found in Ref. [80]. In the following, we review the results of Ref. [80] for the demonstration of how the current operators are related to the correlation functions.

### 4.6.1 Factorization formula for the Temperley-Lieb models

In the following, the expectation value is denoted by $\langle \mathcal{O} \rangle \equiv \langle E|\mathcal{O}|E\rangle$, where $|E\rangle$ is a simultaneous eigenstate of the Hamiltonian (4.23) and the local charges of any representation of the TL algebra, and we assume that $|E\rangle$ is a singlet of the family of the local charges in the TL model. We also assume that $|E\rangle$ simultaneously diagonalizes the fundamental symmetries such as momentum. Let the current mean value be denoted by

$$\psi_{\alpha,\beta} \equiv \langle J_{\alpha+1,\beta+2}(x)\rangle \qquad (\alpha, \beta \geq 0)\,, \tag{4.72}$$

where $\psi_{\alpha,\beta}$ does not depend on the index $x$ because of the site translational invariance $UJ_{\alpha,\beta}(x)U^\dagger = J_{\alpha,\beta}(x+1)$, which comes from $Ue_jU^\dagger = e_{j+1}$. [2] Here we defined $J_{1,\beta} \equiv Q_\beta$.

It is found in Ref. [80] that several short-range correlation functions in TL models are universally factorized in terms of $\psi_{\alpha,\beta}$; the mean value of the local operators constructed from the generator of the TL algebra $e_j$ with respect to an excited eigenstate are expressed as the sum of products of $\psi_{\alpha,\beta}$, and this factorization does not depend on the choice of the representation of the TL algebra and the choice of the eigenstate. These findings, originating from numerical calculations, await a rigorous proof.

We give examples of the factorization formula of short-range correlators below. The simplest one is

$$\langle e_1 e_2\rangle = \frac{1}{d}\left(\frac{1}{2}\psi_{0,2} - 2\psi_{0,0} - \psi_{1,1}\right) + \frac{1}{2}\psi_{0,1}\,. \tag{4.73}$$

[2] We note that in the Potts representation (4.31), the shift of the physical site index and the shift of the Temperley-Lieb index do not coincide with each other as in $Ue_jU^\dagger \neq e_{j+1}$ because of the staggered definition of the TL generator (4.31). The shift of the physical site index increases the Temperley-Lieb index by two: $Ue_jU^\dagger = e_{j+2}$. In this case, we take the staggered average with respect to the TL indices instead of Eq. (4.72), for example, $\langle e_1e_3\rangle \equiv \frac{1}{2}\left(\langle E|e_1e_3|E\rangle + \langle E|e_2e_4|E\rangle\right)$.

This formula can be proved rigorously from the following operator identity:

$$e_1e_2 = \frac{1}{d}\left\{\frac{1}{2}q_4(1) - 2q_2(2) - J_{3,2}(2)\right\} + \frac{1}{2}q_3(1)\,. \tag{4.74}$$

Non-trivial factorizations are observed in more complicated cases; for example

$$\begin{aligned}\langle e_1e_3\rangle =&\frac{1}{d(d^2-1)}\left[(d^2-4)\psi_{0,0} + 2\psi_{0,2} + \frac{d^2-40}{12}\psi_{1,1} + \frac{1}{6}\psi_{1,3} - \frac{1}{4}\psi_{2,2}\right]\\ &+ \frac{1}{d^2-1}\left[\frac{d^2-28}{12}(\psi_{0,1}^2 - \psi_{0,0}\psi_{1,1}) - \frac{1}{4}(\psi_{0,2}^2 - \psi_{0,0}\psi_{2,2})\right.\\ &\left.+\frac{1}{2}(\psi_{0,2}\psi_{1,1} - \psi_{0,1}\psi_{1,2}) + \frac{1}{6}(\psi_{0,1}\psi_{0,3} - \psi_{0,0}\psi_{1,3})\right]\,.\end{aligned} \tag{4.75}$$

Furthermore

$$\begin{aligned}\langle e_1e_2e_3 + e_3e_2e_1\rangle =&\frac{1}{6d\,(d^2-1)}\left[5d^3\psi_{1,1} + 11d^2\left(\psi_{1,0}^2 - \psi_{0,0}\psi_{1,1}\right)\right.\\ &+d(36\psi_{0,0} + 34\psi_{1,1} - 24\psi_{2,0} + 3\psi_{2,2} - 2\psi_{3,1}) + 3\psi_{2,0}(\psi_{2,0} - 2\psi_{1,1})\\ &\left.+2\psi_{1,0}(8\psi_{1,0} + 3\psi_{2,1} - \psi_{3,0}) + \psi_{0,0}(-16\psi_{1,1} - 3\psi_{2,2} + 2\psi_{3,1})\right]\end{aligned} \tag{4.76}$$

and

$$\begin{aligned}\langle e_1e_4\rangle =& \frac{1}{d(d^2-1)(d^2-2)}\left[2(d^2-4)\psi_{0,0} - \frac{11d^2-92}{12}\psi_{0,2} + \frac{d^4+40d^2-392}{36}\psi_{1,1} + \frac{d^2+56}{36}\psi_{1,3}\right.\\ &\left.-\frac{d^2+44}{24}\psi_{2,2} - \frac{1}{12}\psi_{0,4} + \frac{1}{24}\psi_{2,4} - \frac{1}{18}\psi_{3,3}\right]\\ &+ \frac{1}{(d^2-1)(d^2-2)}\left[\frac{5d^4-28d^2-220}{18}\left(\psi_{0,1}^2 - \psi_{0,0}\psi_{1,1}\right)\right.\\ &+\frac{2d^2+16}{9}\left\{2(\psi_{0,1}\psi_{0,3} - \psi_{0,0}\psi_{1,3}) + 3(\psi_{0,0}\psi_{2,2} - \psi_{0,2}^2)\right\}\\ &+\frac{7d^4+226d^2+928}{144}\left(\psi_{0,2}\psi_{1,1} - \psi_{0,1}\psi_{1,2}\right)\\ &+\frac{5d^2+22}{144}\left\{\psi_{0,1}\psi_{1,4} - \psi_{1,1}\psi_{0,4} - 2\psi_{0,1}\psi_{2,3} + 6(\psi_{1,1}\psi_{2,2} - \psi_{1,2}^2)\right\}\\ &+\frac{10d^2+140}{144}\psi_{0,3}\psi_{1,2} + \frac{1}{6}\left(\psi_{0,2}\psi_{0,4} - \psi_{0,0}\psi_{2,4}\right) + \frac{2}{9}\left(\psi_{0,0}\psi_{3,3} - \psi_{0,3}^2\right) - \frac{2}{3}\psi_{0,2}\psi_{1,3}\\ &+\frac{1}{12}\left(\psi_{1,3}\psi_{2,2} - \psi_{1,2}\psi_{2,3}\right) + \frac{1}{18}\left(\psi_{1,1}\psi_{3,3} - \psi_{1,3}^2\right) + \frac{1}{24}\left(\psi_{1,2}\psi_{1,4} - \psi_{1,1}\psi_{2,4}\right)\\ &\left.+\frac{1}{36}\left(\psi_{0,3}\psi_{2,3} - \psi_{0,2}\psi_{3,3}\right) + \frac{1}{48}\left(\psi_{0,2}\psi_{2,4} - \psi_{0,4}\psi_{2,2}\right) + \frac{1}{72}\left(\psi_{0,4}\psi_{1,3} - \psi_{0,3}\psi_{1,4}\right)\right]\,.\end{aligned} \tag{4.77}$$

These factorization formulas hold for any TL representations and for any eigenstates of the TL models. It is conjectured that in more complicated cases, the correlators can be also factorized in the same way [80].

### 4.6.2 Current mean value formula

For the XXZ representation (4.24), it has been proved that the current mean value $\psi_{\alpha,\beta}$ can be written using the corresponding Bethe roots [45, 46, 109]. Here, the Bethe root corresponding to $|E\rangle$ in the XXZ representation is denoted by $\boldsymbol{\lambda} = \{\lambda_1, \ldots, \lambda_N\}$ and $N$ is the number of down spins.

We define the function $\psi(x,y)$ by

$$\psi(x,y) = \mathbf{h}(i\sin(\gamma)x) \cdot G^{-1} \cdot \mathbf{h}(i\sin(\gamma)y) \times (-\sin(\gamma))\,, \tag{4.78}$$

where $\gamma$ is determined from $d = 2\cos\gamma$, $\mathbf{h}(x)$ is a vector of length $N$ with elements $\mathbf{h}_j(x) = h(\lambda_j - x)$ with $h(\lambda)$ given by

$$h(\lambda) = \coth(\lambda - i\gamma/2) - \coth(\lambda + i\gamma/2)\,, \tag{4.79}$$

and $G$ is the Gaudin matrix, defined by

$$G_{jk} = \delta_{jk}\left(L\frac{\sin(\gamma)}{\sinh(\lambda_j + i\gamma/2)\sinh(\lambda_j - i\gamma/2)} - \sum_{l=1}^{N} K(\lambda_{jl})\right) + K(\lambda_{jk})\,, \tag{4.80}$$

where we define $\lambda_{jk} \equiv \lambda_j - \lambda_k$ and

$$K(u) = \frac{\sin(2\gamma)}{\sinh(u + i\gamma)\sinh(u - i\gamma)}\,. \tag{4.81}$$

The remarkable fact is that $\psi(x,y)$ is the generating function for the current mean values [45, 46, 109]:

$$\psi_{a,b} = (\partial_x)^a (\partial_y)^b \psi(x,y)\big|_{x,y=0}\,. \tag{4.82}$$

It was observed that the current mean value in a TL representation other than the XXZ representation is equal to that in the XXZ representation with respect to the eigenstate, which has the same value of the local charges as those in the original representation [80]. This means that the current mean value and the correlators in the TL models can be calculated from the corresponding Bethe root in the XXZ representation.

The formula of the current mean value and the factorization of the correlation functions are demonstrated, especially in the XXZ chain [80]. It is desirable to check whether this framework can be applied to other integrated lattice systems, such as the one-dimensional Hubbard model.

## 4.7 1D Hubbard model: previous result

This section explains the previous progress in the studies of the local charges in the one-dimensional Hubbard model. Here, we consider the spin variable notation (3.24). The difficulty in the Hubbard case is that there is no boost recursive way to obtain the local charges, unlike the XYZ case and the Temperley-Lieb case, as explained in the previous sections. The explicit expressions for the local charges in the one-dimensional Hubbard model are revealed for the first three non-trivial ones [28, 37, 81, 82]. The general expression for the local charges up to the first order of the coupling constant has been obtained in Ref. [28]. However, the complete understanding of the local charges beyond them remained unknown, which will be elucidated in Chapter 5. The following will examine previous studies on local charges in the 1D Hubbard model.

### 4.7.1 Lower-order charges

The first non-trivial charge for the 1D Hubbard model is found by Shastry [37]:

$$
\begin{aligned}
Q_3 = \sum_{j=1}^{L} \big[ &\left(\sigma^x_{j,\uparrow}\sigma^z_{j+1,\uparrow}\sigma^y_{j+2,\uparrow} - \sigma^y_{j,\uparrow}\sigma^z_{j+1,\uparrow}\sigma^x_{j+2,\uparrow}\right) \\
&- U\left(\sigma^x_{j,\uparrow}\sigma^y_{j+1,\uparrow} - \sigma^y_{j,\uparrow}\sigma^x_{j+1,\uparrow}\right)\left(\sigma^z_{j,\downarrow} + \sigma^z_{j+1,\downarrow}\right)\big] + (\uparrow \leftrightarrow \downarrow)\,,
\end{aligned} \tag{4.83}
$$

where $(\uparrow \leftrightarrow \downarrow)$ denotes the terms where the flavor indices are interchanged. In Ref. [37], Shastry utilized the formulation of $Q_3$ to infer the transfer matrix for the 1D Hubbard model and later demonstrated the exact integrability of the 1D Hubbard model by providing the mutual commutativity of the transfer matrix [60].

The next charge was found in Ref. [81, 82]

$$
\begin{aligned}
Q_4 = \sum_{j=1}^{L} \Big[ &\left(\sigma^x_{j,\uparrow}\sigma^z_{j+1}\sigma^z_{j+2,\uparrow}\sigma^x_{j+3,\uparrow} + \sigma^y_{j,\uparrow}\sigma^z_{j+1,\uparrow}\sigma^z_{j+2,\uparrow}\sigma^y_{j+3,\uparrow}\right) \\
&- U\left(\sigma^x_{j,\uparrow}\sigma^z_{j+1,\uparrow}\sigma^x_{j+2,\uparrow} + \sigma^y_{j,\uparrow}\sigma^z_{j+1,\uparrow}\sigma^y_{j+2,\uparrow}\right)\left(\sigma^z_{j,\downarrow} + \sigma^z_{j+1} + \sigma^z_{j+2}\right) \\
&- \frac{U}{2}\left(\sigma^x_{j,\uparrow}\sigma^y_{j+1,\uparrow} - \sigma^y_{j,\uparrow}\sigma^x_{j+1,\uparrow}\right)\left(\sigma^x_{j,\downarrow}\sigma^y_{j+1,\downarrow} - \sigma^y_{j,\downarrow}\sigma^x_{j+1,\downarrow}\right) \\
&- U\left(\sigma^x_{j,\uparrow}\sigma^y_{j+1,\uparrow} - \sigma^y_{j,\uparrow}\sigma^x_{j+1,\uparrow}\right)\left(\sigma^x_{j+1,\downarrow}\sigma^y_{j+2,\downarrow} - \sigma^y_{j+1,\downarrow}\sigma^x_{j+2}\right) \\
&- U\left(\sigma^z_{j,\uparrow}\sigma^z_{j+1,\downarrow}\right) - \frac{U}{2}\left(\sigma^z_{j,\uparrow}\sigma^z_{j,\downarrow}\right) \\
&+U^2\left(\sigma^x_{j,\uparrow}\sigma^x_{j+1,\uparrow} + \sigma^y_{j,\uparrow}\sigma^y_{j+1,\uparrow}\right)\left(\sigma^z_{j,\downarrow}\sigma^z_{j+1,\downarrow}\right) - \frac{U^3}{2}\left(\sigma^z_{j,\uparrow}\sigma^z_{j,\downarrow}\right)\Big] \\
&+ (\uparrow \leftrightarrow \downarrow)\,.
\end{aligned} \tag{4.84}
$$

Grabowski and Mathieu observed that these local charges can be written much more compactly in terms of the tensor products of the densities of the XX charges and they introduced the following notation for the density of the XX charges [28]:

$$
h^{s,+}_{n,j} = \sigma^x_{j,s}\sigma^z_{j+1,s}\sigma^z_{j+2,s}\cdots\sigma^z_{j+n-1,s}\sigma^x_{j+n,s} + \sigma^y_{j,s}\sigma^z_{j+1,s}\sigma^z_{j+2,s}\cdots\sigma^z_{j+n-1,s}\sigma^y_{j+n,s}, \tag{4.85}
$$

$$
h^{s,-}_{n,j} = \sigma^x_{j,s}\sigma^z_{j+1,s}\sigma^z_{j+2,s}\cdots\sigma^z_{j+n-1,s}\sigma^y_{j+n,s} - \sigma^y_{j,s}\sigma^z_{j+1,s}\sigma^z_{j+2,s}\cdots\sigma^z_{j+n-1,s}\sigma^x_{j+n,s}, \tag{4.86}
$$

for $n > 1$ with $s \in \{\uparrow, \downarrow\}$ and

$$h_{1,j}^{s,+} = -\sigma_{j,s}^{z}\,, \quad h_{1,j}^{s,-} = 0\,. \tag{4.87}$$

With these notations, the local charges above are rewritten as

$$H = \sum_{j=1}^{L} \left[ h_{2,j}^{\uparrow,+} + h_{2,j}^{\downarrow,+} + U h_{1,j}^{\uparrow,+} h_{1,j}^{\downarrow,+} \right], \tag{4.88}$$

$$Q_3 = \sum_{j=1}^{L} \left\{ h_{3,j}^{\uparrow,-} + h_{3,j}^{\downarrow,-} + U \left[ h_{2,j}^{\uparrow,-} \left( h_{1,j}^{\downarrow,+} + h_{1,j+1}^{\downarrow,+} \right) + h_{2,j}^{\downarrow,-} \left( h_{1,j}^{\uparrow,+} + h_{1,j+1}^{\uparrow,+} \right) \right] \right\}, \tag{4.89}$$

$$\begin{aligned} Q_4 = \sum_{j=1}^{L} \Big\{ & h_{4,j}^{\uparrow,+} + h_{4,j}^{\downarrow,+} + U \Big[ h_{3,j}^{\uparrow,+} \left( h_{1,j}^{\downarrow,+} + h_{1,j+1}^{\downarrow,+} + h_{1,j+2}^{\downarrow,+} \right), \\ & + h_{3,j}^{\downarrow,+} \left( h_{1,j}^{\uparrow,+} + h_{1,j+1}^{\uparrow,+} + h_{1,j+2}^{\uparrow,+} \right) - h_{2,j}^{\uparrow,-} \left( h_{2,j-1}^{\downarrow,-} + h_{2,j}^{\downarrow,-} + h_{2,j+1}^{\downarrow,-} \right) \\ & - h_{1,j}^{\uparrow,+} \left( h_{1,j-1}^{\downarrow,+} + h_{1,j}^{\downarrow,+} + h_{1,j+1}^{\downarrow,+} \right) \Big] \\ & + U^2 \left[ h_{2,j}^{\uparrow,+} \left( h_{1,j}^{\downarrow,+} h_{1,j+1}^{\downarrow,+} + 1 \right) + h_{2,j}^{\downarrow,+} \left( h_{1,j}^{\uparrow,+} h_{1,j+1}^{\uparrow,+} + 1 \right) \right] - U^3 h_{1,j}^{\uparrow,+} h_{1,j}^{\downarrow,+} \Big\}. \end{aligned} \tag{4.90}$$

They obtained one more higher-order charge [28]:

$$\begin{aligned} Q_5 = \sum_{j=1}^{L} \Big\{ & h_{5,j}^{\uparrow,(-)} + U \Big[ h_{4,j}^{\uparrow,-} \left( h_{1,j}^{\downarrow,+} + h_{1,j+1}^{\downarrow,+} + h_{1,j+2}^{\downarrow,+} + h_{1,j+3}^{\downarrow,+} \right) \\ & + h_{3,j}^{\uparrow,+} \left( h_{2,j-1}^{\downarrow,-} + h_{2,j}^{\downarrow,-} + h_{2,j+1}^{\downarrow,-} + h_{2,j+2}^{\downarrow,-} \right) \\ & - h_{2,j}^{\uparrow,-} \left( h_{1,j-1}^{\downarrow,+} + h_{1,j}^{\downarrow,+} + h_{1,j+1}^{\downarrow,+} + h_{1,j+2}^{\downarrow,+} \right) \Big] \end{aligned} \tag{4.91}$$

$$\begin{aligned} & + U^2 \Big[ h_{3,j}^{\uparrow,-} \left( h_{1,j}^{\uparrow,+} h_{1,j+1}^{\downarrow,+} + h_{1,j}^{\downarrow,+} h_{1,j+2}^{\downarrow,+} + h_{1,j+1}^{\downarrow,+} h_{1,j+2}^{\downarrow,+} \right) \\ & + h_{2,j}^{\uparrow,+} \left( h_{2,j-1}^{\downarrow,-} h_{1,j+1}^{\downarrow,+} + h_{2,j+1}^{\downarrow,-} h_{1,j}^{\downarrow,+} \right) \Big] \\ & - U^3 \left[ h_{2,j}^{\uparrow,-} \left( h_{1,j}^{\downarrow,+} + h_{1,j+1}^{\downarrow,+} \right) \right] \Big\} + (\uparrow \leftrightarrow \downarrow)\,. \end{aligned} \tag{4.92}$$

From the above expressions, Grabowski and Mathieu conjectured that the local charges in the one-dimensional Hubbard model are written as the sum of the terms of the form

$$h_{l_1,j_1}^{(s_1,\epsilon_1)} h_{l_2,j_2}^{(s_2,\epsilon_2)} \cdots h_{l_{p-1},j_{p-1}}^{(s_{p-1},\epsilon_{p-1})} h_{l_p,j_p}^{(s_p,\epsilon_p)}\,, \tag{4.93}$$

where $p$ is an integer and $\epsilon_i \in \{+, -\}$. This is rigorously shown in Chapter 5. They also conjectured that the appearing coefficients are all in the form of

$$\pm U^j\,. \tag{4.94}$$

In fact, this does not hold for higher-order charges beyond $Q_{6\geq k}$, as demonstrated in Chapter 5.

### 4.7.2 General expression of local charges up to $\mathcal{O}(U)$

The authors of Ref. [28] obtained the general expression of the local charges up to the term linear in $U$ in the local charge $Q_k$ [28]:

$$
\begin{aligned}
Q_n =& \sum_{j=1}^{L} [h_{n,j}^{\uparrow,(-)^n} + h_{n,j}^{\downarrow,(-)^n} \\
&+ U \sum_{k=0}^{[n/2]-1} \sum_{m=1}^{n-2k-1} \sum_{\ell=0}^{n-2} (-)^{n+k+m(n-m)+1} h_{n-m-2k,j}^{\uparrow,(-)^{n+m+1}} h_{m,j-m-k+\ell+1}^{\downarrow,(-)^{m+1}}] \\
&+ \mathcal{O}\left(U^2\right) \\
\equiv& Q_n^0 + U Q_n^1 + \mathcal{O}\left(U^2\right) .
\end{aligned}
\tag{4.95}
$$

The explicit expression for all orders had not been obtained before. In Chapter 5, we derive the explicit expression of $Q_k$ beyond the first-order term.

# Chapter 5

# All local conserved quantities in the one-dimensional Hubbard model

In this chapter, I show the explicit expressions for all the local charges in the one-dimensional Hubbard model. The local charges have been found only up to $Q_5$ in the previous studies [28]. Also, there have been no recursive ways to construct them for the one-dimensional (1D) Hubbard model. The local charges in the spin-$1/2$ XYZ chain and the Temperley-Lieb models can be derived from its boost operator. However, the 1D Hubbard model does not have a boost operator, and one needs an alternative approach to the derivation of its local charges.

Here, I identify the operator basis to represent the local charges efficiently in the 1D Hubbard model and derive their explicit expressions. Contrary to the conjecture in the previous study [28], they involve non-trivial coefficients for charges higher than $Q_{k\geq 6}$, and I derive the recursion equation for the coefficients. One also sees that from the strong-coupling limit of these charges, one can derive the mutually commuting family of charges of the special case of the XXC model [110, 111], which is relevant to the $t$-0 model [32].

## 5.1 Setup and notations

In this section, I introduce the Hamiltonian and several notations which will be used to represent the local charges.

### 5.1.1 Hamiltonian and local charges

The Hamiltonian of the one-dimensional Hubbard model is

$$H = -2t \sum_{j=1}^{L} \sum_{\sigma=\uparrow,\downarrow} \left( c_{j,\sigma}^{\dagger} c_{j+1,\sigma} + \text{h.c.} \right) + 4U \sum_{j=1}^{L} \left( n_{j,\uparrow} - \frac{1}{2} \right) \left( n_{j,\downarrow} - \frac{1}{2} \right), \tag{5.1}$$

where $c_{j,\sigma}$ is the fermion annihilation operator on the $j$ th site of flavor $\sigma \in \{\uparrow, \downarrow\}$, $n_{j\sigma} \equiv c^\dagger_{j,\sigma} c_{j,\sigma}$, $U$ is the coupling constant of the fermion on-site interaction, and the periodic boundary condition $c_{j+L,\sigma} = c_{j,\sigma}$ is imposed. In the following, I assume $U \neq 0$ and set $t = 1$.

The Hamiltonian can be rewritten in terms of spin variables via the Jordan-Wigner transformation:

$$c_{j,\uparrow} = \exp\left( \mathrm{i}\pi \sum_{l=1}^{j-1} \sigma^+_{l,\uparrow} \sigma^-_{l,\uparrow} \right) \sigma^-_{j,\uparrow}, \tag{5.2}$$

$$c_{j,\downarrow} = \exp\left( \mathrm{i}\pi \sum_{l=1}^{L} \sigma^+_{l,\uparrow} \sigma^-_{l,\uparrow} \right) \exp\left( \mathrm{i}\pi \sum_{l=1}^{j-1} \sigma^+_{l,\downarrow} \sigma^-_{l,\downarrow} \right) \sigma^-_{j,\downarrow}, \tag{5.3}$$

where $\sigma^a_{j,s}$ $(a \in \{x, y, z\}, s \in \{\uparrow, \downarrow\})$ is the Pauli matrix on the $j$ th site of flavor $s$ and $\sigma^\pm_{j,s} = \frac{1}{2}\left(\sigma^x_{j,s} \pm \mathrm{i}\sigma^y_{j,s}\right)$. Then, the relation $\sigma^z_{j,s} = 2n_{j,s} - 1$ holds, and $\sigma^+_{j,s}\sigma^-_{j,s} = \frac{1}{2}\left(1 + \sigma^z_{j,s}\right) = n_{j,s}$. The Hamiltonian (5.1) is written in terms of the spin variables as

$$H = \sum_{j=1}^{L} \sum_{s=\uparrow,\downarrow} \sum_{a=x,y} \sigma^a_{j,s} \sigma^a_{j+1,s} + U \sum_{j=1}^{L} \sigma^z_{j,\uparrow} \sigma^z_{j,\downarrow}, \tag{5.4}$$

where the boundary condition is twisted as is explained in Eq. (3.28) from the Jordan-Wigner transformation (5.2). Nevertheless, I treat the Hamiltonian in the spin variable notation (5.4) with the periodic boundary condition. Note that the Hamiltonian (5.4) is the form of the two coupled XX chains.

In the following, I will look for the local charges for the fermion Hamiltonian (5.1) and the spin variable Hamiltonian (5.4) in the periodic boundary condition in both cases. I denote the $k$ th conserved quantity, which is a polynomial of $U$ denoted by

$$Q_k = \sum_{j=0}^{j_f} U^j Q^j_k, \tag{5.5}$$

where $j_f = k - 1$ for even $k$, and $j_f = k - 2$ for odd $k$, and $Q^j_k$ is an operator independent of $U$. On the right-hand side, $Q^j_k$ represents the operator for each power when expressing $Q_k$ as a polynomial of $U$, while on the left-hand side, $Q_k$ acts on at most $k$ adjacent sites. I will determine each of $Q^j_k$ to satisfy $[Q_k, H] = 0$.

### 5.1.2 Unit operator and diagram

In this subsection, I introduce notations to simply represent $Q^j_k$. I define a *unit operator* $\psi^{\pm,n}_\sigma(j)$ by

$$\psi^{\pm,n}_\sigma(j) := 2\left(\pm c_{j,\sigma} c^\dagger_{j+n,\sigma} + (-1)^n c^\dagger_{j,\sigma} c_{j+n,\sigma}\right), \tag{5.6}$$

and I will call $n(>0)$ the *length* of the unit operator, and call $\psi_\sigma^{\pm,n}(j)$ $\pm$-type from the $j$ th site with $\sigma(\in\{\uparrow,\downarrow\})$ flavor. In particular, the zero-length unit operators are

$$\psi_\sigma^{-,0}(j) \coloneqq 2n_{j,\sigma}-1, \tag{5.7}$$
$$\psi_\sigma^{+,0}(j) = 0. \tag{5.8}$$

These unit operators satisfy the following relation:

$$\psi_\sigma^{-,n}(j) = \psi_\sigma^{+,n}(j)\cdot\psi_\sigma^{-,0}(j+n) = \psi_\sigma^{-,0}(j)\cdot\psi_\sigma^{+,n}(j). \tag{5.9}$$

Let us represent a unit operator by the spin variables. For this purpose, I employ *doubling product* [41]. In the following, I denote the usual Pauli matrices of $\sigma$ flavor acting on the $j$ th site by $X_{j,\sigma}, Y_{j,\sigma}, Z_{j,\sigma}$. The doubling product of flavor $\sigma$ that starts from the $i$ th site is defined by

$$\overline{A_1\cdots A_k}_{\,\sigma}(i) \coloneqq (A_1)_{i,\sigma}\,(A_1A_2)_{i+1,\sigma}\cdots(A_lA_{l+1})_{l+i,\sigma}\cdots(A_{k-1}A_k)_{i+k-1,\sigma}\,(A_k)_{i+k,\sigma}\,, \tag{5.10}$$

where $A_i\in\{I,X,Y,Z\}$, $A_lA_{l+1}$ is the product of $A_l$ and $A_{l+1}$, and $(\cdot)_{i,\sigma}$ means the operator of the flavor $\sigma$ acting on the $i$ th site. The unit operators are written in terms of the spin variables as

$$\psi_\sigma^{+,n}(j) = \overline{B_1B_2\cdots B_n}_{\,\sigma}(j) + \overline{B'_1B'_2\cdots B'_n}_{\,\sigma}(j) \qquad (n>0)\,, \tag{5.11}$$

where $B_i=X$ and $B'_i=Y$ for odd $i$, while $B_i=Y$ and $B'_i=X$ for even $i$. The zero-length unit operators are written as $\psi_\sigma^{-,0}(j)=Z_{j,\sigma}$, and the non-zero length unit operators of $-$ type are obtained thorough Eq. (5.9). I note that $\sum_{j=1}^{L}\psi_\sigma^{\pm,n}(j)$ is the conserved quantity of the spin Hamiltonian (5.4) in the non-interacting case $U=0$ [28].

I next introduce a graphical notation to represent a unit operator. I represent a unit operator by

$$\psi_\sigma^{+,n}(j) = \overbrace{\bigcirc\bigcirc\cdots\bigcirc}^{n}{}_{\sigma}(j)\ \Big(= \overbrace{|\bigcirc\bigcirc\cdots\bigcirc|}^{n}{}_{\sigma}(j)\Big) \qquad (n>0)\,, \tag{5.12}$$
$$\psi_\sigma^{-,n}(j) = \overbrace{\bigcirc\bigcirc\cdots\bigcirc|}^{n}{}_{\sigma}(j)\ \Big(= \overbrace{|\bigcirc\bigcirc\cdots\bigcirc}^{n}{}_{\sigma}(j)\Big) \qquad (n>0)\,, \tag{5.13}$$
$$\psi_\sigma^{-,0}(j) = |_{\sigma}(j)\,. \tag{5.14}$$

I note from Eq. (5.7) that

$$\bigcirc\cdots\bigcirc|(j)_\sigma = \bigcirc\cdots\bigcirc(j)_\sigma\cdot|(j+n)_\sigma = |\bigcirc\cdots\bigcirc = |(j)_\sigma\cdot\bigcirc\cdots\bigcirc(j)_\sigma\,, \tag{5.15}$$
$$|\bigcirc\cdots\bigcirc|(j)_\sigma = |(j)_\sigma\cdot\bigcirc\cdots\bigcirc(j)_\sigma\cdot|(j+n)_\sigma = \bigcirc\cdots\bigcirc(j)_\sigma\,. \tag{5.16}$$

I next introduce a notation to represent the product of unit operators:

**Definition 5.1.** *I define a diagram for a product of unit operators in which the unit operations of the same flavor have no spatial overlap; such a product is represented as*

$$\prod_{\alpha=1}^{l}\psi_{\sigma_\alpha}^{t_\alpha,n_\alpha}(j_{\alpha,i}), \tag{5.17}$$

*where $j_{\alpha,i} \equiv i + j_\alpha$, $j_1 = 0$, $j_\alpha \le j_{\alpha+1}$, while $j_\alpha$ and $\sigma_\alpha$ are chosen such that if $\sigma_\alpha = \sigma_\beta (\alpha < \beta)$, then $j'_\alpha < j_\beta$ ($j'_\alpha \equiv j_\alpha + n_\alpha$) and if $j_\alpha = j_{\alpha+1}$, then $\sigma_\alpha =\uparrow, \sigma_{\alpha+1} =\downarrow$. Here, $i$ is the starting site of the diagram.*

Here, the term "diagram" refers to an operator, which is a product of units as defined above. Such operators, as one will see later, have a diagrammatic representation.

I represent a diagram starting from the $i$ th site by the symbol $\Psi(i)$. I denote the number of unit operators that constitute the product of $\Psi(i)$ by $l_\Psi$ called the *unit number* of $\Psi(i)$. Note that different unit operators in a diagram mutually commute: $\left[\psi^{t_\alpha,n_\alpha}_{\sigma_\alpha}(j_{\alpha,i}), \psi^{t_\beta,n_\beta}_{\sigma_\beta}(j_{\beta,i})\right] = 0$.

A diagram without a site index denotes the site translation summation:

$$\Psi := \sum_{i=1}^{L} \Psi(i). \tag{5.18}$$

I introduce a graphical representation for diagrams by a two-row sequence. The unit operator $\psi^{t_\alpha,n_\alpha}_{\sigma_\alpha}(j_{\alpha,i})$ is placed in the upper row for $\sigma_\alpha =\uparrow$, while placed in the lower row for $\sigma_\alpha =\downarrow$, with $j_\alpha$ columns being on its left. In the whole system, some parts remain unoccupied by the designated unit operators, and I indicate these parts by filling them with the symbols ◌. For example, the diagram $\Psi(i) = \psi^{-,2}_{\uparrow}(i)\psi^{+,3}_{\downarrow}(i+1)\psi^{-,1}_{\uparrow}(i+4)$ is represented as

$$\bigcirc\bigcirc|_{\uparrow}(i)\cdot \bigcirc\bigcirc\bigcirc_{\downarrow}(i+1)\cdot \bigcirc|_{\uparrow}(i+4) = \begin{smallmatrix}\bigcirc\bigcirc| & ◌ & ◌ & \bigcirc|\\ ◌ & \bigcirc\bigcirc\bigcirc & ◌ & \end{smallmatrix}(i). \tag{5.19}$$

The diagram $\Psi(i) = \psi^{+,2}_{\uparrow}(i)\psi^{-,0}_{\downarrow}(i)\psi^{+,1}_{\downarrow}(i+3)\psi^{-,0}_{\uparrow}(i+6)$ is represented as

$$\bigcirc\bigcirc_{\uparrow}(i)\cdot |_{\downarrow}(i)\cdot \bigcirc_{\downarrow}(i+3)\cdot |_{\uparrow}(i+6) = \begin{smallmatrix}\bigcirc\bigcirc & ◌◌◌◌ & | \\ | & ◌◌◌\bigcirc◌◌◌ & \end{smallmatrix}(i). \tag{5.20}$$

Note that different unit operators on the same row are separated by ◌. In this notation, the interaction term is written as $\Big|(j) = |_{\uparrow}(j) \times |_{\downarrow}(j)$.

I explain the meaning of ◌ in the diagram representation. The part of $n$ consecutive ◌ corresponds to trivial action over $n-1$ consecutive sites. I explain this fact in the following example:

$$\begin{aligned}\bigcirc\bigcirc\overbrace{◌◌◌◌}^{4}\bigcirc\bigcirc\bigcirc(0) &= \psi^{+,2}_{\uparrow}(0)\psi^{+,3}_{\uparrow}(6)\\ &= 4\left(c_{0,\uparrow}c^\dagger_{2,\uparrow} + c^\dagger_{2,\uparrow}c_{0,\uparrow}\right)\left(c_{6,\uparrow}c^\dagger_{9,\uparrow} - c^\dagger_{9,\uparrow}c_{6,\uparrow}\right),\end{aligned} \tag{5.21}$$

where I only show the upper row of the diagram on the left-hand side of the first line, and consequently, the fermion flavors appearing on the right-hand sides of the first and second line are $\uparrow$, and the unit operators with the flavor $\downarrow$ are omitted. The $0$ th to the $2$ th sites and the $6$ th to the $9$ th sites have hopping terms with the spin flavor $\uparrow$; on the other hand, the $4-1=3$ sites, the $3$ th to the $5$ th sites do not have any non-trivial operator with the spin flavor $\uparrow$ on them, which corresponds to the $4$ symbols of ◌ on the left-hand side.

In the following, I refer to a column 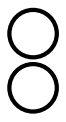 in the graphical representation of a diagram as an *overlap*. I also refer to a column 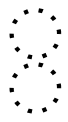 in the graphical representation of a diagram as a *gap*.

I define some integers for the diagram $\Psi$ for Eq. (5.17). First, I define $\{p_i\}$ $(1 \leq i \leq l_\Psi$, $p_i \leq p_{i+1})$ from the set of $\{j_l\}$ and $\{j'_l\}$ by

$$\{p_1, \ldots, p_{2l_\Psi}\} = \{j_1, \ldots, j_{l_\Psi}, j'_1, \ldots, j'_{l_\Psi}\} \qquad (p_i \leq p_{i+1}), \tag{5.22}$$

where $\{p_i\}$ are arranged in ascending order.

Then, the *support* of a diagram $\Psi$ is defined by

$$s_\Psi := j_{l_\Psi} + n_{l_\Psi} + 1, \tag{5.23}$$

which represents the locality of $\Psi$ and $s_\Psi - 1$ corresponds to the total number of columns in the two-row graphical representation of $\Psi$.

The integer *double* of a diagram $\Psi$ is defined by

$$d_\Psi := \sum_{i=1}^{l_\Psi - 1} (p_{2i+1} - p_{2i}), \tag{5.24}$$

which corresponds to the total number of columns of gap and overlap in the graphical representation of $\Psi$. I refer to a diagram with $s_\Psi = s$ and $d_\Psi = d$ as the $(s,d)$-diagram.

The *gap number* of a diagram $\Psi$ is defined by

$$g_\Psi := \sum_{i=1}^{l_\Psi - 1} \delta_i (p_{2i+1} - p_{2i}), \tag{5.25}$$

where $\delta_i = 1$ for $p_{2i+1} \in \{j_1, \ldots, j_{l_\Psi}\}$ and otherwise $0$. The gap number of $\Psi$ is the total number of columns of gap and overlap in the graphical representation of $\Psi$. I note that $s_\Psi > d_\Psi \geq g_\Psi$.

For example, for the diagram (5.19), these integers are $(s_\Psi, d_\Psi, g_\Psi, l_\Psi) = (6, 1, 0, 3)$, while $(7, 3, 3, 4)$ for the diagram (5.20).

Using these diagram representation, the Hamiltonian (5.1) is simply written as

$H =$ 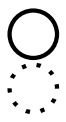 $+$ 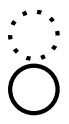 $+ U\Big|$, (5.26)

where the non-interacting term is $H_0 =$ 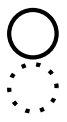 $+$ 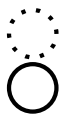 and the interaction term is $H_{\text{int}} = \Big|$. I note that site translation summation is assumed here.

### 5.1.3 Connection of unit operators

In this subsection, I introduce the operation of *connection* of unit operators, which is necessary to define *connected diagram* in the following subsection.

I define the *connection* of unit operators as follows:

**Definition 5.2.** *Two unit operators in a diagram $\Psi$ (5.17), indexed by $\alpha$ and $\beta$ ($\alpha < \beta$), are said to be connected iff the following conditions are satisfied:*

(i) $\sigma_\alpha \neq \sigma_\beta$,

(ii) $j'_\gamma < j'_\alpha$ *or* $j_\beta < j_\gamma$ *holds for* $\forall\gamma(\neq \alpha, \beta)$.

The intuitive comprehension of the connection of unit operators is as follows: connected two unit operators are not separated by any other unit operator, and one is on the upper row, while the other is on the lower row.

For the sake of later convenience, I first prove the following lemma.

**Lemma 5.3.** *For any three mutually different unit operators in a diagram indexed by $\alpha, \beta$ and $\gamma$ ($\sigma_\alpha \neq \sigma_\beta$), either of $j'_\gamma < \max(j_\alpha, j_\beta)$ or $\min(j'_\alpha, j'_\beta) < j_\gamma$ holds.*

*Proof.* I prove this lemma by contradiction. Suppose $\max(j_\alpha, j_\beta) \leq j'_\gamma$ and $j_\gamma \leq \min(j'_\alpha, j'_\beta)$. First, in the case of $\sigma_\alpha = \sigma_\gamma$ and $\alpha < \gamma$, $j'_\alpha < j_\gamma$ should hold from the definition of diagram 5.1, but this contradict with $j_\gamma \leq \min(j'_\alpha, j'_\beta) \leq j'_\alpha$. Second, the case of $\sigma_\alpha = \sigma_\gamma$ and now $\gamma < \alpha$ should yield $j'_\gamma < j_\alpha$. However, this leads to $j_\alpha \leq \max(j_\alpha, j_\beta) < j'_\gamma \leq j_\alpha$, and this is clearly impossible. With the same argument, one can show the contradiction in the case of $\sigma_\gamma = \sigma_\beta \neq \sigma_\alpha$. Thus, the above lemma holds. □

The connection of two unit operators can be categorized into three types: *overlap type*, *adjacent type*, and *gap-sandwiching type*, as will be explained below. Their explanations are first given using the indices of unit operators. Afterward, I describe how these connected unit operators look in the graphical notation of a diagram.

### Overlap type

The first type of connection is *overlap*: the unit operators $\alpha$ and $\beta$ satisfy $j_\alpha < j'_\beta$ and $j_\beta < j'_\alpha$ ($\sigma_\alpha \neq \sigma_\beta, \alpha < \beta$).

Let us first prove that they are actually connected. Lemma 5.3 guarantees that one can see either of $j'_\gamma < j_\beta$ or $\min(j'_\alpha, j'_\beta) < j_\gamma$ holds. In the former case, I have $j'_\gamma < j_\beta < j'_\alpha$, i.e, $j'_\gamma < j'_\alpha$. In the later case, together with the assumption of $j_\beta < j'_\alpha$ and the trivial fact $j_\beta \leq j'_\beta$, I have $j_\beta \leq \min(j'_\alpha, j'_\beta) < j_\gamma$, i.e, $j_\beta < j_\gamma$. Then, from Definition 5.2, one can see that the unit operators $\alpha$ and $\beta$ are connected.

The following are three examples of overlap connections in the graphical representation:

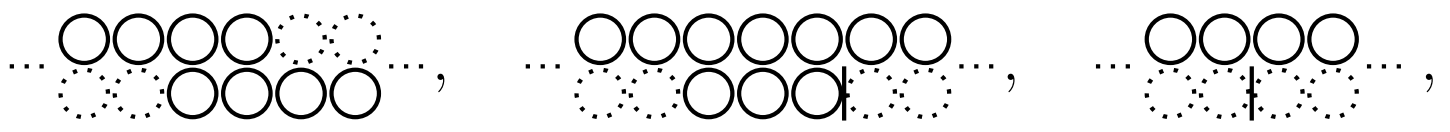 (5.27)

where the two unit operators on the upper and lower rows are connected. In the left and middle examples, the unit operators on the upper and lower rows share a few columns in the overlap region. In the right example, the unit operator in the lower row has zero length.

**Adjacent type**

The second type of connection is *adjacent*: the unit operators $\alpha$ and $\beta$ satisfy $j'_\alpha = j_\beta$ ($\sigma_\alpha \neq \sigma_\beta, \alpha < \beta$).

I again check the connection of the two units. Lemma 5.3 and the assumption above show that any unit operator $\gamma (\neq \alpha, \beta)$ satisfies either $j'_\gamma < j_\beta$ or $j_\beta < j_\gamma$. Thus Definition 5.2 guarantees the connection of the two unit operators $\alpha$ and $\beta$.

The following are two examples of adjacent type connections in the graphical representation of a diagram:

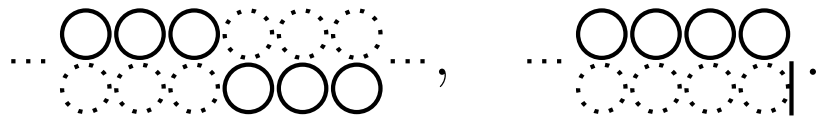

(5.28)

For both examples, two unit operators on the upper and lower row are adjacent. In the right example, the unit operator in the lower row has zero length.

**Gap-sandwiching type**

The third type of connection is *gap-sandwiching*: the unit operators $\alpha$ and $\beta$ satisfy $j'_\alpha < j_\beta$ ($\sigma_\alpha \neq \sigma_\beta$) and either of $j'_\gamma < j'_\alpha$ or $j_\beta < j_\gamma$ for any other unit operator $\gamma (\neq \alpha, \beta)$. One can easily check their connection following the Definition 5.2.

The following are three examples of gap-sandwiching connection:

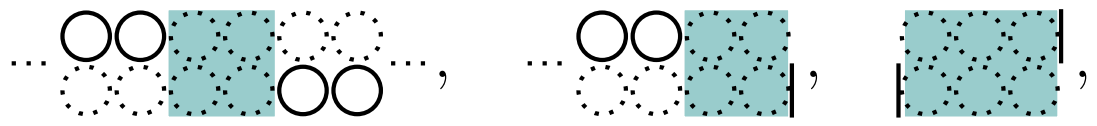

(5.29)

where the columns inside the gap are shaded in teal.

### 5.1.4 Connected diagram

In this subsection, I introduce the concept of *connected diagrams*, which constitute local charges in the 1D Hubbard model.

**Definition 5.4.** *A diagram is called a connected diagram if it satisfies the following two conditions:*

(i) *For any two unit operators $\alpha$ and $\beta$ in the diagram, there exists a sequence of unit operators ($\alpha = \gamma_0, \gamma_1, \ldots, \gamma_N, \gamma_{N+1} = \beta$), in which $\gamma_i$ th and $\gamma_{i+1}$ th unit operators are connected;*

(ii) *The unit operator $\alpha$ is the type $t_\alpha = (-)^{C_\alpha}$, where $C_\alpha$ is the number of unit operators connected with it and $(-)^C = -$ for an odd $C$ and $(-)^C = +$ for an even $C$.*

The diagrams in Eqs. (5.19) and (5.20) are connected. I show further examples of connected diagrams with an explanation below. The first example $\psi_\downarrow^{-,2}(i)\psi_\uparrow^{-,9}(i+4)\psi_\downarrow^{-,3}(i+7)\psi_\downarrow^{-,3}(i+13)$

(the four-unit operator are denoted as $\psi_1, \psi_2, \psi_3, \psi_4$ respectively below) is represented in the graphical notation as

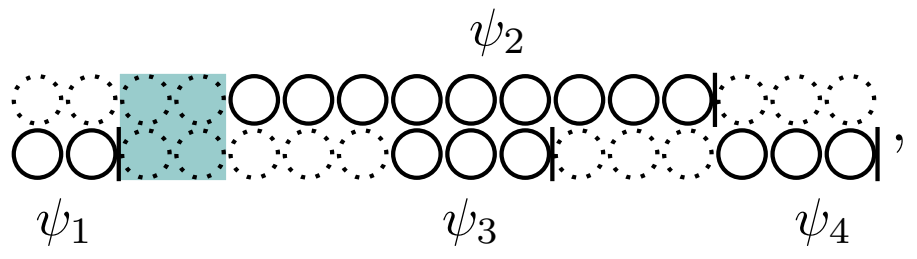

(5.30)

where the columns in the gap are shaded in teal. All the unit operators are of the $-$ type, which are determined by the condition (ii) in Definition 5.4: $C_1 = C_3 = C_4 = 1$ as $\psi_1, \psi_3, \psi_4$ are connected only with $\psi_2$, while $C_2 = 3$ as $\psi_2$ is connected with $\psi_1, \psi_3, \psi_4$. The connection is of the gap-sandwiching type for the $\psi_1$-$\psi_2$ pair, the overlap type for $\psi_2$-$\psi_3$ pair, and the adjacent type for $\psi_2$-$\psi_4$ pair. One should note that the condition (i) in Definition 5.4 is also satisfied for these pairs; for example, the $\psi_1$-$\psi_4$ unit operator pair has the sequence of indices $(1, 2, 4)$, in which $\psi_1$-$\psi_2$ and $\psi_2$-$\psi_4$ pairs are both connected.

The second example of a connected diagram is $\psi_\uparrow^{+,2}(i)\psi_\downarrow^{-,0}(i)\psi_\downarrow^{-,4}(i+3)\psi_\uparrow^{-,0}(i+5)\psi_\uparrow^{+,1}(i+9)\psi_\downarrow^{-,3}(i+11)$ (the six-unit operators are denoted as $\psi_1, \ldots, \psi_6$ respectively below), and this is represented in the graphical notation as

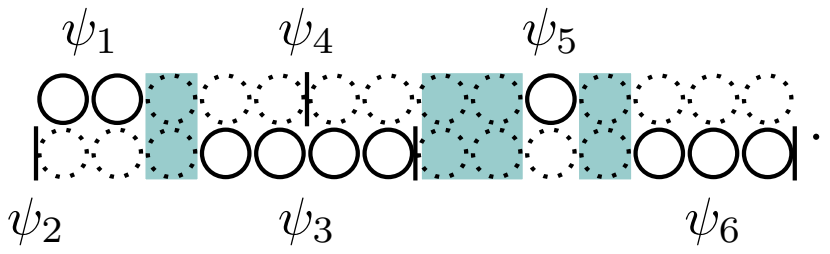

(5.31)

Again, the columns shaded in teal are gaps, and the connection numbers of unit operators are $C_1 = 2, C_2 = 1, C_3 = 3, C_4 = 1, C_5 = 2$ and $C_6 = 1$. The connection for $\psi_1$-$\psi_3$ pair is of the gap-sandwiching type, the connection for $\psi_3$-$\psi_4$ pair is of the overlap type, and the connection for $\psi_1$-$\psi_2$ pair is of the adjacent type. The condition (i) in Definition 5.4 is also satisfied; for example, for the unit operators $\psi_2$ and $\psi_6$, there is a sequence of indices $(2, 1, 3, 5, 6)$, in which the unit operator pairs $\psi_2$-$\psi_1$, $\psi_1$-$\psi_3$, $\psi_3$-$\psi_5$, $\psi_5$-$\psi_6$ are all connected.

I also show two examples of non-connected unit operators that violate the condition (i) or (ii) in Definition 5.4. The first example of a non-connected diagram is $\psi_\downarrow^{-,4}(i)\psi_\uparrow^{-,5}(i+2)\psi_\uparrow^{-,4}(i+9)\psi_\downarrow^{-,6}(i+11)$ represented as

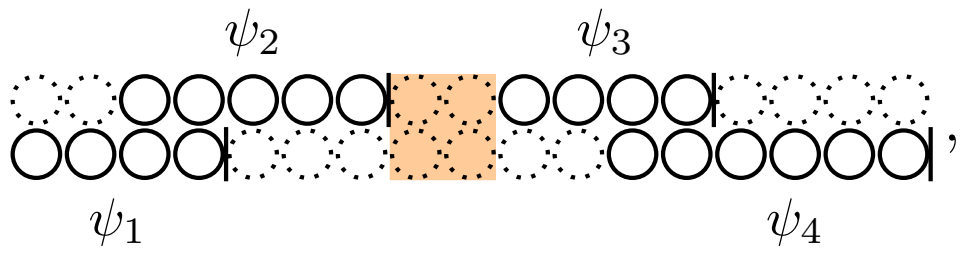

(5.32)

where the columns in the gap are shaded in orange. Now, the condition (i) of Definition 5.4 is violated. In the current example, no sequence of indices connects $\psi_2$ and $\psi_3$ because the connected unit operator pairs are $\psi_1$-$\psi_2$ and $\psi_3$-$\psi_4$. The two unit operators sandwiching the orange gap are not connected.

The second example of a non-connected diagram is $\psi_\downarrow^{-,2}(i)\psi_\uparrow^{+,6}(i+4)\psi_\downarrow^{-,0}(i+6)\psi_\downarrow^{-,4}(i+8)$

represented as

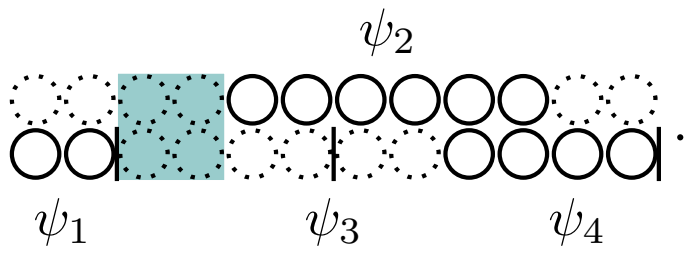

(5.33)

Now, the condition (ii) of Definition 5.4 is violated. The connection number of $\psi_2$ is $C_2 = 3$, while the type of $\psi_2$ is $+ \neq (-)^{c_2}$.

### 5.1.5 List for diagram

In this subsection, I introduce the *list* for a diagram that is necessary to express the coefficients of a diagram in local charges. I first give the definition of a list and then explain it intuitively using the graphical notation of diagrams.

**Definition 5.5.** *The list for a diagram $\Psi$ in Eq. (5.17) with $l_\Psi > 1$ is defined by the sequence of non-negative integers $\boldsymbol{\lambda}_\Psi = \{\lambda_0, \lambda_1, \ldots, \lambda_{l_\Psi - 1}\}$, where I call $\lambda_i$ as crevice. Here, a crevice $\lambda_\alpha$ is defined by $\lambda_\alpha \equiv p_{2\alpha+2} - p_{2\alpha+1} - \eta_\alpha$, where $\eta_\alpha = 1 - (\delta_{\alpha,0} + \delta_{\alpha,l_\Psi-1})$ and $p_i$ is already defined in Eq. (5.22).*

For a given diagram, I define consecutive columns, where one of the rows is ⬚ and the other is ○, such as

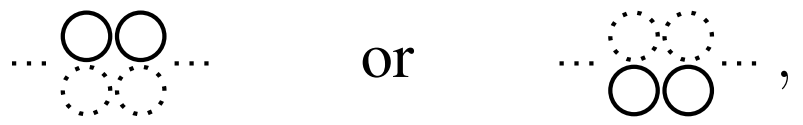

(5.34)

as *coast*. I call the left example a southern coast and the right example a northern coast. A crevice $\lambda_\alpha$ in a list thus corresponds to the length of the coast. I explain this fact using a concrete example below. The list of the connected diagram $\Psi = \psi_\uparrow^{-,3}(i)\psi_\downarrow^{+,4}(i+2)\psi_\uparrow^{-,10}(i+5)\psi_\downarrow^{-,2}(i+10)\psi_\downarrow^{+,6}(i+17)\psi_\uparrow^{-,0}(i+20)$ is $\boldsymbol{\lambda}_\Psi = \{\lambda_0, \lambda_1, \lambda_2, \lambda_3, \lambda_4, \lambda_5\} = \{2, 1, 3, 2, 2, 3\}$, which is depicted below:

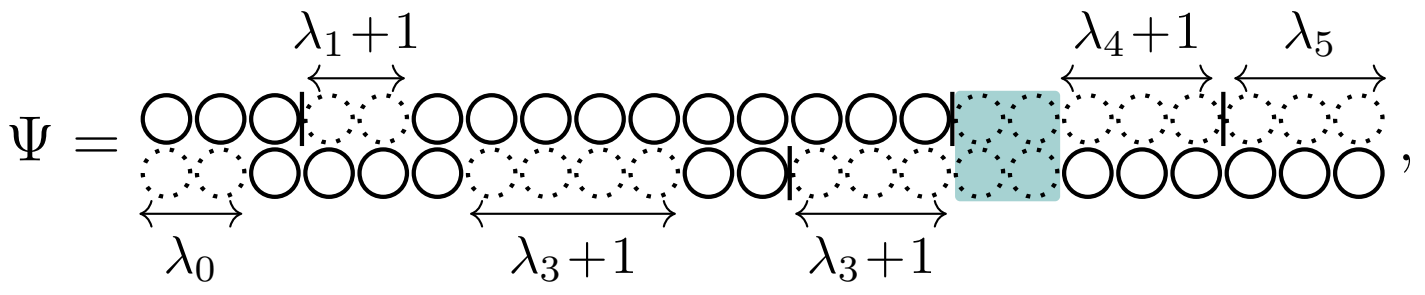

(5.35)

where the lengths of the coasts are indicated by arrows, and the gap is shaded in teal. One should note that $\lambda_0$ and $\lambda_{l_\Psi-1}$ are treated specially ($l_\Psi = 6$, in the above example of Eq. (5.35)); $\lambda_0$ is the length of leftmost coast and $\lambda_{l_\Psi-1}$ is the length of rightmost coast, whereas the length is $\lambda_i + 1$ for the other coasts.

I also define the first and last components of a list as $\lambda_0 \equiv \lambda_L$ and $\lambda_{l_\Psi-1} \equiv \lambda_R$, respectively, define the middle components as $\vec{\lambda} = \{\lambda_1, \ldots, \lambda_{l_\Psi-2}\}$, and then define a list as $\boldsymbol{\lambda}_\Psi = \{\lambda_L; \vec{\lambda}; \lambda_R\}$, where I separate the first and last components by semicolon:

$$\boldsymbol{\lambda}_\Psi = \{\lambda_L; \vec{\lambda}; \lambda_R\} = \{\lambda_0 \equiv \lambda_L; \lambda_1, \ldots, \lambda_{l_\Psi-2}; \lambda_R \equiv \lambda_{l_\Psi-1}\}\,. \tag{5.36}$$

The following is another example of a connected diagram:

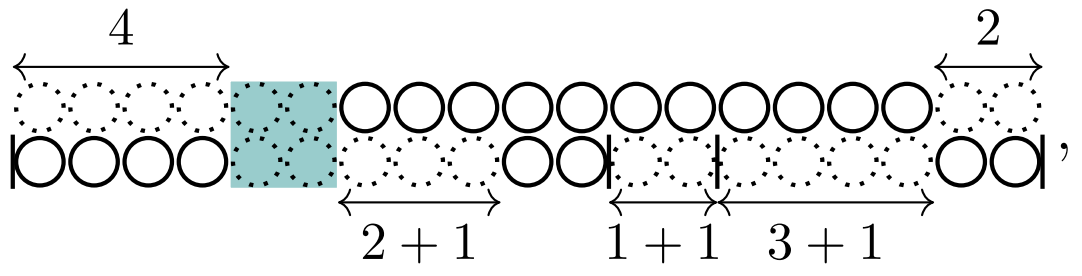

(5.37)

and its list is $\{\lambda_L; \lambda_1, \lambda_2, \lambda_3; \lambda_R\} = \{4; 2, 1, 3; 2\}$.

Two other examples of connected diagrams with $\lambda_L = 0$ are

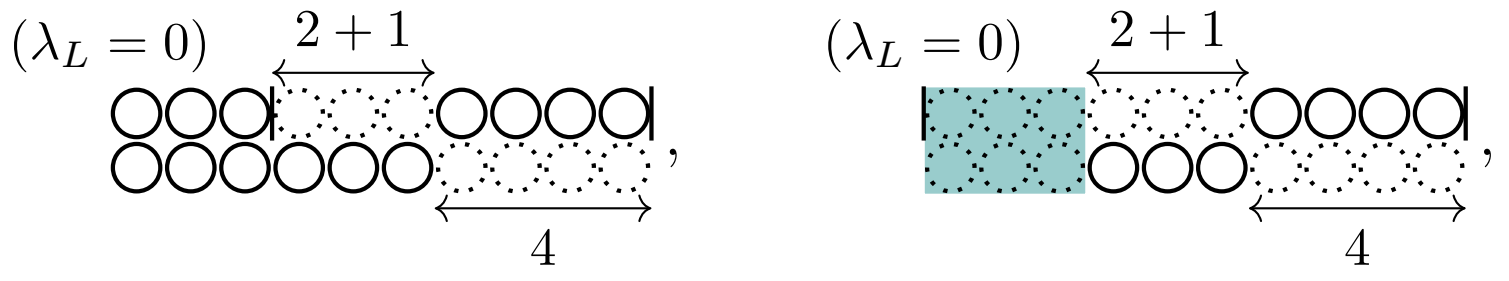

(5.38)

and their lists are both $\{\lambda_L; \lambda_1; \lambda_R\} = \{0; 2; 4\}$.

The following are examples of connected diagrams that have no middle components:

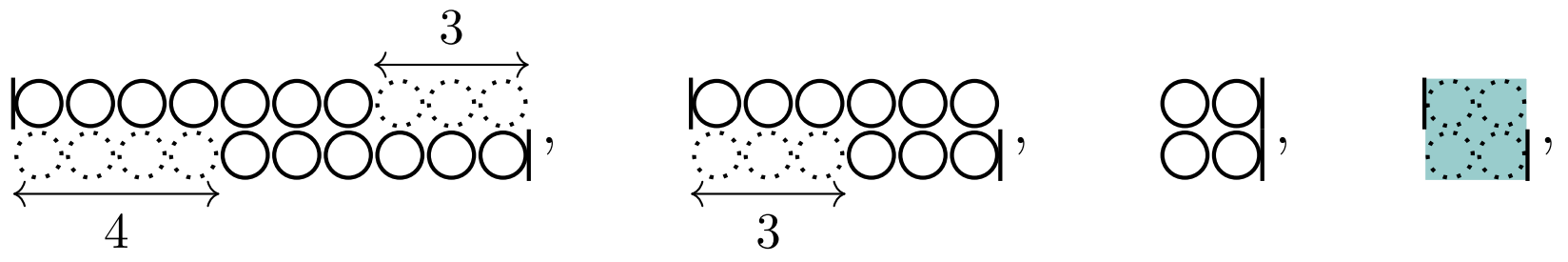

(5.39)

where the leftmost diagram has the list $\{4; 3\}$ and the second left has $\{3; 0\}$, while the rest ones have $\{0; 0\}$.

I next introduce a few operations for the list $\boldsymbol{\lambda}$. The swapped list $\boldsymbol{\lambda}_{a\leftrightarrow b}$ is defined by the configuration in which $\lambda_a$ and $\lambda_b$ are swapped in $\boldsymbol{\lambda}$:

$$\{\dots, \lambda_a, \dots, \lambda_b, \dots\}_{a\leftrightarrow b} = \{\dots, \lambda_b, \dots, \lambda_a, \dots\}\,. \tag{5.40}$$

The removed list $\boldsymbol{\lambda}_{\hat{a}}$ is defined by the configuration in which $\lambda_a$ is removed from $\boldsymbol{\lambda}$:

$$\{\dots, \lambda_{a-1}, \lambda_a, \lambda_{a+1}, \dots\}_{\hat{a}} = \{\dots, \lambda_{a-1}, \lambda_{a+1}, \dots\}\,. \tag{5.41}$$

The replaced list $\boldsymbol{\lambda}_{a\to\widetilde{\lambda}_a}$ is defined by the configuration where $\lambda_a$ in $\boldsymbol{\lambda}$ is replaced by $\widetilde{\lambda}_a$:

$$\{\dots, \lambda_a, \dots\}_{a\to\widetilde{\lambda}_a} = \{\dots, \widetilde{\lambda}_a, \dots\}. \tag{5.42}$$

Two combined operations are also defined: $\boldsymbol{\lambda}_{a:\pm\delta} \equiv \boldsymbol{\lambda}_{a\to\lambda_a\pm\delta}$, and $\mathcal{T}\boldsymbol{\lambda} \equiv \boldsymbol{\lambda}_{L:-1,R:+1} = (\boldsymbol{\lambda}_{L:-1})_{R:+1}$. I also define two operations of adding $0$ in the first or last position of the list:

$$_{0\to}\{\lambda_L; \lambda_1, \dots, \lambda_w; \lambda_R\} = \{0; \lambda_L, \lambda_1, \dots, \lambda_w; \lambda_R\}\ , \tag{5.43}$$

$$\{\lambda_L; \lambda_1, \dots, \lambda_w; \lambda_R\}_{\leftarrow 0} = \{\lambda_L; \lambda_1, \dots, \lambda_w, \lambda_R; 0\}\ . \tag{5.44}$$

The following are some examples of these operations:

$$\{\lambda_L; \lambda_1, \lambda_2, \lambda_3, \lambda_4; \lambda_R\}_{1\leftrightarrow 4} = \{\lambda_L; \lambda_4, \lambda_2, \lambda_3, \lambda_1; \lambda_R\}, \tag{5.45}$$

$$\{\lambda_L; \lambda_1, \lambda_2, \lambda_3, \lambda_4; \lambda_R\}_{L\leftrightarrow R} = \{\lambda_R; \lambda_4, \lambda_2, \lambda_3, \lambda_1; \lambda_L\}, \tag{5.46}$$

$$\{\lambda_L; \lambda_1, \lambda_2, \lambda_3, \lambda_4; \lambda_R\}_{1\to\widetilde{\lambda}_1} = \{\lambda_L; \widetilde{\lambda}_1, \lambda_2, \lambda_3, \lambda_4; \lambda_R\}, \tag{5.47}$$

$$\{\lambda_L; \lambda_1, \lambda_2, \lambda_3, \lambda_4; \lambda_R\}_{\hat{3}} = \{\lambda_L; \lambda_1, \lambda_2, \lambda_4; \lambda_R\}, \tag{5.48}$$

$$\{\lambda_L; \lambda_1; \lambda_R\}_{\hat{1}} = \{\lambda_L; \lambda_R\}, \tag{5.49}$$

$$\{\lambda_L; \lambda_1, \lambda_2, \lambda_3, \lambda_4; \lambda_R\}_{2:+5} = \{\lambda_L; \lambda_1, \lambda_2+5, \lambda_3, \lambda_4; \lambda_R\}, \tag{5.50}$$

$$\{\lambda_L; \lambda_1, \lambda_2, \lambda_3, \lambda_4; \lambda_R\}_{2:+3,4:-2} = \{\lambda_L; \lambda_1, \lambda_2+3, \lambda_3, \lambda_4-2; \lambda_R\}, \tag{5.51}$$

$$\{\lambda_L; \lambda_1, \lambda_2, \lambda_3, \lambda_4; \lambda_R\}_{L:-1} = \{\lambda_L-1; \lambda_1, \lambda_2, \lambda_3, \lambda_4; \lambda_R\}, \tag{5.52}$$

$$\mathcal{T}\{\lambda_L; \lambda_1, \lambda_2, \lambda_3, \lambda_4; \lambda_R\} = \{\lambda_L-1; \lambda_1, \lambda_2, \lambda_3, \lambda_4; \lambda_R+1\}, \tag{5.53}$$

$$_{0\to}\{\lambda_L; \lambda_1, \lambda_2, \lambda_3, \lambda_4; \lambda_R\} = \{0; \lambda_L, \lambda_1, \lambda_2, \lambda_3, \lambda_4; \lambda_R\}, \tag{5.54}$$

$$\{\lambda_L; \lambda_1, \lambda_2, \lambda_3, \lambda_4; \lambda_R\}_{\leftarrow 0} = \{\lambda_L, \lambda_1, \lambda_2, \lambda_3, \lambda_4, \lambda_R; 0\}. \tag{5.55}$$

## 5.2 Explicit expressions for the local conserved quantities

In this section, I show the explicit expressions of the local conserved quantities in the one-dimensional Hubbard model and analyze their structures.

### 5.2.1 Main theorem

I now try to obtain explicit expressions of the local charges $\{Q_k\}$ in the one-dimensional Hubbard model. I will show that each $Q_k$ is a linear combination of connected diagrams.

I first study the non-interacting term ($U^0$ term) in $Q_k$, i.e., $Q_k^0$, which is represented by the $(k,0)$-diagram:

$$Q_k^0 = \psi_\uparrow^{+,k-1} + \psi_\downarrow^{+,k-1} = \tag{5.56}$$

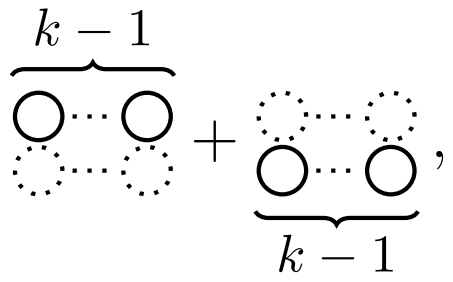

where $\psi_\sigma^{+,k-1} \equiv \sum_{i=1}^{L} \psi_\sigma^{+,k-1}(i)$. One can easily show $[H_0, Q_k^0] = 0$ for all $k$.

Non-trivial terms are $Q_k^j$ for $j > 0$. In order to obtain $Q_k^j$, I define the following integer series:

**Definition 5.6.** *The non-negative integer $C_{n,d}^{j,m}(\boldsymbol{\lambda})$ is defined for the list $\boldsymbol{\lambda} \equiv \{\lambda_L; \lambda_1, \ldots, \lambda_w; \lambda_R\}$ with $w = j-1-2m$ as a solution of the following recursive equation:*

$$\begin{aligned}\Delta C_{n,d}^{j,m}(\boldsymbol{\lambda}) = {} & 2\Delta C_{n-1,d+1}^{j,m}(\boldsymbol{\lambda}) - \Delta C_{n-2,d+2}^{j,m}(\boldsymbol{\lambda}) \\ & + C_{n,d}^{j-1,m-1}(\boldsymbol{\lambda}_{\leftarrow 0}) - C_{n,d}^{j-1,m-1}(_{0\to}(\mathcal{T}\boldsymbol{\lambda})),\end{aligned} \tag{5.57}$$

*where the difference* $\Delta C^{j,m}_{n,d}(\boldsymbol{\lambda})$ *is defined by*

$$\Delta C^{j,m}_{n,d}(\boldsymbol{\lambda}) \equiv C^{j,m}_{n,d}(\boldsymbol{\lambda}) - C^{j,m}_{n,d}(\mathcal{T}\boldsymbol{\lambda}). \tag{5.58}$$

*In the case of* $\lambda_L = -1$, *I define*

$$C^{j,m}_{n,d}(-1;\lambda_1,\dots) := \begin{cases} C^{j,m}_{n,d-1}(1;\lambda_1,\dots) & \text{for } d>0, \\ C^{j,m}_{n-1,1}(0;\lambda_1-1,\dots) + C^{j-1,m}_{n,0}(\lambda_1+1;\dots) & \text{for } d=0. \end{cases} \tag{5.59}$$

*The initial condition is* $C^{j=1,m=0}_{n,d}(\boldsymbol{\lambda}) = 1$. *I also set* $C^{j,m}_{n,d}(\boldsymbol{\lambda}) = 0$ *if one of the three conditions is satisfied: (i)* $\lambda_i < 0 (\exists i \neq L, R)$, *(ii)* $n < 0$, *and (iii)* $m \notin \{1, 2, \dots, \lfloor j/2 \rfloor - 1\}$.

The next theorem gives an explicit expression of the nontrivial terms $Q^{j>0}_k$ of the local charges; this is the main result in this chapter.

**Theorem 5.7.**

$$Q^j_k = \sum_{n=0}^{\lfloor \frac{k-1-j}{2} \rfloor} \sum_{d=0}^{\lfloor \frac{k-1-j}{2} \rfloor - n} \sum_{m=0}^{\lfloor \frac{j-1}{2} \rfloor} \sum_{g=0}^{d} (-1)^{n+m+g} \sum_{\Psi \in \mathcal{S}^{k,j,m}_{n,d,g}} C^{j,m}_{n,d}(\boldsymbol{\lambda}_\Psi)\, \Psi \quad \text{for } j>0, \tag{5.60}$$

*where* $\mathcal{S}^{k,j,m}_{n,d,g}$ *is the set of all* $(k-j-2n-d, d)$*-connected diagrams* $\Psi$ *with* $l_\Psi = j+1-2m$ *and* $g_\Psi = g$.

I will prove Theorem 5.7 in Section 5.4. The positivity of the coefficient $C^{j,m}_{n,d}(\boldsymbol{\lambda}) > 0$ is proved in Appendix A.1.6.

I note that the diagram $\Psi$ in $Q^j_k$ has the coefficient $C^{j,m}_{n,d}(\boldsymbol{\lambda}_\Psi)$, which only depends on the $\Psi$'s list $\boldsymbol{\lambda}$ and indices $j, m, n, d$. Thus, if two connected diagrams $\Psi_1$ and $\Psi_2$ have the same list $\boldsymbol{\lambda}_1 = \boldsymbol{\lambda}_2$, their coefficients are identical up to the sign factor. For example, in $Q^6_{20}$, the following diagrams have the same list $\{1; 1; 2\}$ and their coefficients are all $C^{6,2}_{2,2}(1;1;2) = 1175$:

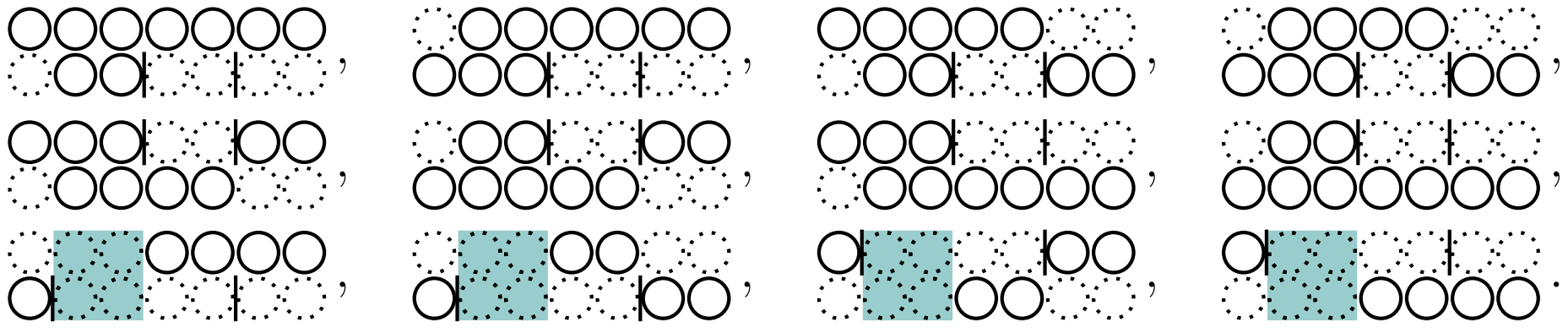

(5.61)

Here, the gaps are again shaded in teal. Consider also the case in which two connected diagrams are almost identical, but a consecutive gap in one diagram is replaced by an overlap of unit operators. Their coefficients are identical up to the sign factor, which comes from the gap numbers. Examples are the following:

coefficient of 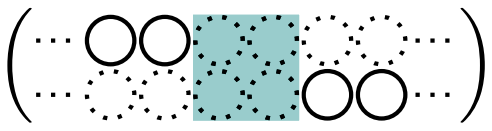 $\times (-1)^2 =$ coefficient of 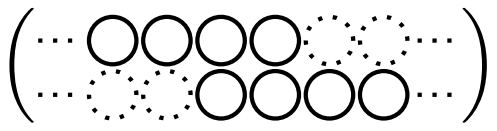 (5.62)

and

coefficient of 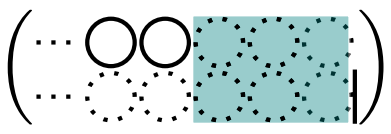 $\times (-1)^3 =$ coefficient of 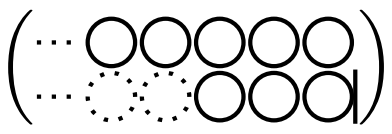 . (5.63)

### 5.2.2 Some expressions of coefficients

In this subsection, I show a general form of $Q_k^j$ for some $j$ using the coefficient obtained by Theorem 5.7.

#### General expressions of $U^1$ terms: previously known result

The general expressions of $Q_k^1$ for all $k$ have been obtained in Ref. [28] as is reviewed in Eq. (4.95). One can represent $Q_k^1$ also in my diagram notation.

Since the coefficients in $Q_k^1$ are trivial as in $C_{n,d}^{j=1,0}(\boldsymbol{\lambda}) = 1$, $Q_k^{j=1}$ is written as

$$
\begin{aligned}
Q_k^{j=1} &= \sum_{\substack{n+d \le \lfloor \frac{k}{2} \rfloor - 1 \\ n,d \ge 0}} \sum_{g=0}^{d} (-1)^{n+g} \sum_{\Psi \in \mathcal{S}_{n,d,g}^{k,1,0}} \Psi \\
&= \sum_{n=0}^{\lfloor \frac{k}{2} \rfloor - 1} (-1)^n \sum_{\substack{u+d=k-2-2n \\ u,d \ge 0}} \sum_{\delta=-(d+n)}^{u+n} \sum_{i=1}^{L} \psi_{\uparrow}^{-,u}(i) \psi_{\downarrow}^{-,d}(i+\delta) \\
&= \sum_{n=0}^{\lfloor \frac{k}{2} \rfloor - 1} (-1)^n \sum_{\substack{u+d=k-2-2n \\ u,d \ge 0}} \sum_{\delta=-(d+n)}^{u+n}
\end{aligned}
$$

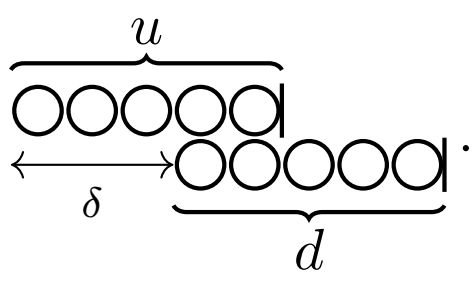

(5.64)

Here, the second line is represented using constructing unit operators, while the last line is represented in the graphical notation. Note that the symbols ◌ are omitted.

In contrast to the previously known cases of $Q_k^1$ [28], I found that some coefficients differ from the trivial value 1 for $Q_k^{j>1}$.

One of the main results of this thesis is that one can construct $Q_k^j$ also for $j > 1$, as shown in Theorem 5.7 and the recursion equation (5.57). One can construct $Q_{k \ge 6}$, which has not been obtained in the previous studies. Moreover, one can solve the recursion equation analytically for some cases, which I explain in the following.

#### Expressions of some coefficients

I obtain an expression of the coefficients $C_{n,d}^{j,m=0}(\boldsymbol{\lambda})$ by solving the recursion equation (5.57).

$$
C_{n,d}^{j,m=0}(\lambda_L; \lambda_1, \ldots, \lambda_{j-1}; \lambda_R) = \sum_{x_1=0}^{\lambda_1} \cdots \sum_{x_{j-1}=0}^{\lambda_{j-1}} \theta\left(n - \sum_{i=1}^{j-1} x_i\right) \quad \text{for } j > 1, \tag{5.65}
$$

where $\theta(x) = 1$ for $x \ge 0$ and $\theta(x) = 0$ otherwise. The coefficients $C_{n,d}^{j,m=0}(\boldsymbol{\lambda})$ are independent of the values of $\lambda_L, \lambda_R$ and $d$. The expression (5.65) seems slightly complicated at a glance.

However, the generating function for Eq. (5.65) is simple:

$$\begin{aligned} G^{j,m=0}(s;\lambda_1,\dots,\lambda_{j-1}) &\coloneqq \sum_{n=0}^{\infty} s^n C^{j,m=0}_{n,d}\left(\lambda_L;\lambda_1,\dots,\lambda_{j-1};\lambda_R\right) \\ &= \sum_{x_1=0}^{\lambda_1}\cdots\sum_{x_{j-1}=0}^{\lambda_{j-1}}\sum_{n=0}^{\infty} s^n\theta\left(n-\sum_{i=1}^{j-1}x_i\right) \\ &= \sum_{x_1=0}^{\lambda_1}\cdots\sum_{x_{j-1}=0}^{\lambda_{j-1}} s^{\sum_{i=1}^{j-1}x_i}(1+s+s^2+\dots) \\ &= \frac{1}{1-s}\prod_{i=1}^{j-1}\left(\sum_{x=0}^{\lambda_i}s^x\right) \\ &= \frac{1}{(1-s)^j}\prod_{i=1}^{j-1}\left(1-s^{\lambda_i+1}\right), \end{aligned} \tag{5.66}$$

where I omit the indices for $\lambda_L, \lambda_R$ and $d$, because this function does not depend on them.

The result in the $n=0$ cases is the following:

$$C^{j,m}_{n=0,d}\left(\boldsymbol{\lambda}\right) = \binom{j-1+d}{m} - \binom{j-1+d}{m-1}, \tag{5.67}$$

which is independent of $\boldsymbol{\lambda}$. I note that Eq. (5.67) is the generalized Catalan number [27, 28, 112]. In Appendix A.2, I demonstrate that the solutions for the $m=0$ case in Eq. (5.65) and for the $n=0$ case in Eq. (5.67) indeed satisfy the recursion equation (5.57) and the identities in Eqs. (5.79)–(5.84).

The expressions are more complicated in more general cases of $n, m > 0$. For example, for $j=3$ and $m=1$, I have

$$\begin{aligned} C^{j=3,m=1}_{n,d}\left(\lambda_L;\lambda_R\right) = &\sum_{\eta=\lambda_L;\lambda_R}\sum_{x=1}^{\eta}\binom{n+3-x}{3} + 2\binom{n+4}{4} \\ &+ (d-1)\left\{2\binom{n+3}{3} - \binom{n+2}{2}\right\}. \end{aligned} \tag{5.68}$$

I have also obtained the explicit expression of $Q^2_k$ and $Q^3_k$ for all $k$ from Eqs. (5.65) and (5.68).

### Expressions for $U^2$ and $U^3$ terms

From Eq. (5.65), one can obtain the explicit expression for $Q^{j=2}_k$ for all $k$. The necessary coefficients are already obtained in Eq. (5.65):

$$C^{j=2,m=0}_{n,d}(\lambda_L;\lambda_1;\lambda_R) = 1 + \min(\lambda_1, n)\,. \tag{5.69}$$

The expression of $Q_k^{j=2}$ is

$$Q_k^{j=2} = \sum_{\substack{n+d \le \lfloor \frac{k-3}{2} \rfloor \\ n,d \ge 0}} \sum_{g=0}^{d} (-1)^{n+g} \sum_{\Psi \in \mathcal{S}_{n,d,g}^{k,2,0}} (1 + \min(\lambda_1, n)) \Psi \, . \tag{5.70}$$

I have also obtained the general explicit form for $Q_k^{j=3}$ for all $k$ from Eqs. (5.65) and (5.68).

With the same argument, if I solve the recursion equation (5.57) explicitly, one can obtain the explicit expressions for more general cases. However, the general solution for the recursion equation, i.e., the explicit expression for $C_{n,d}^{j,m}(\boldsymbol{\lambda})$ in terms of $j, m, n, d$, and $\boldsymbol{\lambda}$, has not been obtained and I leave this issue to the future study. Nonetheless, once I calculate the coefficient appearing in $Q_k$ from the recursion equation (5.57), one can obtain the expression for $Q_k$.

### 5.2.3 Structure of the local conserved quantities

In this subsection, I explain the structure of the local charges obtained in Theorem 5.7.

#### Lower-order charges: previously known results

I first show the explicit expressions of the lower-order charges $Q_3$ [37], $Q_4$ [81, 82] and $Q_5$ [28]. I now represent them in terms of connected diagrams:

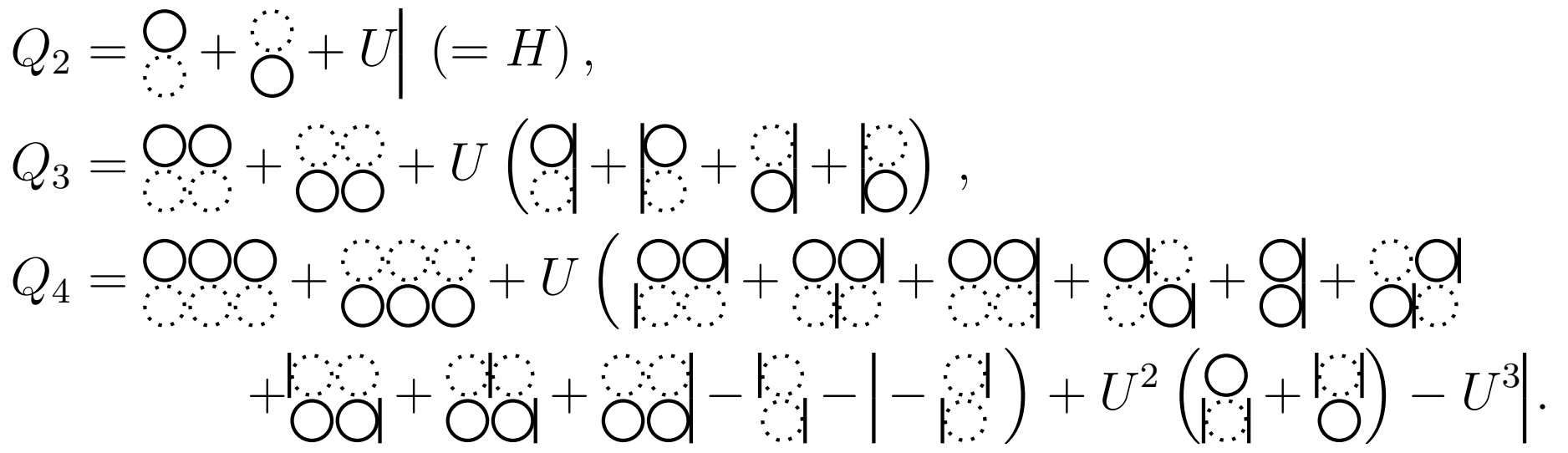

(5.71)

(5.72)

(5.73)

The expression of $Q_5$ [28] is quite long:

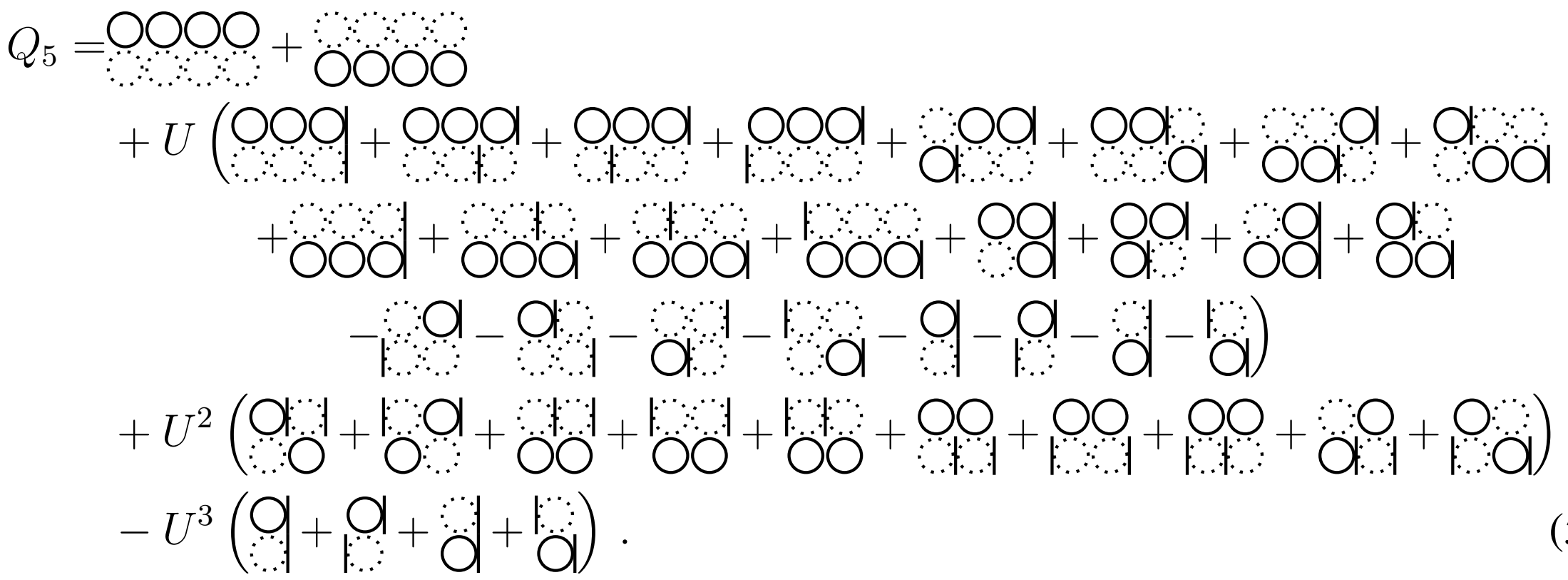

(5.74)

One should note that the numerical coefficients are all $\pm 1$ up to $Q_5$ [28], and this has been known previously.

### The higher-order charges: new results

One can also derive the higher-order charges above $Q_{k>5}$ from Theorem 5.7, which is a new result. To this end, it is useful to rewrite the expression of $Q_k$ in Eq. (5.60) in Theorem 5.7 as

$$Q_k = Q_k^0 + \sum_{j=1}^{j_f} U^j \sum_{n=0}^{\lfloor \frac{k-1-j}{2} \rfloor} \sum_{d=0}^{\lfloor \frac{k-1-j}{2} \rfloor - n} Q_k^j(k-j-2n-d, d), \tag{5.75}$$

where $Q_k^j(s,d)$ is the set of the $(s,d)$-diagrams in $Q_k^j$:

$$Q_k^j(k-j-2n-d, d) = \sum_{m=0}^{\lfloor \frac{j-1}{2} \rfloor} \sum_{g=0}^{d} \sum_{\Psi \in \mathcal{S}_{n,d,g}^{k,j,m}} (-1)^{n+m+g} C_{n,d}^{j,m}\left(\boldsymbol{\lambda}_\Psi\right) \Psi. \tag{5.76}$$

I explain the structure of $Q_6^j$ for each $j$, i.e., the classification of operators in $Q_6^j$ with a non-vanishing coefficient, in Figure 5.1 (a) and all the structure of $Q_6$ in Figure 5.1 (b). The circle at the position $(s,d)$ represents $Q_6^j(s,d)$, i.e, the set of the $(s,d)$-diagrams in $Q_6^j$. The solid arrows indicate how the commutator with $H_0$ generates diagrams at different positions shown by the $\times$ symbol. The vertical dotted arrows indicate the commutator with $H_{\text{int}}$. At the positions shown by the $\times$ symbol, the values of the commutator cancel each other between those coming from different positions.

The explicit expressions of each component in $Q_6$ in Figure 5.1 are

$Q_6^0(6,0)$ 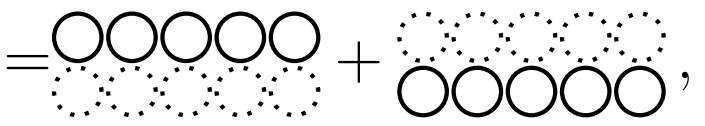

$Q_6^1(5,0)$ 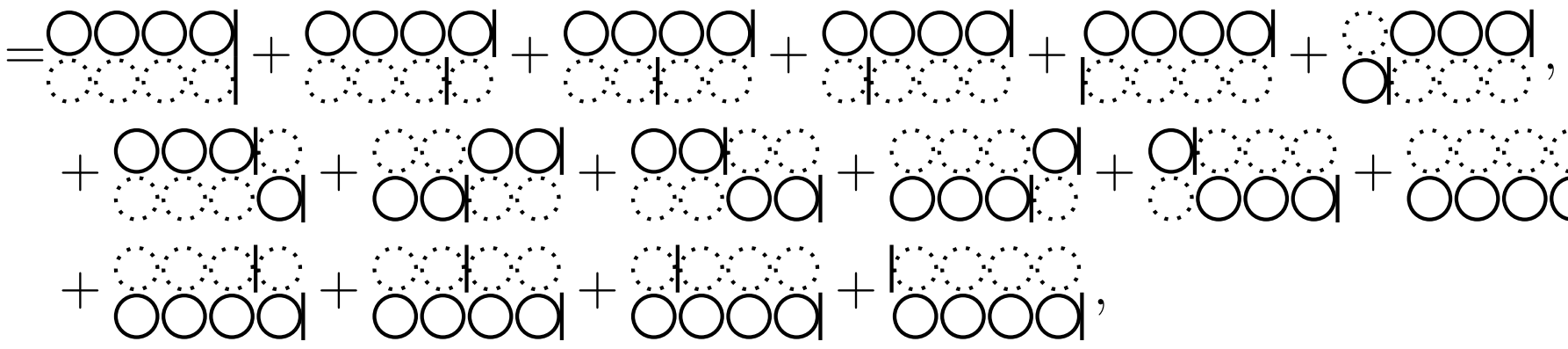

$Q_6^1(4,1)$ 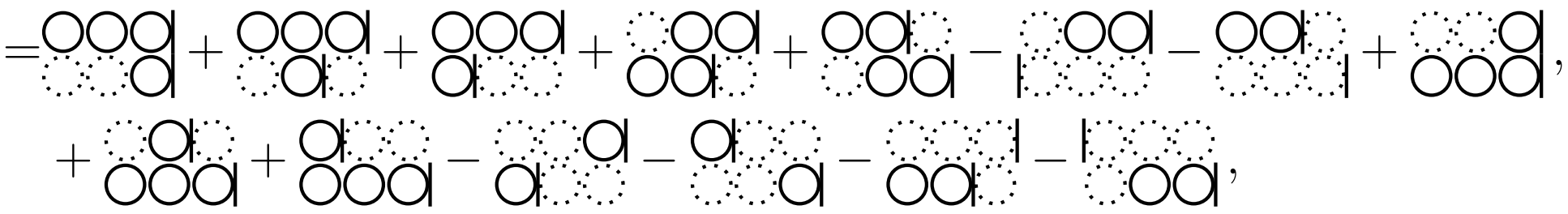

$Q_6^1(3,2)$ 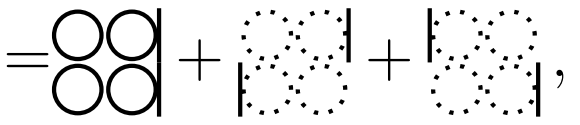

$Q_6^1(3,0)$ 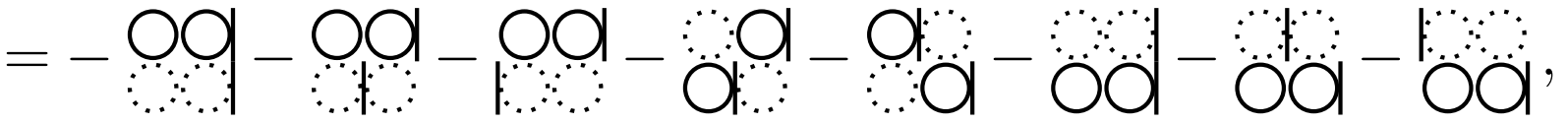

$Q_6^1(2,1)$ 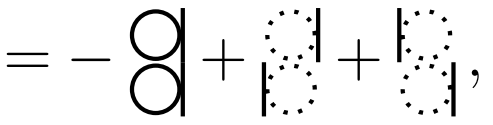

$Q_6^1(1,0)$ 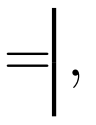

$Q_6^2(4,0)$ 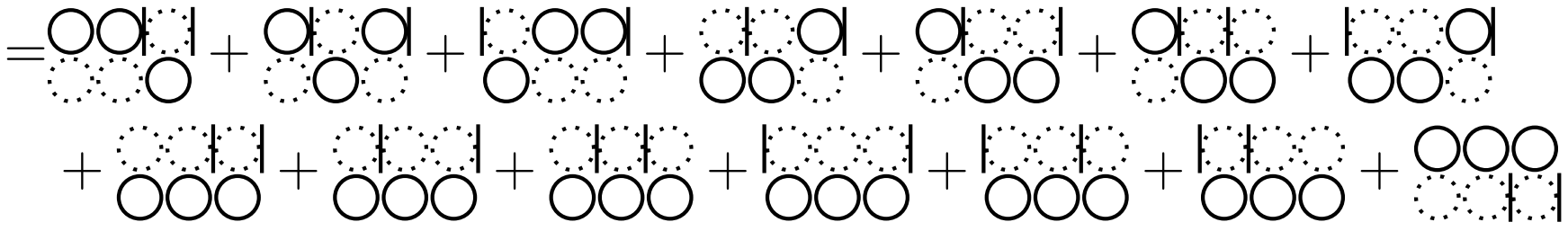

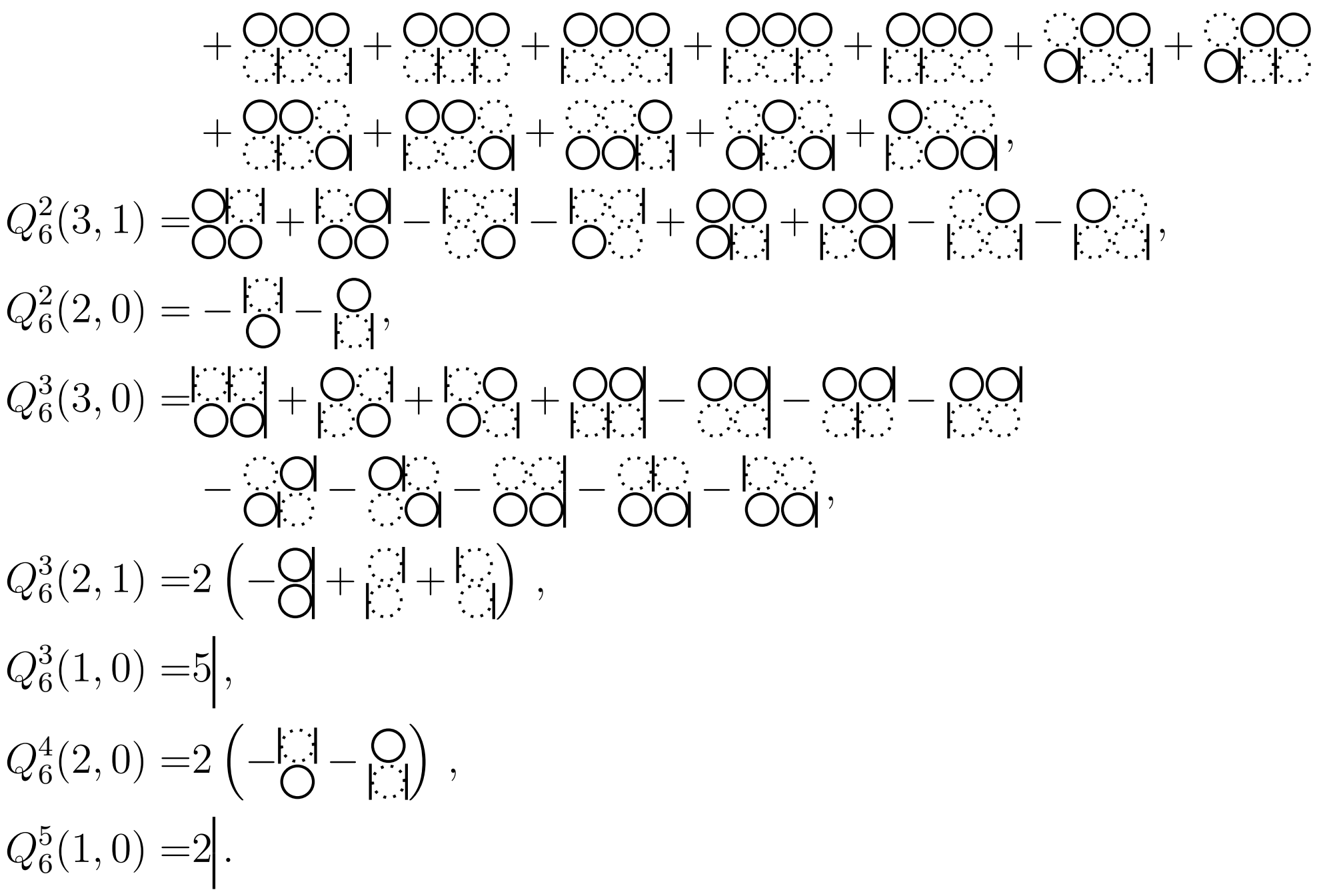

(5.77)

Here, note that $Q_k^0(k,0) \equiv Q_k^0$. It is interesting that some coefficients differ from the trivial values $\pm 1$ in $Q_6^3$, $Q_6^4$ and $Q_6^5$. My diagram representation has an advantage in simplifying expressions compared with the original spin operator representation. This enables us to find the pattern rules in $Q_k$ more easily.

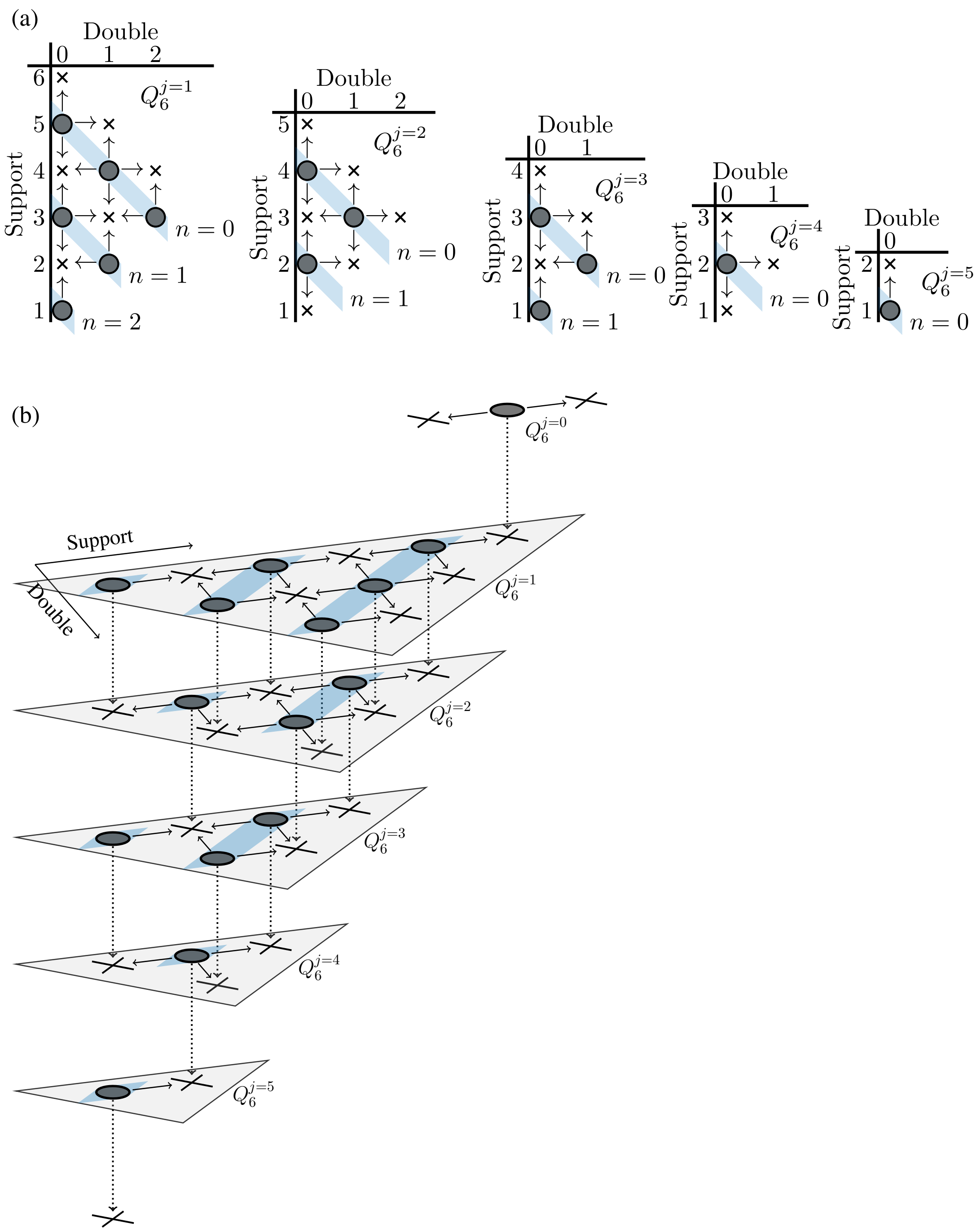

Figure 5.1: The structure of $Q_6^j$. (a) Structure for each $j$. (b) The whole structure in $Q_6$. The solid arrows represent the operation of the commutator $H_0$, while the vertical dotted arrows represent the operation of the commutator with $H_{\text{int}}$.

**Cancellation of diagram**

I now explain the general structure of $Q_k^j$ in Figure 5.2. The diagrams in $Q_k^j$ are classified as $(s,d)$-connected diagrams as shown in Figure 5.2, where the circles represent $(s,d)$-connected diagrams in $Q_k^j$, i.e., $Q_k^j(s,d)$, and the crosses represent diagrams generated by the commutator with the Hamiltonian. The solid arrow represents the commutator of diagrams in the circles at its origin and $H_0$, which generates diagrams in the crosses at the tip of the arrows. The generated diagrams at the crosses are to be canceled together with the contribution from the commutator $Q_k^{j-1}$ and $H_{\text{int}}$ for the conservation law.

I show the basic structure of cancellation of $(s+1,d)$-diagram in $[Q_k, H] = 0$ in Figure 5.3, which represents the following equation:

$$
\begin{aligned}
0 &= [Q_k, H]\big|_{(s+1,d)} \\
&= \left[Q_k^j(s,d), H_0\right]\big|_{(s+1,d)} + \left[Q_k^j(s+1,d-1), H_0\right]\big|_{(s+1,d)} \\
&\qquad + \left[Q_k^j(s+1,d+1), H_0\right]\big|_{(s+1,d)} + \left[Q_k^j(s+2,d), H_0\right]\big|_{(s+1,d)} \\
&\qquad\quad + \left[Q_k^{j-1}(s+1,d), H_{\text{int}}\right] ,
\end{aligned} \tag{5.78}
$$

where I denote the $(s,d)$-diagram in $A$ by $A|_{(s,d)}$. The first four terms in the last line correspond to the four circles in the plane of $Q_k^j$ in Figure 5.3. The last term in the last line corresponds to the circles in the plane of $Q_k^{j-1}$ in Figure 5.3.

I note that the commutator with $H_0$ increases or decreases the support or the double number by one, while the commutator with $H_{\text{int}}$ does not change the support and double number.

## 5.2.4 Identities of the coefficients

The coefficient $C_{n,d}^{j,m}(\boldsymbol{\lambda})$ obtained from the recursion equation (5.57) satisfies the following lemma, which is useful in the proof of Theorem 5.7:

**Lemma 5.8.** *The coefficient $C_{n,d}^{j,m}(\boldsymbol{\lambda})$ defined by Definition 5.6 satisfies the following identities:*

$$
\begin{aligned}
C_{n,d}^{j,m}(\boldsymbol{\lambda}) &= C_{n,d}^{j,m}(\boldsymbol{\lambda}_{L\leftrightarrow R}) , && (5.79)\\
&= C_{n,d}^{j,m}(\boldsymbol{\lambda}_{i_1\leftrightarrow i_2}) \quad (w \geq 2), && (5.80)\\
&= C_{n,d}^{j,m}\left(\boldsymbol{\lambda}_{a\to\min(\lambda_a,n)}\right) , && (5.81)\\
&= C_{n-1,d+2}^{j,m}\left(\boldsymbol{\lambda}_{L(R):-1,i:-1}\right) + C_{n,d+1}^{j-1,m}(\boldsymbol{\lambda}_{L(R)\to\lambda_{L(R)}+\lambda_i,\hat{i}}) \quad (w \geq 1), && (5.82)\\
&= C_{n-1,d+2}^{j,m}(\boldsymbol{\lambda}_{i_1:-1,i_2:-1}) + C_{n,d+1}^{j-1,m}(\boldsymbol{\lambda}_{i_1\to\lambda_{i_1}+\lambda_{i_2},\hat{i_2}}) \quad (w \geq 2), && (5.83)\\
&= C_{n,d}^{j,m}(\boldsymbol{\lambda}_{a:-1,b:+1}) + C_{n-1,d+2}^{j,m}(\boldsymbol{\lambda}_{a:-1,b:-1}) - C_{n-1,d+2}^{j,m}(\boldsymbol{\lambda}_{a:-2}) \quad (\lambda_a \geq 1), && (5.84)
\end{aligned}
$$

*where $w \equiv j-1-2m$, $\boldsymbol{\lambda} = \{\lambda_L \equiv \lambda_0; \lambda_1, \ldots, \lambda_w; \lambda_{w+1} \equiv \lambda_R\}$, $0 \leq a, b \leq w+1$ and $1 \leq i, i_1, i_2 \leq w$.*

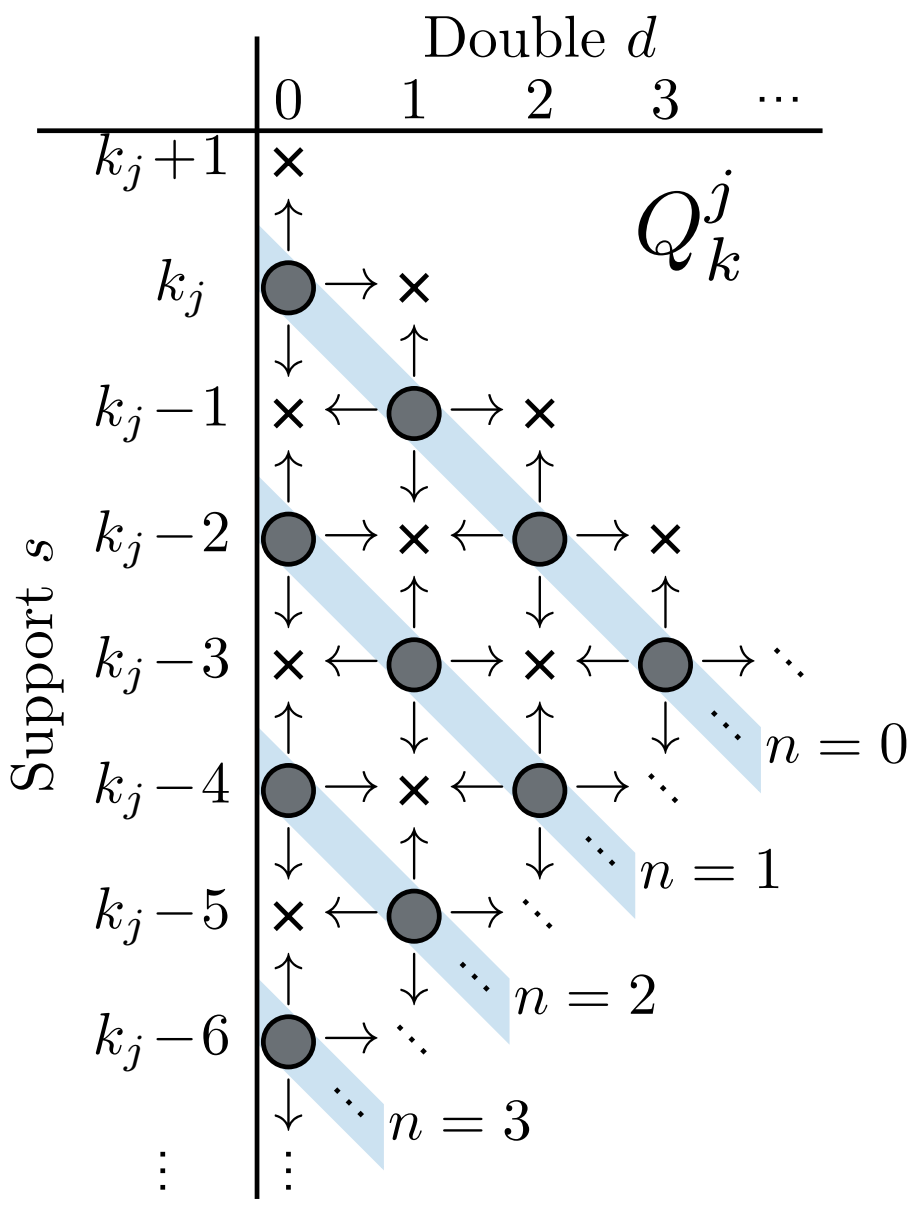

Figure 5.2: Structure of $Q_k^j$, where $k_j \equiv k-j$. Circles at $(s,d)$ represent $(s,d)$-connected diagrams in $Q_k^j$, i.e., $Q_k^j(s,d)$. I note that the circles are in the area of $s>d$ and $s+d\leq k_j$. The commutator of diagrams in the circle at $(s,d)$ with $H_0$ generates the diagrams in the crosses at $(s\pm 1,d)$ and $(s,d\pm 1)$, indicated by the solid arrow tip.

The proof of Lemma 5.8 is given in Appendix A.

In the case of $\lambda_i=0$ in Eq. (5.82), I have

$$C_{n,d}^{j,m}\left(\boldsymbol{\lambda}_{i\to 0}\right)=C_{n,d+1}^{j-1,m}\left(\boldsymbol{\lambda}_{\hat{i}}\right). \tag{5.85}$$

This relationship means the equivalence of the coefficients of connected diagrams in $Q_k$; for example

coefficient of 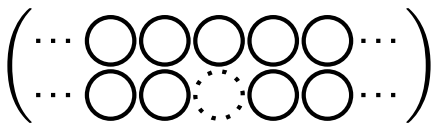 = coefficient of 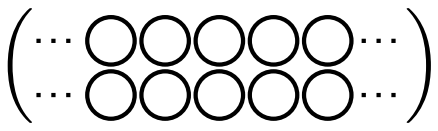 $\times U\,,$ (5.86)

coefficient of 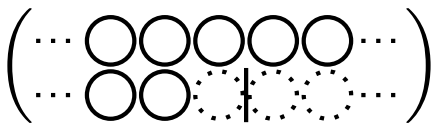 = coefficient of 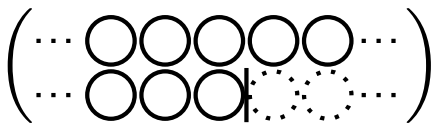 $\times U\,,$ (5.87)

coefficient of 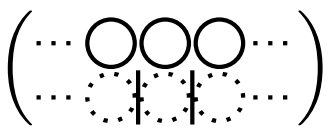 = coefficient of 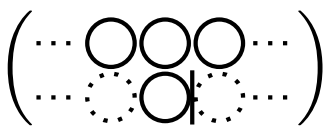 $\times U\,.$ (5.88)

Using Eq. (5.84) repeatedly, I have for $p\geq 1$

$$C_{n,d}^{j,m}\left(\boldsymbol{\lambda}\right)-C_{n,d}^{j,m}\left(\boldsymbol{\lambda}_{a:-p,b:p}\right)=C_{n-1,d+2}^{j,m}\left(\boldsymbol{\lambda}_{a:-1,b:-1}\right)-C_{n-1,d+2}^{j,m}\left(\boldsymbol{\lambda}_{a:-p-1,b:p-1}\right), \tag{5.89}$$

where $\lambda_a\geq p$.

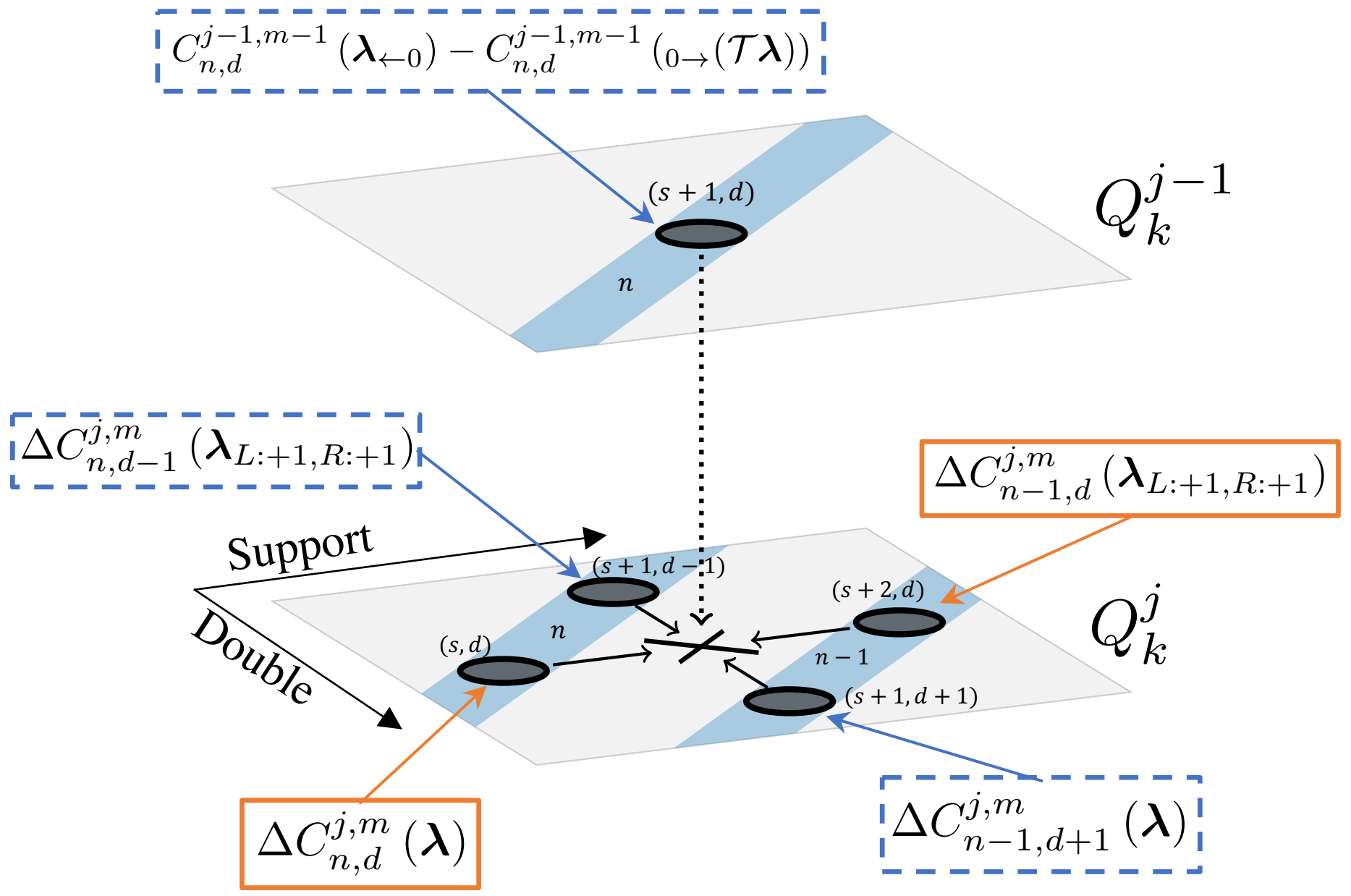

Figure 5.3: Basic structure for the cancellation of diagrams in the commutator with the Hamiltonian. The diagrams generated at the cross of $(s+1,d)$-diagram are to be canceled. The arrows on the plane of $Q^j_k$ represent the commutator with $H_0$: $\left[Q^j_k(s,d),H_0\right]\big|_{(s+1,d)}$ etc. The vertical dotted arrow represents the commutator $\left[Q^{j-1}_k(s+1,d),H_{\text{int}}\right]$. The coefficients, which contribute to the equation for the local cancellation (5.90), are encircled with the orange solid (blue dotted) line and contribute to Eq. (5.90) with the factor $+1(-1)$.

The recursion equation (5.57) reflects the local cancellation in $[Q_k,H]=0$ (see Figure 5.3). To see this, I further rewrite Eq. (5.57) as

$$\begin{aligned}0 =&\Delta C^{j,m}_{n,d}(\boldsymbol{\lambda}) - \Delta C^{j,m}_{n-1,d+1}(\boldsymbol{\lambda}) - \Delta C^{j,m}_{n,d-1}(\boldsymbol{\lambda}_{L:+1,R:+1}) + \Delta C^{j,m}_{n-1,d}(\boldsymbol{\lambda}_{L:+1,R:+1}) \\ &- \left(C^{j-1,m-1}_{n,d}(\boldsymbol{\lambda}_{\leftarrow 0}) - C^{j-1,m-1}_{n,d}(_{0\rightarrow}(\mathcal{T}\boldsymbol{\lambda}))\right),\end{aligned} \tag{5.90}$$

where I used the relations obtained from Eq. (5.84): $\Delta C^{j,m}_{n-1,d+1}(\boldsymbol{\lambda}) = \Delta C^{j,m}_{n,d-1}(\boldsymbol{\lambda}_{L:+1,R:+1})$ and $\Delta C^{j,m}_{n-2,d+2}(\boldsymbol{\lambda}) = \Delta C^{j,m}_{n-1,d}(\boldsymbol{\lambda}_{L:+1,R:+1})$. The first, second, third, and fourth terms in the right-hand side of Eq. (5.90) are the contribution from the $(s,d)$, $(s+1,d+1)$, $(s+1,d-1)$, $(s+2,d)$-diagrams in $Q^j_k$, respectively, with $s=k-j-2n-d$, which are represented by the circles on the $Q^j_k$ plane in Figure 5.3. The commutator of these diagrams in $Q^j_k$ with $H_0$ generates the diagrams in the cross at $(s+1,d)$ on the $Q^j_k$ plane in Figure 5.3. The last term in the right-hand side of Eq. (5.90) is the contribution from the $(s+1,d)$ diagrams in $Q^{j-1}_k$, which are represented by the circle on the $Q^{j-1}_k$ plane in Figure 5.3. The commutator of these diagrams in $Q^{j-1}_k$ with $H_{\text{int}}$ generates the diagrams in the cross at $(s+1,d)$ on the $Q^j_k$ plane in Figure 5.3. The diagrams generated in the cross at $(s+1,d)$ cancel for the conservation law.

I note that for the case $d=0$ in Eq. (5.90) and Figure 5.3, the contribution from $(s+1,1)$-connected diagrams is doubled, as one can see $\Delta C^{j,m}_{n-1,d+1}(\boldsymbol{\lambda}) = \Delta C^{j,m}_{n,d-1}(\boldsymbol{\lambda}_{L:+1,R:+1})$ and the factor 2 in the original recursion equation (5.57). This factor 2 comes from the commutator of Eq. (5.112) as will be explained in Section 5.3.

One can obtain another recursion equation:

$$C_{n,d}^{j,m}(\boldsymbol{\lambda}) = C_{n,d}^{j,m}(\lambda_L - 1; \vec{\lambda}; \lambda_R + 1) + \sum_{\tilde{n}=0}^{n} (n+1-\tilde{n}) \left( C_{\tilde{n},n+d-\tilde{n}}^{j-1,m-1}(\lambda_L; \vec{\lambda}, \lambda_R; 0) - C_{\tilde{n},n+d-\tilde{n}}^{j-1,m-1}(0; \lambda_L - 1, \vec{\lambda}; \lambda_R + 1) \right). \tag{5.91}$$

One can easily derive this as follows. I introduce the notation $\Delta^2 C_{n,d}^{j,m}(\boldsymbol{\lambda}) \equiv \Delta C_{n,d}^{j,m}(\boldsymbol{\lambda}) - \Delta C_{n-1,d+1}^{j,m}(\boldsymbol{\lambda})$. Then using the recursion equation (5.57) repeatedly, Eq. (5.57) can be rewritten as

$$\begin{aligned} \Delta^2 C_{n,d}^{j,m}(\boldsymbol{\lambda}) &= \Delta^2 C_{n-1,d+1}^{j,m}(\boldsymbol{\lambda}) + C_{n,d}^{j-1,m-1}(\boldsymbol{\lambda}_{\leftarrow 0}) - C_{n,d}^{j-1,m-1}({}_{0\rightarrow}(\mathcal{T}\boldsymbol{\lambda})) \\ &= \Delta^2 C_{n-2,d+2}^{j,m}(\boldsymbol{\lambda}) + \sum_{p=0}^{1} \left( C_{n-p,d+p}^{j-1,m-1}(\boldsymbol{\lambda}_{\leftarrow 0}) - C_{n-p,d+p}^{j-1,m-1}({}_{0\rightarrow}(\mathcal{T}\boldsymbol{\lambda})) \right) \\ &\vdots \\ &= \sum_{p=0}^{n} \left( C_{n-p,d+p}^{j-1,m-1}(\boldsymbol{\lambda}_{\leftarrow 0}) - C_{n-p,d+p}^{j-1,m-1}({}_{0\rightarrow}(\mathcal{T}\boldsymbol{\lambda})) \right). \end{aligned} \tag{5.92}$$

Next, I use (5.92) repeatedly, and I obtain

$$\begin{aligned} \Delta C_{n,d}^{j,m}(\boldsymbol{\lambda}) &= \Delta C_{n-1,d+1}^{j,m}(\boldsymbol{\lambda}) + \sum_{p=0}^{n} \left( C_{n-p,d+p}^{j-1,m-1}(\boldsymbol{\lambda}_{\leftarrow 0}) - C_{n-p,d+p}^{j-1,m-1}({}_{0\rightarrow}(\mathcal{T}\boldsymbol{\lambda})) \right) \\ &= \Delta C_{n-2,d+2}^{j,m}(\boldsymbol{\lambda}) + \sum_{q=0}^{1}\sum_{p=0}^{n-q} \left( C_{n-(p+q),d+p+q}^{j-1,m-1}(\boldsymbol{\lambda}_{\leftarrow 0}) - C_{n-(p+q),d+p+q}^{j-1,m-1}({}_{0\rightarrow}(\mathcal{T}\boldsymbol{\lambda})) \right) \\ &\vdots \\ &= \sum_{q=0}^{n}\sum_{p=0}^{n-q} \left( C_{n-(p+q),d+p+q}^{j-1,m-1}(\boldsymbol{\lambda}_{\leftarrow 0}) - C_{n-(p+q),d+p+q}^{j-1,m-1}({}_{0\rightarrow}(\mathcal{T}\boldsymbol{\lambda})) \right) \\ &= \sum_{\tilde{n}=0}^{n} (\tilde{n}+1) \left( C_{n-\tilde{n},d+\tilde{n}}^{j-1,m-1}(\boldsymbol{\lambda}_{\leftarrow 0}) - C_{n-\tilde{n},d+\tilde{n}}^{j-1,m-1}({}_{0\rightarrow}(\mathcal{T}\boldsymbol{\lambda})) \right) \\ &= \sum_{\tilde{n}=0}^{n} (n+1-\tilde{n}) \left( C_{\tilde{n},n+d-\tilde{n}}^{j-1,m-1}(\boldsymbol{\lambda}_{\leftarrow 0}) - C_{\tilde{n},n+d-\tilde{n}}^{j-1,m-1}({}_{0\rightarrow}(\mathcal{T}\boldsymbol{\lambda})) \right). \end{aligned} \tag{5.93}$$

This concludes the proof of Eq. (5.91).

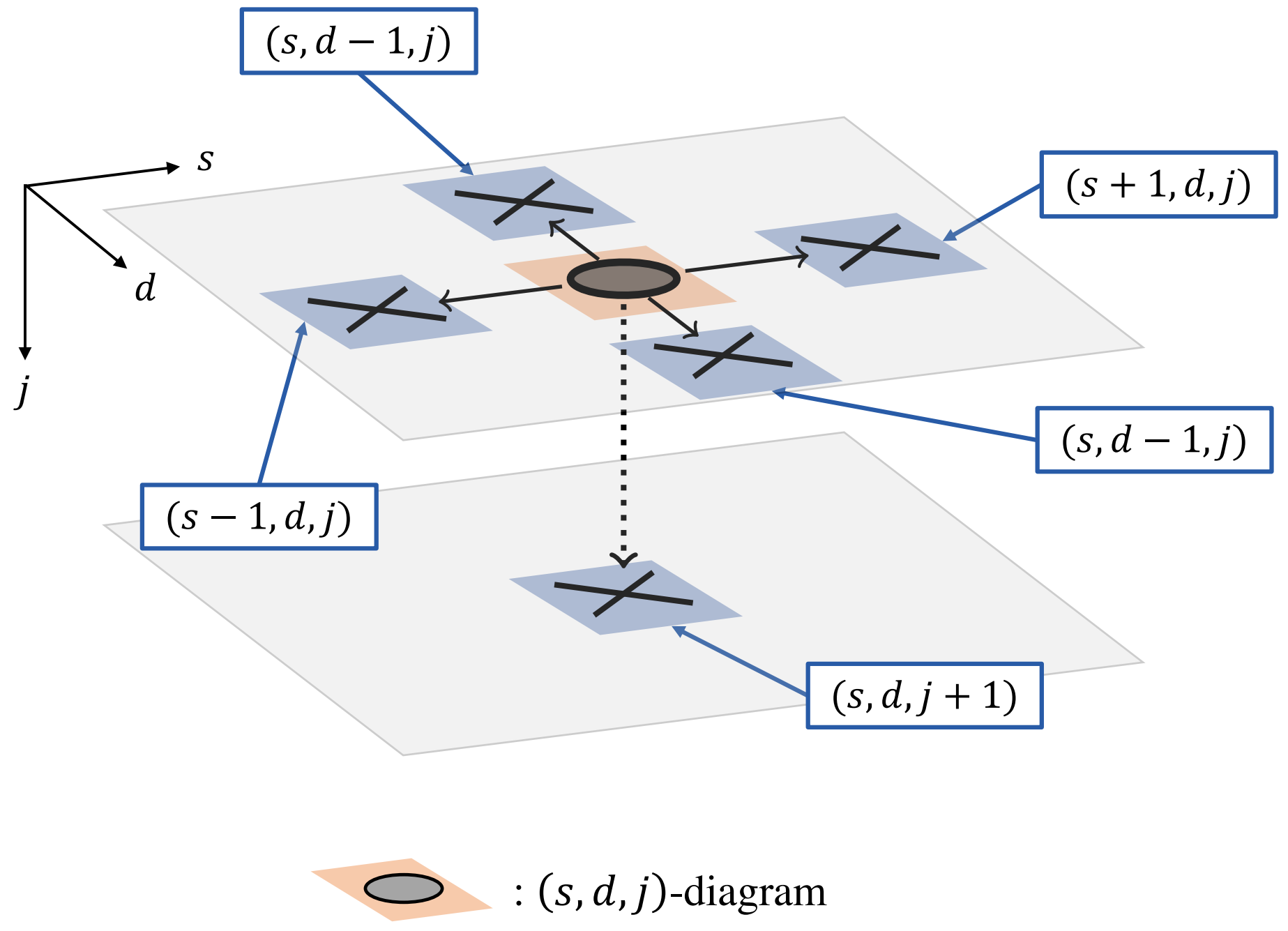

Figure 5.4: Schematic picture of the commutator of $(s, d, j)$-diagram and the Hamiltonian. The commutator generates diagrams at the crosses. The arrows on the upper plane indicate the commutator with $H_0$. The vertical dotted arrow indicates the commutator with $UH_{\text{int}}$.

### 5.2.5 Origin of the result

Before going into the proof of Theorem 5.7, I explain how I arrived at Theorem 5.7. While the reader might think this history is illuminating, it is not essential for understanding the rest of this chapter.

#### Characteristic quantity

First and foremost, a crucial step is identifying *characteristic quantity* to classify the operators constituting the local charges. The characteristic quantity should be chosen so that one can efficiently compute the commutator of local charges and the Hamiltonian and represent conserved quantities as simply as possible.

The first idea that comes to mind as a characteristic quantity would be *locality*; it is literally the quantity that characterizes local charges. Actually, the *locality* is represented by support, introduced in Subsection 5.1.2.

The second characteristic quantity is *double* introduced in Subsection 5.1.2. The characteristic quantity, double, is more evident in my diagram notation; it is hard to come up with it with the usual spin variable notation or the standard fermion notation in the previous studies [28]. One can classify the operators using these characteristic quantities, $(\text{support}, \text{double})$; this is just the classification of $(s, d)$-diagram.

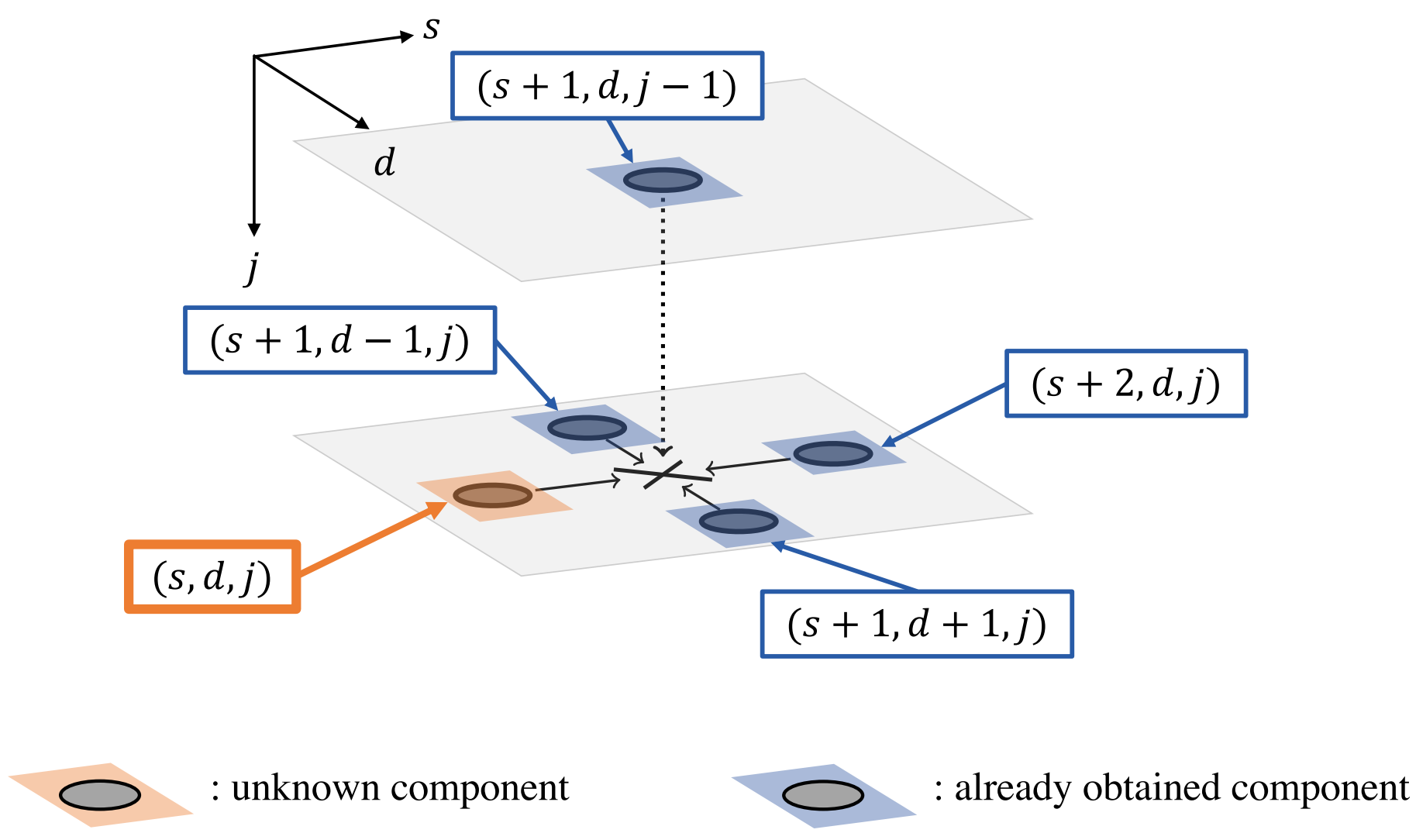

Figure 5.5: Schematic picture of the recursive way to calculate a component of local charges. The unknown diagrams at $(s, d, j)$ shaded with orange are determined from the information of the other diagrams shaded with blue, which have already been obtained.

The third characteristic quantity is the power of the coupling constant $U$, which corresponds to the $z$-coordinate of Figure 5.1 (b). I denote an $(s, d)$-diagram with the coefficient $U^j$ by $(s, d, j)$-diagram; these three coordinate correspond to the three axes of Figure 5.1 (b).

With these characteristic quantities, the commutator of $(s, d, j)$-diagram with the Hamiltonian is easily managed as explained in Figure 5.4. The commutator of $(s, d, j)$-diagram and $H_0$ generates $(s \pm 1, d, j)$-diagram and $(s, d \pm 1, j)$-diagram, which is depicted by the arrows in the upper plane of Figure 5.4. The commutator of $(s, d, j)$-diagram and $UH_{\text{int}}$ generates $(s, d, j+1)$-diagram, which is depicted by the vertical dotted arrow in the upper plane of Figure 5.4. Thus, in any case, one of the indices is changed by one in the commutator with the Hamiltonian.

### Recursive construction of the local charges

The three characteristic quantities follow the recursive construction of the local charges, which is explained in Figure 5.5. The $(s, d, j)$-diagram in $Q_k$ can be obtained by calculating the local cancellation of $(s + 1, d, j)$-diagrams in the commutator $[Q_k, H] = 0$ if I already determined the $(s, d \pm 1, j)$, $(s + 2, d, j)$, and $(s + 1, d, j - 1)$-diagrams in $Q_k$. This local cancellation is the same as (5.78) explained above.

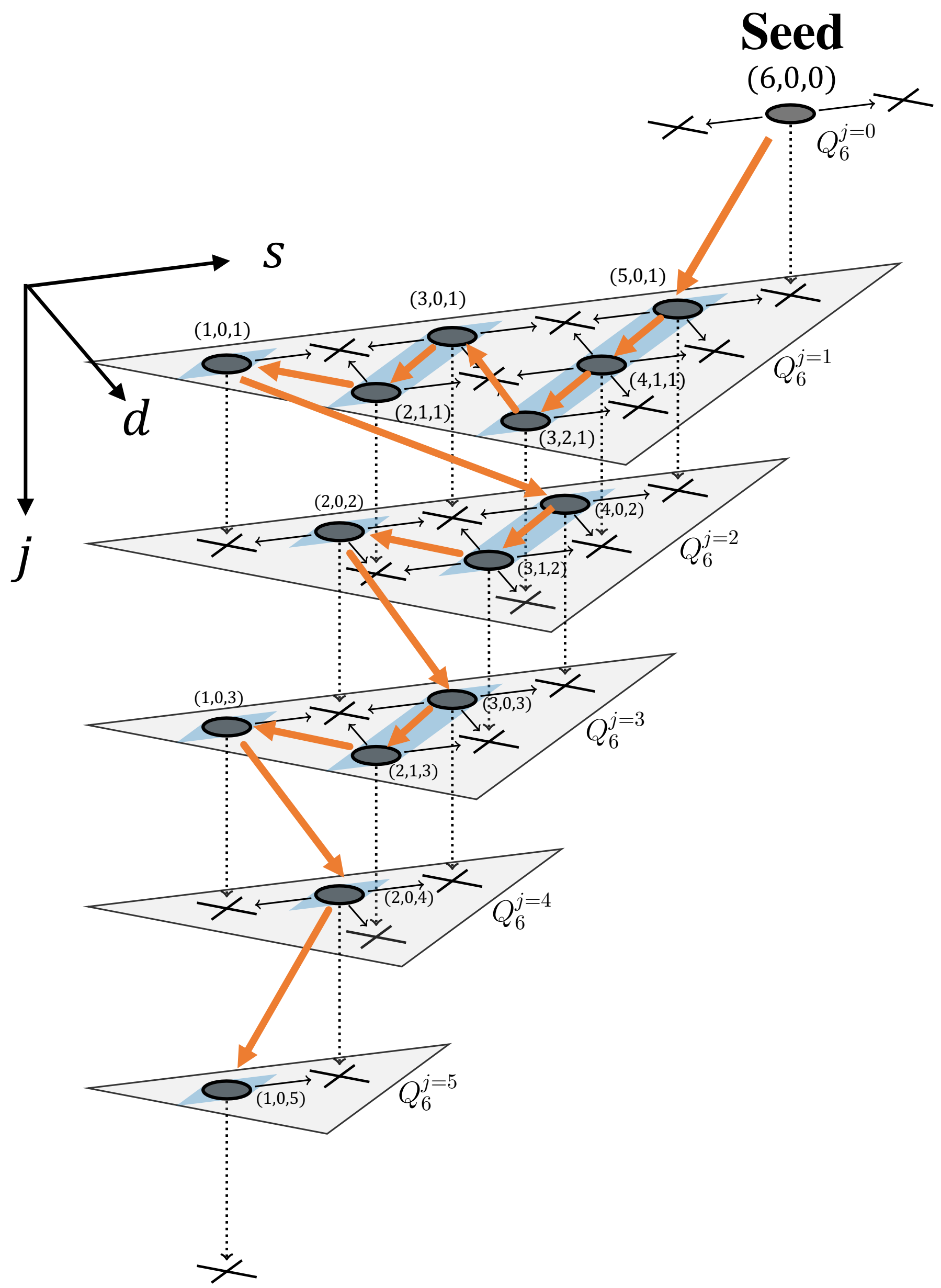

Figure 5.6: Schematic picture of the order in which all the components in $Q_6$ are determined by the recursive way. The components in $Q_6$ are calculated in the order that the orange arrow indicates from the seed at $(6, 0, 0)$.

The recursive way starts from the "seed" component, which is $Q_k^0$ (5.56). Figure 5.6 shows the order in which all the coefficients are determined by the recursive way for $Q_6$ as an example.

I fix the freedom to add lower-order charges so that there are no components of $Q_{k'}^0$ $(k' < k)$ in $Q_k$. With this normalization, $Q_k$ is uniquely determined in this recursive way.

I would like to emphasize that there has never been a recursive way to construct the local charges due to the absence of the boost operator $[B, Q_{k-1}] = Q_k$ [43]. My recursive way does not require the information of the lower order charge $Q_{k-1}$ to obtain $Q_k$, unlike the boost recursion relation.

**Finding pattern in the local charges**

I first computed the first twelve local charges in terms of a diagram, calculating the local cancellation recursively. All the computations for the local charges are conducted in a self-made symbolic algebra program implemented in C++ language.

I first make the ansatz, where the local charges are constructed of a general diagram, including connected and non-connected diagrams. From the data up to $Q_{12}$, I found the following regularities:

- Local charges are constructed of connected diagrams;
- The coefficient of a diagram in local charges depends on the characteristic quantities $(s, d, j)$ and its list.

Using these findings, one can reduce the information that I have to treat in the computation, and I could compute higher-order charges up to $\sim Q_{35}$. Then, I found the identities for the coefficients (5.79)–(5.84).

## 5.3 Commutation relations of diagram with Hamiltonian

This section explains how to calculate the commutation relation of a diagram and the Hamiltonian. I will use this to prove the conservation law of $Q_k$ in the next section. In order to simplify calculations, I here use a definition different from the standard one, and the commutators in the following have the additional factor of $1/2$: $[A, B] \equiv \frac{1}{2}(AB - BA)$.

### 5.3.1 Algebra of unit operators

To explain the operation of the commutators with the Hamiltonian, I first show the algebra of unit operators. I change the notation of the unit operator in the following form:

$$\psi_\sigma^\pm(i,j) := (2-\delta_{i,j})(\pm c_{i,\sigma} c_{j,\sigma}^\dagger + (-1)^{i-j} c_{i,\sigma}^\dagger c_{j,\sigma}), \quad (s \in \{+,-\}, \sigma \in \{\uparrow,\downarrow\}) \tag{5.94}$$

which has the following symmetry:

$$\psi_\sigma^s(i,j) = s(-1)^{i-j+1}\psi_\sigma^s(j,i)\,. \tag{5.95}$$

The original notation of a unit operator (5.6) is now written as

$$\psi_\sigma^{s,n}(j) = \psi_\sigma^s(j, j+n). \tag{5.96}$$

I also note that $\psi_\sigma^+(j,j)=0$.

Using this new notation, the unit operators satisfy the following algebra:

$$\psi_\sigma^s(i,j)\psi_\sigma^{s'}(j,k)=\psi_\sigma^{ss'}(i,k)-\psi_\sigma^{-ss'}(i,k)\psi_\sigma^-(j,j) \qquad (i\neq j\neq k\neq i), \quad (5.97)$$

$$\psi_\sigma^s(i,j)\psi_\sigma^-(j,j)=\psi_\sigma^{-s}(i,j) \qquad (i\neq j), \quad (5.98)$$

$$\psi_\sigma^-(j,j)\psi_\sigma^s(i,j)=-\psi_\sigma^{-s}(i,j) \qquad (i\neq j), \quad (5.99)$$

$$\psi_\sigma^s(i,j)\psi_\sigma^{-s}(j,i)=2[\psi_\sigma^-(i,i)-\psi_\sigma^-(j,j)] \qquad (i\neq j), \quad (5.100)$$

$$\psi_\sigma^s(i,j)\psi_\sigma^s(j,i)=2[1-\psi_\sigma^-(i,i)\psi_\sigma^-(j,j)] \qquad (i\neq j), \quad (5.101)$$

$$\psi_\sigma^-(i,i)^2=1, \quad (5.102)$$

$$[\psi_\sigma^s(i,j),\psi_{\sigma'}^s(k,l)]=0 \qquad (\sigma\neq\sigma' \text{ or } \{i,j\}\cap\{k,l\}=\phi). \quad (5.103)$$

I also have from Eqs. (5.97) and (5.95)

$$\psi_\sigma^{s'}(j,k)\psi_\sigma^s(i,j)=-\psi_\sigma^{ss'}(i,k)-\psi_\sigma^{-ss'}(i,k)\psi_\sigma^-(j,j) \qquad (i\neq j\neq k\neq i)\,. \quad (5.104)$$

Combining these, I obtain the following commutation relations for $i\neq j$:

$$\left[\psi_\sigma^s(i,j),\psi_\sigma^{s'}(j,k)\right]=\psi_\sigma^{ss'}(i,k)\,, \quad (5.105)$$

$$\left[\psi_\sigma^s(i,j),\psi_\sigma^-(j,j)\right]=\psi_\sigma^{-s}(i,j)\,, \quad (5.106)$$

$$\left[\psi_\sigma^s(i,j),\psi_\sigma^-(i,i)\right]=-\psi_\sigma^{-s}(i,j)\,, \quad (5.107)$$

$$\left[\psi_\sigma^s(i,j),\psi_\sigma^{-s}(j,i)\right]=2(\psi_\sigma^-(i,i)-\psi_\sigma^-(j,j))\,. \quad (5.108)$$

These commutators will be used to prove the conservation of local charges $[Q_k,H]=0$, and may also be used to prove their mutual commutativity $[Q_k,Q_l]=0$. Alternatively, one can prove the mutual commutativity indirectly using the completeness of $Q_k$, which I will do in Chapter 6. This completeness guarantees the expansion of my $Q_k$ into a linear combination of the local charges generated from the transfer matrix [37]. These transfer matrix charges are known to commute with each other [60]. Thus, I conclude that my $Q_k$ also mutually commute.

### 5.3.2 Commutators of diagram with $H_0$

First, let us consider the commutator with the non-interacting term in the Hamiltonian. The non-interacting term is written as

$$H_0=\sum_{\sigma=\uparrow,\downarrow}\sum_{j=1}^{L}\bigcirc_\sigma(j)\,. \quad (5.109)$$

This shows that it suffices to calculate the commutator with the density $\bigcirc_\sigma(j)$. From Eqs. (5.105)–(5.108), the following commutators are non-vanishing:

$$\Big[\overbrace{\bigcirc\bigcirc\cdots\bigcirc\bigcirc}^{n}{}_\sigma(j),\bigcirc_\sigma(j')\Big]=\begin{cases}\overbrace{\bigcirc\bigcirc\cdots\bigcirc\bigcirc\bigcirc}^{n+1}{}_\sigma(j) & (j'=j+n,n\geq 1),\\ \overbrace{\bigcirc\bigcirc\bigcirc\cdots\bigcirc\bigcirc}^{n+1}{}_\sigma(j-1)\times(-1) & (j'=j-1,n\geq 1),\\ \overbrace{\bigcirc\bigcirc\cdots\bigcirc}^{n-1}{}_\sigma(j) & (j'=j+n-1,n\geq 2),\\ \overbrace{\bigcirc\cdots\bigcirc\bigcirc}^{n-1}{}_\sigma(j+1)\times(-1) & (j'=j,n\geq 2),\end{cases} \tag{5.110}$$

$$\Big[\overbrace{\bigcirc\bigcirc\cdots\bigcirc\bigcirc\!|}^{n}{}_\sigma(j),\bigcirc_\sigma(j')\Big]=\begin{cases}\overbrace{\bigcirc\bigcirc\cdots\bigcirc\bigcirc\bigcirc\!|}^{n+1}{}_\sigma(j) & (j'=j+n,n\geq 1),\\ \overbrace{\bigcirc\bigcirc\bigcirc\cdots\bigcirc\bigcirc\!|}^{n+1}{}_\sigma(j-1)\times(-1) & (j'=j-1,n\geq 1),\\ \overbrace{\bigcirc\bigcirc\cdots\bigcirc\!|}^{n-1}{}_\sigma(j) & (j'=j+n-1,n\geq 2),\\ \overbrace{\bigcirc\cdots\bigcirc\bigcirc\!|}^{n-1}{}_\sigma(j+1)\times(-1) & (j'=n,n\geq 2),\end{cases} \tag{5.111}$$

$$\Big[\bigcirc\!|_\sigma(j),\bigcirc_\sigma(j)\Big]=2\,|_\sigma(j)-2\,|_\sigma(j-1), \tag{5.112}$$

$$\Big[\,|_\sigma(j),\bigcirc_\sigma(j')\Big]=\begin{cases}\bigcirc\!|_\sigma(j) & (j'=j),\\ \bigcirc\!|_\sigma(j-1)\times(-1) & (j'=j-1),\end{cases} \tag{5.113}$$

and the others are all zero.

I introduce a notation to represent the commutator of a diagram with the Hamiltonian. I now represent a diagram by the two-row representation: 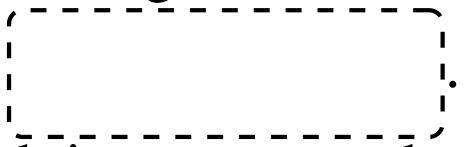. I also define the commutator for the quantity summed over all the sites, which is represented as follows:

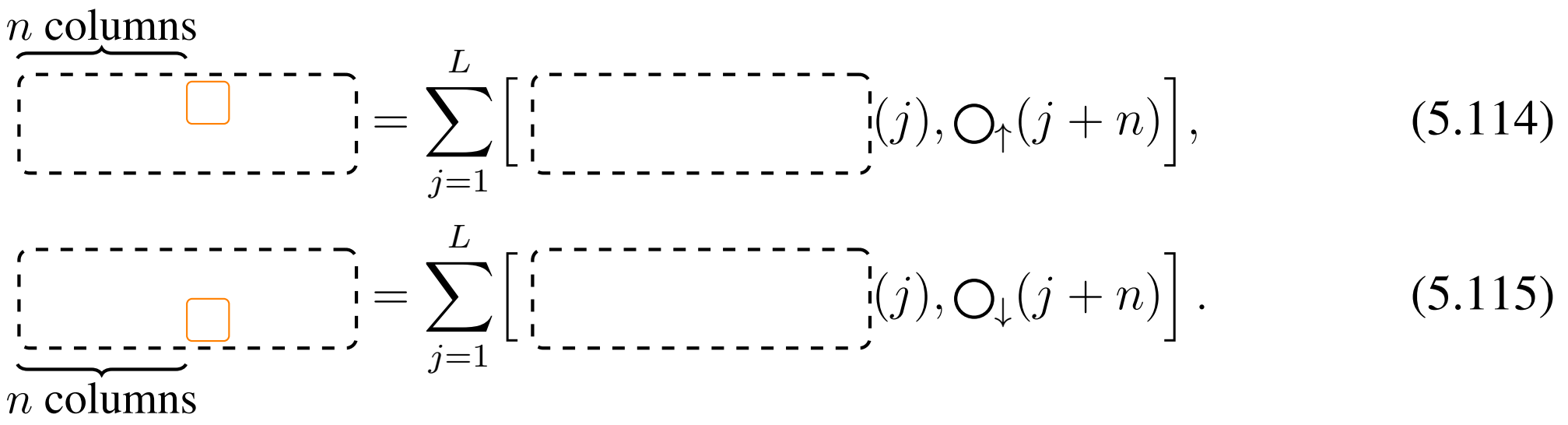

Here, 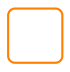 indicates that one should take the commutator of the operator with $\bigcirc_\sigma$ at this position.

In the case in which the commutator increases the support, I use the following notations:

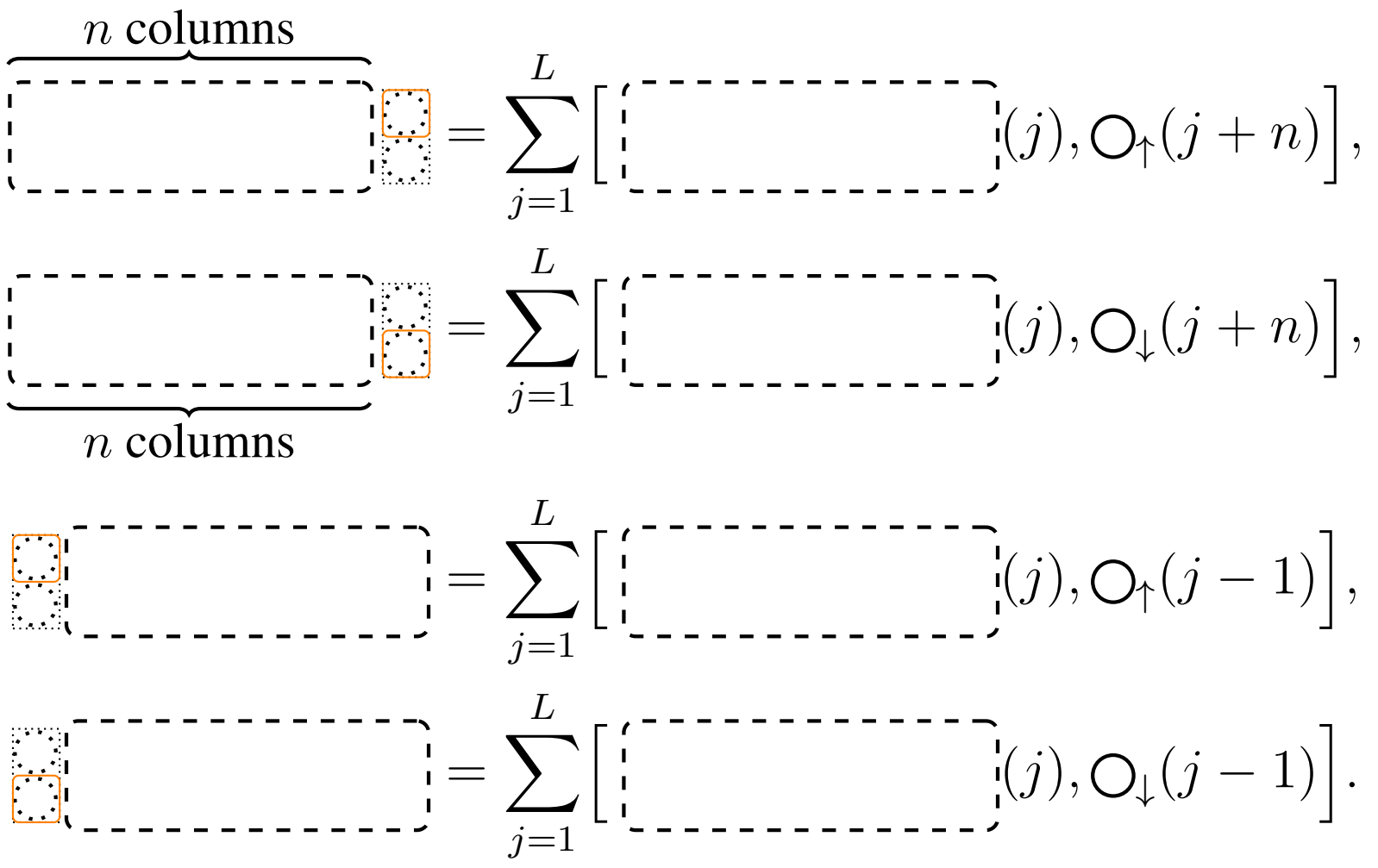

(5.116)

(5.117)

(5.118)

(5.119)

Examples of the explicit non-zero commutators with the density $\bigcirc_\sigma(j)$ in $H_0$ are

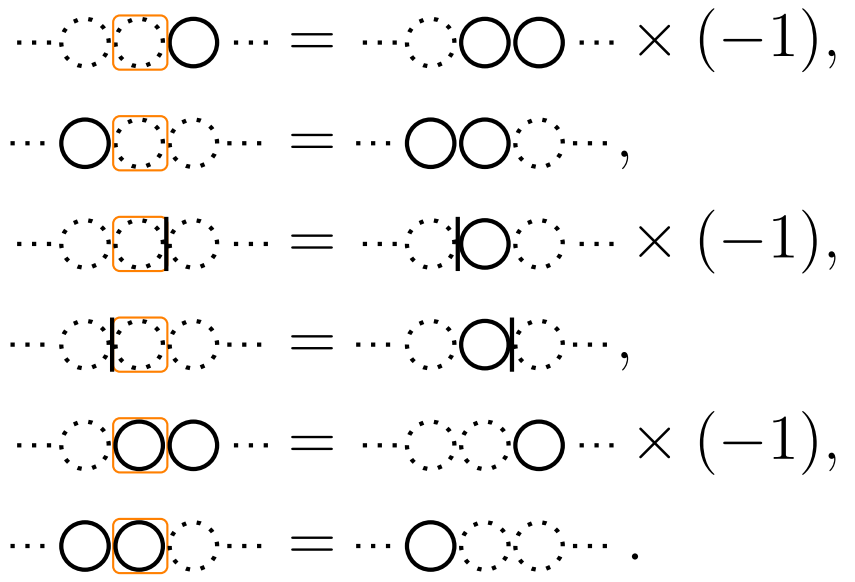

(5.120)

(5.121)

(5.122)

(5.123)

(5.124)

(5.125)

Here, I show only the relevant parts in the diagram rather than the two-row representation used above. The types of unit operators are unchanged upon taking the commutator.

The following commutator generates two terms:

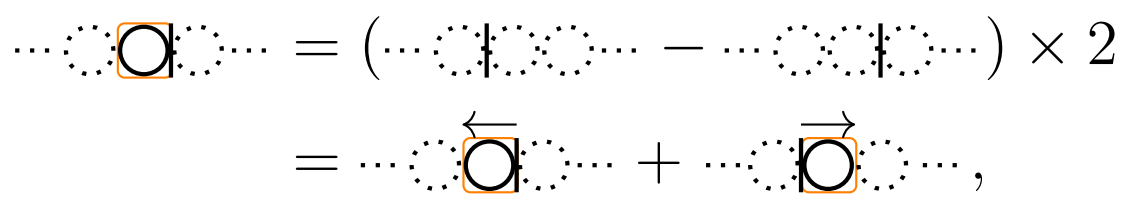

(5.126)

where I have introduced the following notations in the second line:

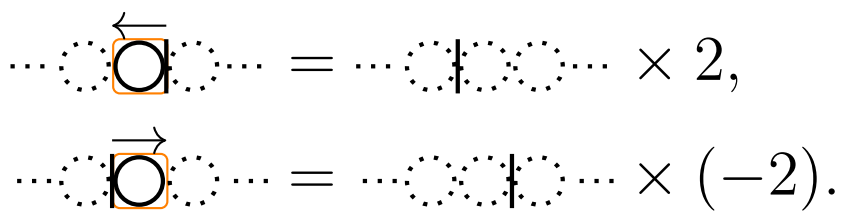

(5.127)

(5.128)

I introduce a unit operator with *hole*:

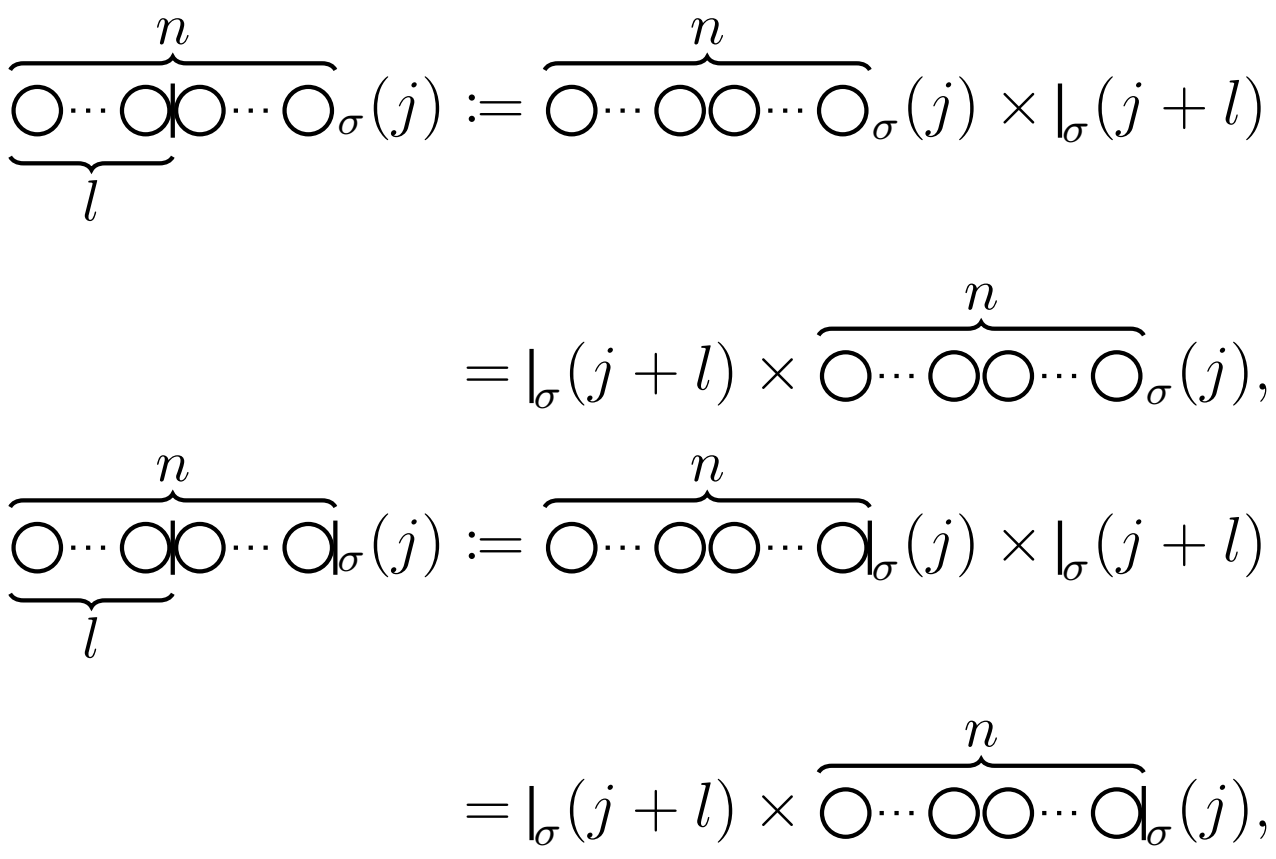

(5.129)

(5.130)

where $1 \leq l \leq n-1$ and $2 \leq n$. I call the part of $\cdots\bigcirc\!|\!\bigcirc\cdots$ as a hole. The types of unit operators with a hole in the first (second) lines are $+$ $(-)$. In the spin variable representation, a hole in a diagram indicates the identity operator at the corresponding site.

The following commutator generates the unit operator with a hole:

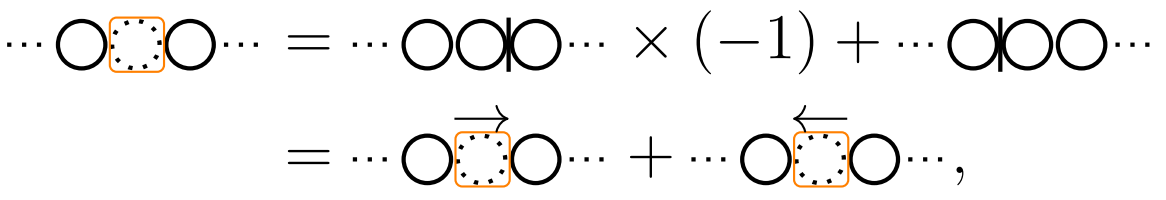

(5.131)

where in the second line, I introduced the following notations:

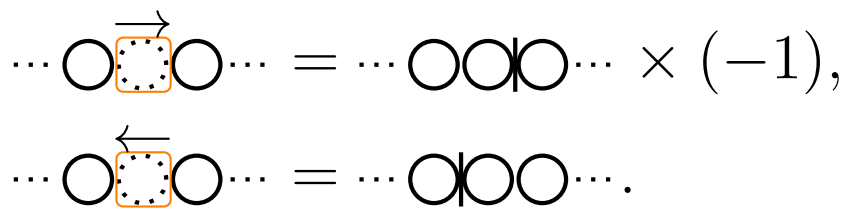

(5.132)

(5.133)

The type of the generated unit operator with a hole is $-t_L t_R$, where $t_L$ is the type of the unit operator on the left and $t_R$ is the type of the unit operator on the right before merging. Other examples of a hole-making commutator are

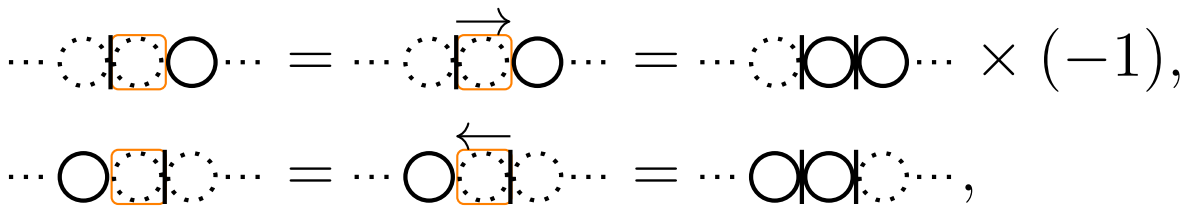

(5.134)

(5.135)

where the type of the generated unit operator is determined with the same argument as above using $t_L = -1$ for Eq. (5.134) and $t_R = -1$ for Eq. (5.135). The following relations will be used afterward:

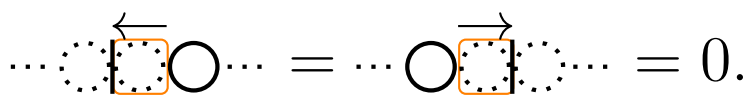

(5.136)

Using the above properties, one can calculate the commutator with $H_0$ for any diagram; for example,

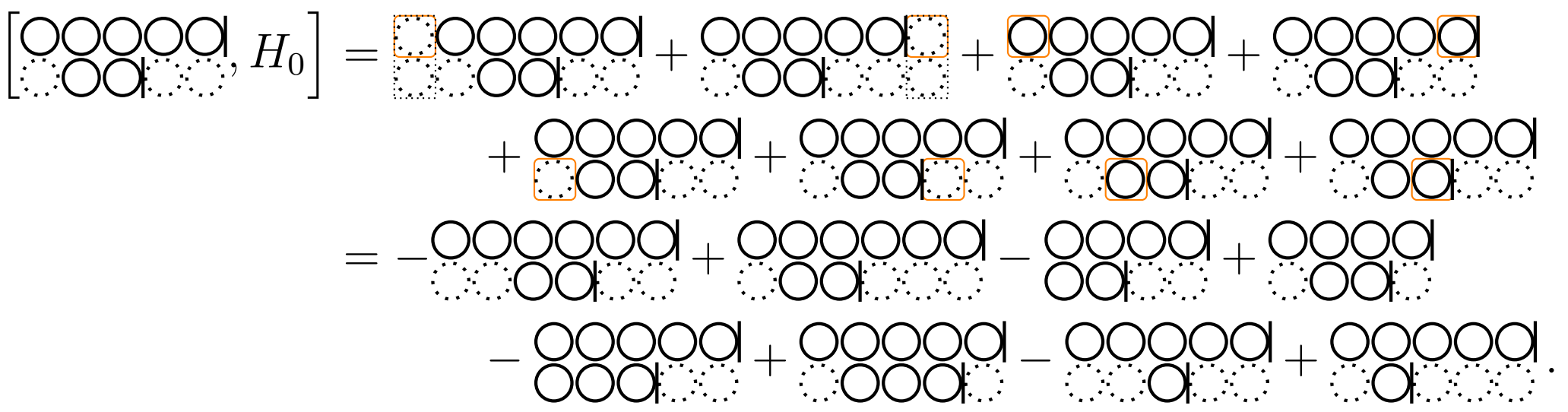

(5.137)

I also note that the list of a connected diagram generated by $[\Psi, H_0]$ differs from the original one $\boldsymbol{\lambda}_\Psi$, in which one element changes by $\pm 1$. For example, in Eq. (5.137), the list of the original diagram is $\{1;2\}$ and the generated diagrams have the lists $\{2;2\}$, $\{1;3\}$, $\{0;2\}$ or $\{1;1\}$.

The following example involves diagrams with a hole and a non-connected diagram:

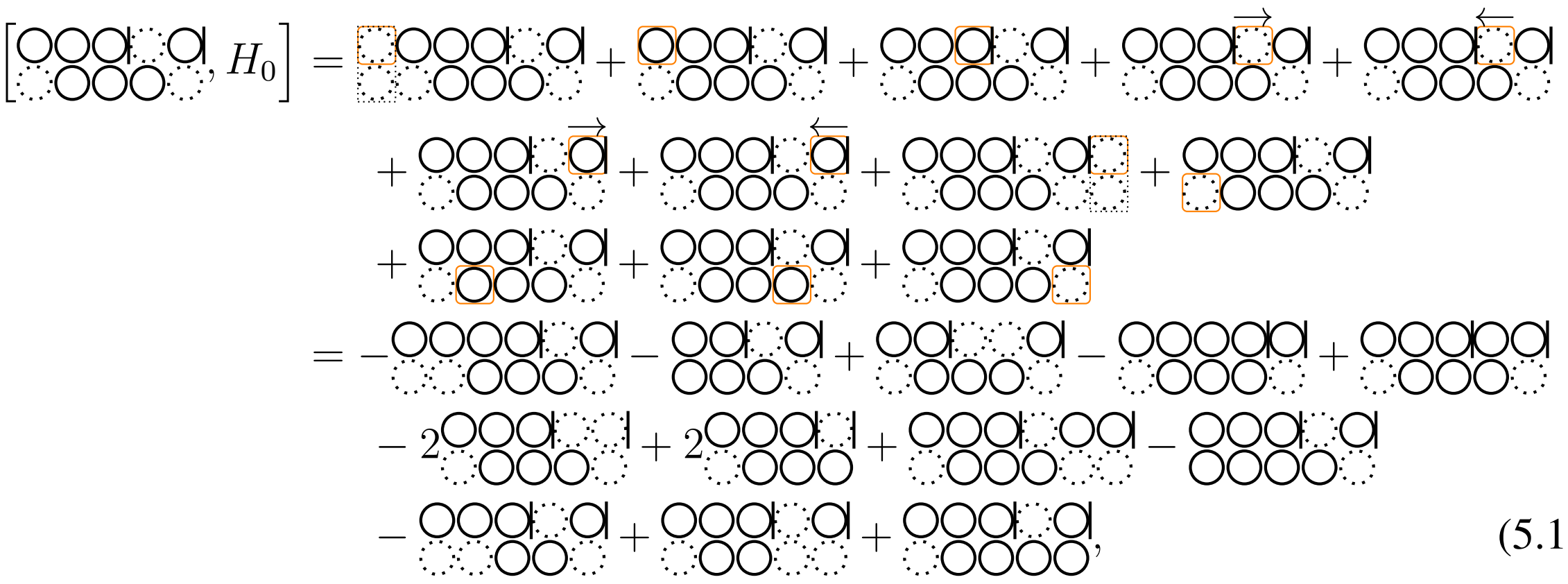

(5.138)

where 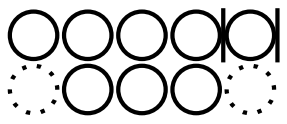 and 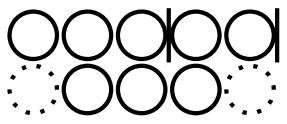 are the diagrams with a hole, while 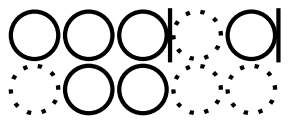 is non-connected.

### 5.3.3 Commutators of diagram with $H_{\text{int}}$

I consider the commutator of a diagram with the interacting term in the Hamiltonian. The interaction term is written as

$$H_{\text{int}} = \sum_{j=1}^{L} \Big| (j) \,. \tag{5.139}$$

From Eq. (5.105)–(5.108), the non-zero commutators of unit operators with the density of the interaction term $|(j)$ are given by

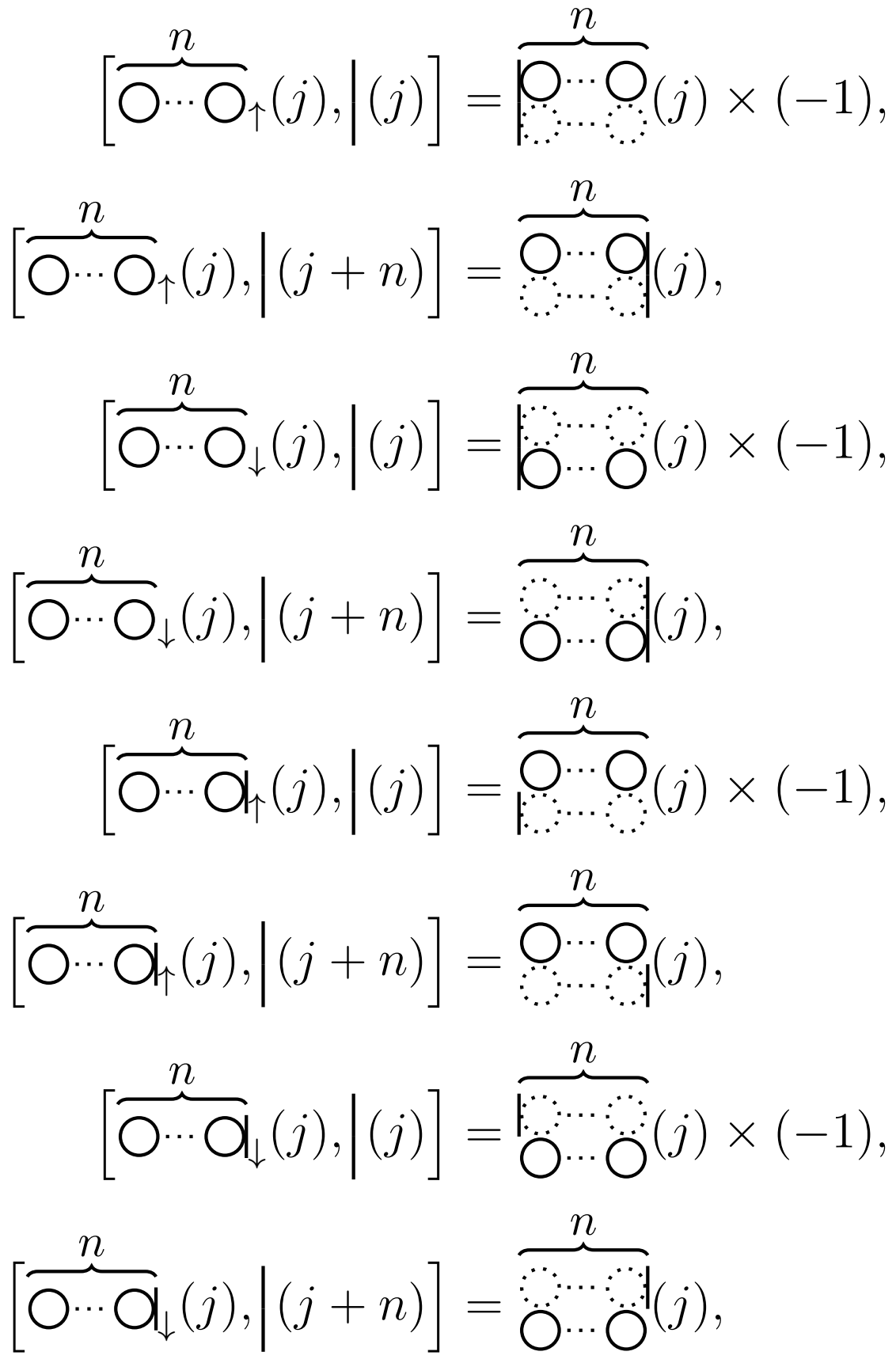

(5.140)
(5.141)
(5.142)
(5.143)
(5.144)
(5.145)
(5.146)
(5.147)

where $n \geq 1$. The other commutators are all zero.

I introduce the notation to represent the commutator of a diagram with the interaction term of the Hamiltonian $H_1$ as follows:

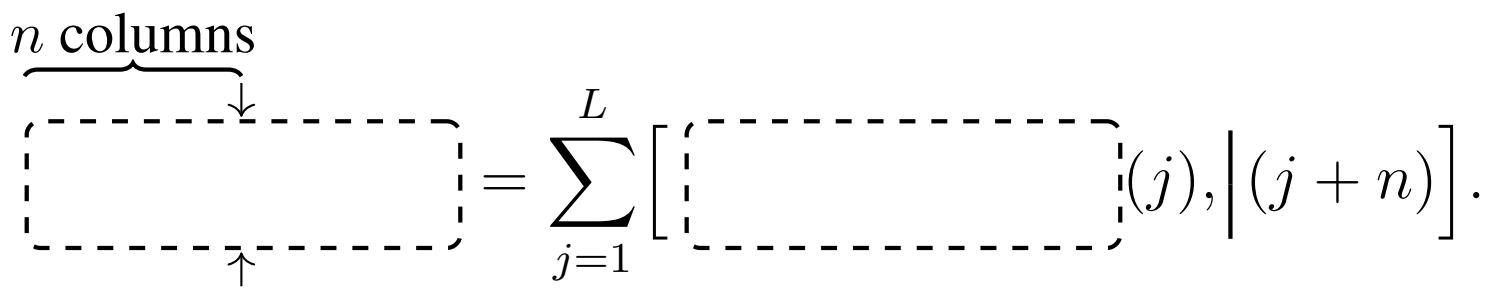

(5.148)

I often omit one of the two arrows in Eq. (5.148) in the following.

Examples of the non-zero commutators with the density of $H_{\rm int}$ are

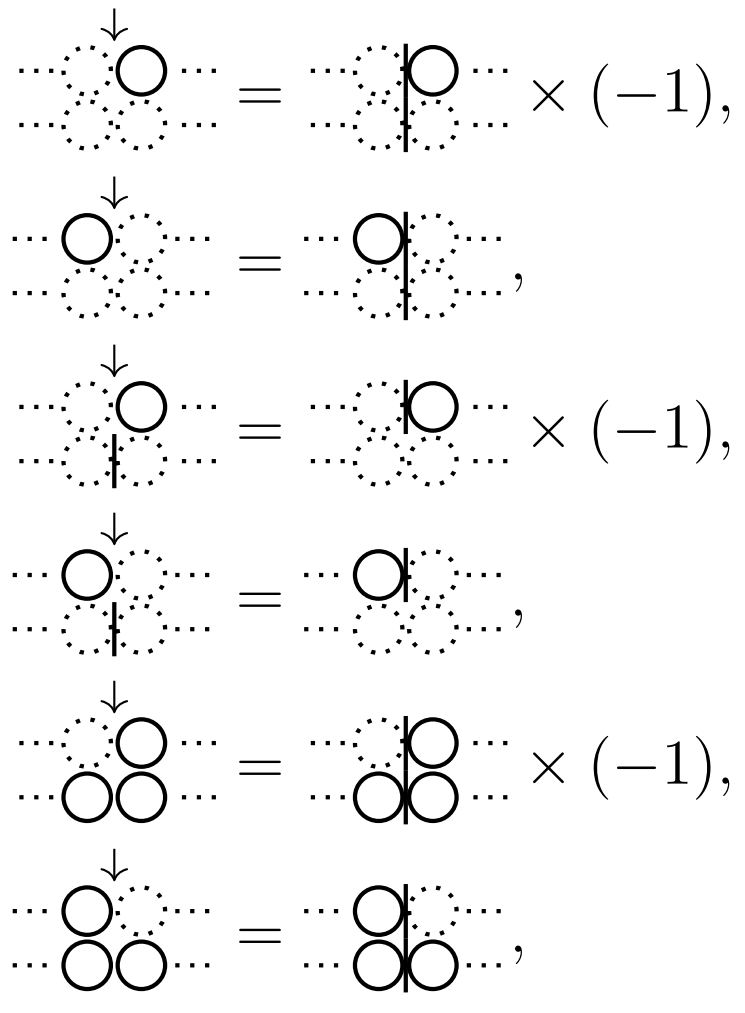

(5.149) (5.150) (5.151) (5.152) (5.153) (5.154)

where the types of the unit operators on the upper row are changed by the commutator. The types of the unit operators on the lower row in Eqs. (5.153) and (5.154) are unchanged by the commutator. I note that the commutator may generate one hole, such as Eqs. (5.153) and (5.154). The equations above hold when I exchange all the upper and lower rows.

I note that the following commutators are zero:

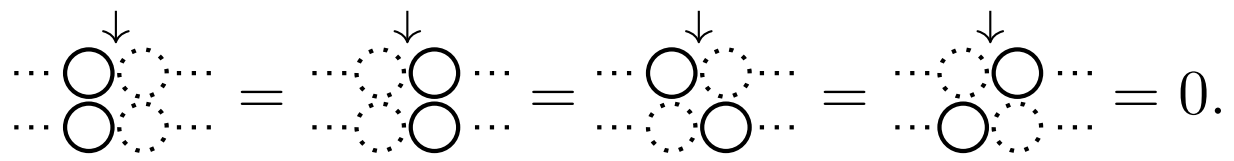

(5.155)

I also note the distinct expressions for a unit operator of $+$ type:

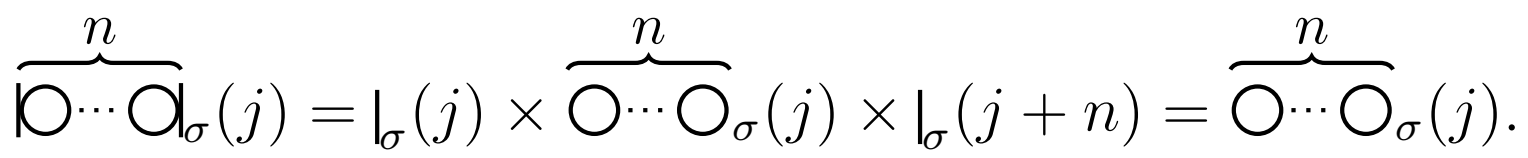

(5.156)

I show examples of the commutator of diagrams with $H_{\rm int}$ below:

(5.157)

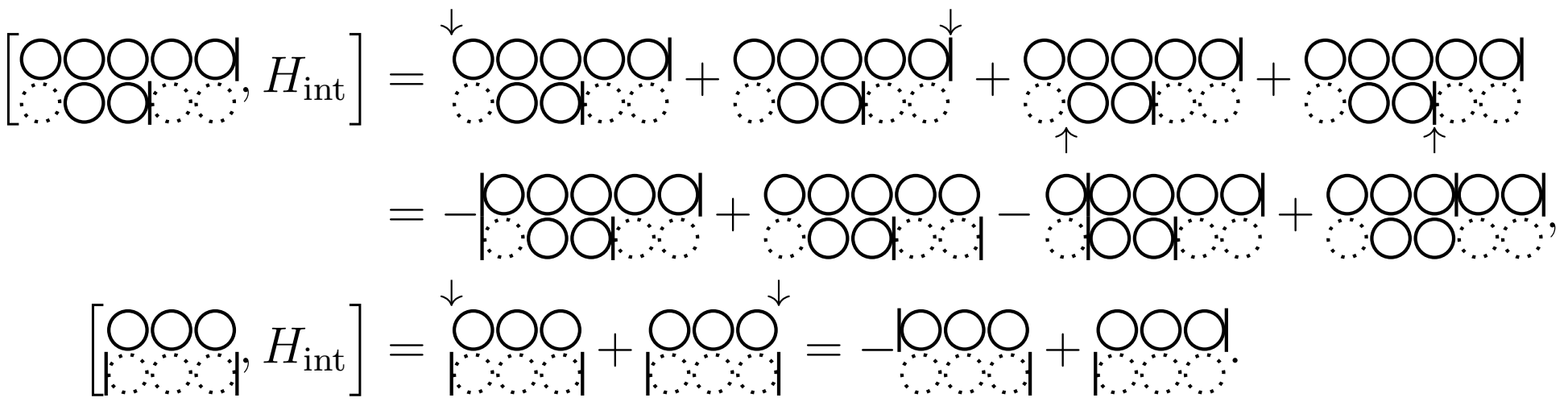

(5.158) (5.159)

## 5.4 Proof of the conservation law of $Q_k$

In this section, I prove Theorem 5.7, i.e. I prove $[Q_k, H] = 0$. I represent the commutator $[Q_k, H]$ as

$$[Q_k, H] = \sum_{j=0}^{j_f+1} U^j \left\{ \left[Q_k^j, H_0\right] + \left[Q_k^{j-1}, H_{\text{int}}\right] \right\}$$
$$= \sum_{j=1}^{j_f+1} U^j \sum_{n=0}^{\lfloor \frac{k-1-j}{2} \rfloor} \sum_{m=0}^{\lceil \frac{j-1}{2} \rceil} \sum_{d=0}^{\lfloor \frac{k-1-j}{2} \rfloor - n} \sum_{g=0}^{d} \sum_{\widetilde{\Psi} \in \mathcal{F}_{n,d,g}^{k,j,m}} \widetilde{D}_{n,d,g}^{k,j,m}(\widetilde{\Psi}) \widetilde{\Psi}, \tag{5.160}$$

where $j_f \equiv 2\lfloor k/2 \rfloor - 1$, $Q_k^{-1} = Q_k^{j_f+1} = 0$, $\mathcal{F}_{n,d,g}^{k,j,m}$ is the set of $(k-j-2n-d+1, d)$-diagrams $\widetilde{\Psi}$ whose unit number is $j+1-2m$, and the gap number is $g$. The $U^0$ terms are canceled because of the conservation law $[Q_k^0, H_0] = 0$ in the non-interacting case.

The symbol $\mathcal{F}_{n,d,g}^{k,j,m}$ denotes the set of the diagrams that can be generated in the cross at $(k - j - 2n - d + 1, d)$ on the plane of $Q_k^j$ in Figure 5.2 by the commutator $[Q_k, H]$. Note that $\mathcal{F}_{n,d,g}^{k,j,m}$ include not only connected diagrams but also non-connected diagrams.

What I have to prove is $\widetilde{D}_{n,d,g}^{k,j,m}(\widetilde{\Psi}) = 0$ for all $\widetilde{\Psi} \in \mathcal{F}_{n,d,g}^{k,j,m}$.

### 5.4.1 Cancellation of connected diagrams

I prove $\widetilde{D}_{n,d,g}^{k,j,m}(\widetilde{\Psi}) = 0$ in the case in which $\widetilde{\Psi} \in \mathcal{F}_{n,d,g}^{k,j,m}$ is a connected diagram and $l_{\widetilde{\Psi}} > 1$, i.e., $j > 2m$. I let the list of $\widetilde{\Psi}$ denoted by $\boldsymbol{\lambda}$.

I rewrite $\widetilde{D}_{n,d,g}^{k,j,m}(\widetilde{\Psi})$ as

$$\widetilde{D}_{n,d,g}^{k,j,m}(\widetilde{\Psi}) = A_0^{\sigma_0^R}(\widetilde{\Psi}) + \sum_{i=1}^{w} A_i^{\sigma_i^L, \sigma_i^R}(\widetilde{\Psi}) + A_{w+1}^{\sigma_{w+1}^L}(\widetilde{\Psi}), \tag{5.161}$$

where $A_i^{\sigma_i^L,\sigma_i^R}(\widetilde{\Psi})$ is the contribution to the cancellation of $\widetilde{\Psi}$ in $[Q_k, H]$ from the diagram in $Q_k^j$ whose list is $\boldsymbol{\lambda}_{i:\pm 1}$. $A_0^{\sigma_0^R}(\widetilde{\Psi})$ is the contribution to the cancellation of $\widetilde{\Psi}$ from diagrams in $Q_k^j$ whose list is $\boldsymbol{\lambda}_{L:\pm 1}$ and diagrams in $Q_k^{j-1}$ whose list is ${}_{0\to}(\boldsymbol{\lambda}_{L:-1})$. In the case of $\lambda_L = 0$, $A_0^{\sigma_0^R}(\widetilde{\Psi})$ also includes the contribution from diagrams in $Q_k^{j-1}$ whose list is $(\boldsymbol{\lambda}_{\hat{L}})_{L:+1}$. $A_{w+1}^{\sigma_0^L}(\widetilde{\Psi})$ is the contribution to the cancellation of $\widetilde{\Psi}$ from diagrams in $Q_k^j$ whose list is $\boldsymbol{\lambda}_{R:\pm 1}$ and diagrams in $Q_k^{j-1}$ whose list is $(\boldsymbol{\lambda}_{R:-1})_{\leftarrow 0}$. In the case of $\lambda_R = 0$, $A_{w+1}^{\sigma_0^L}(\widetilde{\Psi})$ also includes the contribution from diagrams in $Q_k^{j-1}$ whose list is $(\boldsymbol{\lambda}_{\hat{R}})_{R:+1}$. There are no other contributions to the cancellation of $\widetilde{\Psi}$ other than the above contribution. $\sigma_i^L = +$ if there is overlap or gap between the $i$ th coast and the $(i-1)$ th coast and $\sigma_i^L = -$ if there is no overlap or gap between the $i$ th coast and the $(i-1)$ th coast. $\sigma_i^R = +$ if there is overlap or gap between the $i$ th coast and the $(i+1)$ th coast and $\sigma_i^R = -$ if there is no overlap or gap between the $i$ th coast and the $(i+1)$ th coast. Therefore, $\sigma_i^R = \sigma_{i+1}^L$ holds.

I give examples of the case of $\sigma_i^R = \sigma_{i+1}^L = -$, i.e., the case in which there is no overlap or gap between the $i$ th and the $(i+1)$ th coasts:

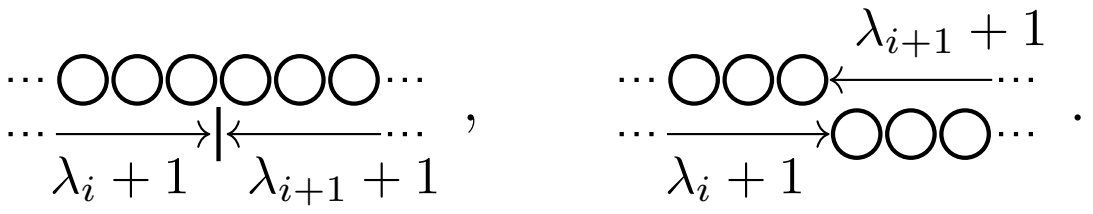

I give the example of the case of $\sigma_i^R = \sigma_{i+1}^L = +$, i.e., the case in which there is some overlap or gap between the $i$ th and the $(i+1)$ th coasts:

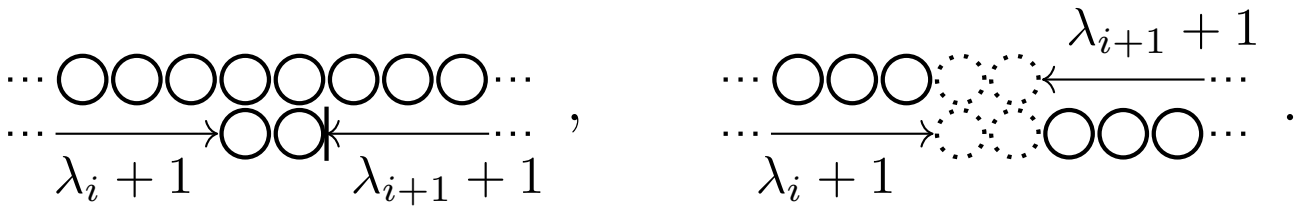

Let us see how $A_i^{\sigma_i^L,\sigma_i^R}(\widetilde{\Psi})$ is calculated. For example, I consider the cancellation of diagram $\widetilde{\Psi} \in \mathcal{F}_{12,5,0}^{42,2,0}$ in $[Q_{42}, H] = 0$. Here $\widetilde{\Psi}$ is

$\widetilde{\Psi} =$ 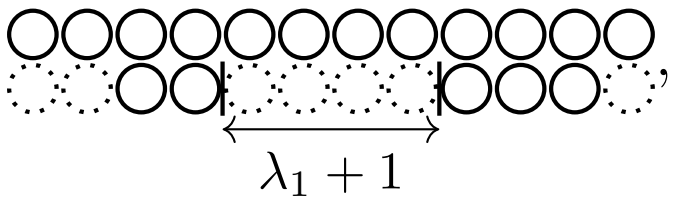 (5.162)

where $\boldsymbol{\lambda} = \{\lambda_L; \lambda_1; \lambda_R\} = \{2; 3; 1\}$. Let us calculate $A_1^{\sigma_1^L,\sigma_1^R}(\widetilde{\Psi})$ for example. Here, I have $\sigma_1^L = \sigma_1^R = +$ because there are overlaps on the left and right of the first coast with the length $\lambda_1 + 1$. The commutators of the connected diagrams in $Q_{42}^2$ whose list is $\{2; 3+1; 1\}$ and the Hamiltonian that can generate $\widetilde{\Psi}$ is

$4$ 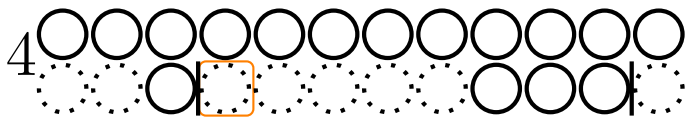 $+\,4$ 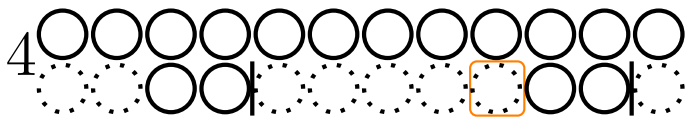 $= 0,$ (5.163)

where the coefficients $4$ come from Eq. (5.69). The two contributions comes from $\mathcal{S}_{12,4,0}^{42,2,0}$ and cancel each other. The commutators of the connected diagrams in $Q_{42}^2$ whose list are $\{2; 3-1; 1\}$ and the Hamiltonian that can generate $\widetilde{\Psi}$ is

$-2$ 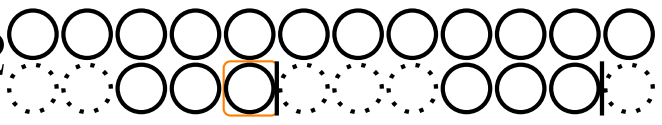 $-\,2$ 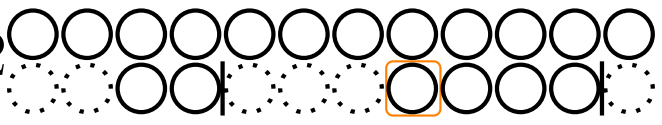 $= 0,$ (5.164)

where the coefficients $2$ come from Eq. (5.69). The two contributions comes from $\mathcal{S}_{12,3,0}^{42,2,0}$ and cancel each other. Therefore, I have $A_1^{++}(\widetilde{\Psi}) = 0$.

In the following, I write $A_0^{\sigma_0^R}(\widetilde{\Psi}), A_i^{\sigma_i^L,\sigma_i^R}(\widetilde{\Psi}), A_{w+1}^{\sigma_{w+1}^L}(\widetilde{\Psi})$ simply as $A_0^{\sigma_0^R}, A_i^{\sigma_i^L,\sigma_i^R}, A_{w+1}^{\sigma_{w+1}^L}$. The following lemma holds:

**Lemma 5.9.** *The contributions to the cancellations $A_i^{\sigma_i^L \sigma_i^R}$ satisfy the following relations:*

$$\begin{aligned}
&A_i^{++} = A_i^{--} = 0, && (5.165)\\
&A_i^{+-} = -A_i^{-+} = (-1)^{n+m+g}\left\{C_{n-1,d+1}^{j,m}\left(\boldsymbol{\lambda}_{i:-1}\right) + C_{n,d-1}^{j,m}\left(\boldsymbol{\lambda}_{i:+1}\right)\right\}, && (5.166)\\
&A_0^{+} = (-1)^{n+m+g}\left\{C_{n-1,d+1}^{j,m}\left(\boldsymbol{\lambda}_{L:-1}\right) - C_{n,d}^{j,m}\left(\boldsymbol{\lambda}_{L:-1}\right)\right.\\
&\qquad\left. + C_{n-1,d}^{j,m}\left(\boldsymbol{\lambda}_{L:+1}\right) - C_{n,d-1}^{j,m}\left(\boldsymbol{\lambda}_{L:+1}\right) + C_{n,d}^{j-1,m-1}\left({}_{0\to}(\boldsymbol{\lambda}_{L:-1})\right)\right\}, && (5.167)\\
&A_{w+1}^{+} = (-1)^{n+m+g+1}\left\{C_{n-1,d+1}^{j,m}\left(\boldsymbol{\lambda}_{R:-1}\right) - C_{n,d}^{j,m}\left(\boldsymbol{\lambda}_{R:-1}\right)\right.\\
&\qquad\left. + C_{n-1,d}^{j,m}\left(\boldsymbol{\lambda}_{R:+1}\right) - C_{n,d-1}^{j,m}\left(\boldsymbol{\lambda}_{R:+1}\right) + C_{n,d}^{j-1,m-1}\left((\boldsymbol{\lambda}_{R:-1})_{\leftarrow 0}\right)\right\}, && (5.168)\\
&A_0^{-} = (-1)^{n+m+g}\left\{2C_{n-1,d+1}^{j,m}\left(\boldsymbol{\lambda}_{L:-1}\right) - C_{n,d}^{j,m}\left(\boldsymbol{\lambda}_{L:-1}\right)\right.\\
&\qquad\left. + C_{n-1,d}^{j,m}\left(\boldsymbol{\lambda}_{L:+1}\right) + C_{n,d}^{j-1,m-1}\left({}_{0\to}(\boldsymbol{\lambda}_{L:-1})\right)\right\}, && (5.169)\\
&A_{w+1}^{-} = (-1)^{n+m+g+1}\left\{2C_{n-1,d+1}^{j,m}\left(\boldsymbol{\lambda}_{R:-1}\right) - C_{n,d}^{j,m}\left(\boldsymbol{\lambda}_{R:-1}\right)\right.\\
&\qquad\left. + C_{n-1,d}^{j,m}\left(\boldsymbol{\lambda}_{R:+1}\right) + C_{n,d}^{j-1,m-1}\left((\boldsymbol{\lambda}_{R:-1})_{\leftarrow 0}\right)\right\}. && (5.170)
\end{aligned}$$

The proof of Lemma 5.9 is given in Appendix B. Using the identity (5.84), I have

$$\begin{aligned}
A_{i_1}^{+-} + A_{i_2}^{-+} &= (-1)^{n+m+g}\left\{C_{n-1,d+1}^{j,m}\left(\boldsymbol{\lambda}_{i_1:-1}\right) + C_{n,d-1}^{j,m}\left(\boldsymbol{\lambda}_{i_1:+1}\right)\right.\\
&\qquad\left. - \left(C_{n-1,d+1}^{j,m}\left(\boldsymbol{\lambda}_{i_2:-1}\right) + C_{n,d-1}^{j,m}\left(\boldsymbol{\lambda}_{i_2:+1}\right)\right)\right\}\\
&= (-1)^{n+m+g}\left\{C_{n-1,d+1}^{j,m}\left(\boldsymbol{\lambda}_{i_1:-1}\right) - C_{n-1,d+1}^{j,m}\left(\boldsymbol{\lambda}_{i_2:-1}\right)\right.\\
&\qquad\left. - \left(C_{n,d-1}^{j,m}\left(\boldsymbol{\lambda}_{i_2:+1}\right) - C_{n,d-1}^{j,m}\left(\boldsymbol{\lambda}_{i_1:+1}\right)\right)\right\}\\
&= 0, \qquad (5.171)
\end{aligned}$$

where I used Eq. (5.84) in the last equality. With the same argument, I also have $A_{i_1}^{-+} + A_{i_2}^{+-} = 0$.

Using the identity (5.84), I have

$$\begin{aligned}
&A_0^{-} + A_i^{-+}\\
=&(-1)^{n+m+g}\left\{2C_{n-1,d+1}^{j,m}\left(\boldsymbol{\lambda}_{L:-1}\right) - C_{n,d}^{j,m}\left(\boldsymbol{\lambda}_{L:-1}\right) + C_{n-1,d}^{j,m}\left(\boldsymbol{\lambda}_{L:+1}\right) + C_{n,d}^{j-1,m-1}\left({}_{0\to}(\boldsymbol{\lambda}_{L:-1})\right)\right.\\
&\qquad\left. - \left(C_{n-1,d+1}^{j,m}\left(\boldsymbol{\lambda}_{i:-1}\right) + C_{n,d-1}^{j,m}\left(\boldsymbol{\lambda}_{i:+1}\right)\right)\right\}\\
=&(-1)^{n+m+g}\left\{C_{n-1,d+1}^{j,m}\left(\boldsymbol{\lambda}_{L:-1}\right) - C_{n,d}^{j,m}\left(\boldsymbol{\lambda}_{L:-1}\right) + C_{n-1,d}^{j,m}\left(\boldsymbol{\lambda}_{L:+1}\right) + C_{n,d}^{j-1,m-1}\left({}_{0\to}(\boldsymbol{\lambda}_{L:-1})\right)\right.\\
&\qquad\left. - \left(C_{n-1,d+1}^{j,m}\left(\boldsymbol{\lambda}_{i:-1}\right) + C_{n,d-1}^{j,m}\left(\boldsymbol{\lambda}_{i:+1}\right) - C_{n-1,d+1}^{j,m}\left(\boldsymbol{\lambda}_{L:-1}\right)\right)\right\}\\
=&(-1)^{n+m+g}\left\{C_{n-1,d+1}^{j,m}\left(\boldsymbol{\lambda}_{L:-1}\right) - C_{n,d}^{j,m}\left(\boldsymbol{\lambda}_{L:-1}\right) + C_{n-1,d}^{j,m}\left(\boldsymbol{\lambda}_{L:+1}\right) + C_{n,d}^{j-1,m-1}\left({}_{0\to}(\boldsymbol{\lambda}_{L:-1})\right)\right.\\
&\qquad\left. - C_{n,d-1}^{j,m}\left(\boldsymbol{\lambda}_{L:+1}\right)\right\}\\
=&A_0^{+}, \qquad (5.172)
\end{aligned}$$

where I used Eq. (5.84) in the third equality. With the same argument, I have

$$A_i^{+-} + A_{w+1}^{-} = A_{w+1}^{+}. \qquad (5.173)$$

Using the identity (5.84), I have

$$
\begin{aligned}
&A_0^{+} + A_i^{+-}\\
=&(-1)^{n+m+g}\Big\{C^{j,m}_{n-1,d+1}\left(\boldsymbol{\lambda}_{L:-1}\right) - C^{j,m}_{n,d}\left(\boldsymbol{\lambda}_{L:-1}\right) + C^{j,m}_{n-1,d}\left(\boldsymbol{\lambda}_{L:+1}\right)\\
&\qquad - C^{j,m}_{n,d-1}\left(\boldsymbol{\lambda}_{L:+1}\right) + C^{j-1,m-1}_{n,d}\left({}_{0\to}(\boldsymbol{\lambda}_{L:-1})\right)\\
&\qquad + C^{j,m}_{n-1,d+1}\left(\boldsymbol{\lambda}_{i:-1}\right) + C^{j,m}_{n,d-1}\left(\boldsymbol{\lambda}_{i:+1}\right)\Big\}\\
=&(-1)^{n+m+g}\left\{C^{j,m}_{n-1,d+1}\left(\boldsymbol{\lambda}_{L:-1}\right) - C^{j,m}_{n,d}\left(\boldsymbol{\lambda}_{L:-1}\right) + C^{j,m}_{n-1,d}\left(\boldsymbol{\lambda}_{L:+1}\right) + C^{j-1,m-1}_{n,d}\left({}_{0\to}(\boldsymbol{\lambda}_{L:-1})\right)\right.\\
&\qquad \left.+C^{j,m}_{n-1,d+1}\left(\boldsymbol{\lambda}_{i:-1}\right) + \left(C^{j,m}_{n,d-1}\left(\boldsymbol{\lambda}_{i:+1}\right) - C^{j,m}_{n,d-1}\left(\boldsymbol{\lambda}_{L:+1}\right)\right)\right\}\\
=&(-1)^{n+m+g}\left\{C^{j,m}_{n-1,d+1}\left(\boldsymbol{\lambda}_{L:-1}\right) - C^{j,m}_{n,d}\left(\boldsymbol{\lambda}_{L:-1}\right) + C^{j,m}_{n-1,d}\left(\boldsymbol{\lambda}_{L:+1}\right) + C^{j-1,m-1}_{n,d}\left({}_{0\to}(\boldsymbol{\lambda}_{L:-1})\right)\right.\\
&\qquad \left.+C^{j,m}_{n-1,d+1}\left(\boldsymbol{\lambda}_{i:-1}\right) + \left(C^{j,m}_{n-1,d+1}\left(\boldsymbol{\lambda}_{L:-1}\right) - C^{j,m}_{n-1,d+1}\left(\boldsymbol{\lambda}_{i:-1}\right)\right)\right\}\\
=&(-1)^{n+m+g}\left\{2C^{j,m}_{n-1,d+1}\left(\boldsymbol{\lambda}_{L:-1}\right) - C^{j,m}_{n,d}\left(\boldsymbol{\lambda}_{L:-1}\right) + C^{j,m}_{n-1,d}\left(\boldsymbol{\lambda}_{L:+1}\right) + C^{j-1,m-1}_{n,d}\left({}_{0\to}(\boldsymbol{\lambda}_{L:-1})\right)\right\}\\
=&A_0^{-},
\end{aligned}
\tag{5.174}
$$

where I used Eq. (5.84) in the third equality. With the same argument, I have

$$
A_i^{-+} + A_{w+1}^{+} = A_{w+1}^{-}. \tag{5.175}
$$

Using the identity (5.84), I have

$$
\begin{aligned}
&A_0^{-} + A_{w+1}^{-}\\
=&(-1)^{n+m+g}\left\{2C^{j,m}_{n-1,d+1}\left(\boldsymbol{\lambda}_{L:-1}\right) - C^{j,m}_{n,d}\left(\boldsymbol{\lambda}_{L:-1}\right) + C^{j,m}_{n-1,d}\left(\boldsymbol{\lambda}_{L:+1}\right) + C^{j-1,m-1}_{n,d}\left({}_{0\to}(\boldsymbol{\lambda}_{L:-1})\right)\right.\\
&\quad \left.-\left(2C^{j,m}_{n-1,d+1}\left(\boldsymbol{\lambda}_{R:-1}\right) - C^{j,m}_{n,d}\left(\boldsymbol{\lambda}_{R:-1}\right) + C^{j,m}_{n-1,d}\left(\boldsymbol{\lambda}_{R:+1}\right) + C^{j-1,m-1}_{n,d}\left((\boldsymbol{\lambda}_{R:-1})_{\leftarrow 0}\right)\right)\right\}\\
=&(-1)^{n+m+g}\Big\{C^{j,m}_{n-1,d+1}\left(\boldsymbol{\lambda}_{L:-1}\right) - C^{j,m}_{n,d-1}\left(\boldsymbol{\lambda}_{L:+1}\right)\\
&\qquad - C^{j,m}_{n,d}\left(\boldsymbol{\lambda}_{L:-1}\right) + C^{j,m}_{n-1,d}\left(\boldsymbol{\lambda}_{L:+1}\right) + C^{j-1,m-1}_{n,d}\left({}_{0\to}(\boldsymbol{\lambda}_{L:-1})\right)\\
&\quad -(C^{j,m}_{n-1,d+1}\left(\boldsymbol{\lambda}_{R:-1}\right) - C^{j,m}_{n,d-1}\left(\boldsymbol{\lambda}_{R:+1}\right)\\
&\qquad - C^{j,m}_{n,d}\left(\boldsymbol{\lambda}_{R:-1}\right) + C^{j,m}_{n-1,d}\left(\boldsymbol{\lambda}_{R:+1}\right) + C^{j-1,m-1}_{n,d}\left((\boldsymbol{\lambda}_{R:-1})_{\leftarrow 0}\right))\Big\}\\
=&A_0^{+} + A_{w+1}^{+},
\end{aligned}
\tag{5.176}
$$

where I used the identity $C^{j,m}_{n-1,d+1}\left(\boldsymbol{\lambda}_{L:-1}\right) - C^{j,m}_{n-1,d+1}\left(\boldsymbol{\lambda}_{R:-1}\right) = C^{j,m}_{n,d-1}\left(\boldsymbol{\lambda}_{R:+1}\right) - C^{j,m}_{n,d-1}\left(\boldsymbol{\lambda}_{L:+1}\right)$ from Eq. (5.84) in the second equality.

From the above arguments, one can see

$$
\begin{aligned}
A_0^{\sigma} + A_i^{\sigma,-\sigma} &= A_0^{-\sigma},\\
A_0^{\sigma} + A_{w+1}^{\sigma} &= A_0^{+} + A_{w+1}^{+}.
\end{aligned}
\tag{5.177}
$$

Therefore, in any case of a connected diagram $\widetilde{\Psi}$, I have

$$
\begin{aligned}
\widetilde{D}^{k,j,m}_{n,d,g}(\widetilde{\Psi}) &= A_0^{\sigma_0^R} + \sum_{i=1}^{w} A_i^{\sigma_i^L,\sigma_i^R} + A_{w+1}^{\sigma_{w+1}^L} \\
&= A_0^{\sigma} + A_{p_1}^{\sigma,-\sigma} + A_{p_2}^{-\sigma,\sigma} + \cdots + A_{p_M}^{(-)^{M-1}\sigma,(-)^M\sigma} + A_{w+1}^{(-)^M\sigma} \\
&= A_0^{-\sigma} + A_{p_2}^{-\sigma,\sigma} + \cdots + A_{p_M}^{(-)^{M-1}\sigma,(-)^M\sigma} + A_{w+1}^{(-)^M\sigma} \\
&= A_0^{\sigma} + A_{p_3}^{\sigma,-\sigma} + \cdots + A_{p_M}^{(-)^{M-1}\sigma,(-)^M\sigma} + A_{w+1}^{(-)^M\sigma} \\
&\ \vdots \\
&= A_0^{(-)^M\sigma} + A_{w+1}^{(-)^M\sigma} \\
&= A_0^{+} + A_{w+1}^{+} \\
&= (-1)^{n+m+g}\big\{C^{j,m}_{n-1,d+1}(\boldsymbol{\lambda}_{L:-1}) - C^{j,m}_{n,d}(\boldsymbol{\lambda}_{L:-1}) \\
&\qquad\qquad + C^{j,m}_{n-1,d}(\boldsymbol{\lambda}_{L:+1}) - C^{j,m}_{n,d-1}(\boldsymbol{\lambda}_{L:+1}) + C^{j-1,m-1}_{n,d}({}_{0\to}(\boldsymbol{\lambda}_{L:-1})) \\
&\qquad - (C^{j,m}_{n-1,d+1}(\boldsymbol{\lambda}_{R:-1}) - C^{j,m}_{n,d}(\boldsymbol{\lambda}_{R:-1}) \\
&\qquad\qquad + C^{j,m}_{n-1,d}(\boldsymbol{\lambda}_{R:+1}) - C^{j,m}_{n,d-1}(\boldsymbol{\lambda}_{R:+1}) + C^{j-1,m-1}_{n,d}((\boldsymbol{\lambda}_{R:-1})_{\leftarrow 0}))\big\} \\
&= (-1)^{n+m+g}\Big[\big\{C^{j,m}_{n,d}(\boldsymbol{\lambda}_{R:-1}) - C^{j,m}_{n,d}(\boldsymbol{\lambda}_{L:-1})\big\} - \big\{C^{j,m}_{n-1,d+1}(\boldsymbol{\lambda}_{R:-1}) - C^{j,m}_{n-1,d+1}(\boldsymbol{\lambda}_{L:-1})\big\} \\
&\qquad - \big\{C^{j,m}_{n,d-1}(\boldsymbol{\lambda}_{L:+1}) - C^{j,m}_{n,d-1}(\boldsymbol{\lambda}_{R:+1})\big\} + \big\{C^{j,m}_{n-1,d}(\boldsymbol{\lambda}_{L:+1}) - C^{j,m}_{n-1,d}(\boldsymbol{\lambda}_{R:+1})\big\} \\
&\qquad\qquad - \big\{C^{j-1,m-1}_{n,d}((\boldsymbol{\lambda}_{R:-1})_{\leftarrow 0}) - C^{j-1,m-1}_{n,d}({}_{0\to}(\boldsymbol{\lambda}_{L:-1}))\big\}\Big] \\
&= (-1)^{n+m+g}\Big[\big\{C^{j,m}_{n,d}(\boldsymbol{\lambda}_{R:-1}) - C^{j,m}_{n,d}(\boldsymbol{\lambda}_{L:-1})\big\} - 2\big\{C^{j,m}_{n-1,d+1}(\boldsymbol{\lambda}_{R:-1}) - C^{j,m}_{n-1,d+1}(\boldsymbol{\lambda}_{L:-1})\big\} \\
&\qquad + \big\{C^{j,m}_{n-2,d+2}(\boldsymbol{\lambda}_{R:-1}) - C^{j,m}_{n-2,d+2}(\boldsymbol{\lambda}_{L:-1})\big\} \\
&\qquad\qquad - \big\{C^{j-1,m-1}_{n,d}((\boldsymbol{\lambda}_{R:-1})_{\leftarrow 0}) - C^{j-1,m-1}_{n,d}({}_{0\to}(\boldsymbol{\lambda}_{L:-1}))\big\}\Big] \\
&= (-1)^{n+m+g}\Big[\Delta C^{j,m}_{n,d}(\widetilde{\boldsymbol{\lambda}}) - 2\Delta C^{j,m}_{n-1,d+1}(\widetilde{\boldsymbol{\lambda}}) + \Delta C^{j,m}_{n-2,d+2}(\widetilde{\boldsymbol{\lambda}}) \\
&\qquad\qquad - \Big\{C^{j-1,m-1}_{n,d}(\widetilde{\boldsymbol{\lambda}}_{\leftarrow 0}) - C^{j-1,m-1}_{n,d}({}_{0\to}(\mathcal{T}\widetilde{\boldsymbol{\lambda}}))\Big\}\Big] \\
&= 0,
\end{aligned}
\tag{5.178}
$$

where $\widetilde{\boldsymbol{\lambda}} \equiv \boldsymbol{\lambda}_{R:-1}$, $\sigma = \sigma_0^R$, $p_i$ is the $i$ th $p$ from the smallest that satisfies $\sigma_p^L = -\sigma_p^R$, $M$ is an integer that satisfies $M \leq w$, in the third-to-last equality, I used the following identities from Eq. (5.84)

$$
C^{j,m}_{n,d-1}(\boldsymbol{\lambda}_{L:+1}) - C^{j,m}_{n,d-1}(\boldsymbol{\lambda}_{R:+1}) = C^{j,m}_{n-1,d+1}(\boldsymbol{\lambda}_{R:-1}) - C^{j,m}_{n-1,d+1}(\boldsymbol{\lambda}_{L:-1}), \tag{5.179}
$$

$$
C^{j,m}_{n-1,d}(\boldsymbol{\lambda}_{L:+1}) - C^{j,m}_{n-1,d}(\boldsymbol{\lambda}_{R:+1}) = C^{j,m}_{n-2,d+2}(\boldsymbol{\lambda}_{R:-1}) - C^{j,m}_{n-2,d+2}(\boldsymbol{\lambda}_{L:-1}), \tag{5.180}
$$

and in the last equality, I used Eq. (5.57). We note that $\mathcal{T}\widetilde{\boldsymbol{\lambda}} = \boldsymbol{\lambda}_{L:-1}$ and $(-)^M\sigma = \sigma_{w+1}^L$.

I next prove $\widetilde{D}^{k,j,m}_{n,d,g}(\widetilde{\Psi}) = 0$ in the case in which $\widetilde{\Psi} \in \mathcal{F}^{k,j,m}_{n,0,0}$ is a connected diagram and $l_{\widetilde{\Psi}} = 1$, i.e., $j = 2m$ and $m > 0$. In this case, the double and the gap number of $\widetilde{\Psi}$ is zero, such

as

$$\widetilde{\Psi} = $$ 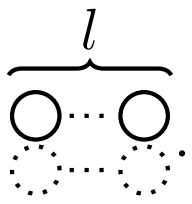 . (5.181)

The contributions to the cancellation of $\widetilde{\Psi}$ in $[Q_k, H]$ come only from $\left[Q_k^{j-1}, H_{\text{int}}\right]$ because $Q_k^{j>0}$ does not have the diagram with a unit number 1, given the normalization that fixes the freedom to add lower-order charges $Q_{k'<k}$. The contributions to the cancellation of $\widetilde{\Psi}$ are

$$(-1)^{n+m} \left\{ -C_{n,0}^{j,m-1}(0;l) \right.$$ 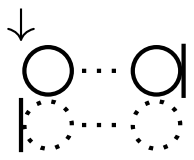 $$- C_{n,0}^{j,m-1}(l;0)$$ 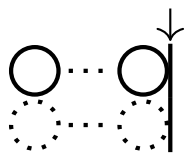 $$\left. \right\}$$

$$=(-1)^{n+m}\left(C_{n,0}^{j,m-1}(0;l) - C_{n,0}^{j,m-1}(l;0)\right)$$ 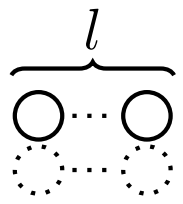

$$=0, \tag{5.182}$$

where I used $C_{n,0}^{j,m-1}(0;l) = C_{n,0}^{j,m-1}(l;0)$ from Eq. (5.79) and I have proved $\widetilde{D}_{n,d,g}^{k,j,m}(\widetilde{\Psi}) = 0$.

### 5.4.2 Cancellation of non-connected diagrams

I next prove $\widetilde{D}_{n,d,g}^{k,j,m}(\widetilde{\Psi}) = 0$ in the case in which $\widetilde{\Psi} \in \mathcal{F}_{n,d,g}^{k,j,m}$ is a non-connected diagram. I consider the case in which there is one hole in the region between the $i$ th and $i+1$ th coasts, such as

$$\widetilde{\Psi} = $$ 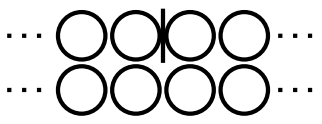 . (5.183)

In this case, the contributions to the cancellation of $\widetilde{\Psi}$ are

$$(-1)^{n+m+g}\left\{ C_{n,d-1}^{j,m}\left(\boldsymbol{\lambda}_{i,(0),i+1}\right) \right.$$ 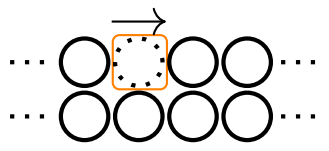 $$+ C_{n,d-1}^{j,m}\left(\boldsymbol{\lambda}_{i,(0),i+1}\right)$$ 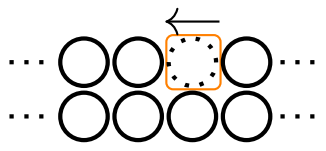 $$\left. \right\}$$

$$=(-1)^{n+m+g+1}\left\{C_{n,d-1}^{j,m}\left(\boldsymbol{\lambda}_{i,(0),i+1}\right) - C_{n,d-1}^{j,m}\left(\boldsymbol{\lambda}_{i,(0),i+1}\right)\right\}\widetilde{\Psi}$$

$$=0, \tag{5.184}$$

where $\boldsymbol{\lambda}_{i,(0),i+1} \equiv \{\ldots, \lambda_i, 0, \lambda_{i+1}, \ldots\}$ and 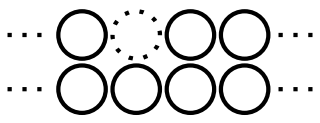 , 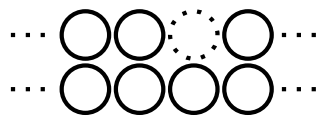 $\in \mathcal{S}_{n,d-1,g}^{k,j,m}$. Thus one can see $\widetilde{D}_{n,d,g}^{k,j,m}(\widetilde{\Psi}) = 0$.

With the same argument, one can prove $\widetilde{D}_{n,d,g}^{k,j,m}(\widetilde{\Psi}) = 0$ in the cases of

$$\widetilde{\Psi} = $$ 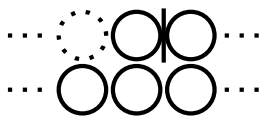 (5.185)

and

$$\widetilde{\Psi} = $$ 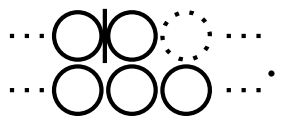 . (5.186)

I next consider the case in which there is one hole on the right of the $i$ th coast, such as

$$\widetilde{\Psi} = \quad (5.187)$$
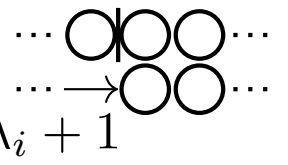

In this case, the contributions to the cancellation of $\widetilde{\Psi}$ are

$$(-1)^{n+m+g}\left\{C^{j,m}_{n,d-1}\left(\boldsymbol{\lambda}_{i,(0),i+1}\right) \cdots + C^{j-1,m}_{n,d}\left(\boldsymbol{\lambda}\right) \cdots \right\}$$
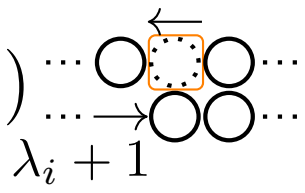
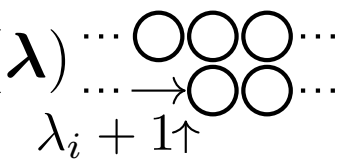

$$=(-1)^{n+m+g}\left\{C^{j,m}_{n,d-1}\left(\boldsymbol{\lambda}_{i,(0),i+1}\right) - C^{j-1,m}_{n,d}\left(\boldsymbol{\lambda}\right)\right\}\widetilde{\Psi}$$
$$=0, \quad (5.188)$$

where $\boldsymbol{\lambda}_{i,(0),i+1} \equiv \{\ldots,\lambda_i,0,\lambda_{i+1},\ldots\}$ and 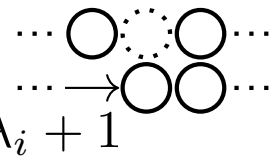 $\in \mathcal{S}^{k,j,m}_{n,d-1,g}$, 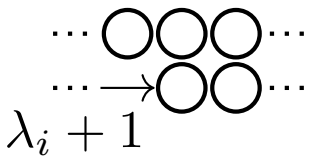 $\in \mathcal{S}^{k,j-1,m}_{n,d,g}$ and I used Eq. (5.85) in the last equality. Thus one can see $\widetilde{D}^{k,j,m}_{n,d,g}(\widetilde{\Psi}) = 0$. With the same argument, one can prove $\widetilde{D}^{k,j,m}_{n,d,g}(\widetilde{\Psi}) = 0$ in the case of

$$\widetilde{\Psi} = \quad . \quad (5.189)$$
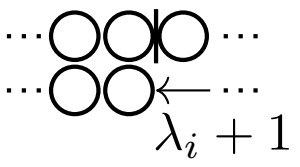

I next consider the case of a non-connected diagram with a gap, such as

$$\widetilde{\Psi} = \quad , \quad (5.190)$$
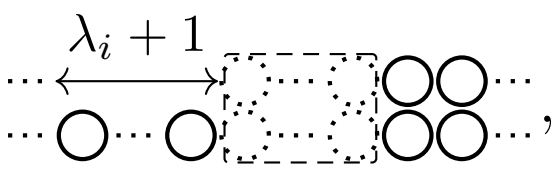

where I enclose the gap with the dashed line. In this case, the contributions to the cancellation of $\widetilde{\Psi}$ are

$$(-1)^{n+m+g}\left\{-C^{j,m}_{n,d-1}\left(\boldsymbol{\lambda}_{i,(0),i+1}\right) \cdots + C^{j,m}_{n,d-1}\left(\boldsymbol{\lambda}_{i,(0),i+1}\right) \cdots \right\}$$
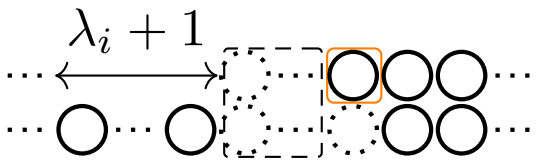
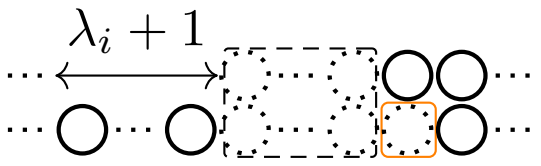

$$=(-1)^{n+m+g}\left\{C^{j,m}_{n,d-1}\left(\boldsymbol{\lambda}_{i,(0),i+1}\right) - C^{j,m}_{n,d-1}\left(\boldsymbol{\lambda}_{i,(0),i+1}\right)\right\}\widetilde{\Psi}$$
$$=0, \quad (5.191)$$

where $\boldsymbol{\lambda}_{i,(0),i+1} \equiv \{\ldots,\lambda_i,1,\lambda_{i+1},\ldots\}$ and 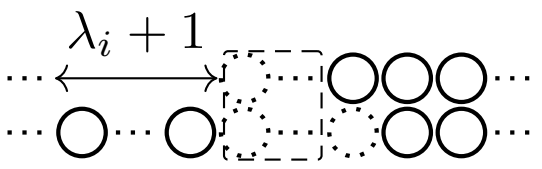 $\in \mathcal{S}^{k,j,m}_{n,d-1,g-1}$, 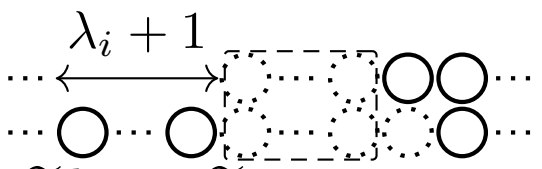 $\in \mathcal{S}^{k,j,m}_{n,d-1,g}$. Thus one can see $\widetilde{D}^{k,j,m}_{n,d,g}(\widetilde{\Psi}) = 0$. With the same argument, one can prove $\widetilde{D}^{k,j,m}_{n,d,g}(\widetilde{\Psi}) = 0$ in the case of

$$\widetilde{\Psi} = \quad . \quad (5.192)$$
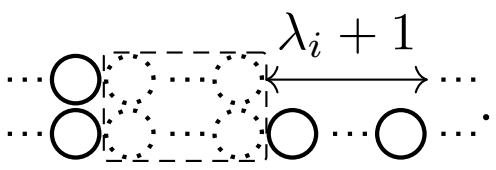

I next consider another case of a non-connected diagram $\widetilde{\Psi}$ with a gap, such as

$$\widetilde{\Psi} =$$ 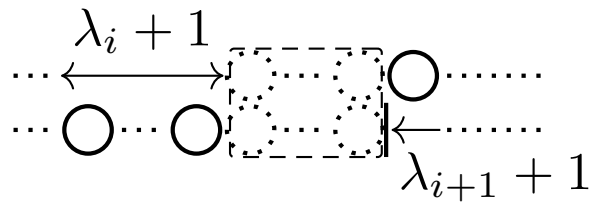 . (5.193)

In this case, the contributions to the cancellation of $\widetilde{\Psi}$ are

$$(-1)^{n+m+g}\left\{-C^{j,m}_{n,d-1}\left(\boldsymbol{\lambda}_{i,(0),i+1}\right)\right.$$ 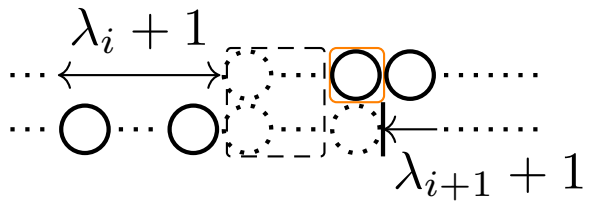 $$+\,C^{j-1,m}_{n,d}\left(\boldsymbol{\lambda}\right)$$ 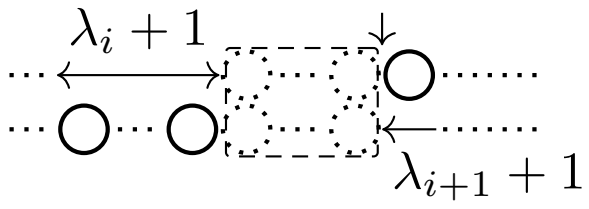 $$\left.\right\}$$

$$=(-1)^{n+m+g}\left\{C^{j,m}_{n,d-1}\left(\boldsymbol{\lambda}_{i,(0),i+1}\right)-C^{j-1,m}_{n,d}\left(\boldsymbol{\lambda}\right)\right\}\widetilde{\Psi}$$
$$=0, \qquad (5.194)$$

where $\boldsymbol{\lambda}_{i,(0),i+1} \equiv \{\ldots,\lambda_i,1,\lambda_{i+1},\ldots\}$, 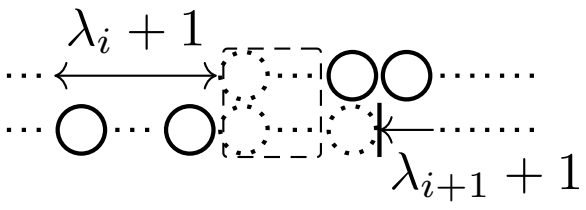 $\in \mathcal{S}^{k,j,m}_{n,d-1,g-1}$, 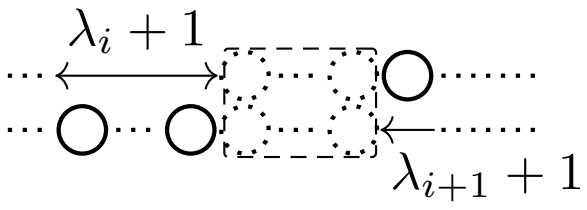 $\in \mathcal{S}^{k,j-1,m}_{n,d,g}$ and I used Eq. (5.85) in the last equality. Thus one can see $\widetilde{D}^{k,j,m}_{n,d,g}(\widetilde{\Psi})=0$.

The same argument holds in the case in which the upper and lower rows of $\widetilde{\Psi}$ above are exchanged.

In the other cases of a non-connected diagram $\widetilde{\Psi}$, the equation $\widetilde{D}^{k,j,m}_{n,d,g}(\widetilde{\Psi})=0$ trivially holds because such $\widetilde{\Psi}$ is not generated in the commutator $[Q_k,H]$. For example, the non-connected diagram $\widetilde{\Psi}$, such as

$$\widetilde{\Psi} =$$ 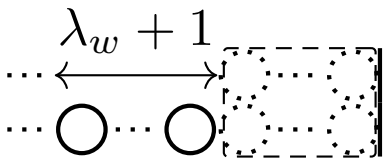 , (5.195)

cannot be generated by the commutator $[Q_k,H]$. At a glance, one may think that this comes from the commutator

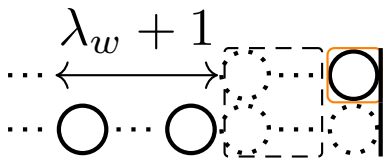 . (5.196)

However, this diagram is non-connected because the condition (ii) in Definition (5.4) is violated.

Thus, I have proved the conservation law $[Q_k,H]=0$.

## 5.5 Strong-coupling limit of the local charges

In this section, I consider the strong-coupling limit of the local charges in the 1D Hubbard model and obtain another family of mutually commuting local charges. These obtained mutual commuting charges are proved to be the local charges of a particular case of Maassarani's model [110, 111, 113–115]. This previously known class of integrable models includes the $\mathrm{SU}(N)$ XX model [116]. These models are proved to be free-fermionic and the general expression for the local charges has been obtained in terms of the Gell-Mann matrices in Ref. [116]. The $\mathrm{SU}(3)$ case corresponds to the $t$-0 model [32]. The $t$-0 model is noteworthy due to the simplicity of calculating the correlation function. A determinant representation for the two-particle Green's functions was obtained in Ref. [117], and finite-temperature spin transport was investigated by considering the spin-spin correlation function in Ref. [118]. In this thesis, I will conjecture the general formula for local charges in a different class of the Massarani model apart from those listed above, from the strong-coupling limit of local charges in the 1D Hubbard model.

The method of obtaining another integrable model through the strong-coupling limit of local charges of an original integrable model was developed in Ref. [119] and first applied to the XXZ model. The Hamiltonian of the folded XXZ chain [120–122] can be derived from the strong-coupling limit of the fourth local charge $Q_4$ in the XXZ chain [119]. The folded XXZ chain is a quantum integrable spin chain with "Hilbert space fragmentation" [123], for which there is an exponentially degenerate eigenspectrum despite the presence of interaction. It corresponds to a special point of the Bariev chain [124–128]. Here, I apply the method developed in Ref. [119] to our local charges $Q_k$ in the 1D Hubbard model.

### 5.5.1 Strong-coupling limit procedure

I first explain the strong-coupling limit strategy to obtain a folded Hamiltonian and its local charges [119]. Consider a mutual commuting family of operators $\{A_k\}$:

$$[A_k, A_l] = 0\,. \tag{5.197}$$

I further assume that $A_k$ depends on a coupling constant $U$, and $A_k$ is expressed as a finite degree polynomial of $U$:

$$A_k = \sum_{j=0}^{p_k} U^j A_k^j\,, \tag{5.198}$$

where $p_k$ is a positive integer and $A_k^j$ is independent of $U$. Substituting (5.198) to (5.197), I have

$$U^{p_k+p_l}\,[A_k^{p_k}, A_l^{p_l}] + \mathcal{O}(U^{p_k+p_l-1}) = 0\,. \tag{5.199}$$

Equation (5.199) must hold for arbitrary $U$. Then, considering the highest-order term, I have

$$[A_k^{p_k}, A_l^{p_l}] = 0\,. \tag{5.200}$$

In the following, I denote $A_k^{p_k}$ as

$$B_k \equiv A_k^{p_k}\,. \tag{5.201}$$

Finally, I obtain another mutual commuting family $\{B_k\}$ from the original family $\{A_k\}$.

I note that there is the freedom for the choice of linear combination for $\{A_k\}$. For example, $A'_k = A_k + U^2 A_{k-1}$ also commutes with all elements in $\{A_l\}$. One should appropriately fix this freedom for $\{A_k\}$ so that the resulting $\{B_k\}$ may be mutually linearly independent.

### 5.5.2 Strong-coupling limit of $Q_k$

I apply the strong-coupling method to the local charges $\{Q_k\}$ in the 1D Hubbard model. To obtain the linearly independent family $\{B_k\}$, I define the lower-order $A_k$ by the linear combinations of $\{Q_k\}$ as

$$A_2 = Q_2, \tag{5.202}$$
$$A_3 = Q_3, \tag{5.203}$$
$$A_4 = Q_4 + U^2 Q_2, \tag{5.204}$$
$$A_5 = Q_5 + U^2 Q_3, \tag{5.205}$$
$$A_6 = Q_6 + 2U^2 Q_4 - 3U^2 Q_2, \tag{5.206}$$
$$A_7 = Q_7 + 2U^2 Q_5 - 4U^2 Q_3, \tag{5.207}$$
$$A_8 = Q_8 + 3U^2 Q_6 + (U^4 - 7U^2) Q_4, \tag{5.208}$$
$$A_9 = Q_9 + 3U^2 Q_7 + (U^4 - 8U^2) Q_5, \tag{5.209}$$
$$A_{10} = Q_{10} + 4U^2 Q_8 + (3U^4 - 11U^2) Q_6 - 7U^4 Q_4 - 15U^4 Q_2, \tag{5.210}$$
$$A_{11} = Q_{11} + 4U^2 Q_9 + (3U^4 - 12U^2) Q_7 - 12U^4 Q_5 + (12U^6 - 18U^4) Q_3 \,. \tag{5.211}$$

With these normalization, I find $p_k = \lfloor \frac{k}{2} \rfloor$, and $\{B_k\}$ is mutually independent up to $k \leq 11$.

I show the corresponding $B_k$ in the following. The explicit expression for the lower-order charges $B_2, B_3, B_4$ are

$B_2 =$ 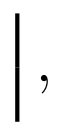 , (5.212)

$B_3 =$ 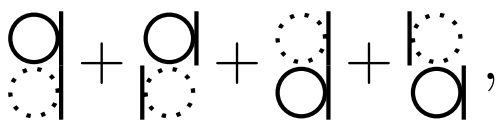 , (5.213)

$B_4 =$ 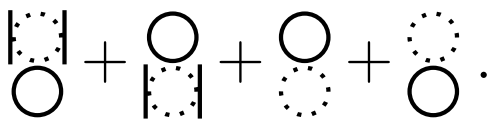 . (5.214)

The expressions in terms of the usual Pauli matrix are as follows:

$$B_2 = \sum_{j=1}^{L} \sigma^z_{j,\uparrow} \sigma^z_{j,\downarrow}, \tag{5.215}$$
$$B_3 = \sum_{j=1}^{L} \left( \sigma^x_{j,\uparrow} \sigma^y_{j+1,\uparrow} - \sigma^y_{j} \sigma^x_{j+1,\uparrow} \right) \left( \sigma^z_{j,\downarrow} + \sigma^z_{j+1,\downarrow} \right) + (\uparrow \leftrightarrow \downarrow) \,, \tag{5.216}$$
$$B_4 = \sum_{j=1}^{L} \left( \sigma^x_{j,\uparrow} \sigma^x_{j+1,\uparrow} + \sigma^y_{j} \sigma^y_{j+1,\uparrow} \right) \left( 1 + \sigma^z_{j,\downarrow} \sigma^z_{j+1,\downarrow} \right) + (\uparrow \leftrightarrow \downarrow) \,. \tag{5.217}$$

One can see that $B_4$ is equal to the Hamiltonian of the special case of the Maassarani model [113].

I study the action of the density of $B_4$. I represent $B_4$ as

$$B_4 = \sum_{j=1}^{L} b^{(4)}_{j,j+1} \,, \tag{5.218}$$

where I denote

$$b^{(4)} = (\sigma^x \otimes \sigma^x + \sigma^y \otimes \sigma^y)(1 + \tau^z \otimes \tau^z) + (\tau^x \otimes \tau^x + \tau^y \otimes \tau^y)(1 + \sigma^z \otimes \sigma^z) \,. \tag{5.219}$$

Here I denote $\sigma^a \equiv \sigma^a_\uparrow$ and $\tau^a \equiv \sigma^a_\downarrow$. I denote the local spin basis as

$$|s_\sigma, s_\tau\rangle \qquad (s_\sigma, s_\tau \in \{\uparrow, \downarrow\}) \,, \tag{5.220}$$

where $s_\sigma$ is the $z$-component of $\sigma^z$ and $s_\tau$ is the $z$-component of $\tau^z$. I further denote the local basis as

$$|1\rangle \equiv |\downarrow, \downarrow\rangle \,, \quad |2\rangle \equiv |\downarrow, \uparrow\rangle \,, \quad |3\rangle \equiv |\uparrow, \downarrow\rangle \,, \quad |4\rangle \equiv |\uparrow, \uparrow\rangle \,. \tag{5.221}$$

The non-transition by the action of $b^{(4)}$ to two local sites are [113]

$$\begin{aligned} |1\rangle \otimes |2\rangle &\leftrightarrow |2\rangle \otimes |1\rangle \,, \\ |1\rangle \otimes |3\rangle &\leftrightarrow |3\rangle \otimes |1\rangle \,, \\ |4\rangle \otimes |2\rangle &\leftrightarrow |2\rangle \otimes |4\rangle \,, \\ |4\rangle \otimes |3\rangle &\leftrightarrow |3\rangle \otimes |4\rangle \,, \end{aligned} \tag{5.222}$$

where $\leftrightarrow$ denotes the non-vanishing transition of two-site basis by $b^{(4)}$, the amplitude is $4$, and all other elements are zero. This representation is helpful in the identification of the $B_4$ and Maassarani's model [111], as explained in the next subsection.

### 5.5.3 Relation with a previously known model

In this subsection, you will see that $B_4$ is a special case of the integrable model that Maassarani found [110, 111].

Maassarani reported the following free-fermionic Hamiltonian [111]:

$$H^{(n_1,n_2)} = \sum_i P^{(n_1,n_2)}_{i,i+1} \,, \tag{5.223}$$

where $P^{(n_1,n_2)}$ is defined by

$$P^{(n_1,n_2)} := \sum_{1\le\alpha\le n_1<\beta\le n} \left( x_{\alpha,\beta} E^{\beta,\alpha} \otimes E^{\alpha,\beta} + x^{-1}_{\alpha,\beta} E^{\alpha,\beta} \otimes E^{\beta,\alpha} \right) \,, \tag{5.224}$$

where $n \equiv n_1 + n_2$, $x_{\alpha,\beta}$ are arbitrary complex parameters, and $E^{\alpha,\beta}$ is the $n$ by $n$ matrix with the element $1$ at row $\alpha$ and column $\beta$ and zeros otherwise. The R-matrix and the algebraic Bethe ansatz solution were found in Ref. [111].

When I set $(n_1 = 1, n_2 = n - 1)$ and $x_{\alpha,\beta} = 1$, I have the $\mathrm{SU}(n)$ XX chain from Eq. (5.224). Note that the $\mathrm{SU}(3)$ XX chain is equal to the $t$-0 model [32], which is also deeply related to the strong-coupling limit of the 1D Hubbard model [32].

From Eq. (5.222), one can see $B_4$ (5.218) corresponds to $(n_1 = 2, n_2 = 2)$ and $x_{\alpha,\beta} = 1$:

$$B_4 = 4 \sum_i P^{(2,2)} \,. \tag{5.225}$$

Thus, one can see the strong-coupling limit of the local charges in the 1D Hubbard model generates the local charges of the special case of Maassarani's model with $(n_1 = 2, n_2 = 2)$ and $x_{\alpha,\beta} = 1$.

### 5.5.4 Strong-coupling limit for the higher-order charges

The expressions for $B_5, B_6, B_7$ and $B_8$ are below:

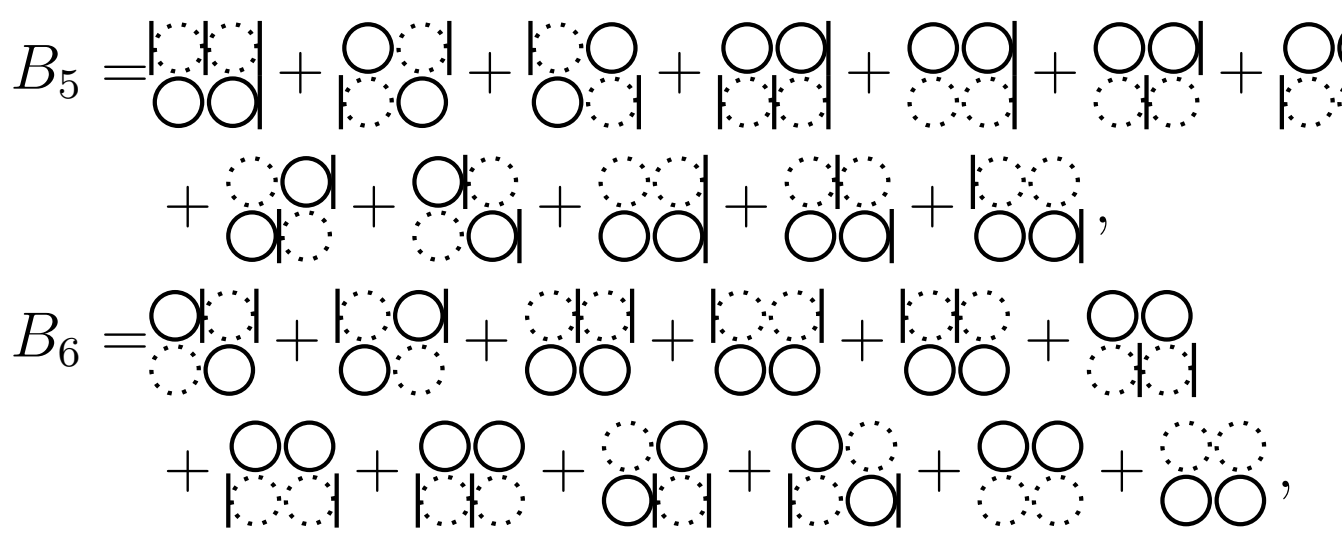

(5.226)

(5.227)

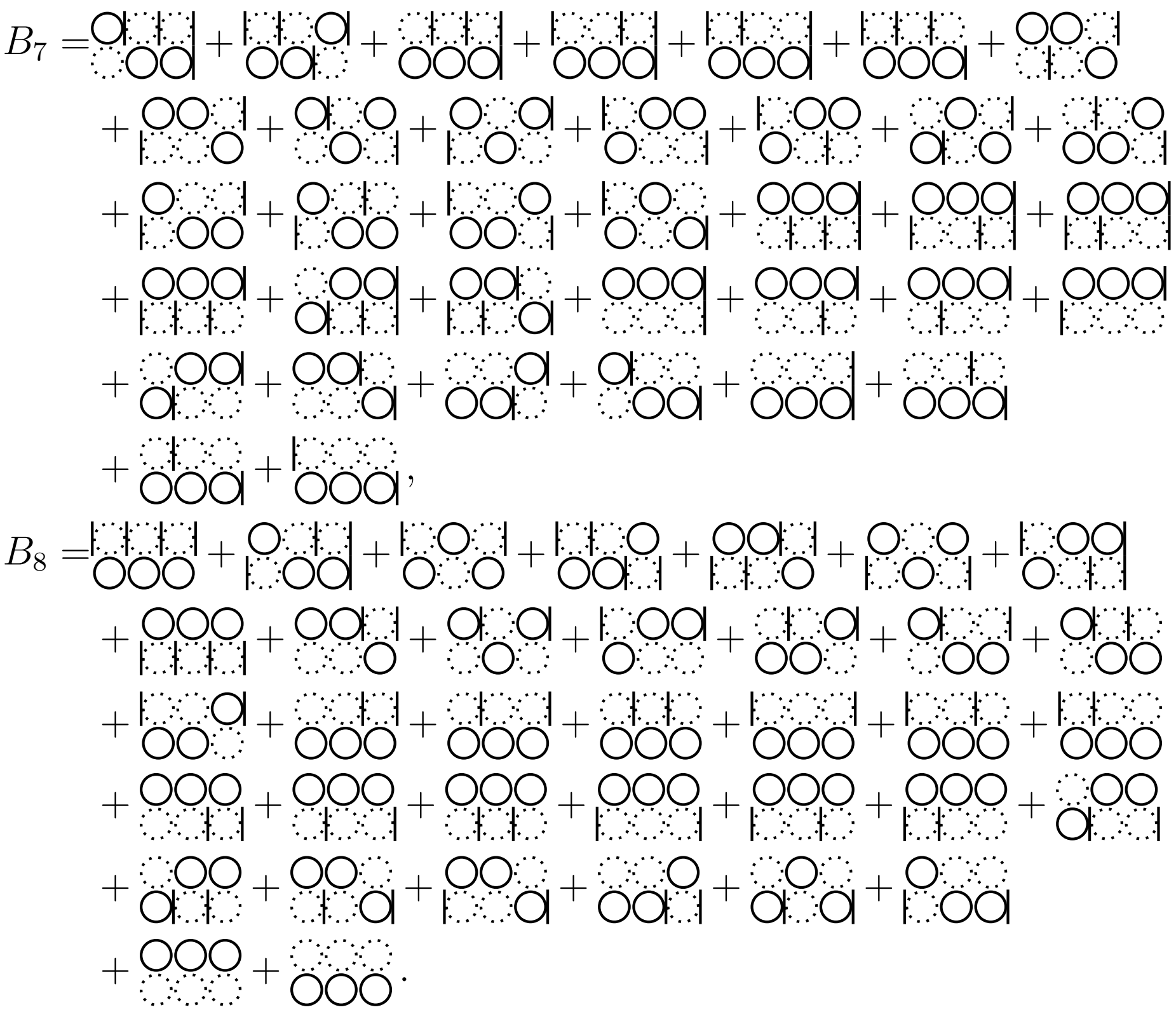

(5.228)

(5.229)

I give the explicit expression for $B_9$,$B_{10}$ and $B_{11}$ in Appendix E.

### 5.5.5 Conjecture for general expression

From the expressions up to $B_{11}$, I made the following conjecture for the general expression for $B_k$.

**Conjecture 5.10.** *The general formula for $B_k$ is*

$$B_k = \sum_{m=0}^{\lfloor k/4 \rfloor} \sum_{\Psi \in A_k^m} \Psi, \tag{5.230}$$

*where $A_k^m$ is the set of connected diagrams whose support is $\lfloor \frac{k+1}{2} \rfloor$ and whose double number is zero and whose unit number is $\lfloor k/2 \rfloor + 1 - 2m$. The $B_k$ defined in Eq. (5.230) satisfies the mutual commutativity $[B_k, B_l] = 0$.*

I note that $B_{2k-1}$ and $B_{2k}$ has the same support. The noteworthy feature of the new local charges $\{B_k\}$ is that the appearing coefficients are trivial; the local charges are just the simple summation of the diagrams. This characteristic is also seen for the local charges in the folded XXZ chain [129]. Then, it is implied that this simplicity of the coefficients in the local charges is a general characteristic of integrable lattice models obtained through the strong-coupling limit procedure [119]. Interestingly, the cumbersome coefficients defined by Eq. (5.57) result in trivial outcomes via the strong-coupling procedure [119].

I note that the conjectured quantities are the local charges of the special case of Maassarani's model, whose local charges are yet to be obtained. Our result in this section can be also seen as progress following the explicit expression of the local charges in the $\mathrm{SU}(n)$ XX model [116].

It should be noted that Eq. (5.230) is still a conjecture; I do not give its rigorous proof. However, I believe this proof may not be so challenging, and thus, I leave it as an exercise for the interested reader.

# Chapter 6

# Completeness of the local conserved quantities in the one-dimensional Hubbard model

In this chapter, I rigorously prove that there are no other local conserved quantities in the one-dimensional (1D) Hubbard model independent of those obtained from Theorem 5.7 in Chapter 5.

From this result, one can see that the local charges are generated by the expansion of the logarithm of the transfer matrix:

$$\widetilde{Q}_k = \frac{\partial^{k-1}}{\partial \lambda^{k-1}} \log T(\lambda)\bigg|_{\lambda=0}, \tag{6.1}$$

which can be written as a linear combination of $Q_k$, such as

$$\widetilde{Q}_k = \sum_{l=1}^{k} \alpha_l^k Q_k, \tag{6.2}$$

where $\alpha_l^k$ is a coefficient, $\alpha_k^k \neq 0$ and $Q_1$ is one-local charges explained below and in Appendix C. Vice versa, $Q_k$ can be also expressed as a linear combination of $\widetilde{Q}_k$ in the same way. Thus, the local charges are exhausted by those generated from the transfer matrix for the one-dimensional Hubbard model.

This proof of the completeness of our charges $\{Q_k\}$ does not need detailed knowledge of $\{Q_k\}$ explained in Chapter 5. It follows the spirit of the proof of the non-integrability of spin chains [41, 42]. I note that our considerations solely focus on ultra-local charges, excluding quasi-local ones [22, 130].

Chapter 6 is organized as follows. In Section 6.1, I explain our setup and present the main theorem of this article. Section 6.2 introduces notations helpful in the proof of the main theorem. In Section 6.3, I provide a rigorous proof of the main theorem. Section 6.4 contains a summary of our work. In Appendix C, I give all the one-support local charges in the one-dimensional Hubbard model.

## 6.1 Main statement

In this section, I present the main result of this chapter in Theorem 6.4 and the subsequent Corollary 6.5.

### 6.1.1 Preliminaries

Here, I restate the Hamiltonian of the one-dimensional Hubbard model:

$$H = -2t\sum_{j=1}^{L}\sum_{\sigma=\uparrow,\downarrow}\left(c^{\dagger}_{j,\sigma}c_{j+1,\sigma}+\text{h.c.}\right)+4U\sum_{j=1}^{L}\left(n_{j,\uparrow}-\frac{1}{2}\right)\left(n_{j,\downarrow}-\frac{1}{2}\right), \tag{6.3}$$

where the periodic boundary condition is imposed, $n_{j\sigma}\equiv c^{\dagger}_{j,\sigma}c_{j,\sigma}$ and $U$ is the coupling constant. I assume $U\neq 0$ and set $t=1$. Let us the first term of Eq. (6.3) denoted by $H_0$, and the second term of Eq. (6.3) denoted by $H_{\text{int}}$. Then, the Hamiltonian is written as $H=H_0+H_{\text{int}}$.

In this chapter, I prove the completeness of $Q_k$ in the fermion notation; however, the same argument can be also made in the case of spin variable notation.

To express the operator basis element constructing the local charges in the one-dimensional Hubbard model, I first define a $k$*-support basis element*:

**Definition 6.1.** *I define the* $k$*-support basis element starting from the* $i$ *th site by*

$$\boldsymbol{e}_i^k \equiv \boldsymbol{e}_{i,\uparrow}^k\boldsymbol{e}_{i,\downarrow}^k, \tag{6.4}$$

$$\boldsymbol{e}_{i,\sigma}^k \equiv e_{i,\sigma}e_{i+1,\sigma}\cdots e_{i+k-1,\sigma}, \tag{6.5}$$

*where* $\sigma\in\{\uparrow,\downarrow\}$*, and*

$$e_{j,\sigma}\in\{c_{j,\sigma},c^{\dagger}_{j,\sigma},z_{j,\sigma},I\}, \tag{6.6}$$

*and* $z_{l,\sigma}\equiv 2n_{l,\sigma}-1$*, and* $I$ *is the identity operator. To ensure that* $\boldsymbol{e}_i^k$ *has strictly* $k$*-support, I impose the following constraint:* $\{e_{i,\uparrow},e_{i,\downarrow}\}\neq\{I,I\}$*, and* $\{e_{i+k-1,\uparrow},e_{i+k-1,\downarrow}\}\neq\{I,I\}$*.*

I also define a $k$*-support operator*:

**Definition 6.2.** *I refer to a linear combination of* $k$*-support basis elements as a* $k$*-support operator.*

I next define a $k$*-local conserved quantity*:

**Definition 6.3.** *A* $k$*-local conserved quantity is a local conserved quantity that is expressed as*

$$F_k=\sum_{l=1}^{k}\sum_{i=1}^{L}\sum_{\boldsymbol{e}_i^l}c_{\boldsymbol{e}_i^l}\boldsymbol{e}_i^l, \tag{6.7}$$

*where the sum of $\mathbf{e}_i^l$ runs over all $l$-support basis elements, $c_{\mathbf{e}_i^l}$ is the coefficient of $\mathbf{e}_i^l$ which may depend on $U$, and there exists a $k$-support basis element $\mathbf{e}_i^k$ such that $c_{\mathbf{e}_i^k} \neq 0$. I call $F_k$ in Eq. (6.7) less than or equal to the $k$-local conserved quantity in the case in which $c_{\mathbf{e}_i^k} = 0$ is allowed.*

Because $F_k$ in Eq. (6.7) is a conserved quantity, $F_k$ commutes with the Hamiltonian in Eq. (6.3): $[F_k, H] = 0$.

### 6.1.2 Recap of $Q_k$

The local conserved quantity $Q_k$ obtained from Theorem 5.7 is a $k$-local conserved quantity. $Q_k$ can be written as

$$Q_k = Q_k^0 + \delta Q_{k-1}(U), \tag{6.8}$$

where $Q_k^0$ is the non-interacting term of local charge $Q_k$, which is a $k$-support operator and which does not depend on $U$:

$$Q_k^0 \equiv 2 \sum_{j=1}^{L} \sum_{\sigma=\uparrow,\downarrow} \left( c_{j,\sigma} c_{j+k-1,\sigma}^{\dagger} + (-1)^{k-1} c_{j,\sigma}^{\dagger} c_{j+k-1,\sigma} \right), \tag{6.9}$$

and $\delta Q_{k-1}(U)$ is the other terms in $Q_k$, which is a $(k-1)$-support operator and depends on $U$:

$$\delta Q_{k-1}(U) = \sum_{j=1}^{j_f} U^j Q_k^j, \tag{6.10}$$

where the explicit expression for $Q_k^j$ is given in Theorem 5.7.

While the exact expression for $Q_k^j$ is given by Theorem 5.7, the exhaustive details of $Q_k^j$ provided there is not necessary for the following argument; the most important fact is that the $k$-support operator in $Q_k$ is $Q_k^0$ in Eq. (6.9).

### 6.1.3 One-local charges

I denote the one-local conserved quantities by $Q_1$, which is not obtained by the expansion of the transfer matrix. The one-local charges $Q_1$ in the one-dimensional Hubbard model are the generators of the $\mathrm{SU}(2) \times \mathrm{SU}(2)$ symmetry, as I reviewed in Chapter 3.

The generators for the fundamental $\mathrm{SU}(2)$ symmetry are

$$S^+ = \sum_{i=1}^{L} c_{j,\uparrow}^{\dagger} c_{j,\downarrow}, \qquad S^- = \sum_{i=1}^{L} c_{j,\downarrow}^{\dagger} c_{j,\uparrow}, \qquad S^z = \frac{1}{2} \sum_{i=1}^{L} \left( n_{j,\uparrow} - n_{j,\downarrow} \right). \tag{6.11}$$

The generators for another $\mathrm{SU}(2)$ symmetry, so called $\eta$-pairing charges [51, 52] are

$$\eta^+ = \sum_{i=1}^{L} (-1)^{i+1} c_{j,\uparrow}^\dagger c_{j,\downarrow}^\dagger, \qquad \eta^- = \sum_{i=1}^{L} (-1)^{i+1} c_{j,\downarrow} c_{j,\uparrow}, \qquad \eta^z = \frac{1}{2}(N - L), \tag{6.12}$$

where $N \equiv \sum_{i=1}^{L}(n_{i,\uparrow} + n_{i,\downarrow})$ is the particle number. I note that $\eta^\pm$ is conserved in the case of even $L$, and is not conserved for odd $L$. The $\mathrm{U}(1)$ charge $\eta^z$ is conserved in both cases of even and odd $L$.

In Appendix C, I prove that one-local conserved quantity $Q_1$ is a linear combination of the $\mathrm{SU}(2)$ charges and $\mathrm{U}(1)$ charge, and also the $\eta$-pairing charges for even $L$. There are no other 1-local charges independent of the generators of $\mathrm{SU}(2) \times \mathrm{SU}(2)/\mathbb{Z}_2 \simeq \mathrm{SO}(4)$.

### 6.1.4 Main theorem: completeness of $Q_k$

Whether or not there exist local charges independent of $\{Q_k\}_{k\geq 1}$ has been a mystery. I prove the family of the local conserved quantities $\{Q_k\}_{k\geq 1}$ is complete: there is no other $k$-local conserved quantities in the one-dimensional Hubbard model independent of $\{Q_k\}_{k\geq 1}$. The proof is based on the following theorem and its corollary:

**Theorem 6.4.** *Let $F_k$ be a $k$-local conserved quantity of the one-dimensional Hubbard model $(k < \lfloor \frac{L}{2} \rfloor)$. A constant $c_k (\neq 0)$ and a less-than-$(k-1)$-local conserved quantity $\Delta_{k-1}$ exist such that $F_k = c_k Q_k + \Delta_{k-1}$.*

When I replace $F_k$ by a less-than-or-equal-to-$k$-local conserved quantity $F_k'$, in Theorem 6.4, $c_k = 0$ is allowed.

I have the following corollary immediately from Theorem 6.4.

**Corollary 6.5.** *Let $F_k$ be a $k$-local conserved quantity of the one-dimensional Hubbard model $(k < \lfloor \frac{L}{2} \rfloor)$. A set of constants $\{c_l\}_{1\leq l\leq k}$ exist such that $F_k = \sum_{l=1}^{k} c_l Q_l$.*

*Proof.* From Theorem 6.4, $F_k$ is written as $F_k = c_k Q_k + F_{k-1}'$, where $F_{k-1}'$ is a less-than-$(k-1)$-local conserved quantity. Using Theorem 6.4 again to $F_{k-1}'$, I have $F_k = c_k Q_k + c_{k-1} Q_{k-1} + F_{k-2}'$, where $F_{k-2}'$ is a less-than-$(k-2)$-local conserved quantity. In the same way, by repeatedly using Theorem 6.4, I have $F_k = \sum_{l=1}^{k} c_l Q_l$ $(c_k \neq 0)$. □

From Corollary 6.5, all local conserved quantities are represented as a linear combination of $\{Q_k\}_{k\geq 1}$, and thereby one can prove the completeness of $\{Q_k\}_{k\geq 1}$. One can also confirm that the local charges generated by the transfer matrix are represented as a linear combination of $\{Q_k\}_{k\geq 1}$. All I have to do next is the proof of Theorem 6.4. I prove Theorem 6.4 in the rest of this Chapter.

## 6.2 Notations

In this section, I introduce useful notations in the proof of Theorem 6.4. In the following, the symbol for the commutator has the additional factor $1/2$: $[A, B] \equiv \frac{1}{2}(AB - BA)$.

### 6.2.1 Notation for the commutator with Hamiltonian

I introduce the notation for the $k$-support basis element

$$\begin{array}{cccc}\bar{a}_1 & \bar{a}_2 & \cdots & \bar{a}_k \\ \bar{b}_1 & \bar{b}_2 & \cdots & \bar{b}_k\end{array}(i) \equiv \boldsymbol{e}_i^k, \tag{6.13}$$

where $\bar{a}_l \equiv \ominus, \oplus, \textcircled{z}, \bigcirc$ for $e_{i+l-1,\uparrow} = c_{i+l-1,\uparrow}, c^\dagger_{i+l-1,\uparrow}, z_{i+l-1,\uparrow}, I$, respectively, and $\bar{b}_l \equiv \ominus, \oplus, \textcircled{z}, \bigcirc$ for $e_{i+l-1,\downarrow} = c_{i+l-1,\downarrow}, c^\dagger_{i+l-1,\downarrow}, z_{i+l-1,\downarrow}, I$, respectively. I note that at least one of $\bar{a}_1$ and $\bar{b}_1$ should not be $I$, and the same is also true for $\bar{a}_k$ and $\bar{b}_k$. I give examples of 5-support basis elements in this notation:

$$\begin{array}{ccccc}\oplus & \ominus & \bigcirc & \bigcirc & \bigcirc \\ \bigcirc & \bigcirc & \bigcirc & \oplus & \ominus\end{array}(i) = c^\dagger_{i,\uparrow} c_{i+1,\uparrow} c^\dagger_{i+3,\downarrow} c_{i+4,\downarrow}, \tag{6.14}$$

$$\begin{array}{ccccc}\oplus & \bigcirc & \ominus & \bigcirc & \textcircled{z} \\ \bigcirc & \ominus & \bigcirc & \oplus & \bigcirc\end{array}(i) = c^\dagger_{i,\uparrow} c_{i+2,\uparrow} z_{i+4,\uparrow} c_{i+1,\downarrow} c^\dagger_{i+3,\downarrow}. \tag{6.15}$$

I refer to $i$ in Eq. (6.13) as the starting site and to the two-row sequence $\begin{array}{cccc}\bar{a}_1 & \bar{a}_2 & \cdots & \bar{a}_k \\ \bar{b}_1 & \bar{b}_2 & \cdots & \bar{b}_k\end{array}$ as a $k$-support *configuration*.

#### Commutator with $H_0$

I introduce the notation for the commutator with the hopping term of the Hamiltonian. I first denote the density of the hopping term as

$$h_l^\uparrow = 2(c_{l,\uparrow} c^\dagger_{l+1,\uparrow} - c^\dagger_{l,\uparrow} c_{l+1,\uparrow}), \tag{6.16}$$

$$h_l^\downarrow = 2(c_{l,\downarrow} c^\dagger_{l+1,\downarrow} - c^\dagger_{l,\downarrow} c_{l+1,\downarrow}), \tag{6.17}$$

and the hopping term is written as $H_0 = \sum_{\sigma\in\{\uparrow,\downarrow\}} \sum_{l=1}^{L} h_l^\sigma$. I represent the commutator of a basis element and $h_l^\sigma$ ($\sigma \in \{\uparrow, \downarrow\}$) by

$$\begin{array}{cccccc}\bar{a}_1 \cdots & \overbrace{\bar{a}_l \quad \bar{a}_{l+1}} & \cdots & \bar{a}_k \\ \bar{b}_1 \cdots & \bar{b}_l \quad \bar{b}_{l+1} & \cdots & \bar{b}_k\end{array}(i) \equiv \left[\begin{array}{cccc}\bar{a}_1 \cdots & \bar{a}_l \quad \bar{a}_{l+1} & \cdots & \bar{a}_k \\ \bar{b}_1 \cdots & \bar{b}_l \quad \bar{b}_{l+1} & \cdots & \bar{b}_k\end{array}(i), h^\uparrow_{i+l-1}\right], \tag{6.18}$$

$$\begin{array}{cccccc}\bar{a}_1 \cdots & \bar{a}_l \quad \bar{a}_{l+1} & \cdots & \bar{a}_k \\ \bar{b}_1 \cdots & \underbrace{\bar{b}_l \quad \bar{b}_{l+1}} & \cdots & \bar{b}_k\end{array}(i) \equiv \left[\begin{array}{cccc}\bar{a}_1 \cdots & \bar{a}_l \quad \bar{a}_{l+1} & \cdots & \bar{a}_k \\ \bar{b}_1 \cdots & \bar{b}_l \quad \bar{b}_{l+1} & \cdots & \bar{b}_k\end{array}(i), h^\downarrow_{i+l-1}\right]. \tag{6.19}$$

When acting on the left or right end of the basis element, I write the action as

$$\begin{matrix} \bar{a}_1 \cdots & \overline{\bar{a}_k \;\; \circ} \\ \bar{b}_1 \cdots & \bar{b}_k \;\; \circ \end{matrix} (i) = \left[ \begin{matrix} \bar{a}_1 \cdots \bar{a}_k \\ \bar{b}_1 \cdots \bar{b}_k \end{matrix} (i), h^{\uparrow}_{i+k-1} \right], \tag{6.20}$$

$$\begin{matrix} \overline{\circ \;\; \bar{a}_1} & \cdots \bar{a}_k \\ \circ \;\; \bar{b}_1 & \cdots \bar{b}_k \end{matrix} (i) = \left[ \begin{matrix} \bar{a}_1 \cdots \bar{a}_k \\ \bar{b}_1 \cdots \bar{b}_k \end{matrix} (i), h^{\uparrow}_{i-1} \right], \tag{6.21}$$

where I add the additional columns with $\circ$ to the right or left end, and the support can be increased by one in this case. The actions of $h^{\downarrow}_{i-1}$ and $h^{\downarrow}_{i+k-1}$ are represented in the same manner.

I show all the possible actions of Eq. (6.18):

$$\cdots \overline{\boxed{\pm}\,\boxed{\phantom{z}}} \cdots = \pm \cdots \boxed{\phantom{z}}\,\boxed{\pm} \cdots , \tag{6.22}$$

$$\cdots \overline{\boxed{\phantom{z}}\,\boxed{\pm}} \cdots = \pm \cdots \boxed{\pm}\,\boxed{\phantom{z}} \cdots , \tag{6.23}$$

$$\cdots \overline{\boxed{\pm}\,\boxed{\mp}} \cdots = \frac{1}{2} \left( \cdots \boxed{\phantom{z}}\,\boxed{z} \cdots - \cdots \boxed{z}\,\boxed{\phantom{z}} \cdots \right), \tag{6.24}$$

$$\cdots \overline{\boxed{z}\,\boxed{\phantom{z}}} \cdots = -2 \left( \cdots \boxed{+}\,\boxed{-} \cdots + \cdots \boxed{-}\,\boxed{+} \cdots \right), \tag{6.25}$$

$$\cdots \overline{\boxed{\phantom{z}}\,\boxed{z}} \cdots = 2 \left( \cdots \boxed{+}\,\boxed{-} \cdots + \cdots \boxed{-}\,\boxed{+} \cdots \right), \tag{6.26}$$

$$\cdots \overline{\boxed{z}\,\boxed{\pm}} \cdots = \pm \cdots \boxed{\pm}\,\boxed{z} \cdots , \tag{6.27}$$

$$\cdots \overline{\boxed{\pm}\,\boxed{z}} \cdots = \pm \cdots \boxed{z}\,\boxed{\pm} \cdots , \tag{6.28}$$

$$\cdots \overline{\boxed{\phantom{z}}\,\boxed{\phantom{z}}} \cdots = \cdots \overline{\boxed{\pm}\,\boxed{\pm}} \cdots = \cdots \overline{\boxed{z}\,\boxed{z}} \cdots = 0, \tag{6.29}$$

where I omit the lower row and the starting site. The action of Eq. (6.19) is also represented in the same way. The symbol $\circ$ in Eqs. (6.20) and (6.21) is treated in the same manner as $\boxed{\phantom{z}}$ in the above actions.

### Commutator with $H_{\text{int}}$

I also introduce the notation for the commutator with the interaction term $H_{\text{int}}$. I first denote the density of the interaction term as

$$h^{\text{int}}_l = U z_{l,\uparrow} z_{l,\downarrow} , \tag{6.30}$$

and the interaction term is written as $H_{\text{int}} = \sum_{l=1}^{L} h^{\text{int}}_l$. I represent the commutator of a basis element and the density of the interaction term by

$$\begin{matrix} & \downarrow & \\ \bar{a}_1 \cdots & \bar{a}_l & \cdots \bar{a}_k \\ \bar{b}_1 \cdots & \bar{b}_l & \cdots \bar{b}_k \\ & \uparrow & \end{matrix} (i) \equiv \left[ \begin{matrix} \bar{a}_1 \cdots \bar{a}_l \cdots \bar{a}_k \\ \bar{b}_1 \cdots \bar{b}_l \cdots \bar{b}_k \end{matrix} (i), h^{\text{int}}_{i+l-1} \right]. \tag{6.31}$$

I show the non-zero action of Eq. (6.31) as

$$\begin{matrix}\downarrow\\ \cdots\boxed{\pm}\cdots\\ \cdots\boxed{\phantom{z}}\cdots\\ \uparrow\end{matrix} = \mp U \begin{matrix}\cdots\boxed{\pm}\cdots\\ \cdots\boxed{z}\cdots\end{matrix}, \qquad \begin{matrix}\downarrow\\ \cdots\boxed{\phantom{z}}\cdots\\ \cdots\boxed{\pm}\cdots\\ \uparrow\end{matrix} = \mp U \begin{matrix}\cdots\boxed{z}\cdots\\ \cdots\boxed{\pm}\cdots\end{matrix}, \tag{6.32}$$

$$\begin{matrix}\downarrow\\ \cdots\boxed{\pm}\cdots\\ \cdots\boxed{z}\cdots\\ \uparrow\end{matrix} = \mp U \begin{matrix}\cdots\boxed{\pm}\cdots\\ \cdots\boxed{\phantom{z}}\cdots\end{matrix}, \qquad \begin{matrix}\downarrow\\ \cdots\boxed{z}\cdots\\ \cdots\boxed{\pm}\cdots\\ \uparrow\end{matrix} = \mp U \begin{matrix}\cdots\boxed{\phantom{z}}\cdots\\ \cdots\boxed{\pm}\cdots\end{matrix}, \tag{6.33}$$

and all other cases are zero. The action of Eq. (6.31) does not change the support.

**Example of commutator**

The commutator of a basis element and the Hamiltonian can be calculated by using the notations introduced above; for example

$$\begin{aligned}\left[\begin{matrix}\boxed{+}&\boxed{-}\\ \boxed{\phantom{z}}&\boxed{z}\end{matrix}(i), H\right] =& \overbrace{\begin{matrix}\dots&\boxed{+}\\ \dots&\boxed{\phantom{z}}\end{matrix}}\begin{matrix}\boxed{-}\\ \boxed{z}\end{matrix}(i) + \overbrace{\begin{matrix}\boxed{+}&\boxed{-}\\ \boxed{\phantom{z}}&\boxed{z}\end{matrix}}(i) + \begin{matrix}\boxed{+}\\ \boxed{\phantom{z}}\end{matrix}\overbrace{\begin{matrix}\boxed{-}&\dots\\ \boxed{z}&\dots\end{matrix}}(i) + \underbrace{\begin{matrix}\boxed{+}&\boxed{-}\\ \boxed{\phantom{z}}&\boxed{z}\end{matrix}}(i) + \begin{matrix}\boxed{+}\\ \boxed{\phantom{z}}\end{matrix}\underbrace{\begin{matrix}\boxed{-}&\dots\\ \boxed{z}&\dots\end{matrix}}(i)\\ &+ \begin{matrix}\downarrow&\\ \boxed{+}&\boxed{-}\\ \boxed{\phantom{z}}&\boxed{z}\\ \uparrow&\end{matrix}(i) + \begin{matrix}&\downarrow\\ \boxed{+}&\boxed{-}\\ \boxed{\phantom{z}}&\boxed{z}\\ &\uparrow\end{matrix}(i)\,,\end{aligned} \tag{6.34}$$

where each term on the right-hand side is calculated as follows:

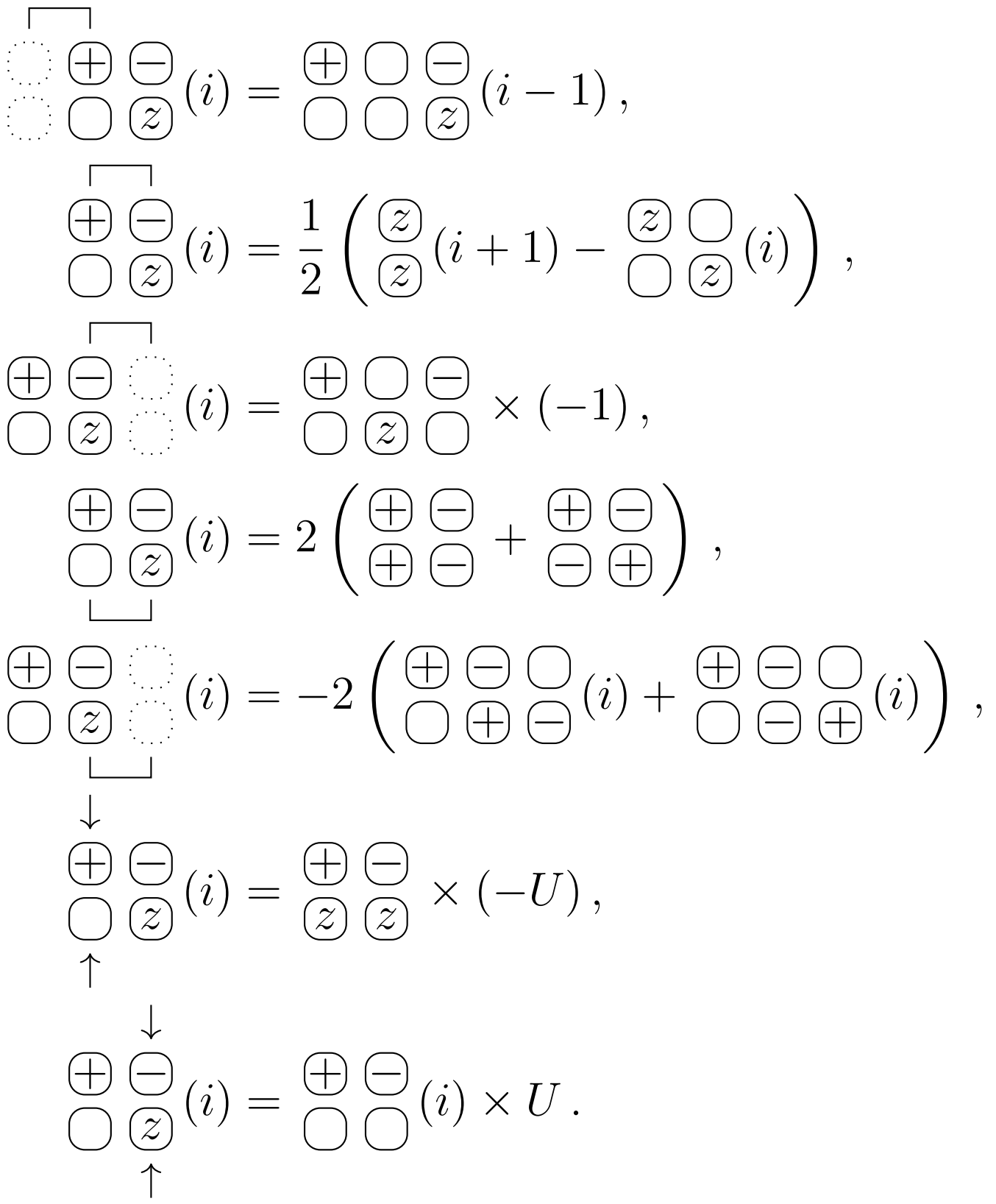

(6.35)

(6.36)

(6.37)

(6.38)

(6.39)

(6.40)

(6.41)

### 6.2.2 Graphical notation for cancellation

I assume that $F_k$ is a $k$-local conserved quantity of the one-dimensional Hubbard model. Configurations are denoted by the symbol $q$ or $p$ in the following. The $k$-local conserved quantity $F_k$ is written as a linear combination of the less-than-or-equal-to-$k$-support basis element:

$$F_k = \sum_{l=1}^{k} \sum_{q \in \mathcal{C}_l} \sum_{i=1}^{L} c_i(q) q(i), \tag{6.42}$$

where I denote the set of all $k$-support configurations by $\mathcal{C}_k$ and $c_i(q)$ is the coefficients of $q(i)$. Because $F_k$ is a conserved quantity, $F_k$ commutes with the Hamiltonian $[F_k, H] = 0$, which gives the equations for $c_i(q)$.

I represent the commutator $[F_k, H]$ by

$$[F_k, H] = \sum_{l=1}^{k+1} \sum_{\widetilde{q} \in \mathcal{C}_l} \sum_{i=1}^{L} d_i(\widetilde{q}) \widetilde{q}(i) , \tag{6.43}$$

where $\widetilde{q}(i)$ is a basis element starting from the $i$ th site, $d_i(\widetilde{q})$ is a linear combination of $\{c_i(q)\}$, which is determined from the commutation relation in Eqs. (6.22)–(6.28), (6.32) and (6.33). The upper limit of the summation over $l$ in the right-hand side of Eq. (6.43) is $k+1$ because the maximum support of the basis elements that can be generated by the commutator of a $k$-local charge and the Hamiltonian is $k+1$. To ensure the conservation of $F_k$, it is necessary that $d_i(\widetilde{q}) = 0$ for all possible $\widetilde{q}$ and $i$. This yields the equations which $\{c_i(q)\}$ must satisfy.

I consider the cancellation of $\widetilde{q}(i)$ in $[F_k, H]$ in Eq. (6.43), i.e., the equation $d_i(\widetilde{q}) = 0$. I first explain how I obtain $d_i(\widetilde{q})$ in terms of $\{c_i(q)\}$. Let $q_1(i_1), \dots, q_m(i_m)$ be the basis elements in $F_k$ such that $[q_l(i_l), H]$ for $l = 1, \dots, m$ generate $\widetilde{q}(i)$ and there is no other contribution to $d_i(\widetilde{q})$. There exists only one $h_{v_l}^{s_l}$ ($s_l \in \{\uparrow, \downarrow, \mathrm{int}\}$, $v_l \in \{1, \dots, L\}$) for $q_l(i_l)$ that satisfies

$$\left[q_l(i_l), h_{v_l}^{s_l}\right] = f_l \widetilde{q}(i) + (\text{rest})\,, \tag{6.44}$$

where $f_l$ is the non-zero factor determined from the commutation relation in Eqs. (6.22)–(6.28), (6.32) and (6.33). The term (rest) is the other basis element generated by the commutator in the case of Eqs. (6.24)–(6.26), where the commutator generates two basis elements. The term $(\text{rest}) = 0$ in the other cases. Then, I have

$$d_i(\widetilde{q}) = \sum_{l=1}^{m} f_l c_{i_l}(q_l)\,, \tag{6.45}$$

and the equation for the cancellation of $\widetilde{q}(i)$ becomes

$$\sum_{l=1}^{m} f_l c_{i_l}(q_l) = 0. \tag{6.46}$$

The condition for the cancellation (6.46) follows from, and will be associated with, the following diagram, which encodes the basis elements $q_1(i_1), \dots, q_m(i_m)$ that yield $\widetilde{q}(i)$ after commutation with the Hamiltonian:

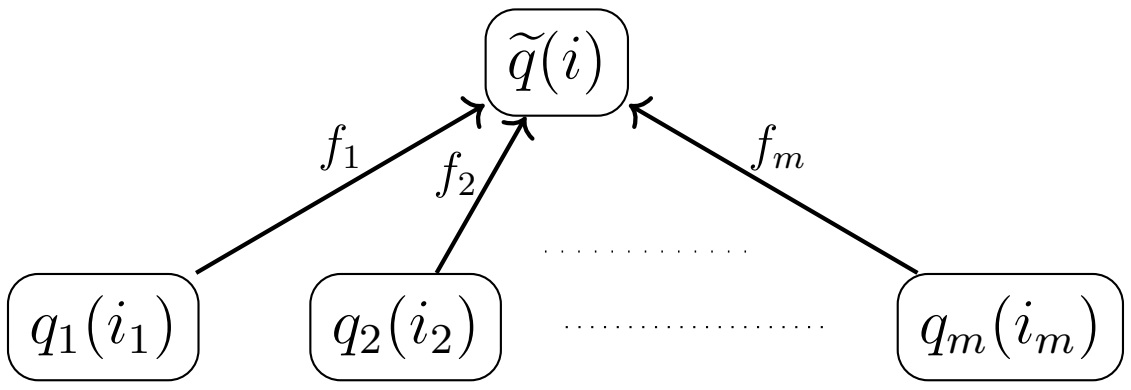

(6.47)

where the node boxes at the starting point of arrows denote the operators that contribute to the cancellation of $\widetilde{q}(i)$ and the factor $f_l$ is depicted in the middle of the arrows. The factor $f_l$ in Eq. (6.47) is often omitted in the following.

in the case of $m = 1$, (6.46) becomes $f_1 c_{i_1}(q_1) = 0$, and I have $c_{i_1}(q_1) = 0$ because of $f_1 \neq 0$. The $m = 1$ case is represented graphically as

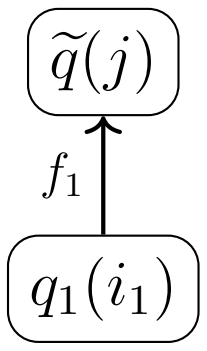

$$\Longrightarrow c_{i_1}(q_1) = 0. \tag{6.48}$$

In the case of $m = 2$, Eq. (6.46) becomes $f_1 c_{i_1}(q_1) + f_2 c_{i_2}(q_2) = 0$, and I have $c_{i_1}(q_1) = f c_{i_2}(q_2)$ ($f = -f_2/f_1$) because of $f_1, f_2 \neq 0$. The case $m = 2$ is represented graphically as

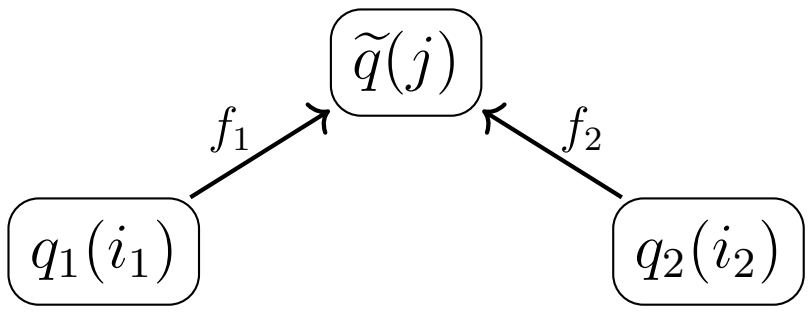

$$\Longrightarrow c_{i_1}(q_1) = f c_{i_2}(q_2). \tag{6.49}$$

In the following, I represent an equation $c_{i_1}(q_1) = f c_{i_2}(q_2)$ for the cancellation in the $m = 2$ case simply as

$$c_{i_1}(q_1) \propto c_{i_2}(q_2)\,. \tag{6.50}$$

Note that if $c_{i_1}(q_1) \propto c_{i_2}(q_2)$ and $c_{i_2}(q_2) = 0$ holds, then one can see $c_{i_1}(q_1) = 0$.

I also introduce an additional notation that represents the node in the graphical notation:

$$\overset{q(i)}{\boxed{\begin{array}{l} \bar{a}_1\ \bar{a}_2 \cdots \bar{a}_k \\ \bar{b}_1\ \bar{b}_2 \cdots \bar{b}_k \end{array}}} \equiv \boxed{\begin{array}{l} \bar{a}_1\ \bar{a}_2 \cdots \bar{a}_k \\ \bar{b}_1\ \bar{b}_2 \cdots \bar{b}_k \end{array} \text{(i)}} \tag{6.51}$$

where $q$ denotes the following configuration:

$$q = \begin{array}{l} \bar{a}_1\ \bar{a}_2 \cdots \bar{a}_k \\ \bar{b}_1\ \bar{b}_2 \cdots \bar{b}_k \end{array}. \tag{6.52}$$

## 6.3 Proof of completeness of $Q_k$

In this section, I prove Theorem 6.4. I write a $k$-local conserved quantity $F_k$ as

$$F_k = F_k^k + F_k^{k-1} + \text{(rest)}, \tag{6.53}$$

where $F_k^k$ is the $k$-support operator in $F_k$, $F_k^{k-1}$ is the $(k-1)$-support operator in $F_k$, and (rest) is the linear combination of the less-than-or-equal-to-$(k-2)$-support operators in $F_k$. For the proof of Theorem 6.4, it suffices to consider only $F_k^k$ and $F_k^{k-1}$. I determine $F_k^k$ so that $F_k$ can satisfy $[F_k, H] = 0$.

From $[F_k, H]\big|_{k+1} = 0$ and $[F_k, H]\big|_k = 0$, I have the following equation for the cancellation of $(k+1)$ and $k$-support operators in $[F_k, H]$:

$$\left[F_k^k, H_0\right]\Big|_{k+1} = 0, \tag{6.54}$$

$$\left[F_k^k, H_0\right]\Big|_k + \left[F_k^{k-1}, H_0\right]\Big|_k + \left[F_k^k, H_{\text{int}}\right] = 0. \tag{6.55}$$

From $[F_k, H]\big|_{k+1} = 0$ and $[F_k, H]\big|_{k} = 0$, I have the following equation for the cancellation of $(k+1)$ and $k$-support operators in $[F_k, H]$:

$$\left[F_k^k, H_0\right]\Big|_{k+1} = 0, \tag{6.56}$$

$$\left[F_k^k, H_0\right]\Big|_{k} + \left[F_k^{k-1}, H_0\right]\Big|_{k} + \left[F_k^k, H_{\text{int}}\right] = 0. \tag{6.57}$$

I note that the following argument is applicable to the maximal support is less than half of the system size: $k < \lfloor \frac{L}{2} \rfloor$, because otherwise there are other contributions to the cancellation of operators beyond what I give below.

### 6.3.1 Cancellation of $(k+1)$-support operator

In the following, I consider the cancellation of $(k+1)$-support operators (6.56). The term $F_k^k$ is written as

$$F_k^k = \sum_{q \in \mathcal{C}_k} \sum_{i=1}^{L} c_i(q) q(i). \tag{6.58}$$

The configuration $q \in \mathcal{C}_k$ is represented as $q = \begin{matrix} \bar{a}_1 & \bar{a}_2 \cdots & \bar{a}_k \\ \bar{b}_1 & \bar{b}_2 \cdots & \bar{b}_k \end{matrix}$, where either of $\bar{a}_1$ or $\bar{b}_1$ and either of $\bar{a}_k$ or $\bar{b}_k$ are not $\bigcirc$. I determine $c_i(q)$ to satisfy the cancellation of the $(k+1)$-support operators (6.56). I note that the nodes at the starting point of the arrows below represent $k$-support operators, as long as I consider the cancellation of Eq. (6.56).

One can see the coefficients are zero if the column at the right end or left end is filled for both upper and lower rows, such as

$$c_i \begin{pmatrix} \oplus \cdots \\ \ominus \cdots \end{pmatrix} = 0\,, \tag{6.59}$$

and

$$c_i \begin{pmatrix} \cdots \textcircled{z} \\ \cdots \oplus \end{pmatrix} = 0\,, \tag{6.60}$$

which is stated in the following Lemma.

**Lemma 6.6.** *If $q \in \mathcal{C}_k$ satisfies $\bigcirc \notin \{\bar{a}_1, \bar{b}_1\}$ or $\bigcirc \notin \{\bar{a}_k, \bar{b}_k\}$, then $c_i(q) = 0$ holds.*

*Proof.* Because either of $\bar{a}_k$ or $\bar{b}_k$ is not $\bigcirc$, I consider the case of $\bar{a}_k \neq \bigcirc$ first. If $\bigcirc \notin \{\bar{a}_1, \bar{b}_1\}$,

I have the following cancellations of $(k+1)$-support basis element $\widetilde{q}(i)$:

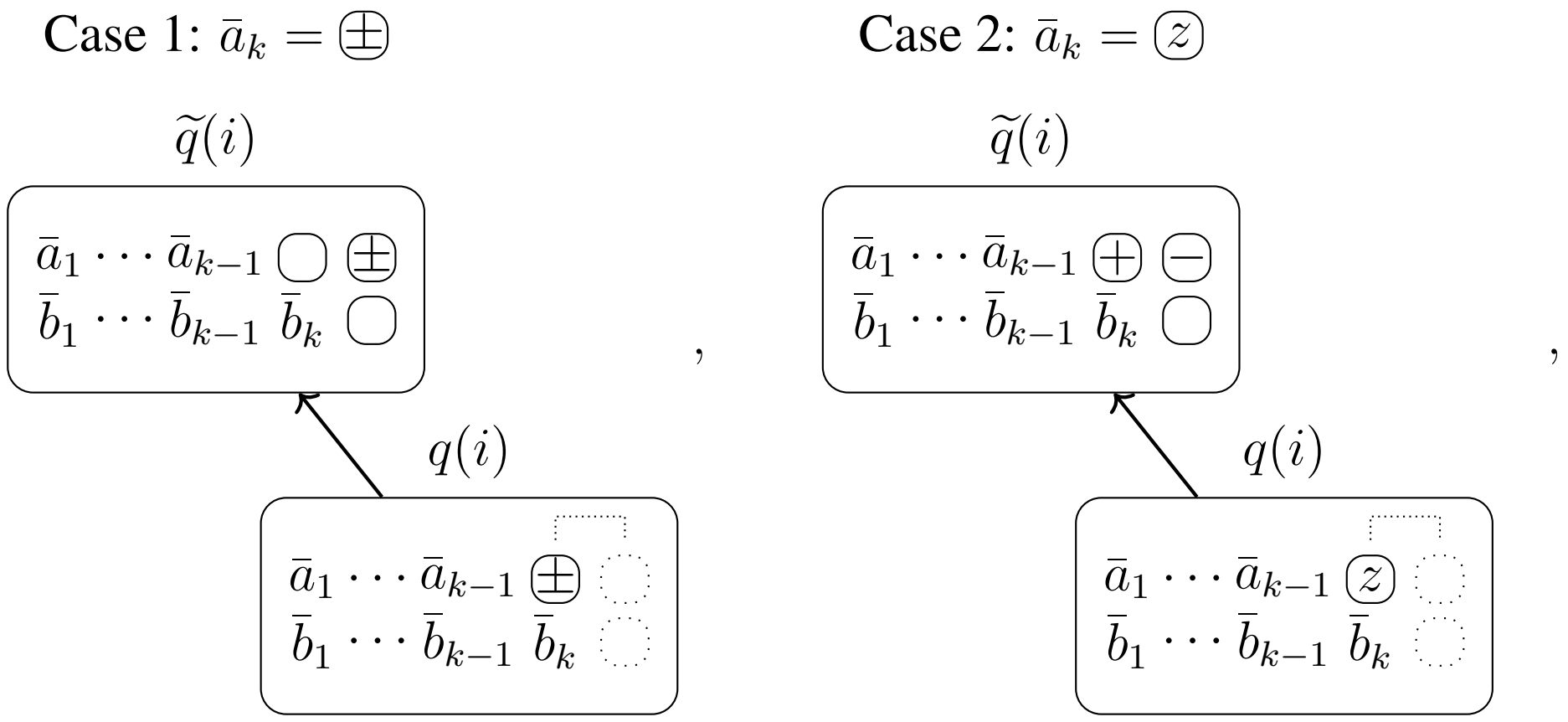

(6.61)

where the commutator that generates $\widetilde{q}$ is indicated by the dotted line as a guide. From Eq. (6.48), I have $c_i(q) = 0$ in both cases.

I have proved $c_i(q) = 0$ if $q$ satisfies $\bigcirc \notin \{\bar{a}_1, \bar{b}_1\}$ and $\bar{a}_k \neq \bigcirc$. Given the same argument holds when the roles of $\bar{a}$ and $\bar{b}$ are interchanged, I have also proved $c_i(q) = 0$ if $q$ satisfies $\bigcirc \notin \{\bar{a}_1, \bar{b}_1\}$ and $\bar{b}_k \neq \bigcirc$. Thus I have proved $c_i(q) = 0$ if $q$ satisfies $\bigcirc \notin \{\bar{a}_1, \bar{b}_1\}$. In the same way, one can also prove $c_i(q) = 0$ if $q$ satisfies $\bigcirc \notin \{\bar{a}_k, \bar{b}_k\}$. $\square$

One can prove that the coefficients are zero if $\textcircled{z}$ is included in the column at the right end or left end, such as

$$c_i \begin{pmatrix} \textcircled{z} \cdots \\ \bigcirc \cdots \end{pmatrix} = 0\,, \tag{6.62}$$

which is stated in the following Lemma.

**Lemma 6.7.** *If $q \in \mathcal{C}_k$ satisfies $\textcircled{z} \in \{\bar{a}_1, \bar{b}_1, \bar{a}_k, \bar{b}_k\}$, then $c_i(q) = 0$ holds.*

*Proof.* Because either of $\bar{a}_k$ or $\bar{b}_k$ is not $\bigcirc$, I consider the case of $\bar{a}_k \neq \bigcirc$ first. If $q \in \mathcal{C}_k$ satisfies $\bar{a}_1 = \textcircled{z}$, the following relation for the cancellation of $(k+1)$-support basis element $\widetilde{q}(i)$ hold:

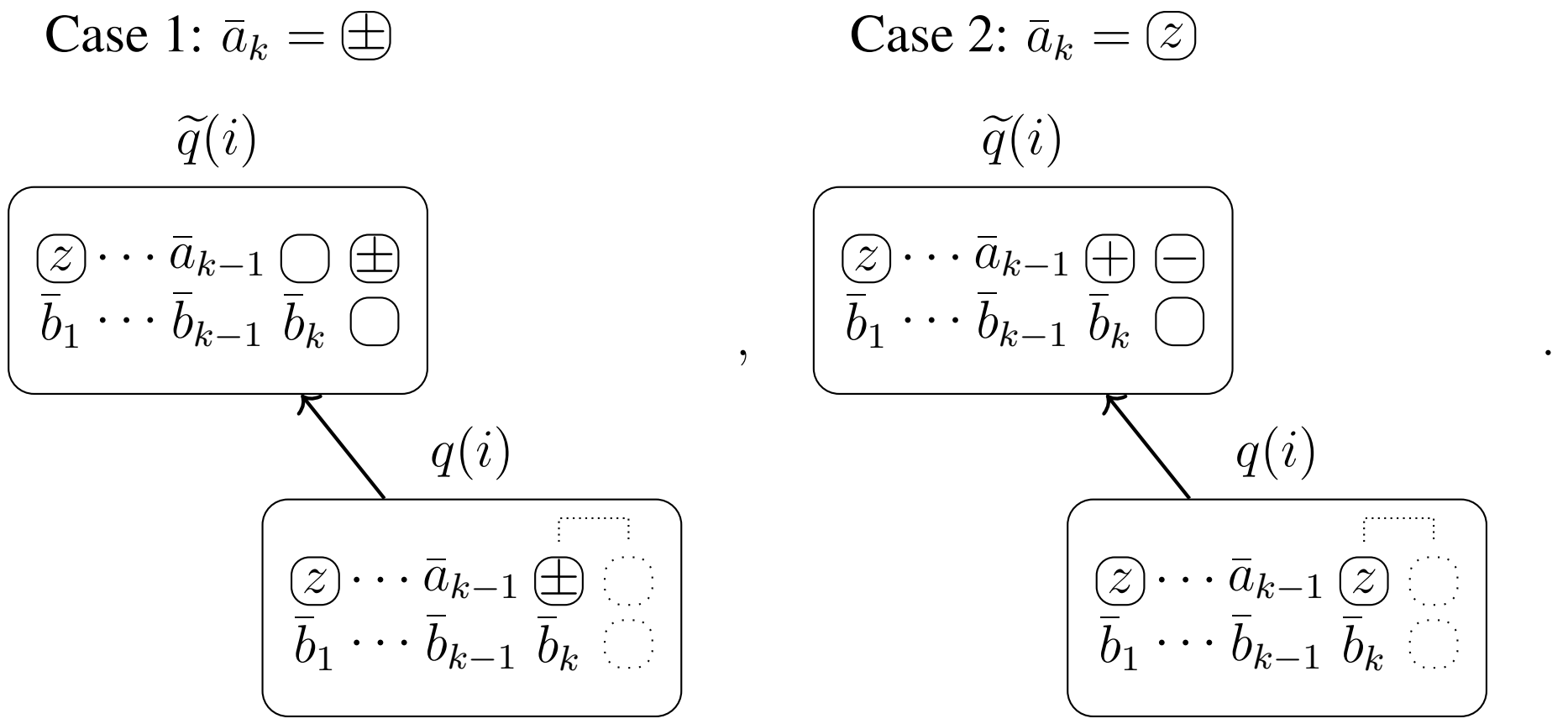

(6.63)

From Eq. (6.48), I have $c_i(q) = 0$ in both cases. In the same way, one can prove $c_i(q) = 0$ if $q$ satisfies $\bar{a}_1 = \textcircled{z}$ and $\bar{b}_k \neq \bigcirc$. Thus I have proved $c_i(q) = 0$ if $q$ satisfies $\bar{a}_1 = \textcircled{z}$.

The same argument holds when the roles of $\bar{a}$ and $\bar{b}$ are interchanged, and then one can also prove $c_i(q) = 0$ if $q$ satisfies $\bar{b}_1 = \textcircled{z}$. Thus I have proved $c_i(q) = 0$ if $q$ satisfies $\textcircled{z} \in \{\bar{a}_1, \bar{b}_1\}$. In the same way, one can also prove $c_i(q) = 0$ if $q$ satisfies $\textcircled{z} \in \{\bar{a}_k, \bar{b}_k\}$. □

One can prove that the coefficients are zero if the second column from the left end is not empty for both rows, such as

$$c_i \begin{pmatrix} \oplus & \bigcirc & \cdots \\ \bigcirc & \ominus & \cdots \end{pmatrix} = 0\,, \tag{6.64}$$

which is stated in the following Lemma.

**Lemma 6.8.** *If $q \in \mathcal{C}_k (k \geq 3)$ satisfies $\{\bar{a}_2, \bar{b}_2\} \neq \{\bigcirc, \bigcirc\}$, then $c_i(q) = 0$ holds.*

*Proof.* If $\textcircled{z} \in \{\bar{a}_1, \bar{b}_1, \bar{a}_k, \bar{b}_k\}$ or $\bigcirc \notin \{\bar{a}_1, \bar{b}_1\}$ or $\bigcirc \notin \{\bar{a}_k, \bar{b}_k\}$, one can see $c_i(q) = 0$ holds from Lemma 6.6 and 6.7. Thus, the nontrivial cases are $\begin{matrix}\bar{a}_1 \\ \bar{b}_1\end{matrix}, \begin{matrix}\bar{a}_k \\ \bar{b}_k\end{matrix} \in \left\{ \begin{matrix}\oplus \\ \bigcirc\end{matrix}, \begin{matrix}\bigcirc \\ \oplus\end{matrix}, \begin{matrix}\ominus \\ \bigcirc\end{matrix}, \begin{matrix}\bigcirc \\ \ominus\end{matrix} \right\}$. First, I consider the cases $\begin{matrix}\bar{a}_1 \\ \bar{b}_1\end{matrix} = \begin{matrix}\oplus \\ \bigcirc\end{matrix}$, $\begin{matrix}\bar{a}_k \\ \bar{b}_k\end{matrix} = \begin{matrix}\oplus \\ \bigcirc\end{matrix}$. If $q \in \mathcal{C}_k$ satisfies $\bar{b}_2 \neq \bigcirc$, I have the following cancellations of the $(k+1)$-support basis element $\widetilde{q}(i)$:

Case 1: $\bar{a}_2 = \bigcirc$ Case 2: $\bar{a}_2 = \textcircled{z}$

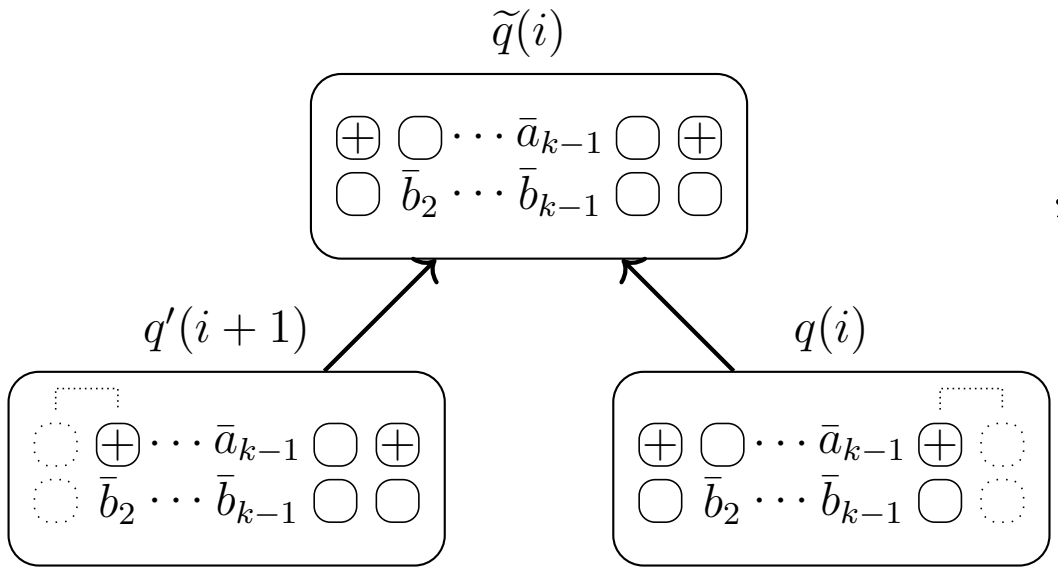

,
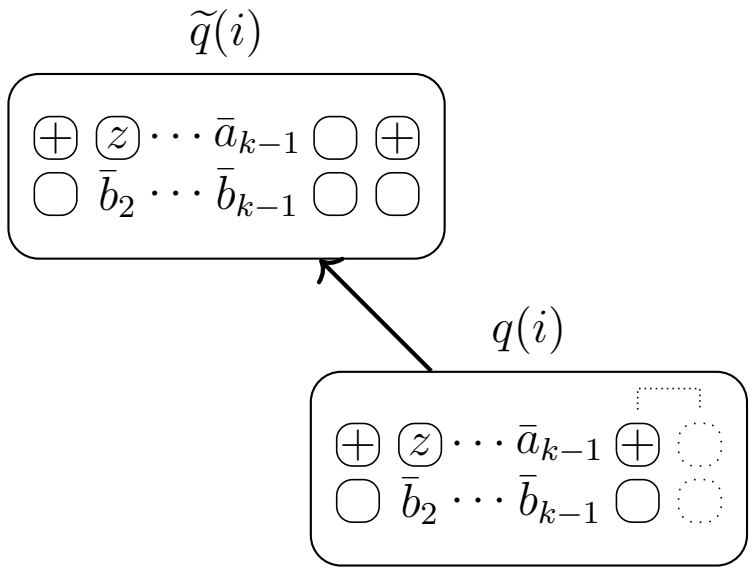

, (6.65)

Case 3: $\bar{a}_2 = \ominus$ Case 4: $\bar{a}_2 = \oplus$

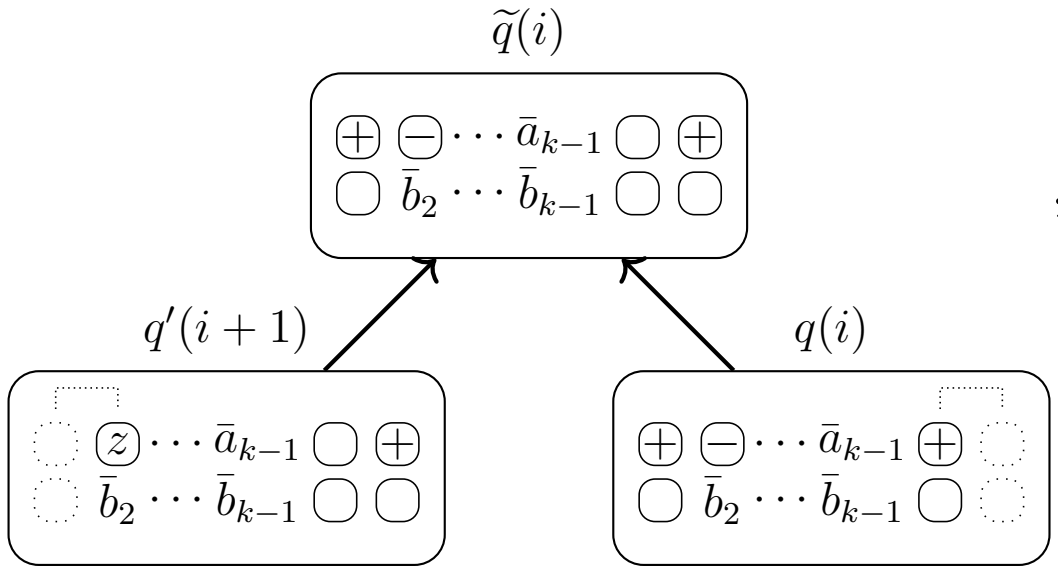

,
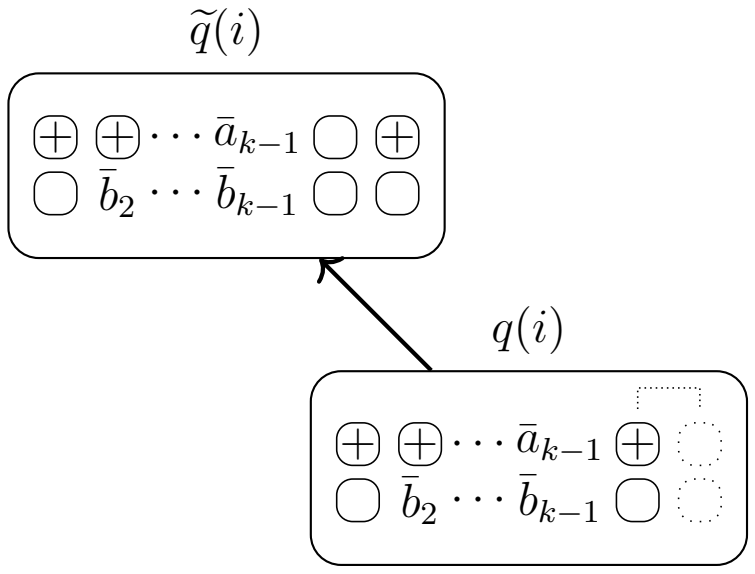

. (6.66)

In the cases 2 and 4, I have $c_i(q) = 0$ from Eq. (6.48). In the cases 1 and 3, I have $c_i(q) \propto c_{i+1}(q') = 0$ from Eq. (6.49) and Lemma 6.6.

One can also do the same argument for the other choices of $\frac{\bar{a}_1}{\bar{b}_1}$ and $\frac{\bar{a}_k}{\bar{b}_k}$. Then I have proved $c_i(q) = 0$ if $\bar{b}_2 \neq \bigcirc$. Given that the same argument holds when the roles of $\bar{a}$ and $\bar{b}$ are interchanged, I have also proved $c_i(q) = 0$ if $\bar{a}_2 \neq \bigcirc$. □

One can prove that the coefficients are zero if a middle column is not empty for both rows, such as

$$c_i \begin{pmatrix} \oplus \cdots \bigcirc \cdots \ominus \\ \bigcirc \cdots \ominus \cdots \bigcirc \end{pmatrix} = 0\,, \tag{6.67}$$

which is stated in the following Lemma.

**Lemma 6.9.** *If there exists an integer $l$ with $2 \leq l \leq k-1$ such that $\bar{a}_l \neq \bigcirc$ or $\bar{b}_l \neq \bigcirc$ for $q \in \mathcal{C}_k$, then $c_i(q) = 0$ holds.*

*Proof.* If $\textcircled{z} \in \{\bar{a}_1, \bar{b}_1, \bar{a}_k, \bar{b}_k\}$ or $\bigcirc \notin \{\bar{a}_1, \bar{b}_1\}$ or $\bigcirc \notin \{\bar{a}_k, \bar{b}_k\}$, one can see $c_i(q) = 0$ holds from Lemma 6.6 and 6.7. Thus, the nontrivial cases are $\frac{\bar{a}_1}{\bar{b}_1}, \frac{\bar{a}_k}{\bar{b}_k} \in \left\{ \frac{\oplus}{\bigcirc}, \frac{\bigcirc}{\oplus}, \frac{\ominus}{\bigcirc}, \frac{\bigcirc}{\ominus} \right\}$. First, I consider the cases $\frac{\bar{a}_1}{\bar{b}_1} = \frac{\oplus}{\bigcirc}, \ \frac{\bar{a}_k}{\bar{b}_k} = \frac{\oplus}{\bigcirc}$. The case $l = 2$ corresponds to the Lemma 6.8, and thus I consider the case $l > 2$ below. If there exist $l (2 < l \leq k-1)$ such that $\{\bar{a}_l, \bar{b}_l\} \neq \{\bigcirc, \bigcirc\}$ and $\bar{a}_m, \bar{b}_m = \bigcirc$ for $(2 \leq m \leq l-1)$, I have the following cancellations of $(k+1)$-support basis element $\widetilde{q}_m(i+m-1)$ $(1 \leq m \leq l-2)$.

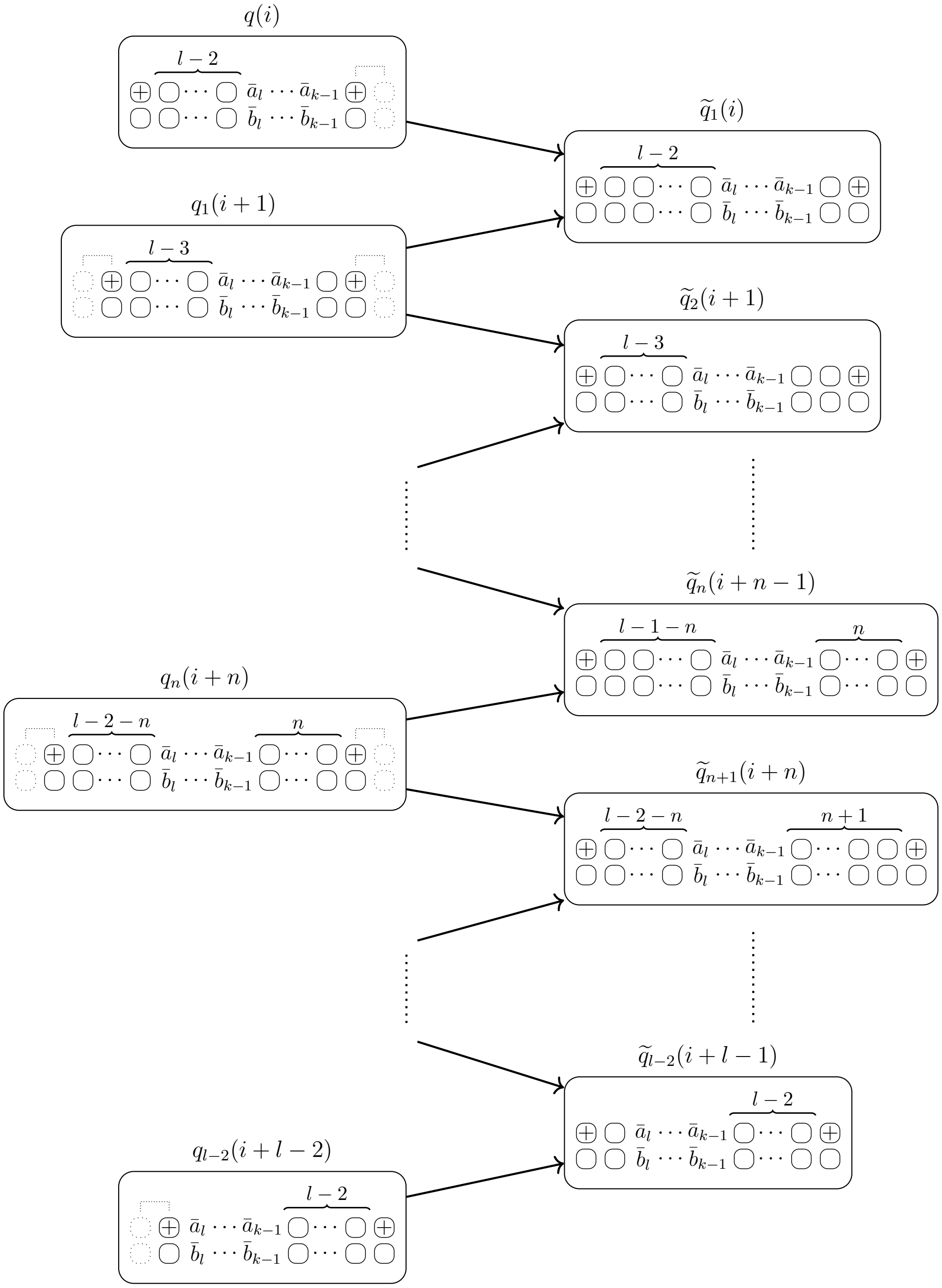

(6.68)

From Eq. (6.49), I have

$$c_i(q) \propto c_{i+1}(q_1) \propto \cdots \propto c_{i+l-2}(q_{l-2}). \tag{6.69}$$

One can see $c_{i+l-2}(q_{l-2}) = 0$ from Lemma 6.8 because of $\left\{\bar{a}_l, \bar{b}_l\right\} \neq \{\bigcirc, \bigcirc\}$. Thus I have $c_i(q) = 0$.

One can also make the same argument for the other choices of $\begin{matrix}\bar{a}_1\\\bar{b}_1\end{matrix}$ and $\begin{matrix}\bar{a}_k\\\bar{b}_k\end{matrix}$. Then I have proved Lemma 6.9. $\square$

From Lemmas 6.6–6.9, one can see $c_i(q) = 0$ $(q \in \mathcal{C}_k)$ unless $q$ is the following form:

$$q = \begin{matrix} \bar{a}_1 & \overbrace{\bigcirc\cdots\bigcirc}^{k-2} & \bar{a}_k \\ \bar{b}_1 & \bigcirc\cdots\bigcirc & \bar{b}_k \end{matrix}, \tag{6.70}$$

where $\begin{matrix}\bar{a}_1\\ \bar{b}_1\end{matrix}, \begin{matrix}\bar{a}_k\\ \bar{b}_k\end{matrix} \in \left\{ \begin{matrix}\boxed{\pm}\\ \bigcirc\end{matrix}, \begin{matrix}\bigcirc\\ \boxed{\pm}\end{matrix} \right\}$. I next study the coefficients of the basis element of these configurations.

**Lemma 6.10.** $c_i\left( \begin{matrix} \boxed{\pm} & \overbrace{\bigcirc\cdots\bigcirc}^{k-2} & \boxed{\mp} \\ \bigcirc & \bigcirc\cdots\bigcirc & \bigcirc \end{matrix} \right)$, $c_i\left( \begin{matrix} \boxed{\pm} & \overbrace{\bigcirc\cdots\bigcirc}^{k-2} & \bigcirc \\ \bigcirc & \bigcirc\cdots\bigcirc & \boxed{\mp} \end{matrix} \right)$, $c_i\left( \begin{matrix} \bigcirc & \overbrace{\bigcirc\cdots\bigcirc}^{k-2} & \bigcirc \\ \boxed{\pm} & \bigcirc\cdots\bigcirc & \boxed{\mp} \end{matrix} \right)$ *and* $c_i\left( \begin{matrix} \bigcirc & \overbrace{\bigcirc\cdots\bigcirc}^{k-2} & \boxed{\mp} \\ \boxed{\pm} & \bigcirc\cdots\bigcirc & \bigcirc \end{matrix} \right)$ *are constants independent of* $i$.

*Proof.* I prove that $c_i\left( \begin{matrix} \boxed{\pm} & \overbrace{\bigcirc\cdots\bigcirc}^{k-2} & \boxed{\mp} \\ \bigcirc & \bigcirc\cdots\bigcirc & \bigcirc \end{matrix} \right)$ is independent of $i$. I consider the following cancellation of the $(k+1)$-support basis element $\widetilde{q}(i)$

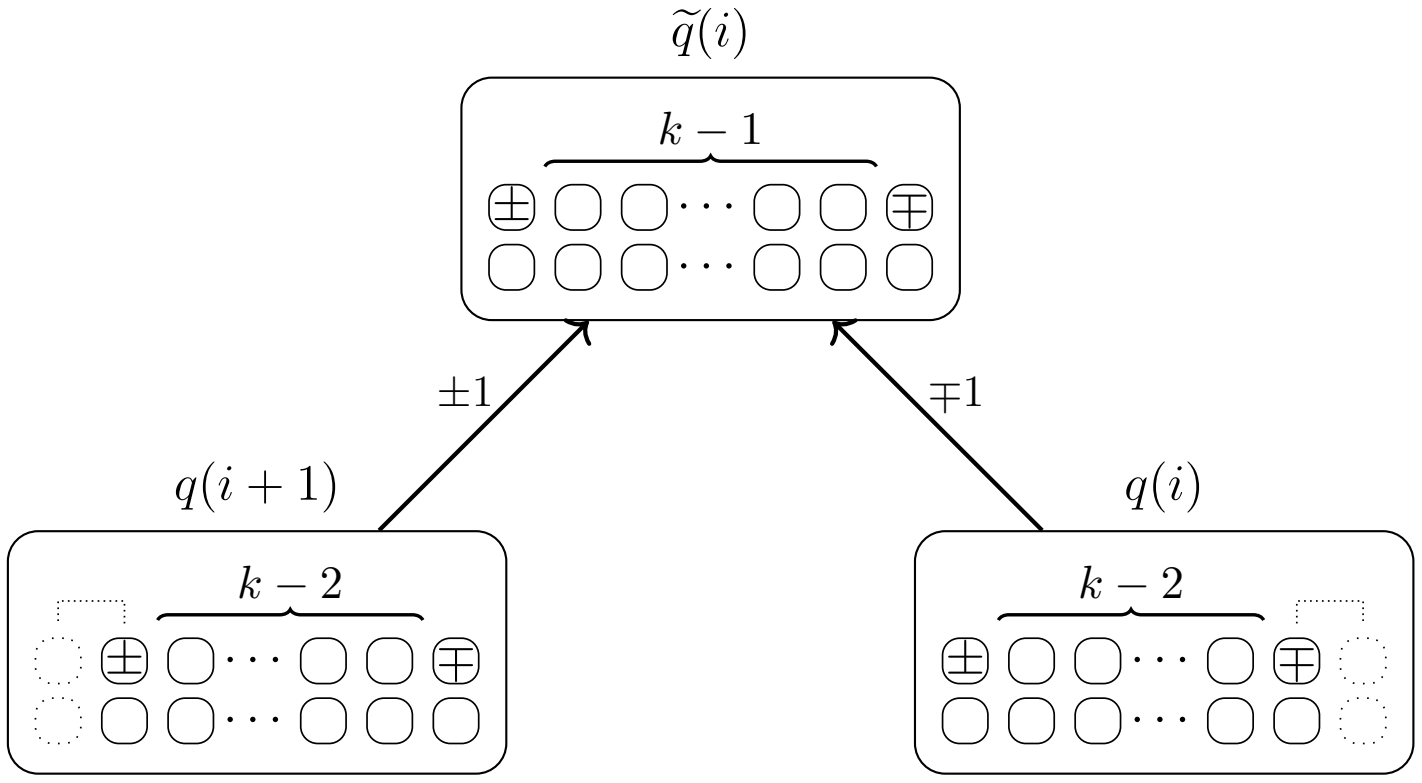

(6.71)

and I have

$$c_i\left( \begin{matrix} \boxed{\pm} & \overbrace{\bigcirc\cdots\bigcirc}^{k-2} & \boxed{\mp} \\ \bigcirc & \bigcirc\cdots\bigcirc & \bigcirc \end{matrix} \right) = c_{i+1}\left( \begin{matrix} \boxed{\pm} & \overbrace{\bigcirc\cdots\bigcirc}^{k-2} & \boxed{\mp} \\ \bigcirc & \bigcirc\cdots\bigcirc & \bigcirc \end{matrix} \right). \tag{6.72}$$

Thus, I have proved that $c_i\left( \begin{matrix} \boxed{\pm} & \overbrace{\bigcirc\cdots\bigcirc}^{k-2} & \boxed{\mp} \\ \bigcirc & \bigcirc\cdots\bigcirc & \bigcirc \end{matrix} \right)$ is independent of $i$. The other cases can be also proved in the same way. □

**Lemma 6.11.** $c_i\left( \begin{matrix} \boxed{\pm} & \overbrace{\bigcirc\cdots\bigcirc}^{k-2} & \boxed{\pm} \\ \bigcirc & \bigcirc\cdots\bigcirc & \bigcirc \end{matrix} \right)$, $c_i\left( \begin{matrix} \boxed{\pm} & \overbrace{\bigcirc\cdots\bigcirc}^{k-2} & \bigcirc \\ \bigcirc & \bigcirc\cdots\bigcirc & \boxed{\pm} \end{matrix} \right)$, $c_i\left( \begin{matrix} \bigcirc & \overbrace{\bigcirc\cdots\bigcirc}^{k-2} & \bigcirc \\ \boxed{\pm} & \bigcirc\cdots\bigcirc & \boxed{\pm} \end{matrix} \right)$ *and* $c_i\left( \begin{matrix} \bigcirc & \overbrace{\bigcirc\cdots\bigcirc}^{k-2} & \boxed{\pm} \\ \boxed{\pm} & \bigcirc\cdots\bigcirc & \bigcirc \end{matrix} \right)$ *are zero for the odd* $L$ *case and are* $(\text{some constant}) \times (-1)^i$ *for the even* $L$ *case.*

*Proof.* I give the proof for $c_i\left(\begin{matrix} \boxplus & \overbrace{\bigcirc \cdots \bigcirc}^{k-2} & \boxplus \\ \bigcirc & \bigcirc \cdots \bigcirc & \bigcirc \end{matrix}\right)$. I consider the following cancellation of the $(k+1)$-support basis element $\widetilde{q}(i)$

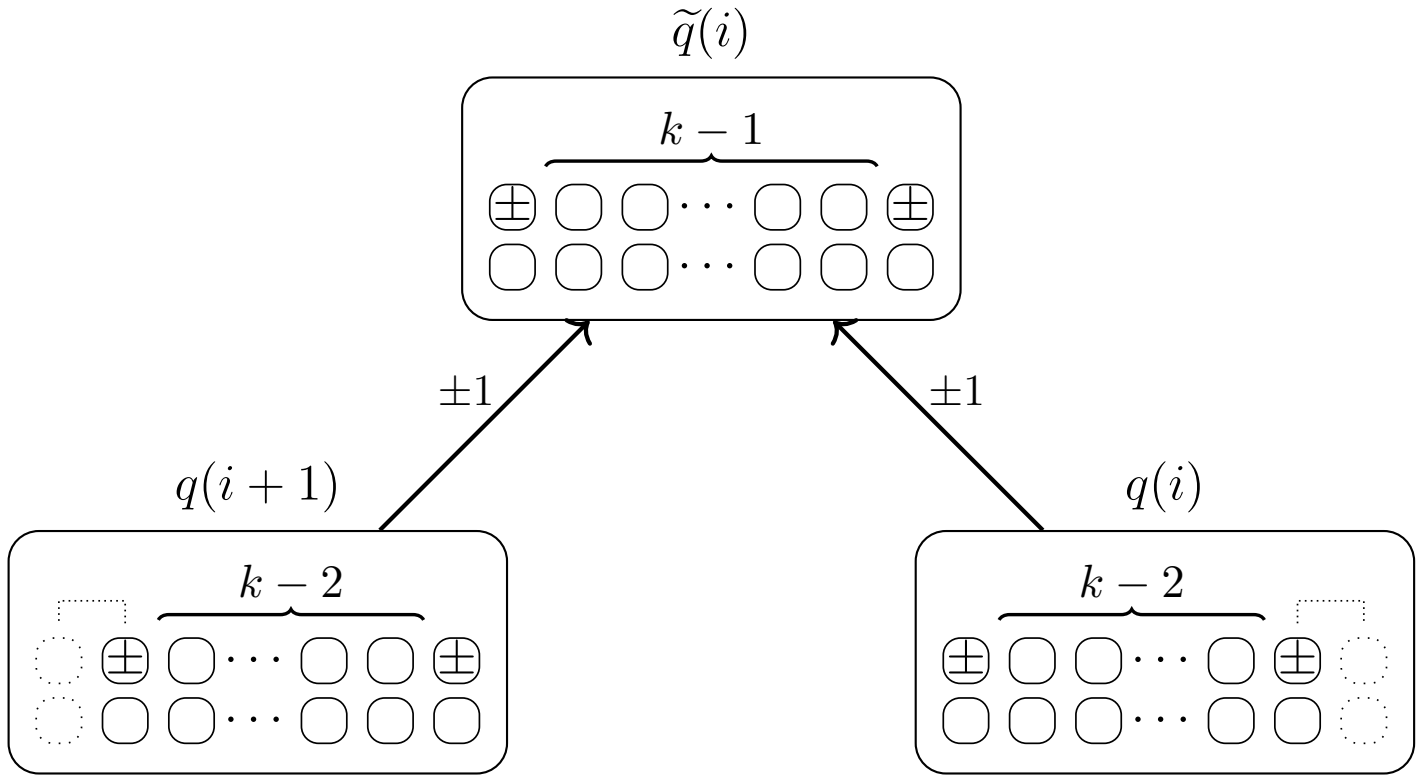

(6.73)

and I have

$$c_i\left(\begin{matrix} \boxplus & \overbrace{\bigcirc \cdots \bigcirc}^{k-2} & \boxplus \\ \bigcirc & \bigcirc \cdots \bigcirc & \bigcirc \end{matrix}\right) = -c_{i+1}\left(\begin{matrix} \boxplus & \overbrace{\bigcirc \cdots \bigcirc}^{k-2} & \boxplus \\ \bigcirc & \bigcirc \cdots \bigcirc & \bigcirc \end{matrix}\right). \tag{6.74}$$

Because of the periodic boundary condition, I have

$$c_i\left(\begin{matrix} \boxplus & \overbrace{\bigcirc \cdots \bigcirc}^{k-2} & \boxplus \\ \bigcirc & \bigcirc \cdots \bigcirc & \bigcirc \end{matrix}\right) = (-1)^L c_i\left(\begin{matrix} \boxplus & \overbrace{\bigcirc \cdots \bigcirc}^{k-2} & \boxplus \\ \bigcirc & \bigcirc \cdots \bigcirc & \bigcirc \end{matrix}\right). \tag{6.75}$$

Thus, I have proved that $c_i\left(\begin{matrix} \boxplus & \overbrace{\bigcirc \cdots \bigcirc}^{k-2} & \boxplus \\ \bigcirc & \bigcirc \cdots \bigcirc & \bigcirc \end{matrix}\right)$ is zero for odd $L$ and is $(\text{some constant}) \times (-1)^i$ for even $L$ case. The other cases can be also proved in the same way. $\square$

From Lemma 6.10 and Lemma 6.11, one can write $F_k^k$ as

$$\begin{aligned} F_k^k = & \sum_{\sigma,\mu \in \{\uparrow,\downarrow\}} \sum_{s\in\{+,-\}} \alpha_{(\sigma,s),(\mu,-s)} \sum_{i=1}^{L} c_{i,\sigma}^{s} c_{i+k-1,\mu}^{-s} \\ & + \sum_{\sigma,\mu \in \{\uparrow,\downarrow\}} \sum_{s\in\{+,-\}} \alpha_{(\sigma,s),(\mu,s)} \sum_{i=1}^{L} (-1)^i c_{i,\sigma}^{s} c_{i+k-1,\mu}^{s}, , \end{aligned} \tag{6.76}$$

where $c_{i,\sigma}^{+} \equiv c_{i,\sigma}^{\dagger}$, $c_{i,\sigma}^{-} \equiv c_{i,\sigma}$, $-s \equiv \mp$ for $s = \pm$, $\alpha_{(\sigma,s),(\mu,-s)}$ is arbitrary constant, $\alpha_{(\sigma,s),(\mu,s)}$ is arbitrary constant for even $L$ and $\alpha_{(\sigma,s),(\mu,s)} = 0$ for odd $L$.

Note that $F_k^k$ commutes with $H_0$: $\left[F_k^k, H_0\right] = 0$. Therefore, the equation for the cancellation of the $k$-support operators (6.57) can be reduced to

$$\left[F_k^{k-1}, H_0\right]\Big|_k + \left[F_k^k, H_{\text{int}}\right] = 0. \tag{6.77}$$

### 6.3.2 Cancellation of $k$-support operators

In the following, I determine $\alpha_{(\sigma,s),(\mu,s')}$ to satisfy the cancellation of $k$-support operators (6.77). I let the $(k-1)$-support configuration denoted by the symbol $p$ below. The term $F_k^{k-1}$ is written as

$$F_k^{k-1} = \sum_{p\in\mathcal{C}_{k-1}} \sum_{i=1}^{L} c_i(p)p(i), \tag{6.78}$$

where $p \in \mathcal{C}_{k-1}$ is denoted as $p = \begin{matrix} \bar{a}_1 & \bar{a}_2 \cdots \bar{a}_{k-1} \\ \bar{b}_1 & \bar{b}_2 \cdots \bar{b}_{k-1} \end{matrix}$ and either of $\bar{a}_1$ or $\bar{b}_1$ and either of $\bar{a}_{k-1}$ or $\bar{b}_{k-1}$ are not $\square$ because $p \in \mathcal{C}_{k-1}$ is $(k-1)$-support configuration.

From Lemma 6.10 and Lemma 6.11, one can calculate explicitly $\left[F_k^k, H_{\text{int}}\right]$ as

$$\begin{aligned}\left[F_k^k, H_{\text{int}}\right] = -U \sum_{s,t\in\{+,-\}} \Bigg\{ & \alpha_{(\uparrow,s),(\uparrow,t)} \left( s \begin{matrix} \boxed{s} & \overbrace{\square\cdots\square}^{k-2} & \boxed{t} \\ \boxed{z} & \square\cdots\square & \square \end{matrix} + t \begin{matrix} \boxed{s} & \overbrace{\square\cdots\square}^{k-2} & \boxed{t} \\ \square & \square\cdots\square & \boxed{z} \end{matrix} \right) \\
& + \alpha_{(\downarrow,s),(\downarrow,t)} \left( s \begin{matrix} \boxed{z} & \overbrace{\square\cdots\square}^{k-2} & \square \\ \boxed{s} & \square\cdots\square & \boxed{t} \end{matrix} + t \begin{matrix} \square & \overbrace{\square\cdots\square}^{k-2} & \boxed{z} \\ \boxed{s} & \square\cdots\square & \boxed{t} \end{matrix} \right) + \alpha_{(\uparrow,s),(\downarrow,t)} \left( s \begin{matrix} \boxed{s} & \overbrace{\square\cdots\square}^{k-2} & \square \\ \boxed{z} & \square\cdots\square & \boxed{t} \end{matrix} + t \begin{matrix} \boxed{s} & \overbrace{\square\cdots\square}^{k-2} & \boxed{z} \\ \square & \square\cdots\square & \boxed{t} \end{matrix} \right) \\
& + \alpha_{(\downarrow,s),(\uparrow,t)} \left( s \begin{matrix} \boxed{z} & \overbrace{\square\cdots\square}^{k-2} & \boxed{t} \\ \boxed{s} & \square\cdots\square & \square \end{matrix} + t \begin{matrix} \square & \overbrace{\square\cdots\square}^{k-2} & \boxed{t} \\ \boxed{s} & \square\cdots\square & \boxed{z} \end{matrix} \right) \Bigg\}.\end{aligned} \tag{6.79}$$

I note that the contributions to the cancellation in Eq. (6.77) from $F_k^k$ are exhausted by the above operators in Eq. (6.79).

One can prove that the coefficients are zero if the spin flavors differ between the right and left ends, such as

$$c_i\left( \begin{matrix} \oplus \cdots \square \\ \square \cdots \ominus \end{matrix} \right) = 0\,, \tag{6.80}$$

which is stated in the following Lemma.

**Lemma 6.12.** $\alpha_{(\sigma,s),(\mu,s')} = 0$ *holds for* $\sigma \neq \mu$*.*

*Proof.* I give the proof in the case of $\sigma = \uparrow$, $\mu = \downarrow$, $s = +$, and $s' = -$. Let $q \in \mathcal{C}_k$ be $q = \begin{matrix} \oplus & \overbrace{\square\cdots\square}^{k-2} & \square \\ \square & \square\cdots\square & \ominus \end{matrix}$ and from Lemma 6.10, I have $c_i(q) = \alpha_{(\uparrow,+),(\downarrow,-)}$. I consider the following cancellation of the $k$-support basis element $\widetilde{q}_l(i+l-1)$ $(1 \le l \le k)$ in Eq. (6.77).

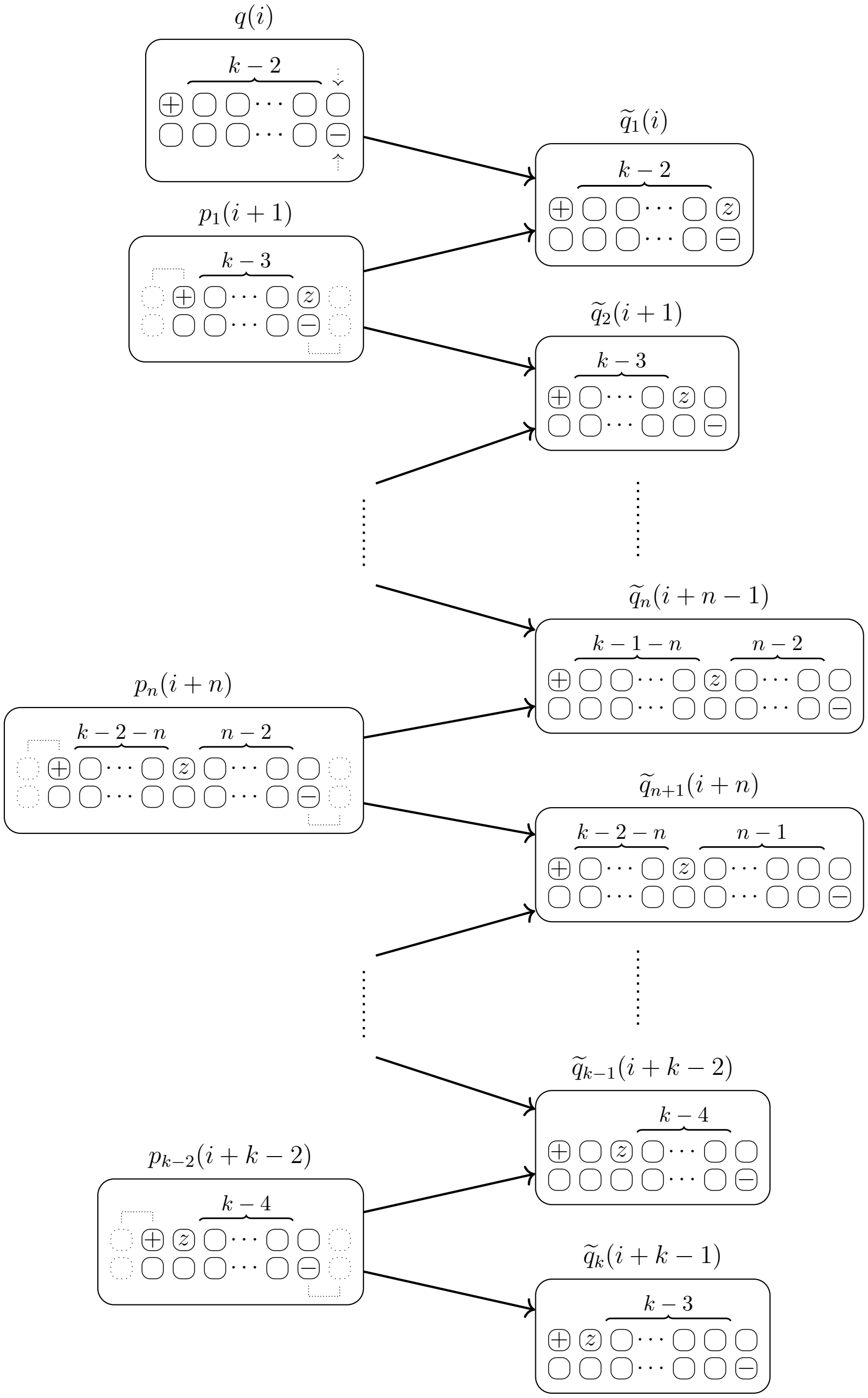

(6.81)

where $p_l \in \mathcal{C}_{k-1}$, $q \in \mathcal{C}_k$, and from Eqs. (6.48) and (6.49), I have

$$\alpha_{(\uparrow,+),(\downarrow,-)} = c_i(q) \propto c_{i+1}(p_1) \propto \cdots \propto c_{i+n}(p_n) \propto \cdots \propto c_{i+k-2}(p_{k-2}) = 0. \tag{6.82}$$

Then, one can see

$$\alpha_{(\uparrow,+),(\downarrow,-)} = 0\,. \tag{6.83}$$

The other cases can be also proved in the same way. □

I will derive a relationship between the coefficients in the following Lemma using the periodic boundary condition.

**Lemma 6.13.** $\alpha_{(\uparrow,s),(\uparrow,-s)} = \alpha_{(\downarrow,s),(\downarrow,-s)}$ *and* $\alpha_{(\sigma,+),(\sigma,-)} = (-1)^{k-1}\alpha_{(\sigma,-),(\sigma,+)}$.

*Proof.* I consider the following cancellation of the $k$-support basis element $\widetilde{q}_l(i+l-1)$ $(1 \leq l \leq$

$k)$ in Eq. (6.77).

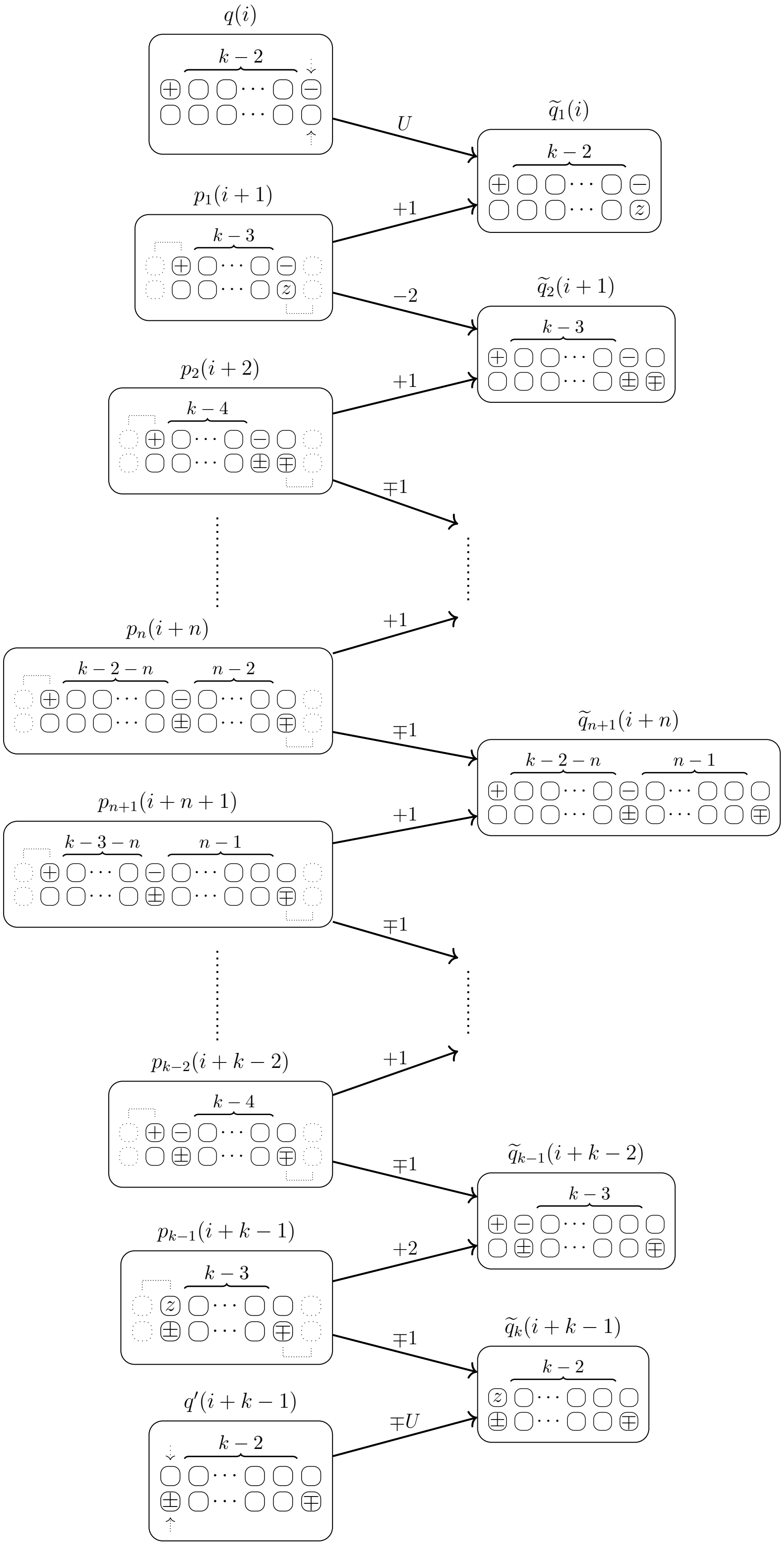

(6.84)

From these graphical notations, I have

$$
\begin{aligned}
c_i(q) &= \left(-\frac{1}{U}\right)\left(-\frac{1}{-2}\right)\left(-\frac{1}{\mp 1}\right)^{k-4}\left(-\frac{2}{\mp 1}\right)\left(-\frac{\mp U}{\mp 1}\right) c_{i+k-1}(q') \\
&= (\pm 1)^{k-3} c_{i+k-1}(q').
\end{aligned}
\tag{6.85}
$$

From Eq. (6.76), I have

$$
c_i(q) = \alpha_{(\uparrow,+),(\uparrow,-)}, \tag{6.86}
$$

and

$$
c_{i+k-1}(q') = \alpha_{(\downarrow,\pm),(\downarrow,\mp)}. \tag{6.87}
$$

Then, one can see

$$
\alpha_{(\uparrow,+),(\uparrow,-)} = (\pm 1)^{k-1}\alpha_{(\downarrow,\pm),(\downarrow,\mp)}\,, \tag{6.88}
$$

and also I have

$$
\alpha_{(\uparrow,+),(\uparrow,-)} = \alpha_{(\downarrow,+),(\downarrow,-)} = (-1)^{k-1}\alpha_{(\downarrow,-),(\downarrow,+)}\,. \tag{6.89}
$$

The remaining case that I have to prove is $\alpha_{(\uparrow,+),(\uparrow,-)} = (-1)^{k-1}\alpha_{(\uparrow,-),(\uparrow,+)}$. This case is also proved in the same way by considering the case in which all spin flavors are reversed (the upper row and the lower row are interchanged) in Eq. (6.84), and I have $\alpha_{(\downarrow,+),(\downarrow,-)} = (\pm 1)^{k-1}\alpha_{(\uparrow,\pm),(\uparrow,\mp)}$, then one can see $\alpha_{(\downarrow,+),(\downarrow,-)} = \alpha_{(\uparrow,+),(\uparrow,-)} = (-1)^{k-1}\alpha_{(\uparrow,-),(\uparrow,+)}$. This concludes the proof. □

One can prove that the coefficients are zero if the columns at the right and left end are the same, such as $c_i\begin{pmatrix} \oplus \cdots \oplus \\ \bigcirc \cdots \bigcirc \end{pmatrix} = 0$, which is stated in the following Lemma.

**Lemma 6.14.** $\alpha_{(\sigma,s),(\sigma,s)} = 0$.

*Proof.* I consider the following cancellation of the $k$-support basis element $\widetilde{q}_l(i+l-1)$ $(1 \leq l \leq$

$k-1)$ in Eq. (6.77).

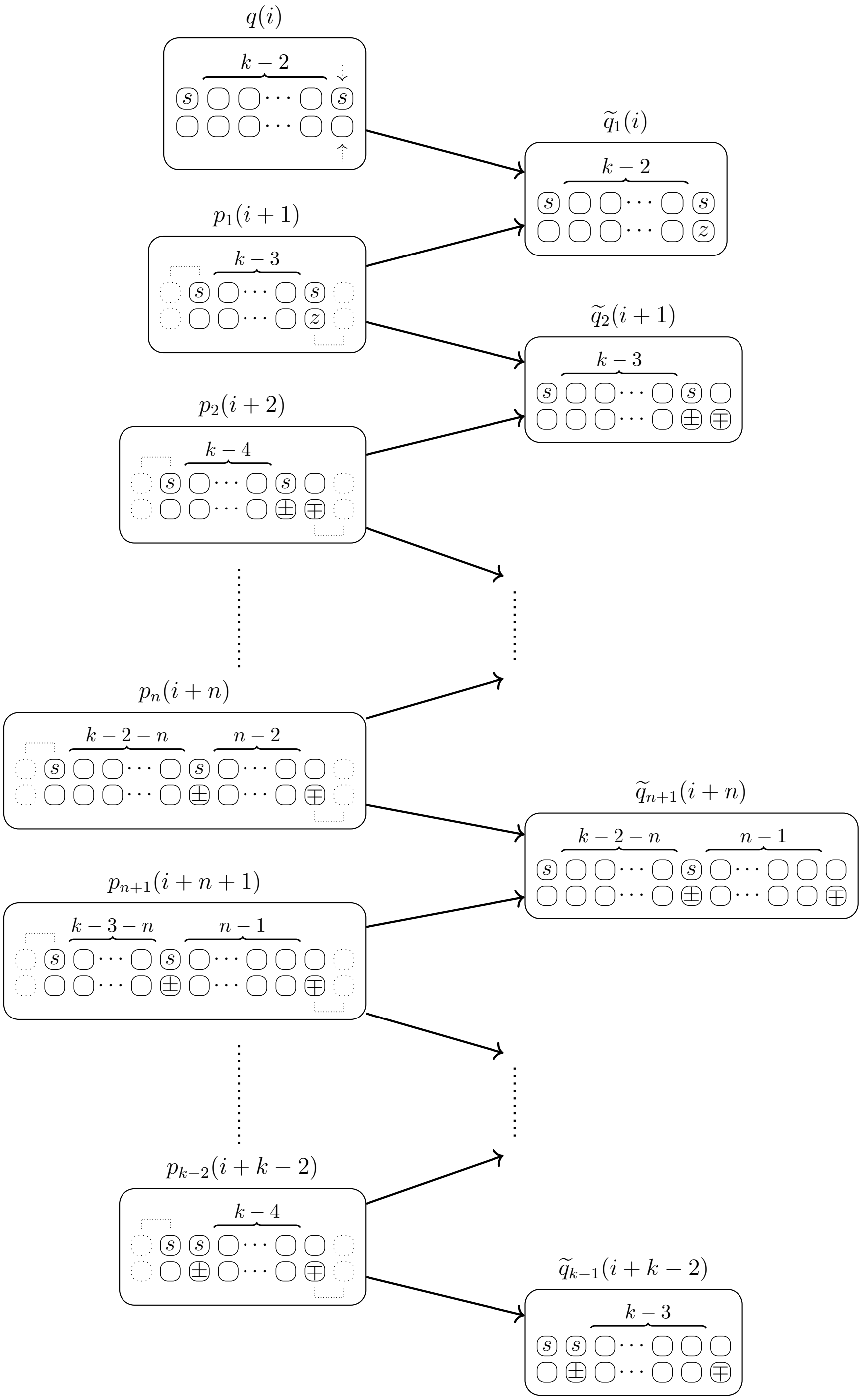

(6.90)

From these graphical notations, I have

$$c_i(q) \propto c_{i+1}(p_1) \propto \cdots \propto c_n(p_n) \propto \cdots \propto c_{k-2}(p_{k-2}) = 0\,. \tag{6.91}$$

From Eq. (6.76), $c_i(q) = \alpha_{(\uparrow,s),(\uparrow,s)}$ and I have $\alpha_{(\uparrow,s),(\uparrow,s)} = 0$.

In the same way, one can prove $\alpha_{(\downarrow,s),(\downarrow,s)} = 0$. This concludes the proof. □

From Lemmas 6.12–6.14, using the notation $\alpha_{(\uparrow,+),(\uparrow,-)} = 2c_k$, I have

$$\begin{aligned} F_k^k &= \sum_{\sigma=\uparrow,\downarrow} \sum_{i=1}^{L} 2c_k \left( c_{i,\sigma} c^\dagger_{i+k-1,\sigma} + (-1)^{k-1} c^\dagger_{i,\sigma} c_{i+k-1,\sigma} \right) \\ &= c_k Q_k^0. \end{aligned} \tag{6.92}$$

One can see that the $k$-support operator in $F_k$ is proportional to $Q_k^0$.

I define

$$\Delta \equiv F_k - c_k Q_k \,, \tag{6.93}$$

where $\Delta$ is also a conserved quantity, which can be seen from $[\Delta, H] = [F_k, H] - c_k [Q_k, H] = 0$. One can prove that $\Delta$ is a less-than-$(k-1)$-local conserved quantity as follows:

$$\begin{aligned} \Delta &\equiv F_k - c_k Q_k \\ &= c_k Q_k^0 + F_k^{k-1} + (\text{rest}) - c_k \left( Q_k^0 + \delta Q_{k-1}(U) \right) \\ &= F_k^{k-1} + (\text{rest}) - c_k \delta Q_{k-1}(U), \end{aligned} \tag{6.94}$$

where the last line is a linear combination of less-than-or-equal-to-$(k-1)$-support operators, and then one can see $\Delta$ is a less-than-$(k-1)$-local conserved quantity, and (rest) is the same as that in Eq. (6.53). Then I have proved that $F_k = c_k Q_k + \Delta$, where $\Delta$ is a less-than-$(k-1)$-local conserved quantity, which concludes the proof of Theorem 6.4.

It should be noted that the aforementioned proof does not hold in the non-interacting case where $U = 0$. In this case, there is no contribution from $F_k^k$ to the cancellation of $k$-support operators in Eqs. (6.77), and (6.76) itself becomes a local charge.

### 6.3.3 Case of $k \geq \lfloor \frac{L}{2} \rfloor$

I briefly explain why the above proof breaks in the case of $k \geq \lfloor \frac{L}{2} \rfloor$. For example, in the case of $k = L/2$ (where I assume even $L$), there is another contribution to the cancellation of the basis element with $(L/2+1)$-support configuration in Eq. (6.71) from the $(L/2)$-support basis

element:

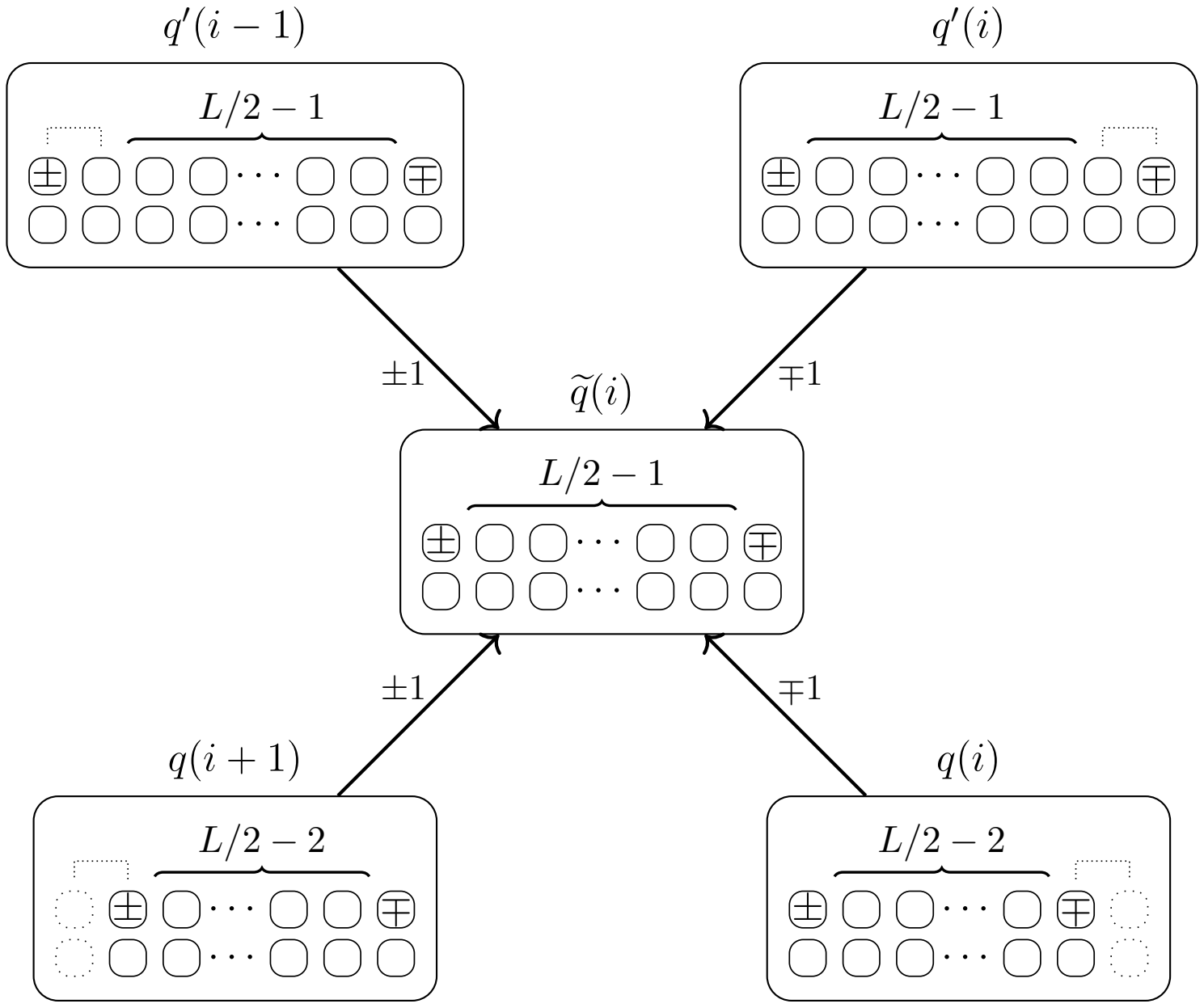

(6.95)

where the configuration $q'$ has $(L/2)$-support because

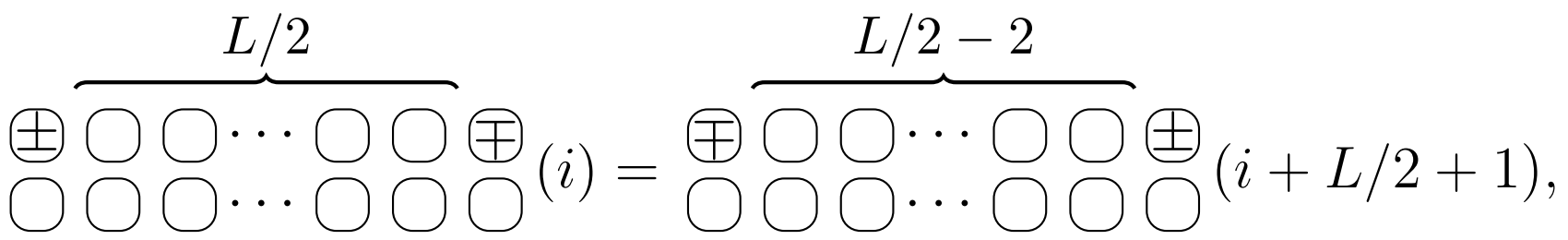

(6.96)

where I used the periodic boundary condition, and one can see this is also a $(L/2)$-support basis element and is included in the linear combination in $F_k^k$.

For a clear understanding of the cancellation (6.95), I show a schematic picture in Figure 6.1. One can confirm that the four contributions to the cancellation of the $(L/2+1)$-support basis element at the center have $(L/2)$-support, which can be seen from the fact that there are $(L/2-2)$ identity operators denoted by the empty circles between the local operators denoted by the filled circles, which are not identities.

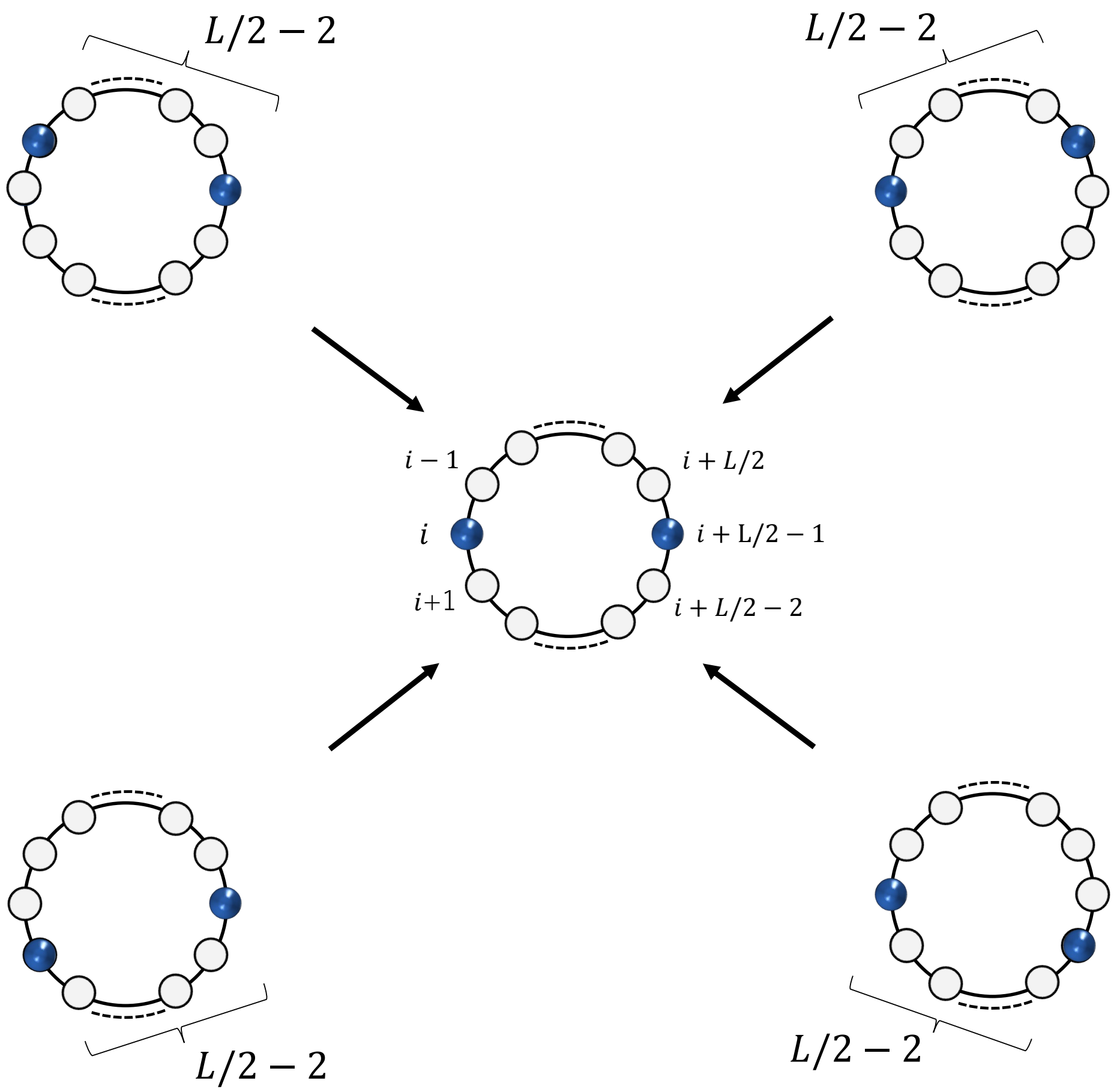

Figure 6.1: Schematic picture of the cancellation in Eq. (6.95). Circles represent the local sites. Empty circles indicate identity operators. Filled circles indicate non-identity operators.

In the same way, in the case of $k \geq \lfloor \frac{L}{2} \rfloor$, there are other contributions to the cancellation apart from what I have considered in the proof of Theorem 6.4.

## 6.4 Summary and remark on non-interacting case

In this chapter, I proved that the $k$-local charges in the 1D Hubbard model are uniquely determined up to the choice of the linear combination of the lower-order local charges. This proof is based on Theorem 6.4, which states that the maximal support components of the local charges are uniquely determined. This strategy has also been employed in the proof of completeness of local charges in the case of the spin-1/2 XYZ chain without a magnetic field [30], which is based on the insights regarding the maximal support components of its local charges [41]. The same strategy may also be immediately applicable to the other quantum integrable lattice models, such as the $\mathrm{SU}(N)$ generalization of the isotropic Heisenberg chain [28, 29], and the Temperley-Lieb models [31], including the spin-$1/2$ XXZ chain.

I note that as for the non-interacting quantum integrable systems, such as the spin-$1/2$ $XY$ chain or free fermion models, $k$-local charges are not uniquely determined. There are more than two independent families of local charges [28, 131–134]. One can also understand that this in Eq. (6.76), where $F_k^k$ is the local charge in the non-interacting case $U = 0$ and there are 16 free parameters, indicating 16 independent families of local charges.

# Chapter 7

# Summary and Discussions

As noted in the introduction, local conservation laws are the foundation of the exact solvability in quantum integrable lattice systems. The existence of the local charges $\{Q_k\}$ is guaranteed by the quantum inverse scattering method [8]. The local charges are obtained from the transfer matrix $T(\lambda)$ as $\partial_\lambda \ln T(\lambda) = \sum_{k\geq 2} \lambda^{k-2} Q_k$, where $\lambda$ is the spectral parameter. An alternative way to calculate $\{Q_k\}$ is achieved by using the Boost operator [43]: $[Q_k, B] = Q_{k+1}$. Despite these known generating procedures, deriving their explicit expressions has remained highly challenging. Though the general expressions are now completely known for some spin chains, there have been no such examples of electron systems. In this thesis, I studied the local conservation laws of the one-dimensional (1D) Hubbard model. This is the first time when the structure of local charges has been clarified for an integrable electron system without the boost operator, i.e., without a recursive way to construct the local charges. This has advanced our understanding of the structure of local charges in quantum integrable systems.

In Chapter 2, I have reviewed the quantum inverse scattering method and seen the Yang-Baxter algebra underlying quantum integrable lattice models. Each integrable model has a corresponding R-matrix, and the transfer matrix is made from the matrix product of the R-matrices. The Yang-Baxter algebra ensures the mutual commutativity of the transfer matrices with different spectral parameters, which results in the mutual commutativity of the local charges. When the R-matrix is of a difference form of spectral parameters, it is possible to construct the Boost operator, and a recursive relationship generates the next-order charge from a lower-order one.

In Chapter 3, I have reviewed the quantum integrability of the one-dimensional Hubbard model. I have briefly reviewed the Lieb-Wu equation [33], which can be obtained from the nested Bethe ansatz, and the spin-charge separation of the low-energy excitation in the one-dimensional Hubbard model [54]. I have also reviewed the R-matrix of the one-dimensional Hubbard model obtained by Shastry [37, 60, 61]. Shastry's R-matrix [37, 60, 61] is not a difference form, and thus the 1D Hubbard model has no boost operator, which makes the study on their local charges difficult.

In Chapter 4, I have reviewed the previous progress on the general expressions of the local charges in quantum integrable spin chains. I have explained the structure of the local charges in the spin-$1/2$ isotropic Heisenberg (XXX) chain [26, 27], its $\mathrm{SU(N)}$ generalization [28], the spin-$1/2$ XYZ chain [30], and the models related to the Temperley-Lieb algebra, which includes the spin-$1/2$ XXZ chain [31]. The critical point in finding a general expression is to discern the patterns in the expressions of the lower-order charges, then formulate a conjecture based on these regularities, and ultimately prove that the conjectured quantities are actually conserved. For the spin-$1/2$ XYZ chain, the pattern emerges when the charges are represented in the doubling-product notation, which was used for the rigorous proof of the non-integrability of the spin-$1/2$ XYZ chain with a non-zero magnetic field [41]. In this way, discerning the regularity and formulating a conjecture for the general expression requires a good representation of the operator constructing the local charges. I also reviewed the procedure to derive the generalized current from the expression of local charges [80], with their demonstrations in the Temperley-Lieb models. Then, I reviewed several examples of the factorization of short-range correlation functions using the generalized currents in the Temperley-Lieb models [80] and the current mean value formula for the XXZ chain [45, 46, 109]. For the one-dimensional Hubbard model, I reviewed the previous studies of its known local conserved quantities [28, 37, 81, 82]. The first three non-trivial charges $Q_k$ $(3 \leq k \leq 5)$ had been obtained there. From those results, Grabowski and Mathieu conjectured that the local charges take the form of a linear combination of a certain kind of the products of local conserved densities of the two uncoupled XX chains [28]. However, the concrete conditions for selecting these special products had been a mystery for decades. The other conjecture is that all the non-vanishing terms should have the coefficient $\pm U^j$, where $U$ is the coupling constant of the on-site Coulomb interaction. This is actually true up to $k = 5$ under an appropriate choice of the normalization for the XX charges. This conjecture turns out to be incorrect for $k > 5$. The coefficients appearing in the linear combination constituting $\{Q_k\}_{k\geq 6}$ involve a rich structure that was not previously expected.

Chapter 5 is the first part of the main results of this thesis. I showed the general expressions for the local conserved quantities in the one-dimensional Hubbard model with any values of the on-site interaction $U$. The results are shown in both the usual fermion representation and the spin-variable representation. I next proved that $Q_k$ is a linear combination of *connected diagrams*. They are notations for the particular kind of products of hopping terms for fermion notation or products of densities of local charges in the XX chain in terms of spin variables. This notation is useful to discern the patterns hidden in the expression of local charges, which was not clarified in previous studies [28]. With this notation, I expressed the higher-order charges $Q_{k\geq 6}$. The connected diagrams constituting them are accompanied by non-trivial coefficients different from $\pm U^j$. Some of them are involved with the generalized Catalan numbers, which also appear in the local charges of the Heisenberg chain [27, 28, 112]. I derived a recursion equation for these coefficients of the connected diagrams in $Q_k$. Obtaining the general explicit formula for the coefficients is a remaining task, and they may be some further generalization of the Catalan number. My result is valid in finite systems with the periodic boundary condition and also in the thermodynamic limit. I also derived another family of mutually commuting charges from the strong-coupling limit of the local charges in the 1D Hubbard model, which is the special case of the Maassarani's model [110, 111].

Chapter 6 is the second part of the main results. I rigorously proved the completeness of the local charges $\{Q_k\}_{k\geq 2}$ obtained in Chapter 5. The proof followed the same idea used for the proof of non-integrability of the spin-$1/2$ XYZ chain [41] and the mixed-field Ising chain [42]. My proof does not require detailed knowledge of $Q_k$ presented in Chapter 5. It immediately follows that the local charges derived from the expansion of the logarithm of the transfer matrix [32, 37, 38, 64], can be expressed as a linear combination of my charges $\{Q_k\}$. I note that my analysis is limited to ultra-local charges and does not apply to quasi-local ones [22, 130].

Finally, I give several discussions and perspectives for future work. One of the applications of my results is the study of the non-equilibrium dynamics of the one-dimensional Hubbard model. An extensive number of the local conserved quantities $\{Q_k\}$ hinders the thermalization of the system, which has been also observed in ultra-cold atom experiments [9]. Instead, it is conjectured that the stationary values of $\{Q_k\}$ are described by the generalized Gibbs ensemble (GGE) [16–18] given by $\rho_{\mathrm{GGE}} \propto e^{-\sum_k \beta_k Q_k}$. The validity of the GGE ansatz has been mainly investigated for integrable spin chains, such as the XXZ chain [19–21, 23]. My result of the exact expression for the local charges may be helpful in investigating the GGE in the Hubbard model, which has not been extensively studied so far. Recently, one method was developed to derive quantum many-body scar models. They are non-integrable systems featuring some non-thermal eigenstates. In Ref. [135], quantum many-body scar models were derived from odd-order charges in integrable lattice models with a non-integrable perturbation. I may also use this method and obtain a quantum many-body scar model from my $Q_{2k+1}$. One can also study the current mean value formula [45, 46, 136], which naturally leads to the fundamental conjectured relation in the generalized hydrodynamics (GHD): the semiclassical expression for the mean value of the current [24, 137]. The formula for current mean value claims that the expectation value of current and generalized current operators with respect to Bethe eigenstates can be written as a simple quadratic form of the single-particle eigenvalues of local charges in terms of the corresponding Bethe roots. This formula has been proved correct for the XXZ chain [45, 46], XYZ chain [138], and the $\mathrm{SU}(3)$-symmetric fundamental model [139]. One can construct current operators of the one-dimensional Hubbard model and study how the current mean value formula holds in this case. Compared to the XXZ chain, a significant difference is that the Hubbard model is solved using the nested Bethe ansatz, and the formulation of the formula for the current mean value for such models is anticipated. It has been known that the current mean values also play an essential role in the factorization of the correlations function in quantum integrable systems [45, 80, 140]. It is interesting to investigate the factorization formula for the correlation functions in the one-dimensional Hubbard model. Similar to the series of studies on the XXZ correlation functions [108], there may be potential for a more profound understanding of the correlation functions in the 1D Hubbard model.

Another interest is concerned with a matrix-product operator (MPO) representation of the local charges $\{Q_k\}$. As I discussed above, their general expressions are very complicated. Recently, for the isotropic Heisenberg chain and its $\mathrm{SU(N)}$ generalization, it was proved that $\{Q_k\}$ are efficiently expressed in the MPO representation [29]. The new coefficients are more straightforward than those appearing in the original study. I expect a similar simplification in the MPO representation for the 1D Hubbard model, which may enable us to obtain a closed form of the general solution of the recursion equation for the coefficients (5.57).

It is also interesting to investigate the boundary terms in the local charges $\{Q_k\}$ when the open boundary condition is imposed. In this case, $\{Q_k\}$ have boundary terms, which are localized near the boundaries. The boundary terms are more complicated than the bulk terms, and their detailed structure is not well understood even for the Heisenberg chain [141]. A method was developed in Ref. [141] to calculate the boundary term from the bulk term for the Heisenberg chain. It is also possible to apply this method to the 1D Hubbard model to calculate its boundary terms of $\{Q_k\}$.

The explicit expressions of local charges are useful in studying non-equilibrium dynamics in integrable systems, such as the GGE and GHD, the quantum many-body scar, and the study of the correlation functions in interacting integrable systems. The direct study of explicit expressions of local charges in quantum integrable lattice models lets us look at integrability from a point of view distinct from the traditional Bethe Ansatz. This new viewpoint may provide a deeper understanding of the essence of quantum integrability. Moreover, it should be emphasized that in theoretical physics, an exact result in quantum many-body systems is inherently significant and stands out as critically crucial on its own.

# Appendix A

# Proof of identities of $C^{j,m}_{n,d}(\boldsymbol{\lambda})$

In Appendix A, I prove the identities of $C^{j,m}_{n,d}(\boldsymbol{\lambda})$ (5.79)–(5.84). Let $P(j,n)$ be the statement (5.79)–(5.84) for $d \geq 0$ and $\lfloor j/2 \rfloor > m \geq 0$. I prove $P(j,n)$ by induction in Section A.1. I also prove the positivity of the coefficient.

I also confirm the analytical solutions of the recursion equation (5.57), (5.65) and (5.67), actually satisfy the identities and the recursion equation (5.57) with direct calculation in Section A.2.

## A.1 Induction step

In this section, I prove $P(j,n)$ by induction. I show that if I suppose $P(j-1,n)$ and $P(j,n-1)$ hold, then $P(j,n)$ also holds. The base cases are $P(j=1,n)$ for any $n \geq 0$, and one can see that $P(j=1,n)$ holds trivially because of $C^{j=1,m=0}_{n,d}(\boldsymbol{\lambda}) = 1$. I define $w \equiv j - 2m - 1$.

### A.1.1 Induction for Eq. (5.79)

Calculating $C^{j,m}_{n,d}(\lambda_L; \vec{\lambda}; \lambda_R) - C^{j,m}_{n,d}(-1; \vec{\lambda}; \lambda_{L+R}+1)$ with Eq. (5.57), I have

$$\begin{aligned}
&C^{j,m}_{n,d}(\lambda_L; \vec{\lambda}; \lambda_R) - C^{j,m}_{n,d}(-1; \vec{\lambda}; \lambda_{L+R}+1) = \sum_{p=0}^{\lambda_L} \Delta C^{j,m}_{n,d}(\mathcal{T}^p \boldsymbol{\lambda}) \\
=&2\left(C^{j,m}_{n-1,d+1}(\lambda_L; \vec{\lambda}; \lambda_R) - C^{j,m}_{n-1,d}(1; \vec{\lambda}; \lambda_{L+R}+1)\right) \\
&\quad - \left(C^{j,m}_{n-1,d}(\lambda_L+1; \vec{\lambda}; \lambda_R+1) - C^{j,m}_{n-1,d}(0; \vec{\lambda}; \lambda_{L+R}+2)\right) \\
&\qquad + \sum_{p=0}^{\lambda_L}\left(C^{j-1,m-1}_{n,d}(p; \lambda_{L+R}-p, \vec{\lambda}; 0) - C^{j-1,m-1}_{n,d}(\lambda_{L+R}-p+1; p-1, \vec{\lambda}; 0)\right),
\end{aligned} \tag{A.1}$$

where $\lambda_{L+R} \equiv \lambda_L + \lambda_R$ and when deriving the term with summation over $p$ in the last line, I used Eqs. (5.79) and (5.80) of $P(j-1,n)$ and variable transformation $p \to \lambda_L - p$. I also used $C_{n-1,d+1}^{j,m}(0;\vec{\lambda};\lambda_{L+R}+1) = C_{n-1,d}^{j,m}(1;\vec{\lambda};\lambda_{L+R}+1)$.

Using Eq. (A.1) interchanging the role of $\lambda_L$ and $\lambda_R$, I have

$$
\begin{aligned}
&C_{n,d}^{j,m}(\lambda_L;\vec{\lambda};\lambda_R) - C_{n,d}^{j,m}(\lambda_R;\vec{\lambda};\lambda_L)\\
=&s\sum_{p=\lambda_-+1}^{\lambda_+}\left(C_{n,d}^{j-1,m}(p;\lambda_{L+R}-p,\vec{\lambda};0) - C_{n,d}^{j-1,m}(\lambda_{L+R}-p+1;p-1,\vec{\lambda};0)\right)\\
=&0,
\end{aligned}
\tag{A.2}
$$

where $\lambda_+ = \max(\lambda_L,\lambda_R)$ and $\lambda_- = \min(\lambda_L,\lambda_R)$ and $s = \pm 1$ for $\lambda_L = \lambda_\pm$, and I used Eq. (5.79) of $P(j,n-1)$ and one can see the last equality holds with the variable transformation $p \to \lambda_{L+R} - p + 1$ in the second term in the summation on the second line. Then, I have proved Eq. (5.79) of $P(j,n)$.

I note that from Eqs. (A.2) and (5.90), I have

$$
\begin{aligned}
&C_{n,d}^{j,m}(\lambda_L;\vec{\lambda};\lambda_R) - C_{n,d}^{j,m}(-1;\vec{\lambda};\lambda_{L+R}+1)\\
=&2\left(C_{n-1,d+1}^{j,m}(\lambda_L;\vec{\lambda};\lambda_R) - C_{n-1,d}^{j,m}(1;\vec{\lambda};\lambda_{L+R}+1)\right)\\
&\quad-\left(C_{n-1,d}^{j,m}(\lambda_L+1;\vec{\lambda};\lambda_R+1) - C_{n-1,d}^{j,m}(0;\vec{\lambda};\lambda_{L+R}+2)\right)\\
&\qquad+\sum_{p=0}^{\lambda_R}\left(C_{n,d}^{j-1,m-1}(p;\lambda_{L+R}-p,\vec{\lambda};0) - C_{n,d}^{j-1,m-1}(\lambda_{L+R}-p+1;p-1,\vec{\lambda};0)\right).
\end{aligned}
\tag{A.3}
$$

### A.1.2 Induction for Eq. (5.80)

Throughout this subsection, I define $F_{n,d}^{j,m}(\boldsymbol{\lambda}) = C_{n,d}^{j,m}(\boldsymbol{\lambda}) - C_{n,d}^{j,m}(\boldsymbol{\lambda}_{i_1\leftrightarrow i_2})$ where $j \geq 2m + 3(w \geq 2)$ and $1 \leq i_1 < i_2 \leq w$ is assumed. I prove $F_{n,d}^{j,m}(\boldsymbol{\lambda}) = 0$ in the following. From the recursion equation (5.57) and Eq. (5.80) of $P(j-1,n)$ and $P(j,n-1)$, I have

$$
\begin{aligned}
&F_{n,d}^{j,m}(\boldsymbol{\lambda}) - F_{n,d}^{j,m}(\mathcal{T}\boldsymbol{\lambda})\\
=&\Delta C_{n,d}^{j,m}(\boldsymbol{\lambda}) - \Delta C_{n,d}^{j,m}(\boldsymbol{\lambda}_{i_1\leftrightarrow i_2})\\
=&2\left[\Delta C_{n-1,d+1}^{j,m}(\boldsymbol{\lambda}) - \Delta C_{n-1,d+1}^{j,m}(\boldsymbol{\lambda}_{i_1\leftrightarrow i_2})\right] - \left[\Delta C_{n-2,d+2}^{j,m}(\boldsymbol{\lambda}) - \Delta C_{n-2,d+2}^{j,m}(\boldsymbol{\lambda}_{i_1\leftrightarrow i_2})\right]\\
&+\left[C_{n,d}^{j-1,m-1}(\boldsymbol{\lambda}_{\leftarrow 0}) - C_{n,d}^{j-1,m-1}((\boldsymbol{\lambda}_{i_1\leftrightarrow i_2})_{\leftarrow 0})\right] - \left[C_{n,d}^{j-1,m-1}({}_{0\rightarrow}(\mathcal{T}\boldsymbol{\lambda})) - C_{n,d}^{j-1,m-1}({}_{0\rightarrow}(\mathcal{T}(\boldsymbol{\lambda}_{i_1\leftrightarrow i_2})))\right]\\
&=0,
\end{aligned}
\tag{A.4}
$$

where in the last equality, I used the relation derived from Eqs. (5.84) and (5.80) of $P(j,n-1)$ to the second term of the third line: $\Delta C_{n-2,d+2}^{j,m}(\boldsymbol{\lambda}) - \Delta C_{n-2,d+2}^{j,m}(\boldsymbol{\lambda}_{i_1\leftrightarrow i_2}) = \Delta C_{n-1,d}^{j,m}(\boldsymbol{\lambda}_{L:+1,R:+1}) -$ $\Delta C^{j,m}_{n-1,d}((\boldsymbol{\lambda}_{L:+1,R:+1})_{i_1\leftrightarrow i_2}) = 0$. Using Eq. (A.4) repeatedly, I have

$$\begin{aligned}
F^{j,m}_{n,d}(\lambda_L;\vec{\lambda};\lambda_R) &= F^{j,m}_{n,d}(-1;\vec{\lambda};\lambda_{L+R}+1) = F^{j,m}_{n,d-1}(1;\vec{\lambda};\lambda_{L+R}+1)\\
&= F^{j,m}_{n,d-1}(-1;\vec{\lambda};\lambda_{L+R}+3) = F^{j,m}_{n,d-1}(1;\vec{\lambda};\lambda_{L+R}+5)\\
&\ \ \vdots\\
&= F^{j,m}_{n,0}(-1;\vec{\lambda};\lambda_{L+R}+1+2d),
\end{aligned} \tag{A.5}$$

and then what I have to prove is $F^{j,m}_{n,0}(-1;\lambda_1,\dots,\lambda_w;A) = 0$ for $A>0$.

For the $1<i_1$ case, I have $F^{j,m}_{n,0}(-1;\lambda_1,\dots,\lambda_w;A)=0$ straightforwardly from the definition of Eqs. (5.59) and (5.80) of $P(j-1,n)$ and $P(j,n-1)$.

For the $i_1=1$ case, I have

$$\begin{aligned}
&F^{j,m}_{n,0}(-1;\lambda_1,\dots,\lambda_{i_2},\dots)\\
=&C^{j,m}_{n,0}(-1;\lambda_1,\dots,\lambda_{i_2},\dots) - C^{j,m}_{n,0}(-1;\lambda_{i_2},\dots,\lambda_1,\dots)\\
=&C^{j,m}_{n-1,1}(0;\lambda_1-1,\dots,\lambda_{i_2},\dots) - C^{j,m}_{n-1,1}(0;\lambda_{i_2}-1,\dots,\lambda_1,\dots)\\
&\quad + C^{j-1,m}_{n,0}(\lambda_1+1;\dots,\lambda_{i_2},\dots) - C^{j-1,m}_{n,0}(\lambda_{i_2}+1;\dots,\lambda_1,\dots)\\
=&C^{j,m}_{n-2,3}(-1;\lambda_1-1,\dots,\lambda_{i_2}-1,\dots) - C^{j,m}_{n-2,3}(-1;\lambda_{i_2}-1,\dots,\lambda_1-1,\dots)\\
&\quad + C^{j-1,m}_{n-1,2}(\lambda_{i_2};\lambda_1-1,\dots,\not{\lambda}_{i_2},\dots) - C^{j-1,m}_{n-1,2}(\lambda_1;\lambda_{i_2}-1,\dots,\not{\lambda}_1,\dots)\\
&\qquad + C^{j-1,m}_{n,0}(\lambda_1+1;\dots,\lambda_{i_2},\dots) - C^{j-1,m}_{n,0}(\lambda_{i_2}+1;\dots,\lambda_1,\dots)\\
=&C^{j-1,m}_{n,0}(\lambda_1+1;\dots,\lambda_{i_2},\dots) - C^{j-1,m}_{n,0}(\lambda_{i_2}+1;\dots,\lambda_1,\dots)\\
&\quad - \left(C^{j-1,m}_{n-1,2}(\lambda_1;\dots,\lambda_{i_2}-1,\dots) - C^{j-1,m}_{n-1,2}(\lambda_{i_2};\dots,\lambda_1-1,\dots)\right)\\
=&0,
\end{aligned} \tag{A.6}$$

where I omitted the irrelevant elements in the list, $\not{\lambda}_i$ denotes the absence of the element at the corresponding position: $\{\dots,\lambda_{i-1},\not{\lambda}_i,\lambda_{i+1},\dots\} = \{\dots,\lambda_{i-1},\lambda_{i+1},\dots\}$, in the third equality, I used Eq. (5.82) of $P(j,n-1)$, in the fourth equality, I used Eq. (5.80) of $P(j,n-2)$, and in the last equality, I used Eq. (5.84) of $P(j-1,n)$.

Then, I have proved Eq. (5.80) of $P(j,n)$.

### A.1.3 Induction for Eq. (5.81)

I next consider Eq. (5.81). It is enough to consider the $\lambda_a>n$ case. I first consider the $a=L$ ($\lambda_L>n$) and $w>0$ case. Using Eq. (A.3), I have

$$\begin{aligned}
&C^{j,m}_{n,d}(\lambda_L;\vec{\lambda};\lambda_R) - C^{j,m}_{n,d}(n;\vec{\lambda};\lambda_R)\\
=&C^{j,m}_{n,d}(-1;\vec{\lambda};\lambda_L+\lambda_R+1) - C^{j,m}_{n,d}(-1;\vec{\lambda};n+\lambda_R+1)\\
=&C^{j,m}_{n,d-1}(1;\vec{\lambda};\lambda_L+\lambda_R+1) - C^{j,m}_{n,d-1}(1;\vec{\lambda};n+\lambda_R+1)\\
=&\left(C^{j,m}_{n,d-1}(\lambda_L+\lambda_R+1;\vec{\lambda};1) - C^{j,m}_{n,d-1}(n;\vec{\lambda};1)\right) - \left(C^{j,m}_{n,d-1}(n+\lambda_R+1;\vec{\lambda};1) - C^{j,m}_{n,d-1}(n;\vec{\lambda};1)\right)
\end{aligned}$$

$$
\begin{aligned}
&= \Big(C^{j,m}_{n,d-1}(-1;\vec{\lambda};\lambda_L+\lambda_R+3) - C^{j,m}_{n,d-1}(-1;\vec{\lambda};n+\lambda_R+3)\Big) \\
&\qquad - \Big(C^{j,m}_{n,d-1}(-1;\vec{\lambda};n+\lambda_R+3) - C^{j,m}_{n,d-1}(-1;\vec{\lambda};n+\lambda_R+3)\Big) \\
&= C^{j,m}_{n,d-1}(-1;\vec{\lambda};\lambda_L+\lambda_R+3) - C^{j,m}_{n,d-1}(-1;\vec{\lambda};n+\lambda_R+3) \\
&\quad \vdots \\
&= C^{j,m}_{n,0}(-1;\vec{\lambda};\lambda_L+\lambda_R+1+2d) - C^{j,m}_{n,0}(-1;\vec{\lambda};n+\lambda_R+1+2d) \\
&= C^{j,m}_{n-1,1}(0;\lambda_1-1,\ldots;\lambda_L+\lambda_R+1+2d) - C^{j,m}_{n-1,1}(0;\lambda_1-1,\ldots;n+\lambda_R+1+2d) \\
&\qquad + C^{j-1,m}_{n,0}(\lambda_1+1;\ldots;\lambda_L+\lambda_R+1+2d) - C^{j-1,m}_{n,0}(\lambda_1+1;\ldots;\vec{\lambda};n+\lambda_R+1+2d) \\
&= 0,
\end{aligned} \tag{A.7}
$$

where in the first equality, I used Eq. (5.81) of $P(j-1,n)$ and $P(j,n-1)$, and then one can see that the contributions from the right-hand side of Eq. (A.3) cancels each other, in the third equality, I used Eq. (5.79) of $P(j,n)$, which I have already proved, in the fourth equality, I used the same argument as the first and second equality for the two terms, and in the last equality, I used Eq. (5.81) of $P(j-1,n)$ and $P(j,n-1)$.

With the same argument, one can prove $C^{j,m}_{n,d}(\boldsymbol{\lambda}) - C^{j,m}_{n,d}(\boldsymbol{\lambda}_{L\to n}) = 0$ for the $w=0$. From Eq. (5.79) of $P(j,n)$ which I have already proved, one can see $C^{j,m}_{n,d}(\boldsymbol{\lambda}) - C^{j,m}_{n,d}(\boldsymbol{\lambda}_{R\to n}) = 0$ also holds for the $\lambda_R > n$ case.

I next consider the case of $a = i$ $(1 \le i \le w)$, $\lambda_i > n$. From Eq. (5.79) of $P(j,n)$ which I proved above, one can set $i=1$ without losing generality. Using Eq. (A.1), I have

$$
\begin{aligned}
&C^{j,m}_{n,d}(\boldsymbol{\lambda}) - C^{j,m}_{n,d}(\boldsymbol{\lambda}_{\lambda_1\to n}) \\
&= C^{j,m}_{n,d}(-1;\lambda_1,\ldots;\lambda_{L+R}+1) - C^{j,m}_{n,d}(-1;n,\ldots;\lambda_{L+R}+1) \\
&= C^{j,m}_{n,d-1}(1;\lambda_1,\ldots;\lambda_{L+R}+1) - C^{j,m}_{n,d-1}(1;n,\ldots;\lambda_{L+R}+1) \\
&= C^{j,m}_{n,d-1}(-1;\lambda_1,\ldots;\lambda_{L+R}+3) - C^{j,m}_{n,d-1}(-1;n,\ldots;\lambda_{L+R}+3) \\
&\quad \vdots \\
&= C^{j,m}_{n,0}(-1;\lambda_1,\ldots;\lambda_{L+R}+1+2d) - C^{j,m}_{n,0}(-1;n,\ldots;\lambda_{L+R}+1+2d) \\
&= C^{j,m}_{n-1,1}(0;\lambda_1-1,\ldots;\lambda_{L+R}+1+2d) - C^{j,m}_{n-1,1}(0;n-1,\ldots;\lambda_{L+R}+1+2d) \\
&\qquad + C^{j-1,m}_{n,0}(\lambda_1+1,\ldots;\lambda_{L+R}+1+2d) - C^{j-1,m}_{n,0}(n+1,\ldots;\lambda_{L+R}+1+2d) \\
&= 0,
\end{aligned} \tag{A.8}
$$

where in the first equality, I used Eq. (5.81) of $P(j-1,n)$ and $P(j,n-1)$ and then one can see the contributions from the right-hand side of Eq. (A.1) cancels each other, and in the last equality, I used Eq. (5.81) of $P(j-1,n)$ and $P(j,n-1)$.

Thus, I have proved Eq. (5.81) of $P(j,n)$.

### A.1.4 Induction for Eqs. (5.82) and (5.83)

Throughout this subsection, I define

$$F^{j,m}_{n,d}(\boldsymbol{\lambda})_{(a,b)} \equiv C^{j,m}_{n,d}(\boldsymbol{\lambda}) - C^{j,m}_{n-1,d+2}(\boldsymbol{\lambda}_{a:-1,b:-1}) - C^{j-1,m}_{n,d+1}(\boldsymbol{\lambda}_{a\to\lambda_a+\lambda_b,\hat{b}}), \tag{A.9}$$

where $0 \le a < b \le w+1$ and $w > 0$ is assumed. First, I prove $F^{j,m}_{n,d}(\boldsymbol{\lambda})_{L(R),i} = 0$ $(1 \le i \le w)$, and complete the proof of Eq. (5.82) of $P(j,n)$. Because I have proved Eqs. (5.79) and (5.80) of $P(j,n)$, one can set $a = L, b = 1$ without losing generality. Using Eq. (A.3), I have

$$\begin{aligned}
&F^{j,m}_{n,d}(\lambda_L;\vec{\lambda};\lambda_R)_{(L,1)}\\
=&C^{j,m}_{n,d}(-1;\vec{\lambda};\lambda_{L+R}+1) - C^{j,m}_{n-1,d+2}(-1;\vec{\lambda}_{1:-1};\lambda_{L+R}) - C^{j-1,m}_{n,d+1}(-1;\vec{\lambda}_{\hat{1}};\lambda_{L+R}+\lambda_1+1)\\
&\quad + 2\left(F^{j,m}_{n-1,d+1}(\boldsymbol{\lambda})_{(L,1)} - F^{j,m}_{n-1,d}(\mathcal{T}\boldsymbol{\eta})_{(L,1)}\right) - \left(F^{j,m}_{n-1,d}(\boldsymbol{\lambda}_{L:+1,R:+1})_{(L,1)} - F^{j,m}_{n-1,d}(\boldsymbol{\eta})_{(L,1)}\right)\\
&\qquad + \sum_{p=0}^{\lambda_R}\left(F^{j-1,m}_{n,d}(\boldsymbol{\xi}^{(p)})_{(1,2)} - F^{j-1,m}_{n,d}(\boldsymbol{\xi}^{(\lambda_{L+R}-p+1)})_{(L,2)}\right)\\
=&C^{j,m}_{n,d-1}(1;\vec{\lambda};\lambda_{L+R}+1) - C^{j,m}_{n-1,d+1}(1;\vec{\lambda}_{1:-1};\lambda_{L+R}) - C^{j-1,m}_{n,d}(1;\vec{\lambda}_{\hat{1}};\lambda_{L+R}+\lambda_1+1)\\
=&F^{j,m}_{n,d-1}(\lambda_{L+R}+1;\vec{\lambda};1)_{(L,1)}\\
=&F^{j,m}_{n,d-2}(\lambda_{L+R}+3;\vec{\lambda};1)_{(L,1)}\\
&\vdots\\
=&F^{j,m}_{n,0}(\lambda_{L+R}+2d-1;\vec{\lambda};1)_{(L,1)}\\
=&C^{j,m}_{n,0}(-1;\vec{\lambda};A+1) - C^{j,m}_{n-1,2}(-1;\vec{\lambda}_{1:-1};A) - C^{j-1,m}_{n,1}(-1;\vec{\lambda}_{\hat{1}};A+\lambda_1+1)\\
=&C^{j,m}_{n-1,1}(0;\vec{\lambda}_{1:-1};A+1) - C^{j,m}_{n-1,1}(1;\vec{\lambda}_{1:-1};A)\\
&\quad + \left(C^{j-1,m}_{n,0}(\lambda_1+1;\vec{\lambda}_{\hat{1}};A+1) - C^{j-1,m}_{n,0}(1;\vec{\lambda}_{\hat{1}};A+\lambda_1+1)\right)\\
=&C^{j,m}_{n-1,1}(0;\vec{\lambda}_{1:-1};A+1) - C^{j,m}_{n-1,1}(1;\vec{\lambda}_{1:-1};A)\\
&\quad + \left(C^{j-1,m}_{n-1,2}(\lambda_1;\vec{\lambda}_{\hat{1}};A) - C^{j-1,m}_{n-1,2}(0;\vec{\lambda}_{\hat{1}};A+\lambda_1)\right)\\
=&C^{j,m}_{n-1,1}(0;\vec{\lambda}_{1:-1};A+1) + \left(-C^{j,m}_{n-1,1}(1;\vec{\lambda}_{1:-1};A) + C^{j-1,m}_{n-1,2}(\lambda_1;\vec{\lambda}_{\hat{1}};A)\right) - C^{j-1,m}_{n-1,2}(0;\vec{\lambda}_{\hat{1}};A+\lambda_1)\\
=&C^{j,m}_{n-1,1}(0;\vec{\lambda}_{1:-1};A+1) - C^{j,m}_{n-2,3}(0;\vec{\lambda}_{1:-2};A) - C^{j-1,m}_{n-1,2}(0;\vec{\lambda}_{\hat{1}};A+\lambda_1)\\
=&0,
\end{aligned} \tag{A.10}$$

where $\boldsymbol{\eta} = \{\lambda_{L+R}+2;\vec{\lambda};0\}$ and $\boldsymbol{\xi}^{(p)} = \{p;\lambda_{L+R}-p,\vec{\lambda};0\}$, $A = \lambda_{L+R}+2d$, in the second equality, I used Eqs. (5.82) and (5.83) of $P(j,n-1)$ and $P(j-1,n)$, in the eighth equality, I used Eq. (5.84) of $P(j-1,n)$, and in the eleven-th and last equalities, I used Eq. (5.82) of $P(j,n-1)$. Thus, I have proved Eq. (5.82) of $P(j,n)$.

Next, I will prove Eq. (5.83) of $P(j,n)$ below. The proof is similar to that of Eq. (5.82) of $P(j,n)$ above. I assume $w \ge 2$. In the following, I will prove $F^{j,m}_{n,d}(\boldsymbol{\lambda})_{(i_1,i_2)} = 0$ $(1 \le i_1 <$ $i_2 \leq w$). Because I have proved Eq. (5.80) of $P(j,n)$, one can set $i_1 = 1, i_2 = 2$ without losing generality. In the similar manner with the proof of Eq. (5.82) of $P(j,n)$, I have

$$\begin{aligned}
&F^{j,m}_{n,d}\left(\lambda_L; \vec{\lambda}; \lambda_R\right)_{(1,2)} \\
=&C^{j,m}_{n,0}(-1;\vec{\lambda};A) - C^{j,m}_{n-1,2}(-1;\vec{\lambda}_{1:-1,2:-1};A) - C^{j-1,m}_{n,1}(-1;\vec{\lambda}_{1:\to\lambda_{1,2},\hat{2}};A) \\
=&\left(C^{j,m}_{n-1,1}(0;\vec{\lambda}_{1:-1};A) + C^{j-1,m}_{n,0}(\lambda_1+1;\vec{\lambda}_{\hat{1}};A)\right) \\
&\qquad - C^{j,m}_{n-1,1}(1;\vec{\lambda}_{1:-1,2:-1};A) - C^{j-1,m}_{n,0}(1;\vec{\lambda}_{1:\to\lambda_{1,2},\hat{2}};A) \\
=&\left(C^{j-1,m}_{n,0}(\lambda_1+1;\lambda_2,\ldots;A) - C^{j-1,m}_{n,0}(1;\lambda_1+\lambda_2,\ldots;A)\right) \\
&\qquad + C^{j,m}_{n-1,1}(0;\vec{\lambda}_{1:-1};A) - C^{j,m}_{n-1,1}(1;\vec{\lambda}_{1:-1,2:-1};A) \\
=&\left(C^{j-1,m}_{n-1,2}(\lambda_1;\lambda_2-1,\ldots;A) - C^{j-1,m}_{n-1,2}(0;\lambda_1+\lambda_2-1,\ldots;A)\right) \\
&\qquad\qquad + C^{j,m}_{n-1,1}(0;\lambda_1-1,\lambda_2,\ldots;A) - C^{j,m}_{n-1,1}(1;\lambda_1-1,\lambda_2-1,\ldots;A) \\
=&\left(C^{j,m}_{n-1,1}(0;\lambda_1-1,\lambda_2,\ldots;A) - C^{j-1,m}_{n-1,2}(0;\lambda_1+\lambda_2-1,\ldots;A)\right) \\
&\qquad\qquad - \left(C^{j,m}_{n-1,1}(1;\lambda_1-1,\lambda_2-1,\ldots;A) - C^{j-1,m}_{n-1,2}(\lambda_1;\lambda_2-1,\ldots;A)\right) \\
=&C^{j,m}_{n-2,3}(0;\lambda_1-2,\lambda_2-1,\ldots;A) - C^{j,m}_{n-2,3}(0;\lambda_1-2,\lambda_2-1,\ldots;A) \\
=&0,
\end{aligned} \tag{A.11}$$

where one can prove the first equality with the same argument as the first to sixth equality in Eq. (A.10), $A = \lambda_{L+R} + 2d + 1$, $\lambda_{1,2} = \lambda_1 + \lambda_2$, the dotted line "$\ldots$" denotes $\lambda_3, \ldots, \lambda_w$, in the fourth equality, I used Eq. (5.89) of $P(j-1,n)$, and in the sixth equality, I used Eqs. (5.83) and (5.82) of $P(j,n-1)$.

Thus, I have proved Eq. (5.83) of $P(j,n)$.

### A.1.5 Induction for Eq. (5.84)

Throughout this subsection, I define

$$F^{j,m}_{n,d}(\boldsymbol{\lambda})_{(a,b)} \equiv C^{j,m}_{n,d}(\boldsymbol{\lambda}) - C^{j,m}_{n,d}(\boldsymbol{\lambda}_{a:-1,b:+1}) - \left\{C^{j,m}_{n-1,d+2}(\boldsymbol{\lambda}_{a:-1,b:-1}) - C^{j,m}_{n-1,d+2}(\boldsymbol{\lambda}_{a:-2})\right\}, \tag{A.12}$$

where $0 \leq a, b \leq w+1$. In the following, I will prove $F^{j,m}_{n,d}(\boldsymbol{\lambda})_{(a,b)} = 0$.

I first consider the case of $a = L$ (or $R$), $b = i (1 \leq i \leq w)$ and $\lambda_L > 0$. Because I have proved Eqs. (5.79) and (5.80) of $P(j,n)$, one can set $a = L, b = 1$ without losing generality. Using Eq. (A.3), I have

$$\begin{aligned}
&F^{j,m}_{n,d}(\lambda_L;\vec{\lambda};\lambda_R)_{(L,1)} \\
=&C^{j,m}_{n,d}(-1;\vec{\lambda};\lambda_{L+R}+1) - C^{j,m}_{n,d}(-1;\vec{\lambda}_{1:+1};\lambda_{L+R}) \\
&\qquad - \left\{C^{j,m}_{n-1,d+2}(-1;\vec{\lambda}_{1:-1};\lambda_{L+R}) - C^{j,m}_{n-1,d+2}(-1;\vec{\lambda};\lambda_{L+R}-1)\right\}
\end{aligned}$$

$$
\begin{aligned}
&+2\left(F^{j,m}_{n-1,d+1}(\boldsymbol{\lambda})_{(L,1)}-F^{j,m}_{n-1,d}(\mathcal{T}\boldsymbol{\eta})_{(L,1)}\right)-\left(F^{j,m}_{n-1,d}(\boldsymbol{\lambda}_{L:+1,R:+1})_{(L,1)}-F^{j,m}_{n-1,d}(\boldsymbol{\eta})_{(L,1)}\right)\\
&+\sum_{p=0}^{\lambda_R}\left(F^{j-1,m}_{n,d}(\boldsymbol{\xi}^{(p)})_{(1,2)}-F^{j-1,m}_{n,d}(\boldsymbol{\xi}^{(\lambda_{L+R}-p+1)})_{(L,2)}\right)\\
=&C^{j,m}_{n,d-1}(1;\vec{\lambda};\lambda_{L+R}+1)-C^{j,m}_{n,d-1}(1;\vec{\lambda}_{1:+1};\lambda_{L+R})\\
&-\left\{C^{j,m}_{n-1,d+1}(1;\vec{\lambda}_{1:-1};\lambda_{L+R})-C^{j,m}_{n-1,d+1}(1;\vec{\lambda};\lambda_{L+R}-1)\right\}\\
=&F^{j,m}_{n,d-1}(\lambda_{L+R}+1;\vec{\lambda};1)_{(L,1)}\\
=&F^{j,m}_{n,d-2}(\lambda_{L+R}+3;\vec{\lambda};1)_{(L,1)}\\
&\vdots\\
=&F^{j,m}_{n,0}(\lambda_{L+R}+2d-1;\vec{\lambda};1)_{(L,1)}\\
=&C^{j,m}_{n,0}(-1;\vec{\lambda};A+1)-C^{j,m}_{n,0}(-1;\vec{\lambda}_{1:+1};A)-\left\{C^{j,m}_{n-1,2}(-1;\vec{\lambda}_{1:-1};A)-C^{j,m}_{n-1,2}(-1;\vec{\lambda};A-1)\right\}\\
=&C^{j,m}_{n-1,1}(0;\lambda_1-1,\ldots;A+1)-C^{j,m}_{n-1,1}(0;\lambda_1,\ldots;A)\\
&+\left(C^{j-1,m}_{n,0}(\lambda_1+1;\ldots;A+1)-C^{j-1,m}_{n,0}(\lambda_1+2;\ldots;A)\right)\\
&-\left\{C^{j,m}_{n-1,1}(1;\lambda_1-1,\ldots;A)-C^{j,m}_{n-1,1}(1;\lambda_1,\ldots;A-1)\right\}\\
=&C^{j,m}_{n-1,1}(0;\lambda_1-1,\ldots;A+1)-C^{j,m}_{n-1,1}(0;\lambda_1,\ldots;A)\\
&+\left(C^{j-1,m}_{n-1,2}(\lambda_1;\ldots;A)-C^{j-1,m}_{n-1,2}(\lambda_1+1;\ldots;A-1)\right)\\
&-\left\{C^{j,m}_{n-1,1}(1;\lambda_1-1,\ldots;A)-C^{j,m}_{n-1,1}(1;\lambda_1,\ldots;A-1)\right\}\\
=&\left(C^{j,m}_{n-1,1}(1;\lambda_1,\ldots;A-1)-C^{j-1,m}_{n-1,2}(\lambda_1+1;\ldots;A-1)\right)\\
&-\left(C^{j,m}_{n-1,1}(0;\lambda_1,\ldots;A)-C^{j-1,m}_{n-1,2}(\lambda_1;\ldots;A)\right)\\
&-\left(C^{j,m}_{n-1,1}(1;\lambda_1-1,\ldots;A)-C^{j,m}_{n-1,1}(0;\lambda_1-1,\ldots;A+1)\right)\\
=&C^{j,m}_{n-2,3}(0;\lambda_1-1,\ldots;A-1)-C^{j,m}_{n-2,3}(-1;\lambda_1-1,\ldots;A)\\
&-\left(C^{j,m}_{n-1,1}(1;\lambda_1-1,\ldots;A)-C^{j,m}_{n-1,1}(0;\lambda_1-1,\ldots;A+1)\right)\\
=&\left(C^{j,m}_{n-1,1}(1;\lambda_1-1,\ldots;A)-C^{j,m}_{n-1,1}(0;\lambda_1-1,\ldots;A+1)\right)\\
&-\left(C^{j,m}_{n-1,1}(1;\lambda_1-1,\ldots;A)-C^{j,m}_{n-1,1}(0;\lambda_1-1,\ldots;A+1)\right)\\
=&0, \qquad\qquad \text{(A.13)}
\end{aligned}
$$

where $\boldsymbol{\eta}=\{\lambda_{L+R}+2;\vec{\lambda};0\}$, $\boldsymbol{\xi}^{(p)}=\{p;\lambda_{L+R}-p;\vec{\lambda};0\}$, $A=\lambda_{L+R}+2d$, in the second equality, I used Eq. (5.84) of $P(j,n-1)$ and $P(j-1,n)$, in the third equality, I used Eq. (5.79) of $P(j,n)$ I have proved above, the dotted line "…" denotes $\lambda_2,\ldots,\lambda_w$, in the eighth equality, I used Eq. (5.84) of $P(j-1,n)$, in the tenth equality, I used Eq. (5.82) of $P(j,n-1)$, and eleventh equality, I used Eq. (5.84) of $P(j,n-1)$.

I next consider the case of $b=L$ (or $R$), $a=i(1\le i\le w)$ and $\lambda_i>0$. Because I have proved Eqs. (5.79) and (5.80) of $P(j,n)$, one can set $b=L, i=1$ without losing generality. Using Eq. (A.3), I have

$$F^{j,m}_{n,d}(\lambda_L;\vec{\lambda};\lambda_R)_{(1,L)}$$

$$
\begin{aligned}
&=C_{n,0}^{j,m}(-1;\vec{\lambda};A)-C_{n,0}^{j,m}(-1;\vec{\lambda}_{1:-1};A+1)-\left\{C_{n-1,2}^{j,m}(-1;\vec{\lambda}_{1:-1};A-1)-C_{n-1,2}^{j,m}(-1;\vec{\lambda}_{i:-2};A)\right\}\\
&=C_{n-1,1}^{j,m}(0;\lambda_1-1,\ldots;A)-C_{n-1,1}^{j,m}(0;\lambda_1-2,\ldots;A+1)\\
&\qquad\qquad+\left(C_{n,0}^{j-1,m}(\lambda_1+1;\ldots;A)-C_{n,0}^{j-1,m}(\lambda_1;\ldots;A+1)\right)\\
&\qquad\qquad\qquad-\left\{C_{n-1,1}^{j,m}(1;\lambda_1-1,\ldots;A-1)-C_{n-1,1}^{j,m}(1;\lambda_1-2,\ldots;A)\right\}\\
&=C_{n-1,1}^{j,m}(0;\lambda_1-1,\ldots;A)-C_{n-1,1}^{j,m}(0;\lambda_1-2,\ldots;A+1)\\
&\qquad\qquad+\left(C_{n-1,2}^{j-1,m}(\lambda_1;\ldots;A-1)-C_{n-1,2}^{j-1,m}(\lambda_1-1;\ldots;A)\right)\\
&\qquad\qquad\qquad-\left\{C_{n-1,1}^{j,m}(1;\lambda_1-1,\ldots;A-1)-C_{n-1,1}^{j,m}(1;\lambda_1-2,\ldots;A)\right\}\\
&=C_{n-1,1}^{j,m}(0;\lambda_1-1,\ldots;A)-C_{n-1,1}^{j,m}(0;\lambda_1-2,\ldots;A+1)\\
&\qquad\qquad+\left(C_{n-1,2}^{j-1,m}(\lambda_1;\ldots;A-1)-C_{n-1,2}^{j-1,m}(\lambda_1-1;\ldots;A)\right)\\
&\qquad\qquad\qquad-\left\{C_{n-1,1}^{j,m}(1;\lambda_1-1,\ldots;A-1)-C_{n-1,1}^{j,m}(1;\lambda_1-2,\ldots;A)\right\}\\
&=C_{n-1,1}^{j,m}(1;\lambda_1-2,\ldots;A)-C_{n-1,1}^{j,m}(0;\lambda_1-2,\ldots;A+1)\\
&\qquad\qquad+\left(C_{n-1,1}^{j,m}(0;\lambda_1-1,\ldots;A)-C_{n-1,2}^{j-1,m}(\lambda_1-1;\ldots;A)\right)\\
&\qquad\qquad\qquad-\left\{C_{n-1,1}^{j,m}(1;\lambda_1-1,\ldots;A-1)-C_{n-1,2}^{j-1,m}(\lambda_1;\ldots;A-1)\right\}\\
&=C_{n-1,1}^{j,m}(1;\lambda_1-2,\ldots;A)-C_{n-1,1}^{j,m}(0;\lambda_1-2,\ldots;A+1)\\
&\qquad\qquad-\left(C_{n-2,3}^{j,m}(0;\lambda_1-2,\ldots;A-1)-C_{n-2,3}^{j,m}(-1;\lambda_1-2,\ldots;A)\right)\\
&=C_{n-1,1}^{j,m}(1;\lambda_1-2,\ldots;A)-C_{n-1,1}^{j,m}(0;\lambda_1-2,\ldots;A+1)\\
&\qquad\qquad-\left(C_{n-1,1}^{j,m}(1;\lambda_1-2,\ldots;A)-C_{n-1,1}^{j,m}(0;\lambda_1-2,\ldots;A+1)\right)\\
&=0,
\end{aligned}
\tag{A.14}
$$

where one can prove that the first equality with the same argument as the first to sixth equality in Eq. (A.13), $A=\lambda_{L+R}+2d+1$, the dotted line "…" denotes $\lambda_2,\ldots,\lambda_w$, in the third equality, I used Eq. (5.84) of $P(j-1,n)$, and in the sixth equality, I used Eq. (5.82) of $P(j,n-1)$ and in the seventh equality, I used Eq. (5.84) of $P(j,n-1)$.

I next consider the case of $(a,b)=(L,R),(R,L)$, and $w=0$ $(j=2m+1)$. Because I have proved Eq. (5.79) of $P(j,n)$, one can set $a=L,b=R$ without losing generality. Using Eq. (A.1), I have

$$
\begin{aligned}
&F_{n,d}^{j,m}(\lambda_L;\lambda_R)_{(L,R)}\\
=&2F_{n-1,d+1}^{j,m}(\lambda_L;\lambda_R)_{(L,R)}-F_{n-1,d}^{j,m}(\lambda_L+1;\lambda_R+1)_{(L,R)}\\
&\qquad+\left(C_{n,d}^{j-1,m-1}(\lambda_L;\lambda_R;0)-C_{n,d}^{j-1,m-1}(\lambda_R+1;\lambda_L-1;0)\right)\\
&\qquad\qquad-\left(C_{n-1,d+2}^{j-1,m-1}(\lambda_L-1;\lambda_R-1;0)-C_{n-1,d+2}^{j-1,m-1}(\lambda_R;\lambda_L-2;0)\right)\\
=&0,
\end{aligned}
\tag{A.15}
$$

where in the first equality, I apply Eq. (A.1) to every term in $F_{n,d}^{j,m}(\lambda_L;\lambda_R)_{(L,R)}$ and one can see there are trivial cancellations with straightforward calculation, and in the second equality, I used Eq. (5.84) of $P(j,n-1)$ and used Eq. (5.89) which is derived from Eq. (5.84) of $P(j-1,n)$.

The other cases of Eq. (5.84) of $P(j,n)$ can be easily proved from the above results. In the case of $w \geq 2$ and $a = i_1, b = i_2 (1 \leq i_1, i_2 \leq w)$ and $\lambda_a > 0$, I have

$$F^{j,m}_{n,d}(\boldsymbol{\lambda})_{(i_1,i_2)} = F^{j,m}_{n,d}(\boldsymbol{\lambda})_{(i_1,L)} + F^{j,m}_{n,d}(\boldsymbol{\lambda}_{L:+1,i_1:-1})_{(L,i_2)} = 0, \tag{A.16}$$

where I used the above result in Eqs. (A.13) and (A.14). In the case of $w > 0$ and $a = L, b = R$, I have

$$F^{j,m}_{n,d}(\boldsymbol{\lambda})_{(L,R)} = F^{j,m}_{n,d}(\boldsymbol{\lambda})_{(L,i)} + F^{j,m}_{n,d}(\boldsymbol{\lambda}_{i:+1,L:-1})_{(i,R)} = 0, \tag{A.17}$$

where I used the above results (A.13) and (A.14).

Thus, I have proved Eq. (5.84) of $P(j,n)$.

Therefore, I have proved Eqs. (5.79)–(5.84) by induction.

### A.1.6 Proof of positivity of $C^{j,m}_{n,d}(\boldsymbol{\lambda})$

In this subsection, I prove the positivity of the coefficient: $C^{j,m}_{n,d}(\boldsymbol{\lambda}) > 0$. I prove this positivity by induction. I show $C^{j,m}_{n,d}(\boldsymbol{\lambda}) > 0$ if I suppose $C^{j-1,m}_{n,d}(\boldsymbol{\lambda}) > 0$ and $C^{j,m}_{n-1,d}(\boldsymbol{\lambda}) > 0$ hold. The base cases trivially hold from $C^{1,m}_{n,d} = 1$ and from Eq. (5.67). From Eqs. (5.79) and (5.81), I can set $0 \leq \lambda_L \leq \lambda_R \leq n$ without losing generality.

First, I prove $\Delta C^{j,m}_{n,d}(\boldsymbol{\lambda}) > 0$. From Eq. (5.91), I have

$$\begin{aligned}
\Delta C^{j,m}_{n,d}(\boldsymbol{\lambda}) &= \sum_{\tilde{n}=0}^{n} (n+1-\tilde{n}) \left( C^{j-1,m-1}_{\tilde{n},n+d-\tilde{n}}(\boldsymbol{\lambda}_{\leftarrow 0}) - C^{j-1,m-1}_{\tilde{n},n+d-\tilde{n}}({}_{0\rightarrow}(\mathcal{T}\boldsymbol{\lambda})) \right) \\
&= \sum_{\tilde{n}=0}^{n} (n+1-\tilde{n}) \left( C^{j-1,m-1}_{\tilde{n},n+d-\tilde{n}}\left(0; \vec{\lambda}, \lambda_R; \lambda_L\right) - C^{j-1,m-1}_{\tilde{n},n+d-\tilde{n}}\left(0; \vec{\lambda}, \lambda_L - 1; \lambda_R + 1\right) \right) \\
&= \sum_{\tilde{n}=0}^{n} (n+1-\tilde{n}) \left( C^{j-1,m-1}_{\tilde{n}-\lambda_L,n+d-\tilde{n}+2\lambda_L}\left(0; \vec{\lambda}, \lambda_R - \lambda_L; 0\right) \right. \\
&\qquad\qquad \left. - C^{j-1,m-1}_{\tilde{n}-\lambda_L,n+d-\tilde{n}+2\lambda_L}\left(0; \vec{\lambda}, -1; \lambda_R - \lambda_L + 1\right) \right) \\
&= \sum_{\tilde{n}=0}^{n} (n+1-\tilde{n}) C^{j-1,m-1}_{\tilde{n}-\lambda_L,n+d-\tilde{n}+2\lambda_L}\left(0; \vec{\lambda}, \lambda_R - \lambda_L; 0\right) \\
&\geq C^{j-1,m-1}_{n-\lambda_L,d+2\lambda_L}\left(0; \vec{\lambda}, \lambda_R - \lambda_L; 0\right) \\
&> 0,
\end{aligned} \tag{A.18}$$

where in the third equality, I used Eq. (5.84) repeatedly, and the last and second-to-last inequalities hold due to the assumption of induction. Then I proved $\Delta C^{j,m}_{n,d}(\boldsymbol{\lambda}) > 0$.

From the above argument, I have

$$C^{j,m}_{n,d}(\boldsymbol{\lambda}) = \Delta C^{j,m}_{n,d}(\boldsymbol{\lambda}) + C^{j,m}_{n,d}(\lambda_L - 1; \vec{\lambda}; \lambda_R + 1)$$

$$
\begin{aligned}
&> C^{j,m}_{n,d}(\lambda_L - 1; \vec{\lambda}; \lambda_R + 1) \\
&\vdots \\
&> C^{j,m}_{n,d}(-1; \vec{\lambda}; \lambda_{L+R} + 1) \\
&= C^{j,m}_{n,d-1}(1; \vec{\lambda}; \lambda_{L+R} + 1) \\
&> C^{j,m}_{n,d-2}(1; \vec{\lambda}; \lambda_{L+R} + 3) \\
&\vdots \\
&> C^{j,m}_{n,0}(1; \vec{\lambda}; \lambda_{L+R} + 2d - 1) \\
&> C^{j,m}_{n,0}(-1; \vec{\lambda}; \lambda_{L+R} + 2d + 1) \\
&= C^{j,m}_{n-1,1}(0; \lambda_1 - 1, \ldots; \lambda_{L+R} + 2d + 1) + C^{j-1,m}_{n,0}(\lambda_1 + 1, \ldots; \lambda_{L+R} + 2d + 1) \\
&> 0,
\end{aligned}
\tag{A.19}
$$

where I used the definition (5.59) repeatedly, and the last inequality holds due to the assumption of induction. Thus, I have proved the positivity of the coefficient $C^{j,m}_{n,d}(\boldsymbol{\lambda}) > 0$.

## A.2 Demonstration of identities in special cases

In this section, I demonstrate the identities (5.79)–(5.84) actually holds for the cases where the general explicit forms are obtained, i.e., $C^{j,m}_{n=0,d}$ and $C^{j,m=0}_{n,d}$, though I have already proved these identities generally above.

The identities in Eqs. (5.79) and (5.80) hold trivially for the $m = 0$ case because $C^{j,m=0}_{n,d}(\boldsymbol{\lambda})$ does not depend on $\lambda_L$ and $\lambda_R$ and is symmetric under the permutation of $\lambda_i$'s $(1 \leq i \leq j-1)$ from Eq. (5.65). Thus, I demonstrate the other cases in this section.

### A.2.1 Demonstration of identities for $C^{j,m}_{n=0,d}(\boldsymbol{\lambda})$

In this subsection, I directly prove that $P(j, n = 0)$ actually holds from the general explicit form (5.67).

Using the generalized Catalan number [27, 28]:

$$
C_{n,k} \coloneqq \binom{n+k}{k} - \binom{n+k}{k-1}, \tag{A.20}
$$

the explicit solution (5.67) is written as

$$
C^{j,m}_{n=0,d}(\boldsymbol{\lambda}) = C_{j-1-m+d,m}. \tag{A.21}
$$

The generalized Catalan number satisfies the following equations.

$$
C_{n,k} = C_{n,k-1} + C_{n-1,k} \qquad \text{for } 1 < k < n, \tag{A.22}
$$

$$
C_{n,n} = C_{n,n-1} \qquad \text{for } n \geq 2. \tag{A.23}
$$

### Demonstration of recursion equation for $n=0$ case

I first confirm the explicit solution (5.67) actually satisfies the recursion equation (5.57). The recursion equation (5.57) for the $n=0$ case is

$$\Delta C^{j,m}_{0,d}(\boldsymbol{\lambda}) = C^{j-1,m-1}_{0,d}(\boldsymbol{\lambda}_{\leftarrow 0}) - C^{j-1,m-1}_{0,d}\left({}_{0\rightarrow}(\mathcal{T}\boldsymbol{\lambda})\right), \tag{A.24}$$

where $\boldsymbol{\lambda} = \{\lambda_L;\lambda_1,\ldots,\lambda_w;\lambda_R\}$ ($w \equiv j-1-2m$), and $\boldsymbol{\lambda}_{\leftarrow 0} = \{\lambda_L;\lambda_1,\ldots,\lambda_w,\lambda_R;0\}$, and ${}_{0\rightarrow}(\mathcal{T}\boldsymbol{\lambda}) = \{0;\lambda_L-1,\lambda_1,\ldots,\lambda_w;\lambda_R+1\}$. I note that $\Delta C^{j,m}_{-1,d+1}(\boldsymbol{\lambda}) = \Delta C^{j,m}_{-2,d+2}(\boldsymbol{\lambda}) = 0$.

In the case of $\lambda_L > 0$, Eq. (A.24) trivially holds; the left-hand side of Eq. (A.24) is

$$\begin{aligned}\Delta C^{j,m}_{0,d}(\boldsymbol{\lambda}) &= C^{j,m}_{0,d}(\lambda_L;\lambda_1,\ldots,\lambda_w;\lambda_R) - C^{j,m}_{0,d}(\lambda_L-1;\lambda_1,\ldots,\lambda_w;\lambda_R+1)\\ &= C_{j-1-m+d,m} - C_{j-1-m+d,m}\\ &= 0,\end{aligned} \tag{A.25}$$

and the left-hand side of Eq. (A.24) is

$$\begin{aligned}&C^{j-1,m-1}_{0,d}(\lambda_L;\lambda_1,\ldots,\lambda_w,\lambda_R;0) - C^{j-1,m-1}_{0,d}(0;\lambda_L-1,\lambda_1,\ldots,\lambda_w;\lambda_R+1)\\ =&C_{j-1-m+d,m-1} - C_{j-1-m+d,m-1}\\ =&0\,.\end{aligned} \tag{A.26}$$

Thus, one can see both sides of Eq. (A.24) are zero.

In the case of $\lambda_L = 0$, the left-hand side of Eq. (A.24) becomes

$$\begin{aligned}\Delta C^{j,m}_{0,d}(\boldsymbol{\lambda}) &= C^{j,m}_{0,d}(0;\lambda_1,\ldots,\lambda_w;\lambda_R) - C^{j,m}_{0,d}(-1;\lambda_1,\ldots,\lambda_w;\lambda_R+1)\\ &= \begin{cases} C^{j,m}_{0,d}(0;\lambda_1,\ldots) - C^{j,m}_{0,d-1}(1;\lambda_1,\ldots) & \text{for } d>0\\ C^{j,m}_{0,0}(0;\lambda_1,\ldots) - C^{j,m}_{-1,1}(0;\lambda_1-1,\ldots) - C^{j-1,m}_{0,0}(\lambda_1+1;\ldots) & \text{for } d=0\end{cases}\\ &= \begin{cases} C_{j-1-m+d,m} - C_{j-2-m+d,m} & \text{for } d>0\\ C_{j-1-m,m} - C_{j-2-m,m} & \text{for } d=0\end{cases}\\ &= C_{j-1-m+d,m} - C_{j-2-m+d,m}\\ &= C_{j-1-m+d,m-1},\end{aligned} \tag{A.27}$$

where I remark that $C^{j,m}_{-1,1}(0;\lambda_1-1,\ldots)=0$, in the last equality, I used Eq. (A.22), and omitted the irrelevant part of lists which is denoted by the dotted line "...". The right-hand side of Eq. (A.24) becomes

$$\begin{aligned}&C^{j-1,m-1}_{0,d}(0;\lambda_1,\ldots,\lambda_w,\lambda_R;0) - C^{j-1,m-1}_{0,d}(0;-1,\lambda_1,\ldots,\lambda_w;\lambda_R+1)\\ &= C_{j-1-m+d,m-1},\end{aligned} \tag{A.28}$$

where I remark that $C^{j-1,m-1}_{0,d}(0;-1,\lambda_1,\ldots,\lambda_w;\lambda_R+1)=0$. Thus, one can see that both sides of Eq. (A.24) become $C_{j-1-m+d,m-1}$ in the case of $\lambda_L = 0$.

Thus, I have proved that the solution (5.67) actually satisfies the recursion equation (5.57).

**Demonstration of** $P(j, n = 0)$

I demonstrate Eq. (A.24) actually satisfies the identities (5.79)-(5.84). From Eq. (A.24), $C_{n=0,d}^{j,m}(\boldsymbol{\lambda})$ does not depend on $\boldsymbol{\lambda}$, and then one can see that Eqs. (5.79), (5.80), (5.81) and (5.84) trivially hold.

I demonstrate Eq. (5.82) for $n = 0$ case:

$$\begin{aligned} C_{n=-1,d+2}^{j,m}\left(\boldsymbol{\lambda}_{L(R):-1,i:-1}\right) + C_{n=0,d+1}^{j-1,m}\left(\boldsymbol{\lambda}_{\lambda_{L(R)}\to\lambda_{L(R)}+\lambda_{i_1},\hat{\imath}}\right) =& C_{(j-1)-1-m+(d+1),m} \\ =& C_{j-1-m+d,m}, \end{aligned} \tag{A.29}$$

where I note that $C_{n=-1,d+2}^{j,m}\left(\boldsymbol{\lambda}_{L(R):-1,i:-1}\right) = 0$, one can see that the last line is the very $C_{n=0,d}^{j,m}(\boldsymbol{\lambda})$ from Eq. (5.67), and then I have demonstrated Eq. (5.82) for the $n = 0$ case. With the same argument, one can also demonstrate Eq. (5.83) for the $n = 0$ case.

Thus, I have demonstrated $P(j, n = 0)$ actually holds.

### A.2.2 Demonstration of identities for $C_{n,d}^{j,m=0}(\boldsymbol{\lambda})$

In this subsection, I demonstrate the explicit solution of Eq. (5.65) actually satisfies the recursion equation (5.57) and the identities (5.79)–(5.84).

The identities (5.79) and (5.80) trivially hold from Eq. (5.65); thus, I demonstrate the other identities.

**Demonstration of recursion equation for** $m = 0$ **case**

I first demonstrate the explicit solution (5.65) actually satisfies the recursion equation (5.57). The recursion equation (5.57) for the $m = 0$ case is

$$\Delta C_{n,d}^{j,0}(\boldsymbol{\lambda}) - 2\Delta C_{n-1,d+1}^{j,0}(\boldsymbol{\lambda}) + \Delta C_{n-2,d+2}^{j,0}(\boldsymbol{\lambda}) = 0. \tag{A.30}$$

one can show this recursion equation (A.30) trivially holds due to $\Delta C_{n,d}^{j,0}(\boldsymbol{\lambda}) = 0$, as is shown in the following.

From Eq. (5.65), $C_{n,d}^{j,m=0}(\lambda_L; \lambda_1, \ldots, \lambda_w; \lambda_R)$ does not depend on $\lambda_L$ and $\lambda_R$, and one can see $\Delta C_{n,d}^{j,0}(\boldsymbol{\lambda}) = 0$ for the $\lambda_L > 0$ cases.

I next consider the case of $\lambda_L = 0$ and $d > 0$. In this case, $\Delta C_{n,d}^{j,0}(\boldsymbol{\lambda})$ becomes

$$\begin{aligned} \Delta C_{n,d}^{j,0}(\boldsymbol{\lambda}) &= C_{n,d}^{j,0}(0; \lambda_1, \ldots, \lambda_{j-1}; \lambda_R) - C_{n,d-1}^{j,0}(1; \lambda_1, \ldots, \lambda_{j-1}; \lambda_R + 1) \\ &= 0, \end{aligned} \tag{A.31}$$

where I used the fact that $C_{n,d}^{j,m=0}(\lambda_L; \lambda_1, \ldots, \lambda_w; \lambda_R)$ also does not depend on $d$ from Eq. (5.65).

I next consider the case of $\lambda_L = 0$ and $d = 0$. In this case, $\Delta C_{n,d}^{j,0}(\boldsymbol{\lambda})$ becomes

$$\Delta C_{n,d}^{j,0}(\boldsymbol{\lambda}) = C_{n,0}^{j,0}(0; \lambda_1, \ldots, \lambda_{j-1}; \lambda_R) - C_{n,0}^{j,0}(-1; \lambda_1, \ldots, \lambda_{j-1}; \lambda_R + 1), \tag{A.32}$$

and the second term on the right-hand side is

$$
\begin{aligned}
&C_{n,0}^{j,0}(-1;\lambda_1,\ldots,\lambda_w;\lambda_R+1)\\
=&C_{n-1,1}^{j,0}(0;\lambda_1-1,\ldots,\lambda_{j-1};\lambda_R+1)+C_{n,0}^{j-1,0}(\lambda_1+1;\lambda_2,\ldots,\lambda_{j-1};\lambda_R+1)\\
=&\sum_{x_1=0}^{\lambda_1-1}\sum_{x_2=0}^{\lambda_2}\cdots\sum_{x_{j-1}=0}^{\lambda_{j-1}}\theta\left(n-1-\sum_{l=1}^{j-1}x_l\right)+\sum_{x_2=0}^{\lambda_2}\cdots\sum_{x_{j-1}=0}^{\lambda_{j-1}}\theta\left(n-\sum_{l=2}^{j-1}x_l\right)\\
=&\sum_{x_1=1}^{\lambda_1}\sum_{x_2=0}^{\lambda_2}\cdots\sum_{x_{j-1}=0}^{\lambda_{j-1}}\theta\left(n-\sum_{l=1}^{j-1}x_l\right)+\sum_{x_2=0}^{\lambda_2}\cdots\sum_{x_{j-1}=0}^{\lambda_{j-1}}\theta\left(n-\sum_{l=2}^{j-1}x_l\right)\\
=&\sum_{x_1=0}^{\lambda_1}\sum_{x_2=0}^{\lambda_2}\cdots\sum_{x_{j-1}=0}^{\lambda_{j-1}}\theta\left(n-\sum_{l=1}^{j-1}x_l\right)\\
=&C_{n,0}^{j,0}(0;\lambda_1,\ldots,\lambda_{j-1};\lambda_R),
\end{aligned}
\tag{A.33}
$$

where I performed the variable transformation of $x_1 \to x_1+1$ in the first term in the third equality. Then, one can see $\Delta C_{n,d}^{j,0}(\boldsymbol{\lambda})=0$.

Thus, I have confirmed $\Delta C_{n,d}^{j,0}(\boldsymbol{\lambda})=0$ for all cases and demonstrated the recursion equation (A.30) trivially holds.

### Demonstration of Eq. (5.81) for $m=0$ case

I demonstrate Eq. (5.81) actually holds for the $m=0$ case. $C_{n,d}^{j,m=0}$ does not depend on $\lambda_L$ and $\lambda_R$, and $a=L,R$ case is trivially holds. I consider the case of $a=i(1\le i\le j-1)$ and $\lambda_a\ge n$.

I introduce the following notation $\overset{\hat{i}}{\cdots}$, which represents the absence of the summation over $x_i$, such as

$$
\sum_{x_1}\overset{\hat{i}}{\cdots}\sum_{x_n}\equiv\sum_{x_1}\cdots\sum_{x_{i-1}}\sum_{x_{i+1}}\cdots\sum_{x_n}.
\tag{A.34}
$$

$C_{n,d}^{j,m=0}(\boldsymbol{\lambda})$ can be transformed as follows:

$$
\begin{aligned}
C_{n,d}^{j,m=0}(\boldsymbol{\lambda})&=\sum_{x_1=0}^{\lambda_1}\cdots\sum_{x_{j-1}=0}^{\lambda_{j-1}}\theta\left(n-\sum_{l=1}^{j-1}x_l\right)\\
&=\sum_{x_1=0}^{\lambda_1}\overset{\hat{i}}{\cdots}\sum_{x_{j-1}=0}^{\lambda_{j-1}}\sum_{x_i=0}^{\lambda_i}\theta\left(n-\sum_{l=1}^{j-1}x_l\right)\\
&=\sum_{x_1=0}^{\lambda_1}\overset{\hat{i}}{\cdots}\sum_{x_{j-1}=0}^{\lambda_{j-1}}\left\{\sum_{x_i=0}^{n}\theta\left(n-\sum_{l=1}^{j-1}x_l\right)+\sum_{x_i=n+1}^{\lambda_i}\theta\left(n-\sum_{l=1}^{j-1}x_l\right)\right\}
\end{aligned}
$$

$$
\begin{aligned}
&= \sum_{x_1=0}^{\lambda_1} \overset{\hat{i}}{\cdots} \sum_{x_{j-1}=0}^{\lambda_{j-1}} \sum_{x_i=0}^{\min(n,\lambda_i)} \theta\left(n - \sum_{l=1}^{j-1} x_l\right) \\
&= \sum_{x_1=0}^{\lambda_1} \overset{\hat{i}}{\cdots} \sum_{x_{j-1}=0}^{\lambda_{j-1}} \sum_{x_i=0}^{\min(n,\lambda_i)} \theta\left(n - \sum_{l=1}^{j-1} x_l\right) \\
&= C^{j,m=0}_{n,d}\left(\boldsymbol{\lambda}_{\lambda_i \to \min(\lambda_i,n)}\right),
\end{aligned} \tag{A.35}
$$

where I used $\sum_{x_i=n+1}^{\lambda_i} \theta\left(n - \sum_{l=1}^{j-1} x_l\right) = 0$ in the fourth equality.

Thus, I have demonstrated (5.81) for the $m = 0$ case.

### Demonstration of Eq. (5.82) for $m = 0$ case

I demonstrate Eq. (5.82) actually holds in the case $m = 0$.

$$
\begin{aligned}
&C^{j,m=0}_{n-1,d+2}\left(\boldsymbol{\lambda}_{L(R):-1,i:-1}\right) + C^{j-1,m=0}_{n,d+1}\left(\boldsymbol{\lambda}_{\lambda_{L(R)} \to \lambda_{L(R)} + \lambda_{i_1}, \hat{i}}\right) \\
&= \sum_{x_1=0}^{\lambda_1} \overset{\hat{i}}{\cdots} \sum_{x_{j-1}=0}^{\lambda_{j-1}} \sum_{x_i=0}^{\lambda_i-1} \theta\left(n - 1 - \sum_{l=1}^{j-1} x_l\right) + \sum_{x_1=0}^{\lambda_1} \overset{\hat{i}}{\cdots} \sum_{x_{j-1}=0}^{\lambda_{j-1}} \theta\left(n - \sum_{\substack{1 \le l \le j-1 \\ l \ne i}} x_l\right) \\
&= \sum_{x_1=0}^{\lambda_1} \overset{\hat{i}}{\cdots} \sum_{x_{j-1}=0}^{\lambda_{j-1}} \left\{ \sum_{x_i=0}^{\lambda_i-1} \theta\left(n - 1 - \sum_{l=1}^{j-1} x_l\right) + \theta\left(n - \sum_{\substack{1 \le l \le j-1 \\ l \ne i}} x_l\right) \right\} \\
&= \sum_{x_1=0}^{\lambda_1} \overset{\hat{i}}{\cdots} \sum_{x_{j-1}=0}^{\lambda_{j-1}} \left\{ \sum_{x_i=1}^{\lambda_i} \theta\left(n - x_i - \sum_{\substack{1 \le l \le j-1 \\ l \ne i}} x_l\right) + \theta\left(n - \sum_{\substack{1 \le l \le j-1 \\ l \ne i}} x_l\right) \right\} \\
&= \sum_{x_1=0}^{\lambda_1} \overset{\hat{i}}{\cdots} \sum_{x_{j-1}=0}^{\lambda_{j-1}} \sum_{x_i=0}^{\lambda_i} \theta\left(n - \sum_{l=1}^{j-1} x_l\right) \\
&= \sum_{x_1=0}^{\lambda_1} \cdots \sum_{x_{j-1}=0}^{\lambda_{j-1}} \theta\left(n - \sum_{l=1}^{j-1} x_l\right) \\
&= C^{j,m=0}_{n,d}(\boldsymbol{\lambda}).
\end{aligned} \tag{A.36}
$$

Then, I have proved (5.82) for the base case $m = 0$.

### Demonstration of Eq. (5.83) for $m = 0$ case

I prove Eq. (5.83) for $m = 0$ base case.

$$
C^{j,m=0}_{n-1,d+2}\left(\boldsymbol{\lambda}_{i_1:-1,i_2:-1}\right) + C^{j-1,m=0}_{n,d+1}\left(\boldsymbol{\lambda}_{\lambda_{i_1} \to \lambda_{i_1} + \lambda_{i_2}, \hat{i}_2}\right)
$$

$$
\begin{aligned}
&=\sum_{x_1=0}^{\lambda_1}\overset{\hat{i}_1,\hat{i}_2}{\cdots}\sum_{x_{j-1}=0}^{\lambda_{j-1}}\sum_{x_{i_1}=0}^{\lambda_{i_1}-1}\sum_{x_{i_2}=0}^{\lambda_{i_2}-1}\theta\left(n-1-\sum_{l=1}^{j-1}x_l\right)+\sum_{x_1=0}^{\lambda_1}\overset{\hat{i}_1,\hat{i}_2}{\cdots}\sum_{x_{j-1}=0}^{\lambda_{j-1}}\sum_{x_p=0}^{\lambda_{i_1}+\lambda_{i_2}}\theta\left(n-x_p-\sum_{\substack{1\le l\le j-1\\ l\ne i_1,i_2}}x_l\right)\\
&=\sum_{x_1=0}^{\lambda_1}\overset{\hat{i}_1,\hat{i}_2}{\cdots}\sum_{x_{j-1}=0}^{\lambda_{j-1}}\left\{\sum_{x_{i_1}=0}^{\lambda_{i_1}-1}\sum_{x_{i_2}=0}^{\lambda_{i_2}-1}\theta\left(n-1-x_{i_1}-x_{i_2}-\sum_{\substack{1\le l\le j-1\\ l\ne i_1,i_2}}x_l\right)+\sum_{x_p=0}^{\lambda_{i_1}+\lambda_{i_2}}\theta\left(n-x_p-\sum_{\substack{1\le l\le j-1\\ l\ne i_1,i_2}}x_l\right)\right\}\\
&=\sum_{x_1=0}^{\lambda_1}\overset{\hat{i}_1,\hat{i}_2}{\cdots}\sum_{x_{j-1}=0}^{\lambda_{j-1}}\left\{\sum_{x_{i_1}=0}^{\lambda_{i_1}-1}\sum_{x_{i_2}=1}^{\lambda_{i_2}}\theta\left(n-x_{i_1}-x_{i_2}-\sum_{\substack{1\le l\le j-1\\ l\ne i_1,i_2}}x_l\right)+\sum_{x_{i_1}=0}^{\lambda_{i_1}-1}\theta\left(n-x_{i_1}-\sum_{\substack{1\le l\le j-1\\ l\ne i_1,i_2}}x_l\right)\right.\\
&\qquad\qquad\left.+\sum_{x_p=\lambda_{i_1}}^{\lambda_{i_1}+\lambda_{i_2}}\theta\left(n-x_p-\sum_{\substack{1\le l\le j-1\\ l\ne i_1,i_2}}x_l\right)\right\}\\
&=\sum_{x_1=0}^{\lambda_1}\overset{\hat{i}_1,\hat{i}_2}{\cdots}\sum_{x_{j-1}=0}^{\lambda_{j-1}}\left\{\sum_{x_{i_1}=0}^{\lambda_{i_1}-1}\sum_{x_{i_2}=0}^{\lambda_{i_2}}\theta\left(n-x_{i_1}-x_{i_2}-\sum_{\substack{1\le l\le j-1\\ l\ne i_1,i_2}}x_l\right)+\sum_{x_p=\lambda_{i_1}}^{\lambda_{i_1}+\lambda_{i_2}}\theta\left(n-x_p-\sum_{\substack{1\le l\le j-1\\ l\ne i_1,i_2}}x_l\right)\right\}\\
&=\sum_{x_1=0}^{\lambda_1}\overset{\hat{i}_1,\hat{i}_2}{\cdots}\sum_{x_{j-1}=0}^{\lambda_{j-1}}\left\{\sum_{x_{i_1}=0}^{\lambda_{i_1}-1}\sum_{x_{i_2}=0}^{\lambda_{i_2}}\theta\left(n-x_{i_1}-\sum_{\substack{1\le l\le j-1\\ l\ne i_1}}x_l\right)+\sum_{x_{i_2}=0}^{\lambda_{i_2}}\theta\left(n-\lambda_{i_1}-x_{i_2}-\sum_{\substack{1\le l\le j-1\\ l\ne i_1,i_2}}x_l\right)\right\}\\
&=\sum_{x_1=0}^{\lambda_1}\overset{\hat{i}_1,\hat{i}_2}{\cdots}\sum_{x_{j-1}=0}^{\lambda_{j-1}}\sum_{x_{i_2}=0}^{\lambda_{i_2}}\left\{\sum_{x_{i_1}=0}^{\lambda_{i_1}-1}\theta\left(n-x_{i_1}-\sum_{\substack{1\le l\le j-1\\ l\ne i_1}}x_l\right)+\theta\left(n-\lambda_{i_1}-\sum_{\substack{1\le l\le j-1\\ l\ne i_1}}x_l\right)\right\}\\
&=\sum_{x_1=0}^{\lambda_1}\overset{\hat{i}_1}{\cdots}\sum_{x_{j-1}=0}^{\lambda_{j-1}}\left\{\sum_{x_{i_1}=0}^{\lambda_{i_1}}\theta\left(n-x_{i_1}-\sum_{\substack{1\le l\le j-1\\ l\ne i_1}}x_l\right)\right\}\\
&=\sum_{x_1=0}^{\lambda_1}\cdots\sum_{x_{j-1}=0}^{\lambda_{j-1}}\theta\left(n-\sum_{l=1}^{j-1}x_l\right)\\
&=C_{n,d}^{j,m=0}(\boldsymbol{\lambda}).
\end{aligned}
\tag{A.37}
$$

Thus, I have proved Eq. (5.83) for the case $m=0$.

**Demonstration of Eq. (5.84) for $m = 0$ case**

I prove Eq. (5.84) for the case $m = 0$. Equation (5.84) in the case $a = R(L), b = L(R)$ holds trivially because $C_{n,d}^{j,m=0}(\boldsymbol{\lambda})$ does not depend on $d$ and $\lambda_L$ and $\lambda_R$.

In the case $a = L(R), b = i (1 \leq i \leq j-1)$, I have

$$\begin{aligned}
&C_{n,d}^{j,m=0}\left(\boldsymbol{\lambda}_{L(R):-1,i:+1}\right) + C_{n-1,d+2}^{j,m=0}\left(\boldsymbol{\lambda}_{L(R):-1,i:-1}\right) - C_{n-1,d+2}^{j,m=0}\left(\boldsymbol{\lambda}_{L(R):-2}\right)\\
=&\sum_{x_1=0}^{\lambda_1}\overset{\hat{i}}{\cdots}\sum_{x_{j-1}=0}^{\lambda_{j-1}}\sum_{x_i=0}^{\lambda_i+1}\theta\left(n-\sum_{l=1}^{j-1}x_l\right)+\sum_{x_1=0}^{\lambda_1}\overset{\hat{i}}{\cdots}\sum_{x_{j-1}=0}^{\lambda_{j-1}}\sum_{x_i=0}^{\lambda_i-1}\theta\left(n-1-\sum_{l=1}^{j-1}x_l\right)\\
&\qquad-\sum_{x_1=0}^{\lambda_1}\overset{\hat{i}}{\cdots}\sum_{x_{j-1}=0}^{\lambda_{j-1}}\sum_{x_i=0}^{\lambda_i}\theta\left(n-1-\sum_{l=1}^{j-1}x_l\right)\\
=&\sum_{x_1=0}^{\lambda_1}\overset{\hat{i}}{\cdots}\sum_{x_{j-1}=0}^{\lambda_{j-1}}\sum_{x_i=0}^{\lambda_i+1}\theta\left(n-\sum_{l=1}^{j-1}x_l\right)-\sum_{x_1=0}^{\lambda_1}\overset{\hat{i}}{\cdots}\sum_{x_{j-1}=0}^{\lambda_{j-1}}\theta\left(n-1-\lambda_i-\sum_{\substack{1\leq l\leq j-1\\ l\neq i}}x_l\right)\\
=&\sum_{x_1=0}^{\lambda_1}\overset{\hat{i}}{\cdots}\sum_{x_{j-1}=0}^{\lambda_{j-1}}\sum_{x_i=0}^{\lambda_i}\theta\left(n-\sum_{l=1}^{j-1}x_l\right)\\
=&\sum_{x_1=0}^{\lambda_1}\cdots\sum_{x_{j-1}=0}^{\lambda_{j-1}}\theta\left(n-\sum_{l=1}^{j-1}x_l\right)\\
=&C_{n,d}^{j,m=0}\left(\boldsymbol{\lambda}\right).
\end{aligned}\tag{A.38}$$

I next consider $a = i, b = L(R) (1 \leq i \leq j-1)$ case:

$$\begin{aligned}
&C_{n,d}^{j,m=0}\left(\boldsymbol{\lambda}_{i:-1,L(R):+1}\right) + C_{n-1,d+2}^{j,m=0}\left(\boldsymbol{\lambda}_{i:-1,L(R):-1}\right) - C_{n-1,d+2}^{j,m=0}\left(\boldsymbol{\lambda}_{i:-2}\right)\\
=&\sum_{x_1=0}^{\lambda_1}\overset{\hat{i}}{\cdots}\sum_{x_{j-1}=0}^{\lambda_{j-1}}\sum_{x_i=0}^{\lambda_i-1}\theta\left(n-\sum_{l=1}^{j-1}x_l\right)+\sum_{x_1=0}^{\lambda_1}\overset{\hat{i}}{\cdots}\sum_{x_{j-1}=0}^{\lambda_{j-1}}\sum_{x_i=0}^{\lambda_i-1}\theta\left(n-1-\sum_{l=1}^{j-1}x_l\right)\\
&\qquad-\sum_{x_1=0}^{\lambda_1}\overset{\hat{i}}{\cdots}\sum_{x_{j-1}=0}^{\lambda_{j-1}}\sum_{x_i=0}^{\lambda_i-2}\theta\left(n-1-\sum_{l=1}^{j-1}x_l\right)\\
=&\sum_{x_1=0}^{\lambda_1}\overset{\hat{i}}{\cdots}\sum_{x_{j-1}=0}^{\lambda_{j-1}}\sum_{x_i=0}^{\lambda_i-1}\theta\left(n-\sum_{l=1}^{j-1}x_l\right)+\sum_{x_1=0}^{\lambda_1}\overset{\hat{i}}{\cdots}\sum_{x_{j-1}=0}^{\lambda_{j-1}}\theta\left(n-1-(\lambda_i-1)-\sum_{\substack{1\leq l\leq j-1\\ l\neq i}}x_l\right)\\
=&\sum_{x_1=0}^{\lambda_1}\overset{\hat{i}}{\cdots}\sum_{x_{j-1}=0}^{\lambda_{j-1}}\sum_{x_i=0}^{\lambda_i}\theta\left(n-\sum_{l=1}^{j-1}x_l\right)\\
=&\sum_{x_1=0}^{\lambda_1}\cdots\sum_{x_{j-1}=0}^{\lambda_{j-1}}\theta\left(n-\sum_{l=1}^{j-1}x_l\right)
\end{aligned}$$

$$=C_{n,d}^{j,m=0}(\boldsymbol{\lambda}). \tag{A.39}$$

I next consider $a=i_1, b=i_2 (1\le i_1, i_2\le j-1)$ case:

$$\begin{aligned}
&C_{n,d}^{j,m=0}(\boldsymbol{\lambda}_{i_1:-1,i_2:+1})+C_{n-1,d+2}^{j,m=0}(\boldsymbol{\lambda}_{i_1:-1,i_2:-1})-C_{n-1,d+2}^{j,m=0}(\boldsymbol{\lambda}_{i_1:-2})\\
&=\sum_{x_1=0}^{\lambda_1}\overset{\hat{i}_1,\hat{i}_2}{\cdots}\sum_{x_{j-1}=0}^{\lambda_{j-1}}\sum_{x_{i_1}=0}^{\lambda_{i_1}-1}\sum_{x_{i_2}=0}^{\lambda_{i_2}+1}\theta\left(n-\sum_{l=1}^{j-1}x_l\right)+\sum_{x_1=0}^{\lambda_1}\overset{\hat{i}_1,\hat{i}_2}{\cdots}\sum_{x_{j-1}=0}^{\lambda_{j-1}}\sum_{x_{i_1}=0}^{\lambda_{i_1}-1}\sum_{x_{i_2}=0}^{\lambda_{i_2}-1}\theta\left(n-1-\sum_{l=1}^{j-1}x_l\right)\\
&\qquad-\sum_{x_1=0}^{\lambda_1}\overset{\hat{i}_1,\hat{i}_2}{\cdots}\sum_{x_{j-1}=0}^{\lambda_{j-1}}\sum_{x_{i_1}=0}^{\lambda_{i_1}-2}\sum_{x_{i_2}=0}^{\lambda_{i_2}}\theta\left(n-1-\sum_{l=1}^{j-1}x_l\right)\\
&=\sum_{x_1=0}^{\lambda_1}\overset{\hat{i}_1,\hat{i}_2}{\cdots}\sum_{x_{j-1}=0}^{\lambda_{j-1}}\sum_{x_{i_1}=0}^{\lambda_{i_1}-1}\sum_{x_{i_2}=0}^{\lambda_{i_2}+1}\theta\left(n-\sum_{l=1}^{j-1}x_l\right)+\sum_{x_1=0}^{\lambda_1}\overset{\hat{i}_1,\hat{i}_2}{\cdots}\sum_{x_{j-1}=0}^{\lambda_{j-1}}\sum_{x_{i_1}=0}^{\lambda_{i_1}-1}\sum_{x_{i_2}=0}^{\lambda_{i_2}-1}\theta\left(n-1-\sum_{l=1}^{j-1}x_l\right)\\
&\quad-\sum_{x_1=0}^{\lambda_1}\overset{\hat{i}_1,\hat{i}_2}{\cdots}\sum_{x_{j-1}=0}^{\lambda_{j-1}}\sum_{x_{i_1}=0}^{\lambda_{i_1}-1}\sum_{x_{i_2}=0}^{\lambda_{i_2}}\theta\left(n-1-\sum_{l=1}^{j-1}x_l\right)+\sum_{x_1=0}^{\lambda_1}\overset{\hat{i}_1,\hat{i}_2}{\cdots}\sum_{x_{j-1}=0}^{\lambda_{j-1}}\sum_{x_{i_2}=0}^{\lambda_{i_2}}\theta\left(n-\lambda_{i_1}-\sum_{l=1}^{j-1}x_l\right)\\
&=\sum_{x_1=0}^{\lambda_1}\overset{\hat{i}_1,\hat{i}_2}{\cdots}\sum_{x_{j-1}=0}^{\lambda_{j-1}}\sum_{x_{i_1}=0}^{\lambda_{i_1}-1}\sum_{x_{i_2}=0}^{\lambda_{i_2}+1}\theta\left(n-\sum_{l=1}^{j-1}x_l\right)-\sum_{x_1=0}^{\lambda_1}\overset{\hat{i}_1,\hat{i}_2}{\cdots}\sum_{x_{j-1}=0}^{\lambda_{j-1}}\sum_{x_{i_1}=0}^{\lambda_{i_1}-1}\theta\left(n-1-\lambda_{i_2}-\sum_{\substack{1\le l\le j-1\\ l\ne i_2}}x_l\right)\\
&\quad+\sum_{x_1=0}^{\lambda_1}\overset{\hat{i}_1,\hat{i}_2}{\cdots}\sum_{x_{j-1}=0}^{\lambda_{j-1}}\sum_{x_{i_2}=0}^{\lambda_{i_2}}\theta\left(n-\lambda_{i_1}-\sum_{l=1}^{j-1}x_l\right)\\
&=\sum_{x_1=0}^{\lambda_1}\overset{\hat{i}_1,\hat{i}_2}{\cdots}\sum_{x_{j-1}=0}^{\lambda_{j-1}}\sum_{x_{i_1}=0}^{\lambda_{i_1}-1}\sum_{x_{i_2}=0}^{\lambda_{i_2}}\theta\left(n-\sum_{l=1}^{j-1}x_l\right)+\sum_{x_1=0}^{\lambda_1}\overset{\hat{i}_1,\hat{i}_2}{\cdots}\sum_{x_{j-1}=0}^{\lambda_{j-1}}\sum_{x_{i_2}=0}^{\lambda_{i_2}}\theta\left(n-\lambda_{i_1}-\sum_{l=1}^{j-1}x_l\right)\\
&=\sum_{x_1=0}^{\lambda_1}\overset{\hat{i}_1,\hat{i}_2}{\cdots}\sum_{x_{j-1}=0}^{\lambda_{j-1}}\sum_{x_{i_1}=0}^{\lambda_{i_1}}\sum_{x_{i_2}=0}^{\lambda_{i_2}}\theta\left(n-\sum_{l=1}^{j-1}x_l\right)\\
&=\sum_{x_1=0}^{\lambda_1}\cdots\sum_{x_{j-1}=0}^{\lambda_{j-1}}\theta\left(n-\sum_{l=1}^{j-1}x_l\right)\\
&=C_{n,d}^{j,m=0}(\boldsymbol{\lambda}).
\end{aligned} \tag{A.40}$$

Thus, I have demonstrated (5.84) for the $m=0$ case.

# Appendix B

# Proof of identities of $A_i^{\sigma\mu}$

In Appendix B, I prove Lemma 5.9, i.e., I prove the identity of $A_i^{\sigma,\nu}(\widetilde{\Psi})$ (5.165)–(5.170). I note that the diagrams generated from the commutator of $H$ and connected diagrams can be non-connected but satisfy rule (ii) for the connected diagram. Thus, the types of such diagrams are determined uniquely. In the following, I do not explicitly represent the type of unit unless I mention it explicitly.

I assume $\widetilde{\Psi} \in \mathcal{F}_{n,d,g}^{k,j,m}$, and I use the notation $A_i^{\sigma,\nu}(\widetilde{\Psi}) = A_i^{\sigma,\nu}$ for simplicity in the following.

I omit 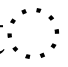 in the coast and indicate the length of the coast by an arrow, for example: 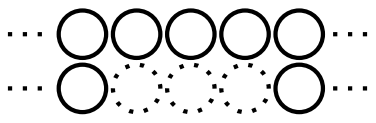 $=$ 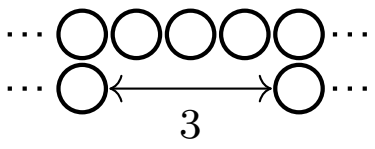 .

## B.1 Proof of Eq. (5.165)

In this subsection, I prove Eq. (5.165):

$$A_i^{++} = A_i^{--} = 0\,. \tag{B.1}$$

I firstly prove $A_i^{++} = 0$. Let $\widetilde{\Psi}$ be a diagram where the $i$ th and $(i \pm 1)$ th coasts are not adjacent.

I consider the case in which the left and right sides of the $i$ th coast are overlaps, such as

$\widetilde{\Psi} =$ 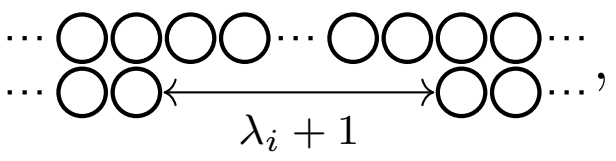 , (B.2)

where 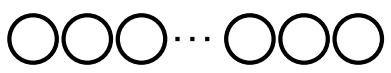 denotes the sequence of ○. Here, one can see $\sigma_i^L = \sigma_i^R = +$, and I have

$A_i^{++}\widetilde{\Psi}$

$=(-1)^{n+m+g} \Big\{ -C_{n-1,d+1}^{j,m}(\boldsymbol{\lambda}_{i:-1})$ 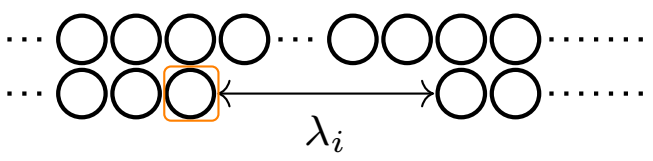

$-C^{j,m}_{n-1,d+1}\left(\boldsymbol{\lambda}_{i:-1}\right)$ 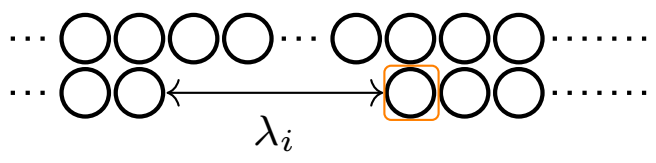 $+\,C^{j,m}_{n,d-1}\left(\boldsymbol{\lambda}_{i:+1}\right)$ 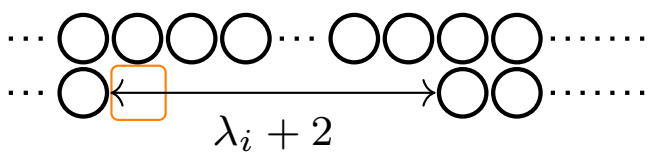

$+C^{j,m}_{n,d-1}\left(\boldsymbol{\lambda}_{i:+1}\right)$ 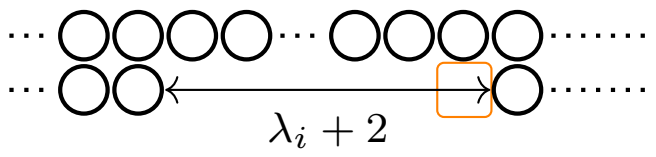 $\Big\}$

$$=(-1)^{n+m+g}\left\{-C^{j,m}_{n-1,d+1}\left(\boldsymbol{\lambda}_{i:-1}\right)+C^{j,m}_{n-1,d+1}\left(\boldsymbol{\lambda}_{i:-1}\right)+C^{j,m}_{n,d-1}\left(\boldsymbol{\lambda}_{i:+1}\right)-C^{j,m}_{n,d-1}\left(\boldsymbol{\lambda}_{i:+1}\right)\right\}\widetilde{\Psi}$$
$$=0. \tag{B.3}$$

One can understand from where the contributions to $A_i^{\sigma,\mu}$ come by observing the changes in the support, double, gap number, and unit number between the $\widetilde{\Psi}$ and the diagram in $Q_k^j$ that contribute to $A_i^{\sigma,\mu}$. For instance, for the above case (B.3), I have 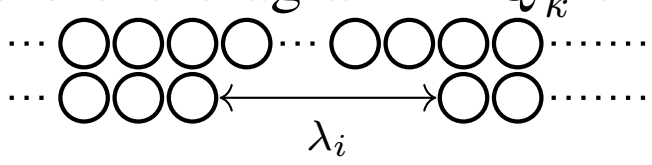 , 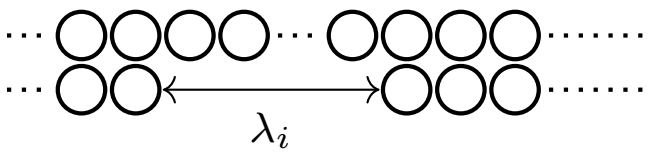 $\in \mathcal{S}^{k,j,m}_{n-1,d+1,g}$ , 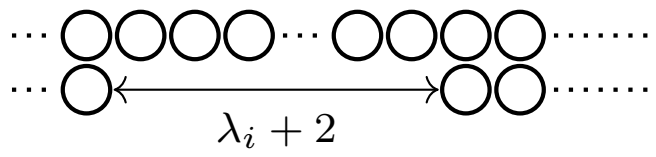 , 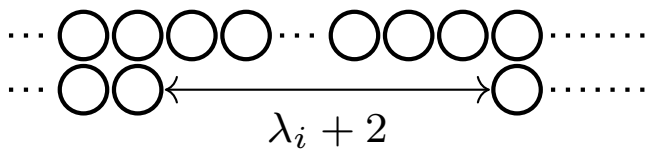 $\in \mathcal{S}^{k,j,m}_{n,d-1,g}$, and the sign factor of each term on the right-hand side on the first line of Eq. (B.3) is determined by where the contribution comes from. If the diagram $\Psi\in\mathcal{S}^{k,j,m}_{n,d,g}$ contribute to $A_i^{\sigma,\nu}$, $\Psi$ is accompanied by the sign $(-1)^{n+m+g}$, which can be seen from Theorem 5.7. In the following, I omit the explanation for where the contribution comes from unless mentioned.

I next consider the case in which the left and right sides of the $i$ th coast are both gaps, such as

$\widetilde{\Psi}=$ 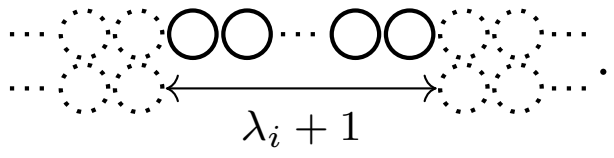 . (B.4)

In this case, one can see $\sigma_i^L=\sigma_i^R=+$, and I obtain the value of $A_i^{\sigma_i^L,\sigma_i^R}=A_i^{++}$ from the following calculation:

$A_i^{++}\widetilde{\Psi}$

$=(-1)^{n+m+g}\Big\{-C^{j,m}_{n-1,d+1}\left(\boldsymbol{\lambda}_{i:-1}\right)$ 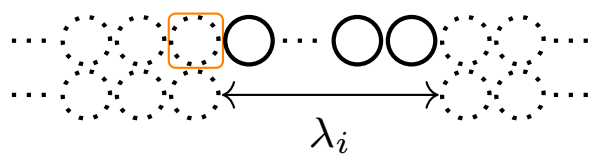 $-\,C^{j,m}_{n-1,d+1}\left(\boldsymbol{\lambda}_{i:-1}\right)$ 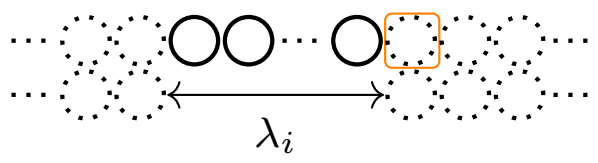

$-C^{j,m}_{n,d-1}\left(\boldsymbol{\lambda}_{i:+1}\right)$ 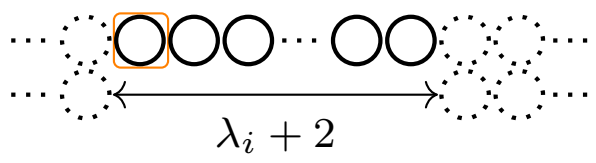 $-\,C^{j,m}_{n,d-1}\left(\boldsymbol{\lambda}_{i:+1}\right)$ 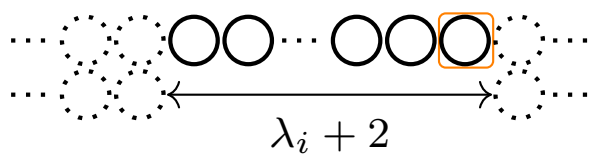 $\Big\}$

$$=(-1)^{n+m+g}\left\{C^{j,m}_{n-1,d+1}\left(\boldsymbol{\lambda}_{i:-1}\right)-C^{j,m}_{n-1,d+1}\left(\boldsymbol{\lambda}_{i:-1}\right)+C^{j,m}_{n,d-1}\left(\boldsymbol{\lambda}_{i:-1}\right)-C^{j,m}_{n,d-1}\left(\boldsymbol{\lambda}_{i:-1}\right)\right\}\widetilde{\Psi}$$
$$=0. \tag{B.5}$$

I next consider the case in which the left side of the $i$ th coast is a gap, and the right side is an overlap, such as

$\widetilde{\Psi}=$ 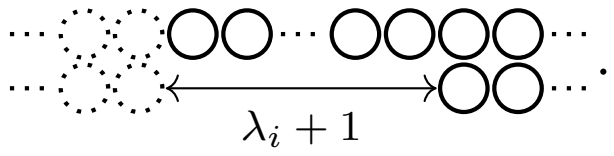 . (B.6)

In this case, one can see $\sigma_i^L = \sigma_i^R = +$, and I obtain the value of $A_i^{\sigma_i^L,\sigma_i^R} = A_i^{++}$ from the following calculation:

$$A_i^{++}\widetilde{\Psi}$$

$$=(-1)^{n+m+g}\Big\{C^{j,m}_{n-1,d+1}\left(\boldsymbol{\lambda}_{i:-1}\right)$$
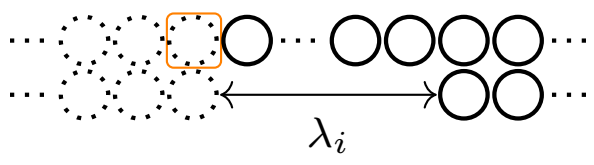

$$+\,C^{j,m}_{n-1,d+1}\left(\boldsymbol{\lambda}_{i:-1}\right)$$
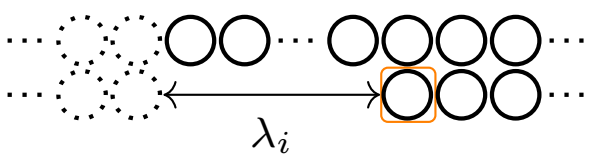

$$-C^{j,m}_{n,d-1}\left(\boldsymbol{\lambda}_{i:+1}\right)$$
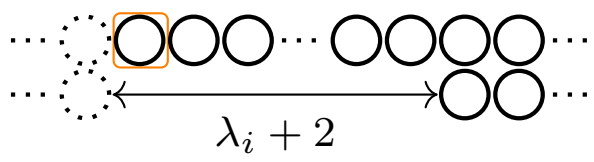

$$+\,C^{j,m}_{n,d-1}\left(\boldsymbol{\lambda}_{i:+1}\right)$$
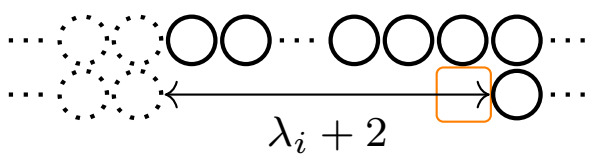

$$\Big\}$$

$$=(-1)^{n+m+g}\left\{-C^{j,m}_{n-1,d+1}\left(\boldsymbol{\lambda}_{i:-1}\right)+C^{j,m}_{n-1,d+1}\left(\boldsymbol{\lambda}_{i:-1}\right)+C^{j,m}_{n,d-1}\left(\boldsymbol{\lambda}_{i:-1}\right)-C^{j,m}_{n,d-1}\left(\boldsymbol{\lambda}_{i:-1}\right)\right\}\widetilde{\Psi}$$

$$=0. \tag{B.7}$$

With the same argument, in the case in which the right side of the $i$ th coast is a gap, and the left side of that is an overlap, such as

$$\widetilde{\Psi}=$$
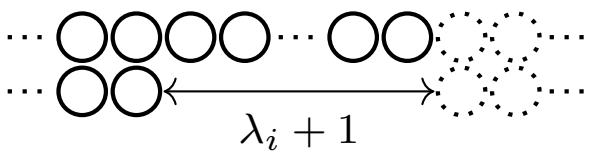

$$, \tag{B.8}$$

One can see $\sigma_i^L = \sigma_i^R = +$, and $A_i^{\sigma_i^L,\sigma_i^R}\widetilde{\Psi} = A_i^{+,+}\widetilde{\Psi} = 0$.

Thus, I have proved $A_i^{++} = 0$ for all possible cases.

I next prove $A_i^{--} = 0$. Let $\widetilde{\Psi}$ be a diagram where the $i$ th and $(i \pm 1)$ th coasts are adjacent. I first consider the case in which the $i$ th and $(i \pm 1)$ th coasts are adjacent, and they are looking in the same direction, such as

$$\widetilde{\Psi}=$$
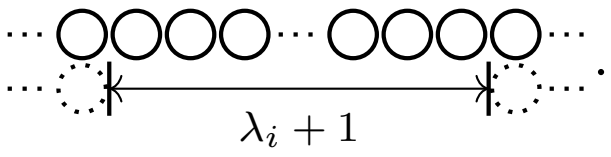

$$. \tag{B.9}$$

In this case, one can see $\sigma_i^L = \sigma_i^R = -$, and I have

$$A_i^{--}\widetilde{\Psi}$$

$$=(-1)^{n+m+g}\Big\{-C^{j,m}_{n-1,d+1}\left(\boldsymbol{\lambda}_{i:-1}\right)$$
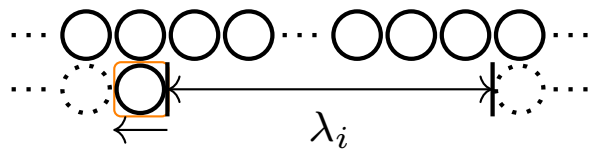

$$-\,C^{j,m}_{n-1,d+1}\left(\boldsymbol{\lambda}_{i:-1}\right)$$
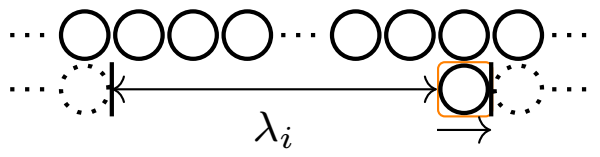

$$\Big\}$$

$$=(-1)^{n+m+g}2\left\{C^{j,m}_{n-1,d+1}\left(\boldsymbol{\lambda}_{i:-1}\right)-C^{j,m}_{n-1,d+1}\left(\boldsymbol{\lambda}_{i:-1}\right)\right\}$$

$$=0. \tag{B.10}$$

I consider the case in which the $i$ th and $(i \pm 1)$ th coasts are adjacent, the $i$ th and $(i+1)$ th coasts are looking in the same direction, and the $i$ th and $(i-1)$ th coasts are not looking in the same direction, such as

$$\widetilde{\Psi}=$$
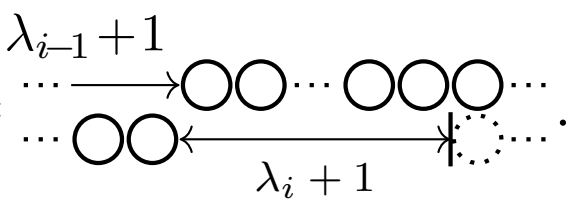

$$. \tag{B.11}$$

In this case, one can see $\sigma_i^L = \sigma_i^R = -$, and I have

$$A_i^{--}\widetilde{\Psi}$$

$$=(-1)^{n+m+g}\Big\{-C^{j,m}_{n-1,d+1}\left(\boldsymbol{\lambda}_{i:-1}\right)$$ 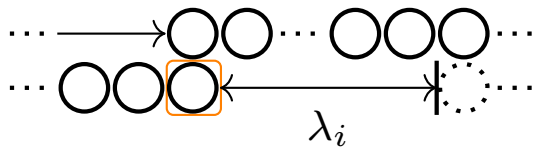 $$-\,C^{j,m}_{n-1,d+1}\left(\boldsymbol{\lambda}_{i:-1}\right)$$ 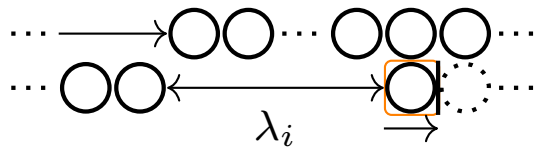

$$+C^{j,m}_{n-1,d+1}\left(\boldsymbol{\lambda}_{i:-1}\right)$$ 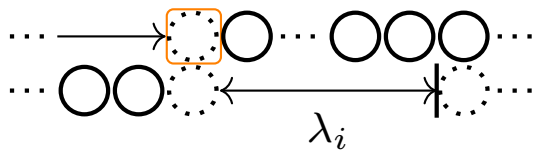 $$\Big\}$$

$$=(-1)^{n+m+g}\left\{-C^{j,m}_{n-1,d+1}\left(\boldsymbol{\lambda}_{i:-1}\right)+2C^{j,m}_{n-1,d+1}\left(\boldsymbol{\lambda}_{i:-1}\right)-C^{j,m}_{n-1,d+1}\left(\boldsymbol{\lambda}_{i:-1}\right)\right\}\widetilde{\Psi}$$

$$=0. \tag{B.12}$$

With the same argument, in the case in which the $i$ th and $(i \pm 1)$ th coasts are adjacent, the $i$ th and $(i-1)$ th coasts are looking in the same direction, and the $i$ th and $(i+1)$ th coasts are not looking in the same direction, such as

$$\widetilde{\Psi} =$$ 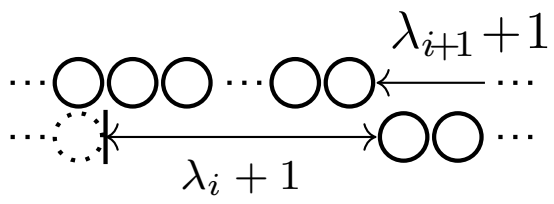 $$, \tag{B.13}$$

One can see $\sigma_i^L = \sigma_i^R = -$, and I have $A_i^{\sigma_i^L,\sigma_i^R} = A_i^{--} = 0$.

I consider the case in which the $i$ th and $(i \pm 1)$ th coasts are adjacent, and they are not looking in the same direction, such as

$$\widetilde{\Psi} =$$ 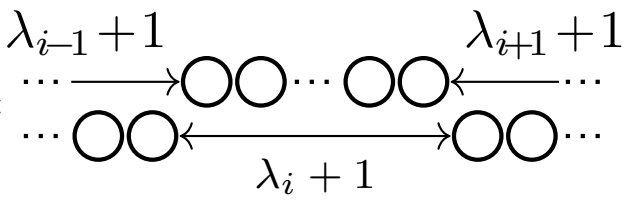 $$. \tag{B.14}$$

In this case, one can see $\sigma_i^L = \sigma_i^R = -$, and I have

$$A_i^{--}\widetilde{\Psi}$$

$$=(-1)^{n+m+g}\Big\{-C^{j,m}_{n-1,d+1}\left(\boldsymbol{\lambda}_{i:-1}\right)$$ 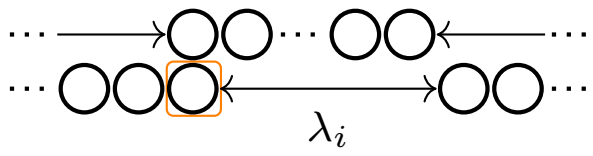 $$-\,C^{j,m}_{n-1,d+1}\left(\boldsymbol{\lambda}_{i:-1}\right)$$ 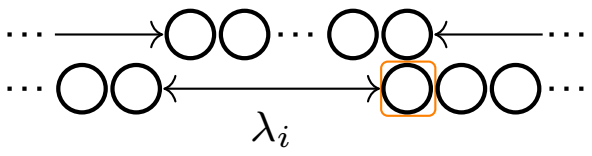

$$+C^{j,m}_{n-1,d+1}\left(\boldsymbol{\lambda}_{i:-1}\right)$$ 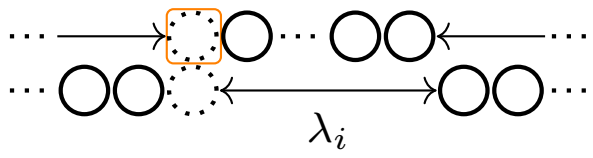 $$+\,C^{j,m}_{n-1,d+1}\left(\boldsymbol{\lambda}_{i:-1}\right)$$ 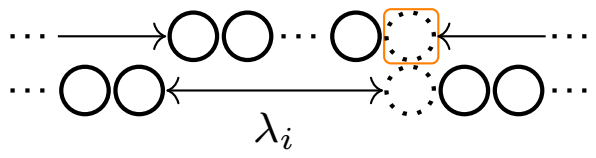 $$\Big\}$$

$$=(-1)^{n+m+g}\left\{-C^{j,m}_{n-1,d+1}\left(\boldsymbol{\lambda}_{i:-1}\right)+C^{j,m}_{n-1,d+1}\left(\boldsymbol{\lambda}_{i:-1}\right)+C^{j,m}_{n-1,d+1}\left(\boldsymbol{\lambda}_{i:-1}\right)-C^{j,m}_{n-1,d+1}\left(\boldsymbol{\lambda}_{i:-1}\right)\right\}\widetilde{\Psi}$$

$$=0. \tag{B.15}$$

Thus, I have proved $A_i^{--} = 0$ for all possible cases. This concludes the proof of Eq. (5.165).

## B.2 Proof of Eq. (5.166)

In this subsection, $R$ is denoted by

$$R \equiv (-1)^{n+m+g} \left\{ C^{j,m}_{n-1,d+1}(\boldsymbol{\lambda}_{i:-1}) + C^{j,m}_{n,d-1}(\boldsymbol{\lambda}_{i:+1}) \right\}, \tag{B.16}$$

and let $\widetilde{\Psi}$ be a diagram in which the $i$ th and $(i-1)$ th coasts are adjacent, and the $i$ th and $(i+1)$ th coasts are not adjacent. I prove Eq. (5.166), i.e., $A_i^{+-} = -A_i^{-+} = R$ in the following.

I first consider the case in which the $i$ th and $(i+1)$ th coasts are adjacent and looking in the same direction, and the left side of the $i$ th coast is an overlap, such as

$\widetilde{\Psi} =$ 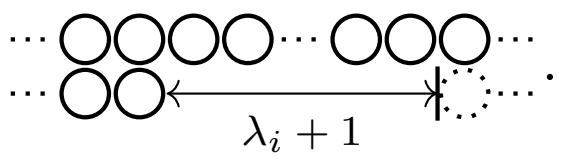. (B.17)

In this case, one can see $\sigma_i^L = +$ and $\sigma_i^R = -$, and I have

$A_i^{+-}\widetilde{\Psi}$

$=(-1)^{n+m+g}\Big\{ -C^{j,m}_{n-1,d+1}(\boldsymbol{\lambda}_{i:-1})$ 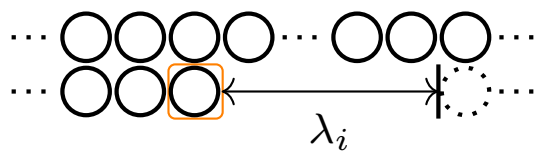 $- C^{j,m}_{n-1,d+1}(\boldsymbol{\lambda}_{i:-1})$ 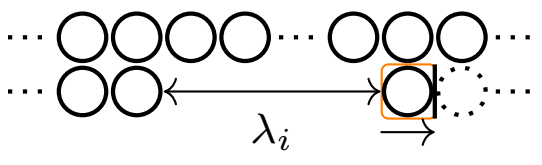

$+C^{j,m}_{n,d-1}(\boldsymbol{\lambda}_{i:+1})$ 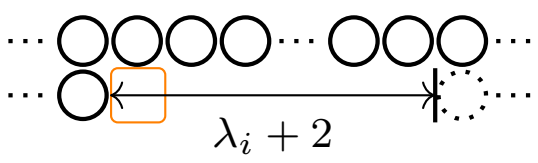 $\Big\}$

$$=(-1)^{n+m+g}\left\{ -C^{j,m}_{n-1,d+1}(\boldsymbol{\lambda}_{i:-1}) + 2C^{j,m}_{n-1,d+1}(\boldsymbol{\lambda}_{i:-1}) + C^{j,m}_{n,d-1}(\boldsymbol{\lambda}_{i:+1}) \right\}\widetilde{\Psi}$$

$$=R\widetilde{\Psi}. \tag{B.18}$$

I next consider the case in which the $i$ th and $(i+1)$ th coasts are adjacent and looking in the same direction, and the left side of the $i$ th coast is a gap, such as

$\widetilde{\Psi} =$ 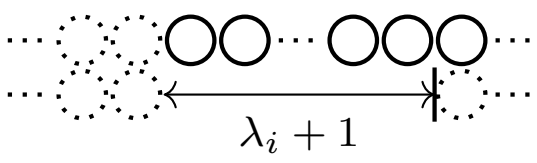. (B.19)

In this case, one can see $\sigma_i^L = +$ and $\sigma_i^R = -$, and I have

$A_i^{+-}\widetilde{\Psi}$

$=(-1)^{n+m+g}\Big\{ C^{j,m}_{n-1,d+1}(\boldsymbol{\lambda}_{i:-1})$ 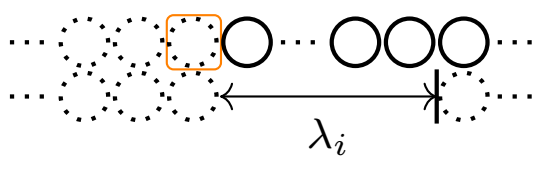 $- C^{j,m}_{n-1,d+1}(\boldsymbol{\lambda}_{i:-1})$ 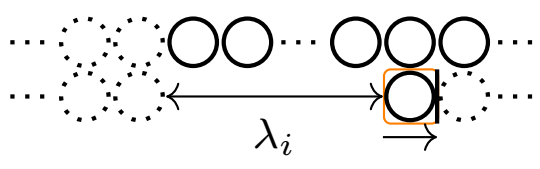

$-C^{j,m}_{n,d-1}(\boldsymbol{\lambda}_{i:+1})$ 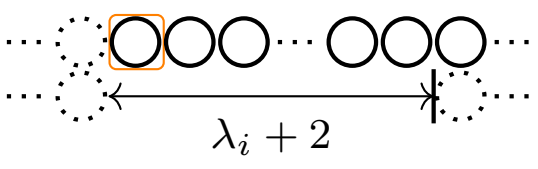 $\Big\}$

$$=(-1)^{n+m+g}\left\{ -C^{j,m}_{n-1,d+1}(\boldsymbol{\lambda}_{i:-1}) + 2C^{j,m}_{n-1,d+1}(\boldsymbol{\lambda}_{i:-1}) + C^{j,m}_{n,d-1}(\boldsymbol{\lambda}_{i:+1}) \right\}\widetilde{\Psi}$$

$$=R\widetilde{\Psi}. \tag{B.20}$$

I next consider the case in which the $i$ th and $(i+1)$ th coasts are adjacent and looking in different directions, and the left side of the $i$ th coast is overlap, such as

$$\widetilde{\Psi} =$$ 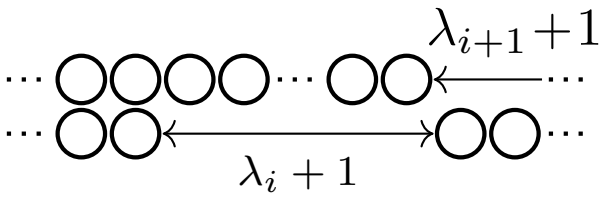

. (B.21)

In this case, one can see $\sigma_i^L = +$ and $\sigma_i^R = -$, and I have

$$A_i^{+-}\widetilde{\Psi}$$

$$=(-1)^{n+m+g}\Big\{-C_{n-1,d+1}^{j,m}(\boldsymbol{\lambda}_{i:-1})$$ 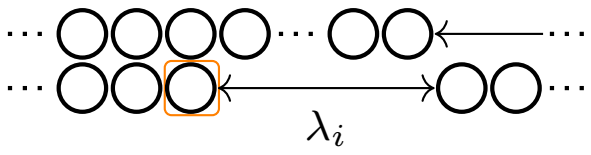

$$-\,C_{n-1,d+1}^{j,m}(\boldsymbol{\lambda}_{i:-1})$$ 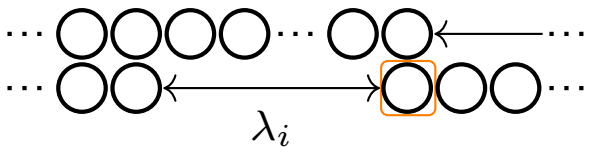

$$+C_{n-1,d+1}^{j,m}(\boldsymbol{\lambda}_{i:-1})$$ 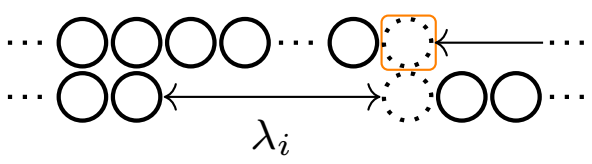

$$+\,C_{n,d-1}^{j,m}(\boldsymbol{\lambda}_{i:+1})$$ 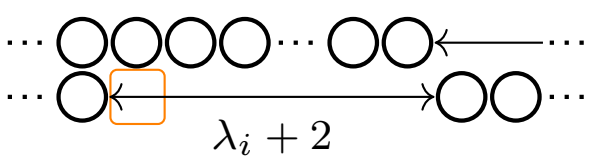

$$\Big\}$$

$$=(-1)^{n+m+g}\left\{-C_{n-1,d+1}^{j,m}(\boldsymbol{\lambda}_{i:-1})+C_{n-1,d+1}^{j,m}(\boldsymbol{\lambda}_{i:-1})+C_{n-1,d+1}^{j,m}(\boldsymbol{\lambda}_{i:-1})+C_{n,d-1}^{j,m}(\boldsymbol{\lambda}_{i:+1})\right\}\widetilde{\Psi}$$
$$=R\widetilde{\Psi}. \tag{B.22}$$

I next consider the case in which the $i$ th and $(i+1)$ th coasts are adjacent and looking in different directions, and the left side of the $i$ th coast is a gap, such as

$$\widetilde{\Psi} =$$ 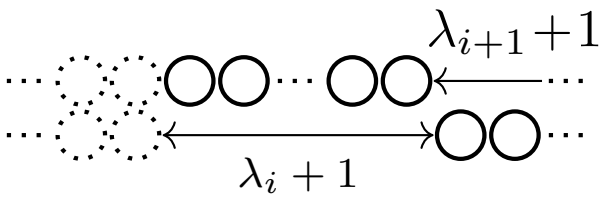

. (B.23)

In this case, $\sigma_i^L = +$ and $\sigma_i^R = -$, and I have

$$A_i^{+-}\widetilde{\Psi}$$

$$=(-1)^{n+m+g}\Big\{C_{n-1,d+1}^{j,m}(\boldsymbol{\lambda}_{i:-1})$$ 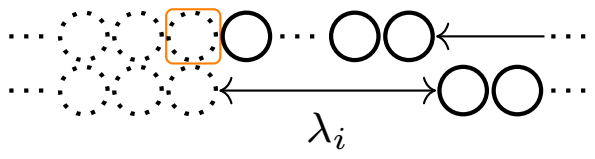

$$+\,C_{n-1,d+1}^{j,m}(\boldsymbol{\lambda}_{i:-1})$$ 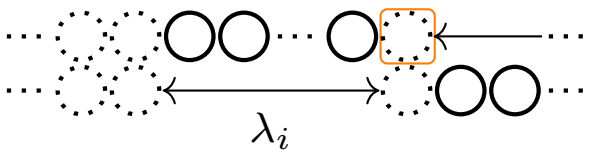

$$-C_{n,d-1}^{j,m}(\boldsymbol{\lambda}_{i:+1})$$ 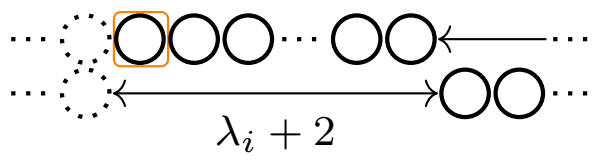

$$-\,C_{n-1,d+1}^{j,m}(\boldsymbol{\lambda}_{i:+1})$$ 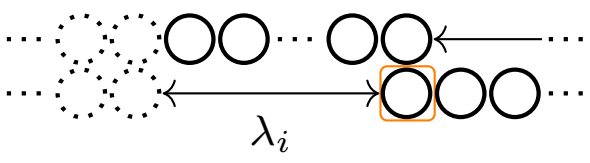

$$\Big\}$$

$$=(-1)^{n+m+g}\left\{-C_{n-1,d+1}^{j,m}(\boldsymbol{\lambda}_{i:-1})+C_{n-1,d+1}^{j,m}(\boldsymbol{\lambda}_{i:-1})+C_{n,d-1}^{j,m}(\boldsymbol{\lambda}_{i:+1})-C_{n-1,d+1}^{j,m}(\boldsymbol{\lambda}_{i:-1})\right\}\widetilde{\Psi}$$
$$=R\widetilde{\Psi}. \tag{B.24}$$

Thus, I have proved $A_i^{+-} = R$ for all possible cases. With the same argument, one can prove $A_i^{-+} = -R$ for all possible cases.

This concludes the proof of Eq. (5.166).

## B.3 Proof of Eqs. (5.167) and (5.168)

In this subsection, $R$ is denoted by

$$\begin{aligned} R \equiv &(-1)^{n+m+g}\{C^{j,m}_{n-1,d+1}(\boldsymbol{\lambda}_{L:-1}) - C^{j,m}_{n,d}(\boldsymbol{\lambda}_{L:-1}) + C^{j,m}_{n-1,d}(\boldsymbol{\lambda}_{L:+1}) \\ &- C^{j,m}_{n,d-1}(\boldsymbol{\lambda}_{L:+1}) + C^{j-1,m-1}_{n,d}({}_{0\rightarrow}(\boldsymbol{\lambda}_{L:-1}))\} \end{aligned} \tag{B.25}$$

and let $\widetilde{\Psi}$ be a diagram where the leftmost coast (the $0$ th coast, corresponding to $\lambda_0$) and the $1$ th coast, corresponding to $\lambda_1$, are not adjacent. In this subsection, I prove Eq. (5.167):

$$A_0^+ = R\,. \tag{B.26}$$

Equation (5.168) is proved with the same argument.

I consider the case in which the right side of the leftmost coast is an overlap and $\lambda_L > 0$, such as

$\widetilde{\Psi} =$ 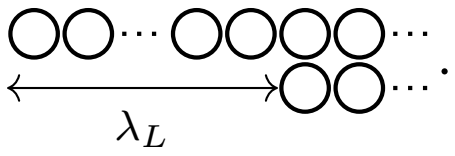 . (B.27)

In this case, one can see $\sigma_0^R = +$, and I have

$A_0^+\widetilde{\Psi}$

$=(-1)^{n+m+g}\Big\{ C^{j,m}_{n,d}(\boldsymbol{\lambda}_{L:-1})$ 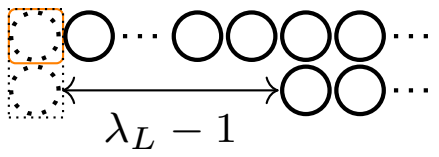 $- C^{j,m}_{n-1,d+1}(\boldsymbol{\lambda}_{L:-1})$ 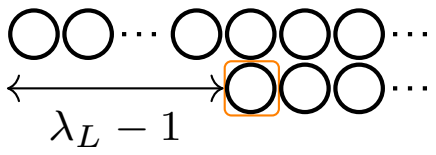

$-C^{j,m}_{n-1,d}(\boldsymbol{\lambda}_{L:+1})$ 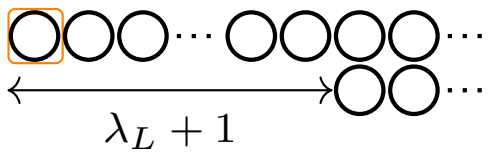 $+ C^{j,m}_{n,d-1}(\boldsymbol{\lambda}_{L:+1})$ 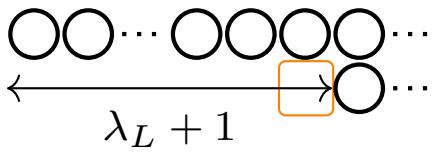

$-C^{j-1,m-1}_{n,d}({}_{0\rightarrow}(\boldsymbol{\lambda}_{L:-1}))$ 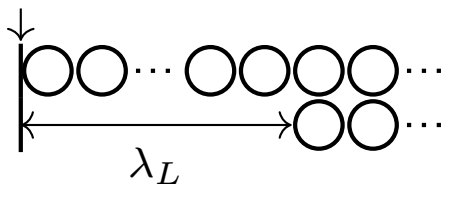 $\Big\}$

$=R\widetilde{\Psi},$ (B.28)

where I note that 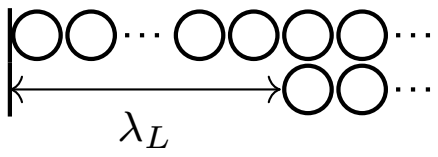 $\in \mathcal{S}^{k,j-1,m-1}_{n,d,g}$.

I consider the case in which $\lambda_L > 0$ and the right side of the leftmost coast is a gap, such as,

$\widetilde{\Psi} =$ 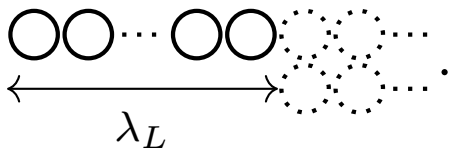 . (B.29)

In this case, one can see $\sigma_0^R = +$, and I have

$A_0^+\widetilde{\Psi}$

$$=(-1)^{n+m+g}\Big\{-C^{j,m}_{n-1,d}\left(\boldsymbol{\lambda}_{L:+1}\right)$$ 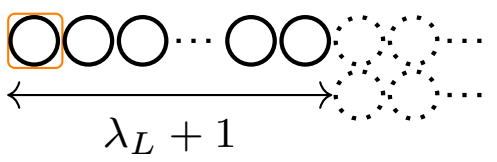 $$+\,C^{j,m}_{n,d}\left(\boldsymbol{\lambda}_{L:-1}\right)$$ 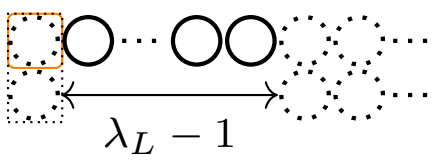

$$+C^{j,m}_{n-1,d+1}\left(\boldsymbol{\lambda}_{L:-1}\right)$$ 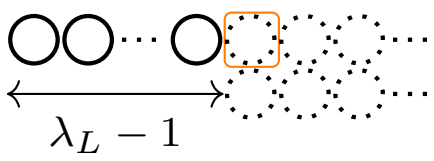 $$-\,C^{j,m}_{n,d-1}\left(\boldsymbol{\lambda}_{L:+1}\right)$$ 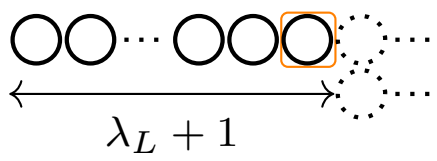

$$-C^{j-1,m-1}_{n,d}\left({}_{0\to}\left(\boldsymbol{\lambda}_{L:-1}\right)\right)$$ 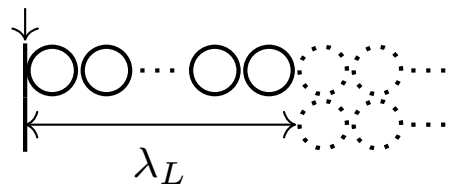 $$\Big\}$$

$$=R\widetilde{\Psi}. \tag{B.30}$$

I consider the case in which $\lambda_L = 0$ and the leftmost of the diagram is an overlap, such as

$$\widetilde{\Psi} =$$ 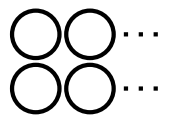 $$. \tag{B.31}$$

In this case, one can see $\sigma_0^R = +$ and I have

$$A_0^+\widetilde{\Psi} =(-1)^{n+m+g}\Big\{-C^{j,m}_{n-1,d}\left(\boldsymbol{\lambda}_{L:+1}\right)$$ 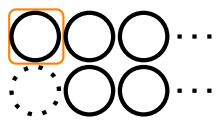 $$-\,C^{j,m}_{n-1,d}\left(\boldsymbol{\lambda}_{L:+1}\right)$$ 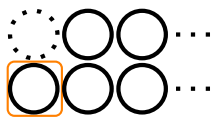

$$+C^{j,m}_{n,d-1}\left(\boldsymbol{\lambda}_{L:+1}\right)$$ 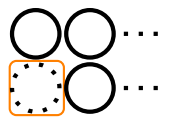 $$+\,C^{j,m}_{n,d-1}\left(\boldsymbol{\lambda}_{L:+1}\right)$$ 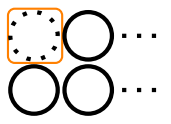 $$\Big\}$$

$$=(-1)^{n+m+g}2\left\{C^{j,m}_{n-1,d}\left(\boldsymbol{\lambda}_{L:+1}\right)-C^{j,m}_{n,d-1}\left(\boldsymbol{\lambda}_{L:+1}\right)\right\}\widetilde{\Psi}$$
$$=R\widetilde{\Psi}, \tag{B.32}$$

where I used $C^{j,m}_{n-1,d}\left(\boldsymbol{\lambda}_{L:+1}\right) = C^{j,m}_{n-1,d+1}\left(\boldsymbol{\lambda}_{L:-1}\right)$ and $C^{j,m}_{n,d-1}\left(\boldsymbol{\lambda}_{L:+1}\right) = C^{j,m}_{n,d}\left(\boldsymbol{\lambda}_{L:-1}\right)$ for the case $\lambda_L = 0$, and $C^{j-1,m-1}_{n,d}\left({}_{0\to}\left(\boldsymbol{\lambda}_{L:-1}\right)\right) = 0$ for the case $\lambda_L = 0$.

I consider the case in which $\lambda_L = 0$ and the leftmost of the diagram is the zero-length unit on either the upper or lower row and its right is a gap, such as

$$\widetilde{\Psi} =$$ 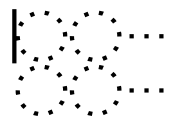 $$. \tag{B.33}$$

In this case, one can see $\sigma_0^R = +$ and I have

$$A_0^+\widetilde{\Psi} =(-1)^{n+m+g}\Big\{-C^{j,m}_{n-1,d}\left(\boldsymbol{\lambda}_{L:+1}\right)$$ 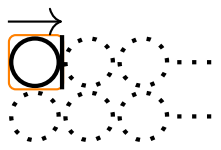 $$-\,C^{j,m}_{(n-1)-(g-1),d-1}\left(\boldsymbol{\lambda}_{L:+1}\right)$$ 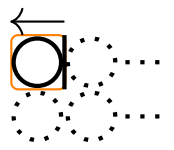 $$\Big\}$$

$$=(-1)^{n+m+g}2\left\{C^{j,m}_{n-1,d}\left(\boldsymbol{\lambda}_{L:+1}\right)-C^{j,m}_{n,d-1}\left(\boldsymbol{\lambda}_{L:+1}\right)\right\}\widetilde{\Psi}$$
$$=R\widetilde{\Psi}, \tag{B.34}$$

where I used $C^{j,m}_{n-1,d}\left(\boldsymbol{\lambda}_{L:+1}\right) = C^{j,m}_{n-1,d+1}\left(\boldsymbol{\lambda}_{L:-1}\right)$, $C^{j,m}_{n,d-1}\left(\boldsymbol{\lambda}_{L:+1}\right) = C^{j,m}_{n,d}\left(\boldsymbol{\lambda}_{L:-1}\right)$ and $C^{j-1,m-1}_{n,d}\left({}_{0\to}\boldsymbol{\lambda}\right) = 0$ for the case $\lambda_L = 0$.

Thus, I have proved Eq. (5.167) for all possible cases. With the same argument, one can prove Eq. (5.168) for all possible cases.

## B.4 Proof of Eqs. (5.169) and (5.170)

In this subsection, $R$ is denoted by

$$R \equiv (-1)^{n+m+g}\{2C^{j,m}_{n-1,d+1}(\boldsymbol{\lambda}_{L:-1}) - C^{j,m}_{n,d}(\boldsymbol{\lambda}_{L:-1}) + C^{j,m}_{n-1,d}(\boldsymbol{\lambda}_{L:+1}) + C^{j-1,m-1}_{n,d}({}_{0\to}(\boldsymbol{\lambda}_{L:-1}))\}\,, \tag{B.35}$$

and let $\widetilde{\Psi}$ be a diagram where the leftmost coast (the $0$ th coast) and the $1$ th coast are adjacent. In this subsection, I prove Eq. (5.169):

$$A_0^- = R\,. \tag{B.36}$$

I note that Eq.(5.170) is proved in the same argument.

First, I consider the case in which the leftmost coast (the $0$ th coast) and the $1$ th coast are looking in the same direction and $\lambda_L > 0$, such as

$$\widetilde{\Psi} = $$ 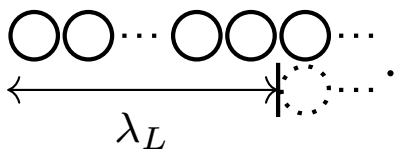

. (B.37)

In this case, one can see $\sigma_0^R = -$, and I have

$$A_0^+\widetilde{\Psi}$$
$$=(-1)^{n+m+g}\Big\{C^{j,m}_{n,d}(\boldsymbol{\lambda}_{L:-1})$$ 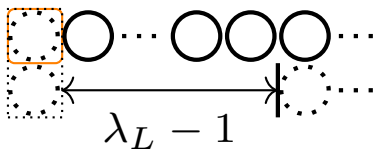

$$- C^{j,m}_{n-1,d+1}(\boldsymbol{\lambda}_{L:-1})$$ 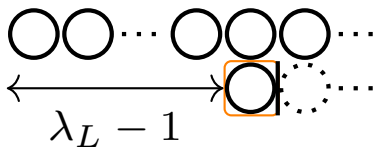

$$-C^{j,m}_{n-1,d}(\boldsymbol{\lambda}_{L:+1})$$ 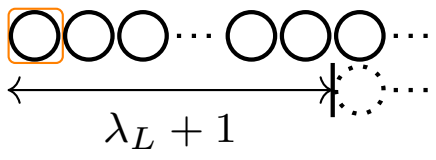

$$- C^{j-1,m-1}_{n,d}({}_{0\to}(\boldsymbol{\lambda}_{L:-1}))$$ 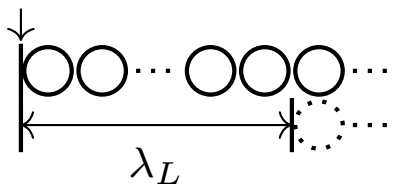

$$\Big\}$$
$$=R\widetilde{\Psi}. \tag{B.38}$$

I next consider the case in which the leftmost coast (the $0$ th coast) and the $1$ th coast are adjacent and looking in the different directions and $\lambda_L > 0$, such as

$$\widetilde{\Psi} = $$ 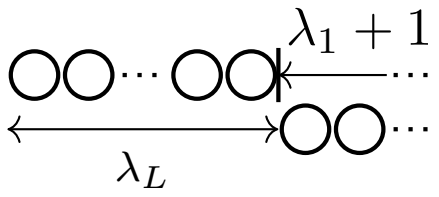

. (B.39)

In this case, one can see $\sigma_0^R = -$,and I have

$$A_0^-\widetilde{\Psi}$$
$$=(-1)^{n+m+g}\Big\{C^{j,m}_{n,d}(\boldsymbol{\lambda}_{L:-1})$$ 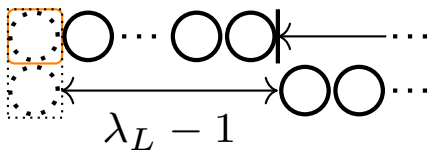

$$- C^{j,m}_{n-1,d+1}(\boldsymbol{\lambda}_{L:-1})$$ 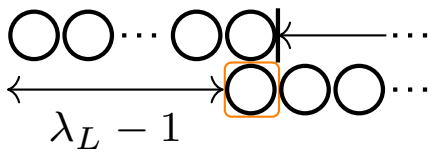

$$-C^{j,m}_{n-1,d}\left(\boldsymbol{\lambda}_{L:+1}\right)$$ 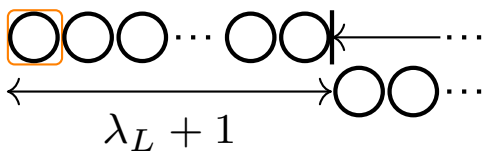 $$+\,C^{j,m}_{n-1,d+1}\left(\boldsymbol{\lambda}_{L:+1}\right)$$ 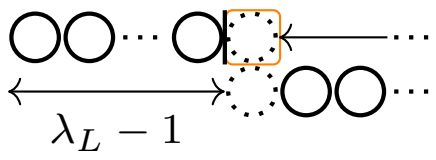

$$-C^{j-1,m-1}_{n,d}\left(_{0\rightarrow}(\boldsymbol{\lambda}_{L:-1})\right)$$ 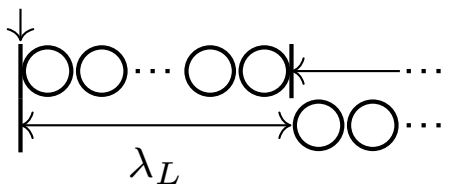 $$\Bigg\}$$

$$=R\widetilde{\Psi}, \tag{B.40}$$

where I note that 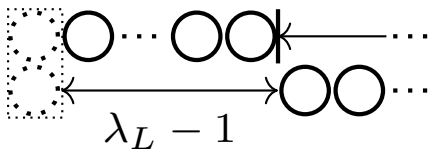 $\in \mathcal{S}^{k,j,m}_{n,d,g}$, 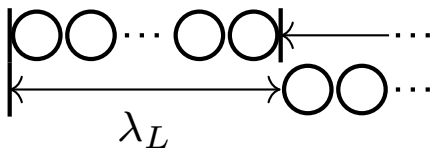 $\in \mathcal{S}^{k,j-1,m-1}_{n,d,g}$.

In the following, I consider the case $\lambda_L = 0$. Here, I also consider the contribution from the diagram in $Q_k^j$ with a list $\boldsymbol{\lambda}_{1:-1} = \{0; \lambda_1 - 1, \ldots\}$, which is 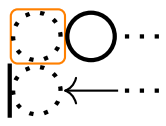 in Eq. (B.42) below. According to the original definition of $A_i^{\sigma,\mu}$, a diagram with a list $\{0; \lambda_1 - 1, \ldots\}$ should contribute to $A_1^{-,\sigma_i^R}$. Nonetheless, I incorporate this contribution into $A_L^-$ rather than $A_1^{-,\sigma_i^R}$ for later convenience. This special treatment does not conflict with the earlier arguments because this contribution has not been considered so far, and It is ensured that there is no double counting and retaining overall consistency.

I consider the case in which $\lambda_L = 0$ and the $0$ th and $1$ th coasts are adjacent and $w > 0$, such as

$$\widetilde{\Psi} = $$ 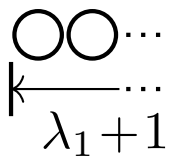 $$. \tag{B.41}$$

In this case, one can see $\sigma_0^R = -$, and I have

$$A_0^-\widetilde{\Psi} =(-1)^{n+m+g}\Bigg\{-C^{j,m}_{n-1,d}\left(\boldsymbol{\lambda}_{L:+1}\right)$$ 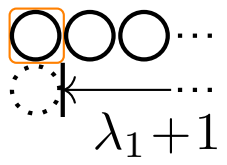 $$-\,C^{j,m}_{n-1,d}\left(\boldsymbol{\lambda}_{L:+1}\right)$$ 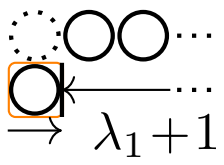

$$+C^{j,m}_{n-1,d+1}\left(\boldsymbol{\lambda}_{1:-1}\right)$$ 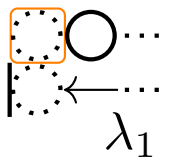 $$+\,C^{j-1,m}_{n,d}\left(\boldsymbol{\lambda}_{\lambda_L\rightarrow 1+\lambda_1,\hat{1}}\right)$$ 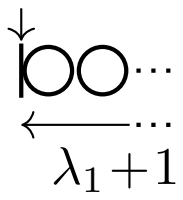 $$\Bigg\}$$

$$\begin{aligned}
&=(-1)^{n+m+g}\left\{3C^{j,m}_{n-1,d}\left(\boldsymbol{\lambda}_{L:+1}\right) - C^{j,m}_{n-1,d+1}\left(\boldsymbol{\lambda}_{1:-1}\right) - C^{j-1,m}_{n,d}\left(\boldsymbol{\lambda}_{\lambda_L\rightarrow 1+\lambda_1,\hat{1}}\right)\right\}\widetilde{\Psi}\\
&=(-1)^{n+m+g}\left\{C^{j,m}_{n-1,d}\left(\boldsymbol{\lambda}_{L:+1}\right) + C^{j,m}_{n-1,d+1}\left(\boldsymbol{\lambda}_{L:-1}\right)\right.\\
&\qquad\left. + \left(C^{j,m}_{n-1,d+1}\left(\boldsymbol{\lambda}_{L:-1}\right) - C^{j,m}_{n-1,d+1}\left(\boldsymbol{\lambda}_{1:-1}\right)\right) - C^{j-1,m}_{n,d}\left(\boldsymbol{\lambda}_{\lambda_L\rightarrow 1+\lambda_1,\hat{1}}\right)\right\}\widetilde{\Psi}\\
&=(-1)^{n+m+g}\left\{C^{j,m}_{n-1,d}\left(\boldsymbol{\lambda}_{L:+1}\right) + C^{j,m}_{n-1,d+1}\left(\boldsymbol{\lambda}_{L:-1}\right)\right.\\
&\qquad\left. + \left(C^{j,m}_{n,d-1}\left(\boldsymbol{\lambda}_{i:+1}\right) - C^{j,m}_{n,d-1}\left(\boldsymbol{\lambda}_{L:+1}\right)\right) - C^{j-1,m}_{n,d}\left(\boldsymbol{\lambda}_{\lambda_L\rightarrow 1+\lambda_1,\hat{1}}\right)\right\}\widetilde{\Psi}\\
&=(-1)^{n+m+g}\left\{C^{j,m}_{n-1,d}\left(\boldsymbol{\lambda}_{L:+1}\right) + C^{j,m}_{n-1,d+1}\left(\boldsymbol{\lambda}_{L:-1}\right) - C^{j,m}_{n,d-1}\left(\boldsymbol{\lambda}_{L:+1}\right)\right.\\
&\qquad\left. + \left(C^{j,m}_{n,d-1}\left(\boldsymbol{\lambda}_{i:+1}\right) - C^{j-1,m}_{n,d}\left(\boldsymbol{\lambda}_{\lambda_L\rightarrow 1+\lambda_1,\hat{1}}\right)\right)\right\}\widetilde{\Psi}\\
&=(-1)^{n+m+g}\left\{C^{j,m}_{n-1,d}\left(\boldsymbol{\lambda}_{L:+1}\right) + C^{j,m}_{n-1,d+1}\left(\boldsymbol{\lambda}_{L:-1}\right) - C^{j,m}_{n,d}\left(\boldsymbol{\lambda}_{L:-1}\right) + C^{j,m}_{n-1,d+1}\left(\boldsymbol{\lambda}_{L:-1}\right)\right\}\widetilde{\Psi}\\
&=R\widetilde{\Psi},
\end{aligned} \tag{B.42}$$

where 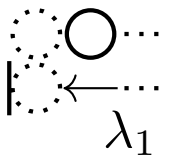 $\in \mathcal{S}^{k,j,m}_{n,d,g+1}$, 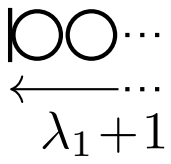 $\in \mathcal{S}^{k,j-1,m}_{n,d,g}$, and I used $C^{j-1,m-1}_{n,d}\left({}_{0\rightarrow}(\boldsymbol{\lambda}_{L:-1})\right) = 0$ and $C^{j,m}_{n-1,d}\left(\boldsymbol{\lambda}_{L:+1}\right) = C^{j,m}_{n-1,d+1}\left(\boldsymbol{\lambda}_{L:-1}\right)$ with $\lambda_L = 0$ and, $C^{j,m}_{n-1,d+1}\left(\boldsymbol{\lambda}_{L:-1}\right) - C^{j,m}_{n-1,d+1}\left(\boldsymbol{\lambda}_{1:-1}\right) = C^{j,m}_{n,d-1}\left(\boldsymbol{\lambda}_{i:+1}\right) - C^{j,m}_{n,d-1}\left(\boldsymbol{\lambda}_{L:+1}\right)$ from Eq. (5.84), and $C^{j,m}_{n,d-1}\left(\boldsymbol{\lambda}_{i:+1}\right) - C^{j-1,m}_{n,d}\left(\boldsymbol{\lambda}_{\lambda_L\rightarrow 1+\lambda_1,\hat{1}}\right) = C^{j,m}_{n-1,d+1}\left(\boldsymbol{\lambda}_{L:-1}\right)$ which can be derived from Eq. (5.82).

I consider the case in which $\lambda_L = 0$, $d = 0$, and $w = 0$ ($j = 2m+1$), such as

$$\widetilde{\Psi} = $$ 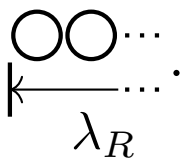 (B.43)

Here, I have $\sigma_0^R = -$, and one can see

$$A_0^- \widetilde{\Psi}$$

$$=(-1)^{n+m+g}\left\{-C^{j,m}_{n-1,0}\left(\boldsymbol{\lambda}_{L:+1}\right)\right.$$ 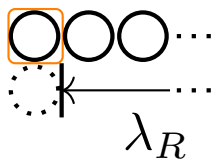 $$- C^{j,m}_{n-1,0}\left(\boldsymbol{\lambda}_{L:+1}\right)$$ 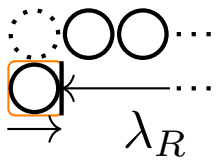 $$\left. + C^{j,m}_{n-1,1}\left(\boldsymbol{\lambda}_{R:-1}\right)\right.$$ 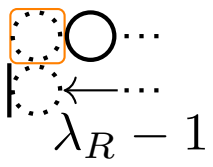 $$\left.\right\}$$

$$=(-1)^{n+m+g}\left\{3C^{j,m}_{n-1,0}\left(\boldsymbol{\lambda}_{L:+1}\right) - C^{j,m}_{n-1,1}\left(\boldsymbol{\lambda}_{R:-1}\right)\right\}\widetilde{\Psi}$$

$$=(-1)^{n+m+g}\left\{2C^{j,m}_{n-1,1}\left(\boldsymbol{\lambda}_{L:-1}\right) + C^{j,m}_{n-1,0}\left(\boldsymbol{\lambda}_{L:+1}\right) - C^{j,m}_{n,0}\left(\boldsymbol{\lambda}_{L:-1}\right)\right\}\widetilde{\Psi}$$

$$=R\widetilde{\Psi}, \tag{B.44}$$

where I used $C^{j-1,m-1}_{n,d}\left({}_{0\rightarrow}(\boldsymbol{\lambda}_{L:-1})\right) = 0$ for $\lambda_L = 0$, and I used $C^{j,m}_{n-1,d}\left(\boldsymbol{\lambda}_{L:+1}\right) = C^{j,m}_{n-1,d+1}\left(\boldsymbol{\lambda}_{L:-1}\right)$, and $C^{j,m}_{n-1,1}\left(\boldsymbol{\lambda}_{R:-1}\right) = C^{j,m}_{n,0}\left(\boldsymbol{\lambda}_{L:-1}\right)$ for $\lambda_L = 0$.

Then, I have proved Eq. (5.169), $A_L^- = R$ for all possible cases. With the same argument, one can prove Eq. (5.170).

This concludes all the proof of Lemma 5.9.

# Appendix C

# One-local conserved quantities

In Appendix C, I prove that there are no one-local conserved quantities independent of the $\mathrm{SU}(2)$ charges and $\mathrm{U}(1)$ charge and $\eta$-pairing charges, which is the another $\mathrm{SU}(2)$ symmetry for even $L$ [51, 52].

A one-local conserved quantity $F_1$ is denoted by

$$F_1 = \sum_{\substack{\bar{a},\bar{b}\in\{\bigcirc,\oplus,\ominus,\textcircled{z}\} \\ \{\bar{a},\bar{b}\}\neq\{\bigcirc,\bigcirc\}}} \sum_{i=1}^{L} c_i \left( \begin{matrix} \bar{a} \\ \bar{b} \end{matrix} \right) \begin{matrix} \bar{a} \\ \bar{b} \end{matrix} (i), \tag{C.1}$$

where $c_i \left( \begin{matrix} \bar{a} \\ \bar{b} \end{matrix} \right)$ is the coefficient of $\begin{matrix} \bar{a} \\ \bar{b} \end{matrix} (i)$ and $\{\bar{a}, \bar{b}\} \neq \{\bigcirc, \bigcirc\}$ is the normalization where $F_1$ is traceless.

I determine $F_1$ so that $[F_1, H] = [F_1, H_0] + [F_1, H_1] = 0$. Because $[F_1, H_0]$ is a two-support operator and $[F_1, H_1]$ is a one-support operator, the two terms must independently vanish, then I have the equations:

$$[F_1, H_0] = 0, \tag{C.2}$$

$$[F_1, H_1] = 0. \tag{C.3}$$

From Eq. (C.3), one can see

$$[F_1, H_1] = \sum_{\substack{\bar{a},\bar{b}\in\{\bigcirc,\oplus,\ominus,\textcircled{z}\} \\ \{\bar{a},\bar{b}\}\neq\{\bigcirc,\bigcirc\}}} \sum_{i=1}^{L} c_i \left( \begin{matrix} \bar{a} \\ \bar{b} \end{matrix} \right) \begin{matrix} \downarrow \\ \bar{a} \\ \bar{b} \\ \uparrow \end{matrix} (i) = 0. \tag{C.4}$$

$\begin{matrix} \downarrow \\ \bar{a} \\ \bar{b} \\ \uparrow \end{matrix} (i)$ is the one-support basis at $i$ th site. $\begin{matrix} \downarrow \\ \bar{a}_1 \\ \bar{b}_1 \\ \uparrow \end{matrix} (i)$ and $\begin{matrix} \downarrow \\ \bar{a}_2 \\ \bar{b}_2 \\ \uparrow \end{matrix} (j)$ are independent for $i \neq j$ and $\bar{a}_1 \neq \bar{a}_2$ and $\bar{b}_1 \neq \bar{b}_2$. Then I have

$$c_i \begin{pmatrix} \bar{a} \\ \bar{b} \end{pmatrix} \begin{matrix} \downarrow \\ \bar{a} \\ \bar{b} \\ \uparrow \end{matrix} (i) = 0. \tag{C.5}$$

If $\begin{matrix}\bar{a}\\ \bar{b}\end{matrix} \in \left\{ \begin{matrix}\boxed{\pm}\\ \boxed{\phantom{z}}\end{matrix}, \begin{matrix}\boxed{\pm}\\ \boxed{z}\end{matrix}, \begin{matrix}\boxed{\phantom{z}}\\ \boxed{\pm}\end{matrix}, \begin{matrix}\boxed{z}\\ \boxed{\pm}\end{matrix} \right\}$, one can see $\begin{matrix} \downarrow \\ \bar{a} \\ \bar{b} \\ \uparrow \end{matrix} \neq 0$ and then I have $c_i \begin{pmatrix} \bar{a} \\ \bar{b} \end{pmatrix} = 0$. Therefore $[F_1, H_1] = 0$ is satisfied if $F_1$ has the following form:

$$F_1 = \sum_{i=1}^{L} \left[ \alpha_i^{+-} \begin{matrix}\boxed{+}\\ \boxed{-}\end{matrix}(i) + \alpha_i^{-+} \begin{matrix}\boxed{-}\\ \boxed{+}\end{matrix}(i) + \alpha_i^{++} \begin{matrix}\boxed{+}\\ \boxed{+}\end{matrix}(i) + \alpha_i^{--} \begin{matrix}\boxed{-}\\ \boxed{-}\end{matrix}(i) + \alpha_i^{z\uparrow} \begin{matrix}\boxed{z}\\ \boxed{\phantom{z}}\end{matrix}(i) + \alpha_i^{z\downarrow} \begin{matrix}\boxed{\phantom{z}}\\ \boxed{z}\end{matrix}(i) + \alpha_i^{zz} \begin{matrix}\boxed{z}\\ \boxed{z}\end{matrix}(i) \right]. \tag{C.6}$$

In the following, by considering the cancellation of two-support operators ($[F_1, H_0] = 0$), I prove $\alpha_i^{\pm\mp}$, $\alpha_i^{z\uparrow}$ and $\alpha_i^{z\downarrow}$ are constants independent of $i$ and $\alpha_i^{\pm\mp}$ is proportional to $(-1)^i$ for even $L$ and is zero for odd $L$ and $\alpha_i^{zz} = 0$.

Considering the following cancellation of the two-support basis

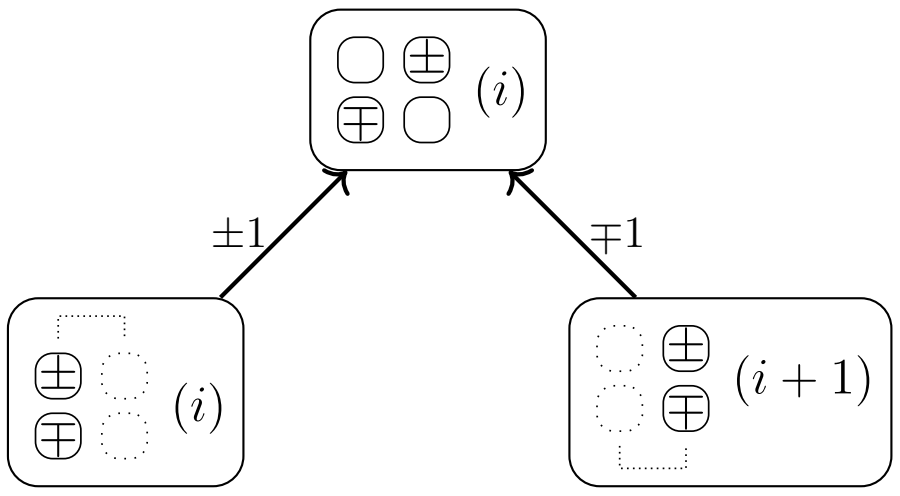

, (C.7)

and I have $\alpha_i^{\pm\mp} = \alpha_{i+1}^{\pm\mp}$. Then one can see that $\alpha_i^{\pm\mp}$ is independent of $i$.

Considering the following cancellation of the two-support basis

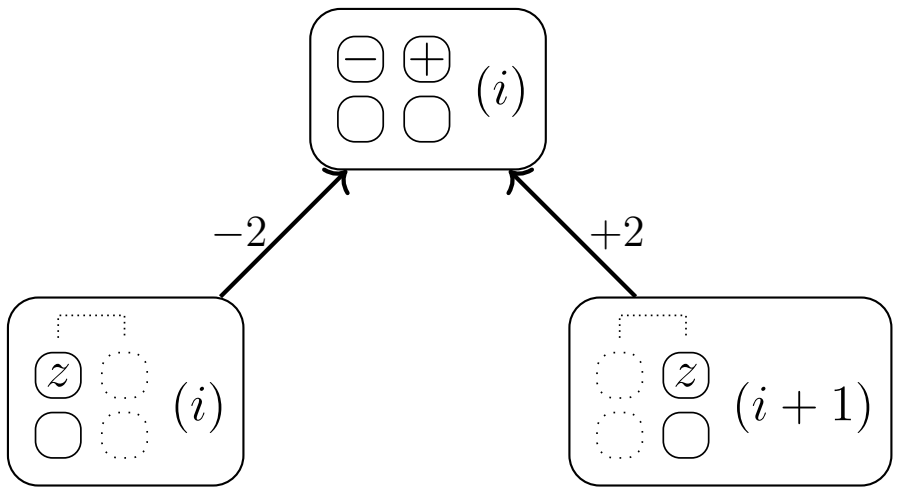

, (C.8)

and one can see

$$\alpha_i^{z\uparrow} = \alpha_{i+1}^{z\uparrow}. \tag{C.9}$$

Then, I have proved $\alpha_i^{z\uparrow}$ is independent of $i$. With the same argument, one can prove that $\alpha_i^{z\downarrow}$ is independent of $i$.

Considering the following cancellation of the two support basis

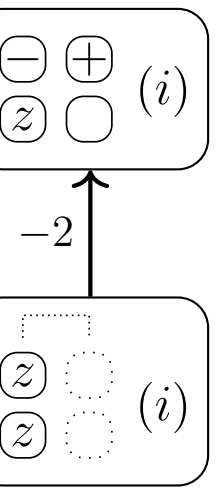

(C.10)

and I have $\alpha_i^{zz} = 0$.

Considering the following cancellation of the two-support basis

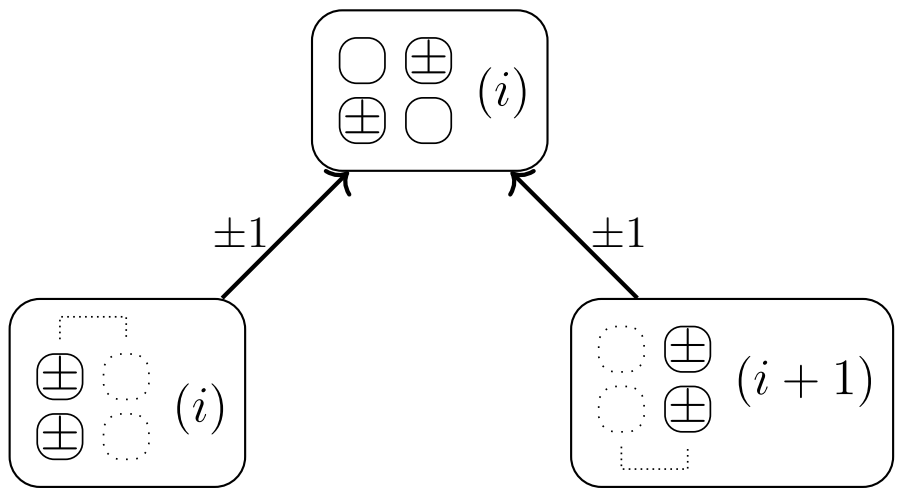

(C.11)

and I have

$$\alpha_i^{\pm\pm} = -\alpha_{i+1}^{\pm\pm} . \tag{C.12}$$

From the periodic boundary condition, I have

$$\alpha_i^{\pm\pm} = (-1)^L \alpha_{i+L}^{\pm\pm} = (-1)^L \alpha_i^{\pm\pm} . \tag{C.13}$$

Then, one can see $\alpha_i^{\pm\pm} = (-1)^i \alpha^{\pm\pm}$ where $\alpha^{\pm\pm}$ is a constant and $\alpha^{\pm\pm} = 0$ for odd $L$.

From the above argument, $[F_1, H_0] = 0$ is satisfied if

$$\begin{aligned} F_1 &= \alpha^{+-} \sum_{i=1}^{L} \begin{smallmatrix}\oplus\\ \ominus\end{smallmatrix}(i) + \alpha^{-+} \sum_{i=1}^{L} \begin{smallmatrix}\ominus\\ \oplus\end{smallmatrix}(i) + \alpha^{z\uparrow} \sum_{i=1}^{L} \begin{smallmatrix}\textcircled{z}\\ \bigcirc\end{smallmatrix}(i) + \alpha^{z\downarrow} \sum_{i=1}^{L} \begin{smallmatrix}\bigcirc\\ \textcircled{z}\end{smallmatrix}(i) \\ &\quad + \alpha^{++} \sum_{i=1}^{L} (-1)^i \begin{smallmatrix}\oplus\\ \oplus\end{smallmatrix}(i) - \alpha^{--} \sum_{i=1}^{L} (-1)^i \begin{smallmatrix}\ominus\\ \ominus\end{smallmatrix}(i) \\ &= \alpha^{+-} S^+ - \alpha^{-+} S^- + 2(\alpha^{z\uparrow} - \alpha^{z\downarrow}) S^z + (\alpha^{z\uparrow} + \alpha^{z\downarrow}) N - \alpha^{++} \eta^+ + \alpha^{--} \eta^- \\ &= c_+ S^+ + c_- S^- + c_z S^z + c_N N + a_+ \eta^+ + a_- \eta^- , \end{aligned} \tag{C.14}$$

where I redefined each coefficient in the last line and

$$S^+ = \sum_{i=1}^{L} c_{j,\uparrow}^\dagger c_{j,\downarrow}, \quad S^- = \sum_{i=1}^{L} c_{j,\downarrow}^\dagger c_{j,\uparrow}, \quad S^z = \frac{1}{2} \sum_{i=1}^{L} (n_{j,\uparrow} - n_{j,\downarrow}) , \tag{C.15}$$

are the $\mathrm{SU}(2)$ charges and $N = \sum_i^L (n_{j,\uparrow} + n_{j,\downarrow} - 1)$ is the $\mathrm{U}(1)$ charge, and

$$\eta^+ = \sum_{i=1}^{L} (-1)^{i+1} c_{j,\uparrow}^\dagger c_{j,\downarrow}^\dagger, \quad \eta^- = \sum_{i=1}^{L} (-1)^{i+1} c_{j,\downarrow} c_{j,\uparrow}, \tag{C.16}$$

are the $\eta$-pairing charges and $a_{\pm} = 0$ for odd $L$.

Therefore, I have proved that one-local conserved quantities are always written as the linear combination of $\mathrm{SU}(2)$ charges and $\mathrm{U}(1)$ charge and $\eta$-pairing charges.

# Appendix D

# Examples of higher-order local charges

In Appendix D, I show the expressions for $Q_7$ and $Q_8$.

## D.1 Expression for $Q_7$

I show all the expressions for the components of $Q_7$. I explain the structure of $Q_7^j$ for each $j$ in Figure. D.1 (a) and all the structure of $Q_7$ in Figure. D.1 (b). The circle at $(s, d)$ in Figure. D.1 represents $Q_7^j(s, d)$.

$Q_{2k+1}$ does not have $(s, s-1)$-connected diagrams because the space reflection of a $(s, s-1)$-connected diagram is even, whereas $Q_{2k+1}$ is odd under space reflection. The unit number of a $(s, s-1)$-connected diagram is always 2. The types of the two units are both $-$, all the region is an overlap or gap, the length of the two units are the same, such as 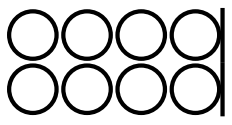 and 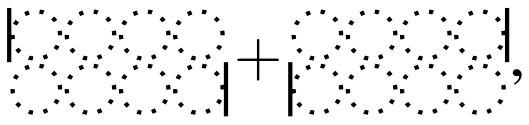, and these diagrams are even under space reflection.

Each component of $Q_7$ is as follows:

$Q_7^0(7, 0) =$ 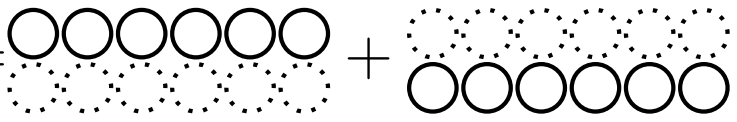

$Q_7^1(6, 0) =$ 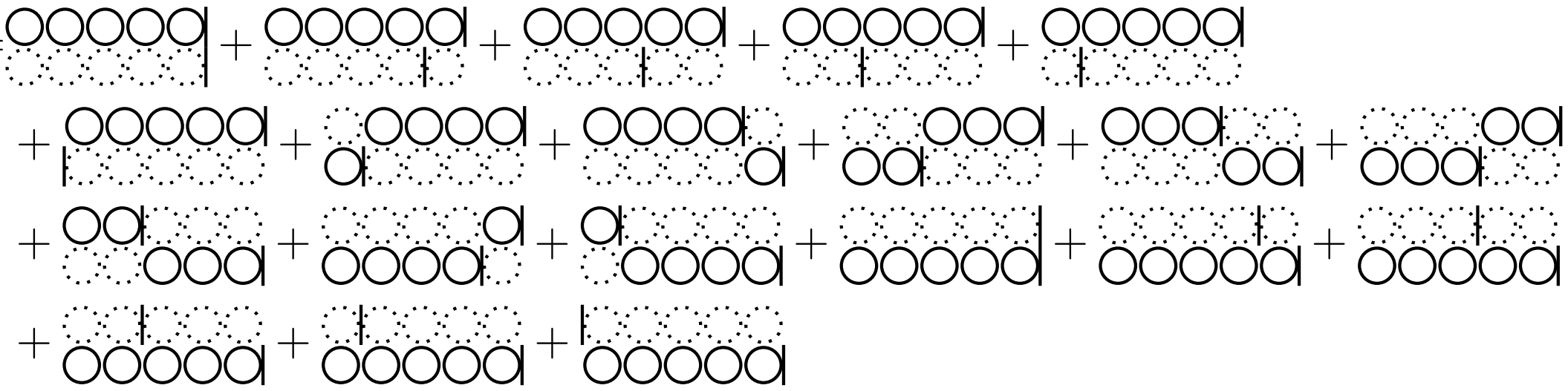

$Q_7^1(5, 1) =$ 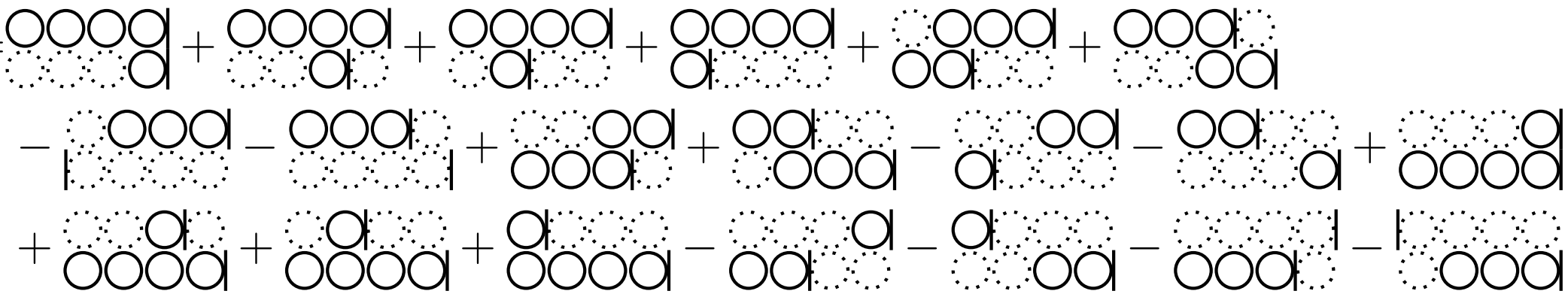

$Q_7^1(4, 2) =$ 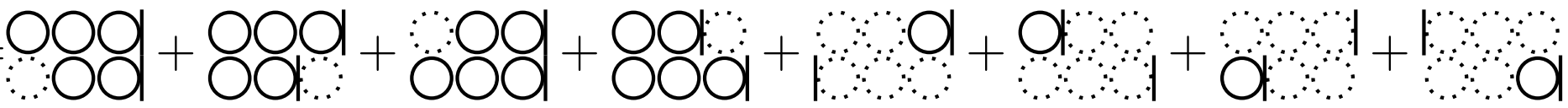

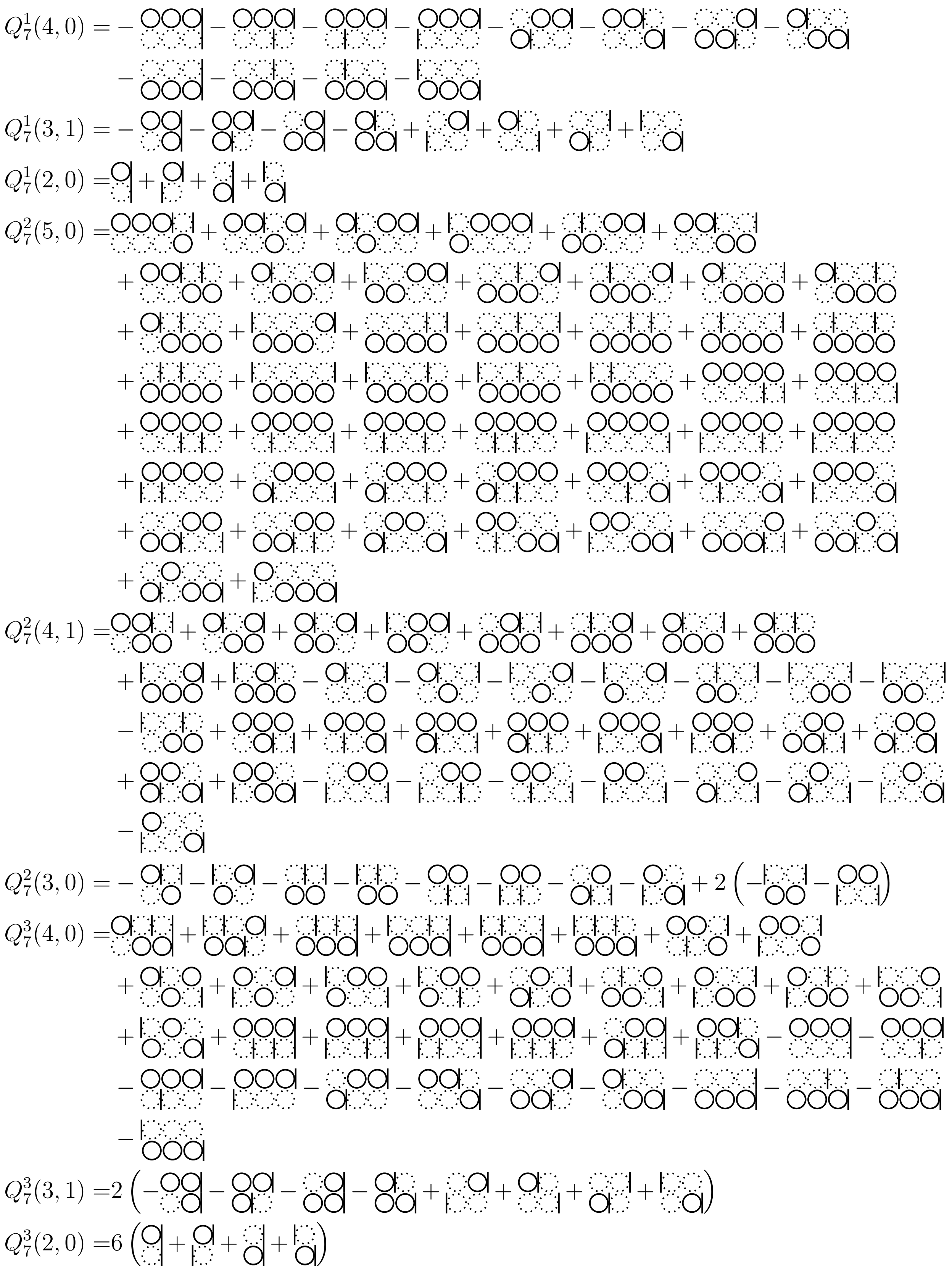
$Q_7^1(4,0) =$
$Q_7^1(3,1) =$
$Q_7^1(2,0) =$
$Q_7^2(5,0) =$
$Q_7^2(4,1) =$
$Q_7^2(3,0) =$
$Q_7^3(4,0) =$
$Q_7^3(3,1) =2$
$Q_7^3(2,0) =6$

$Q_7^4(3,0) =$ 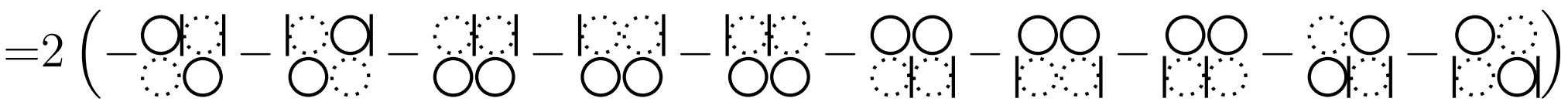

$Q_7^5(2,0) =$ 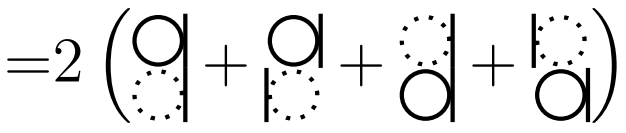

(D.1)

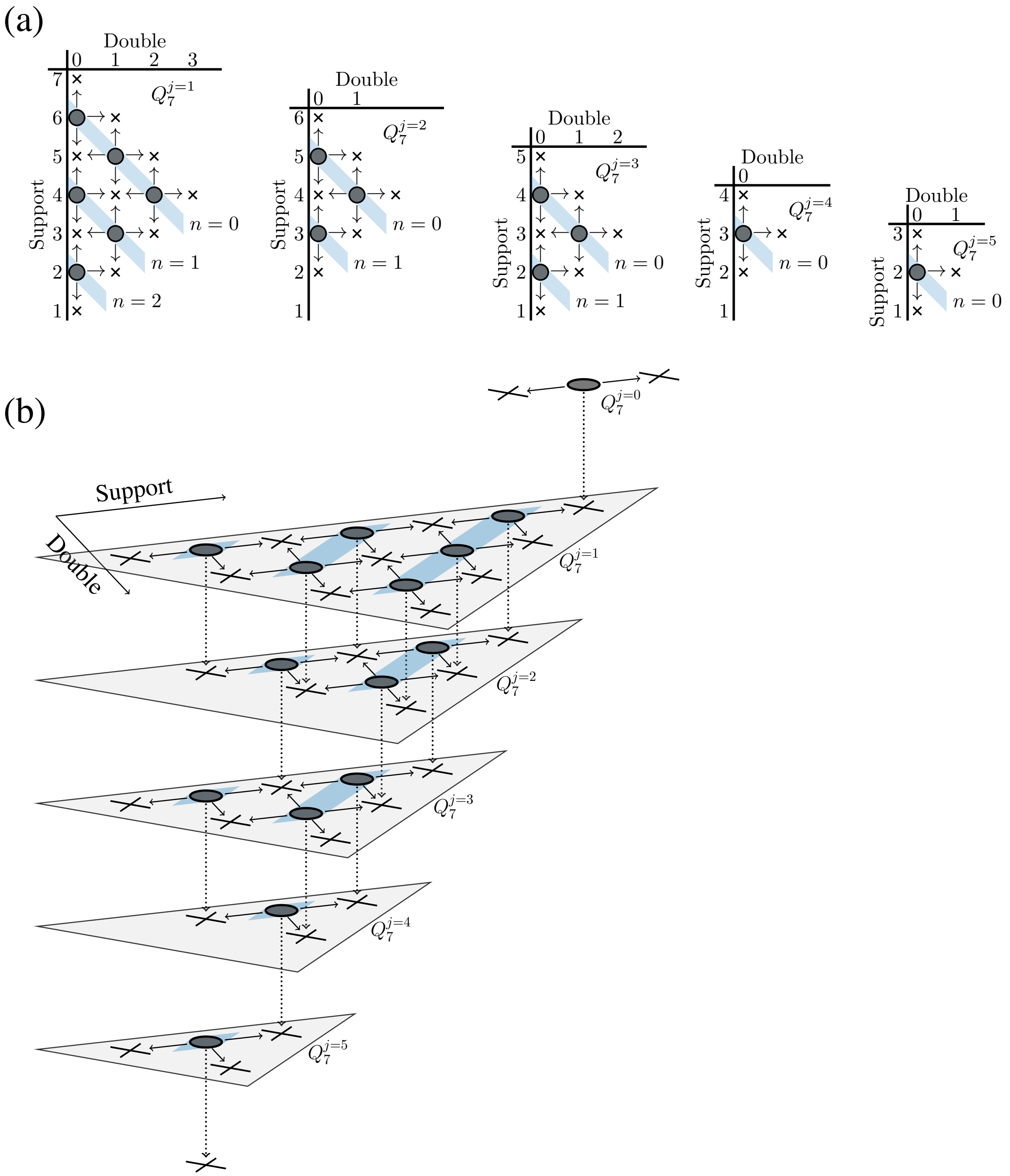

Figure D.1: The structure of $Q_7^j$ for each $j$ (a) and all the structure of $Q_7$ (b). In (b), each plane represents the structure of $Q_6^j$ in (a), and the axis of support and double are omitted. The solid arrow in planes represents the commutator of diagrams with $H_0$, and the vertical dotted arrow represents the commutator of diagrams with $H_{\rm int}$.

## D.2 Expression for $Q_8$

I show all the expressions for the components of $Q_8$. I explain the structure of $Q_8^j$ for each $j$ in Figure. D.2 (a) and all the structure of $Q_8$ in Figure. D.2 (b). The circle at $(s,d)$ in Figure. D.2 represents $Q_8^j(s,d)$. From $Q_8$, the general structure of the cancellation of diagrams (Figure. D.2 (c)) appears. Each component of $Q_8$ is as follows:

$Q_8^0(8,0) =$
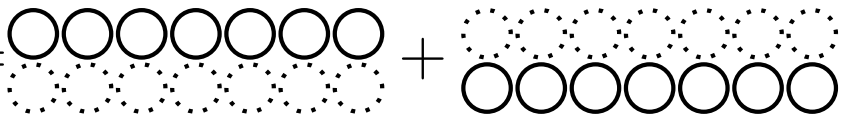

$Q_8^1(7,0) =$
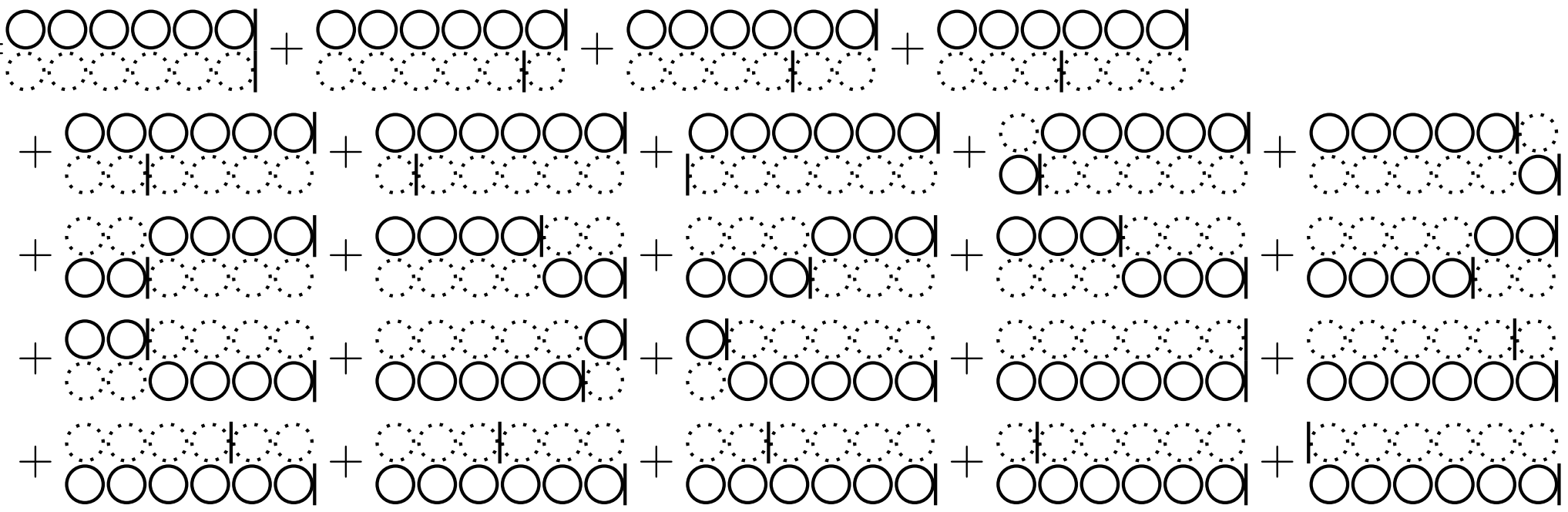

$Q_8^1(6,1) =$
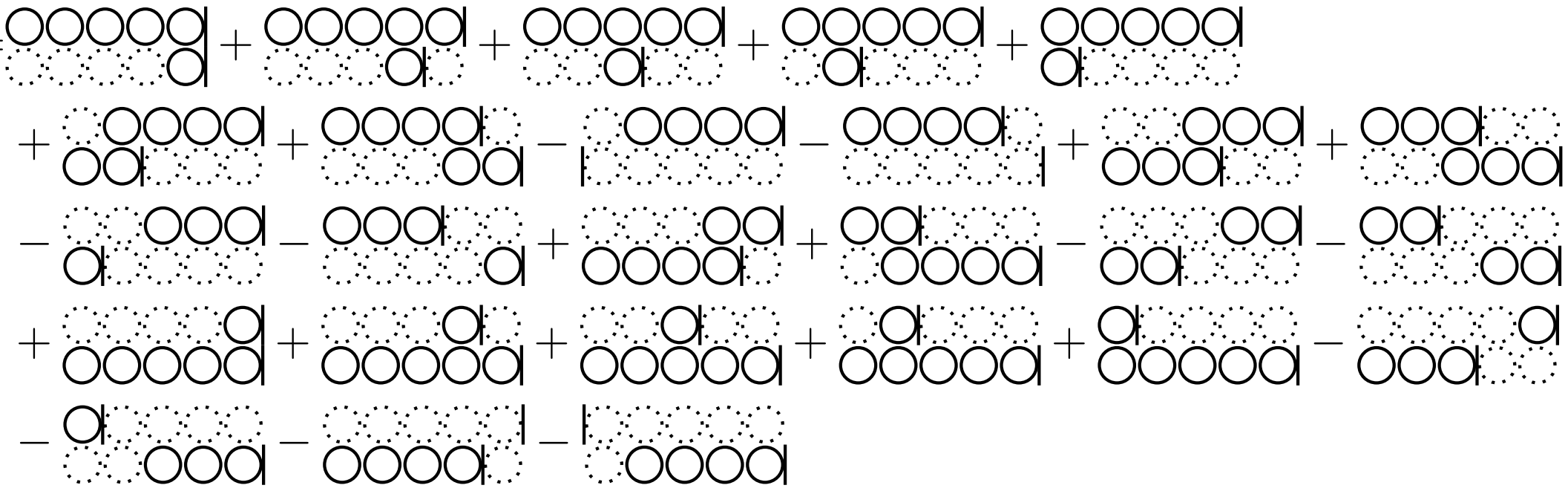

$Q_8^1(5,2) =$
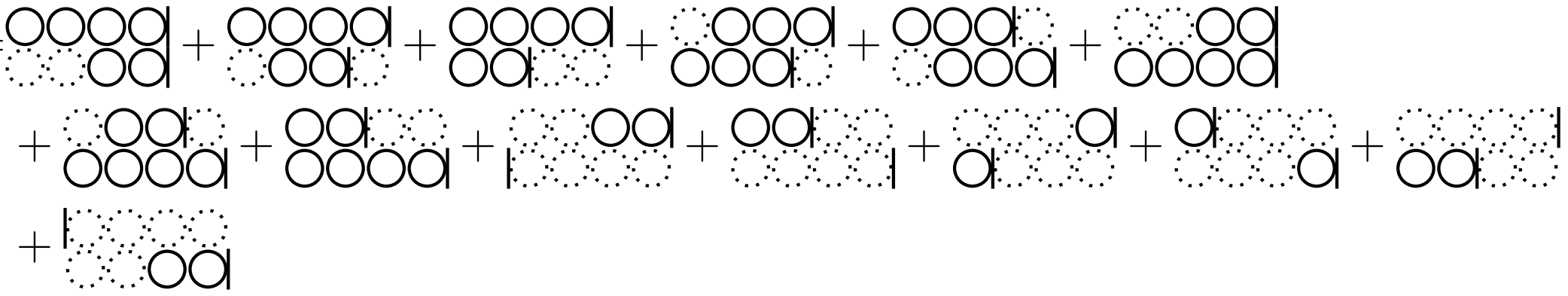

$Q_8^1(4,3) =$
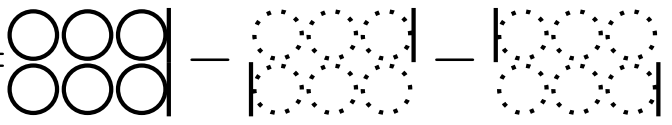

$Q_8^1(5,0) =$
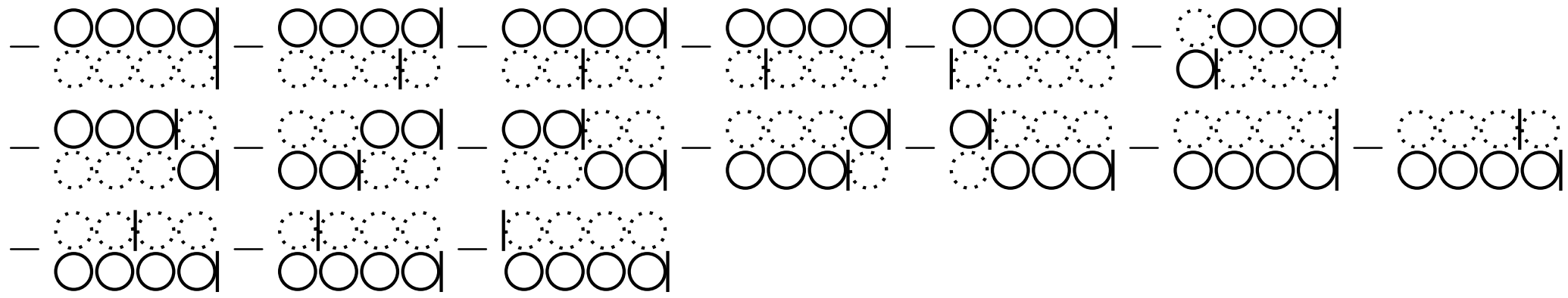

$Q_8^1(4,1) =$
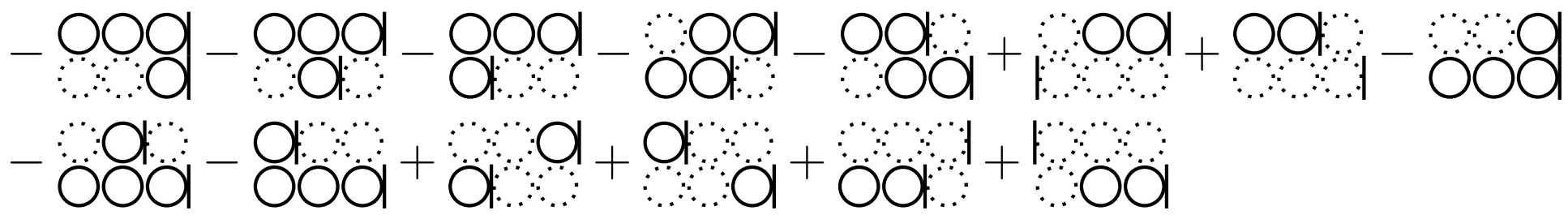

$Q_8^1(3,2) =$
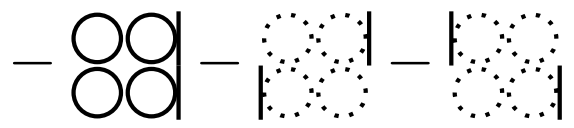

$Q^1_8(3,0) =$ 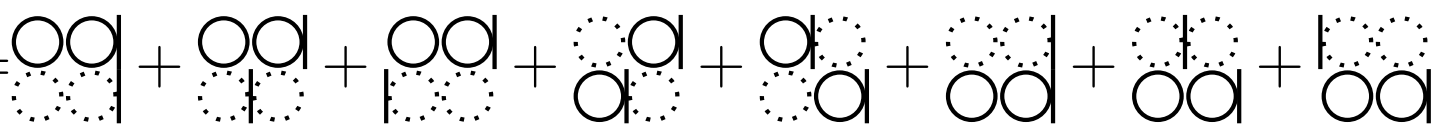

$Q^1_8(2,1) =$ 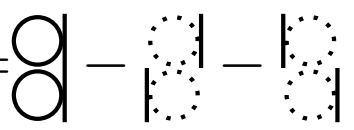

$Q^1_8(1,0) = -$ 

$Q^2_8(6,0) =$ 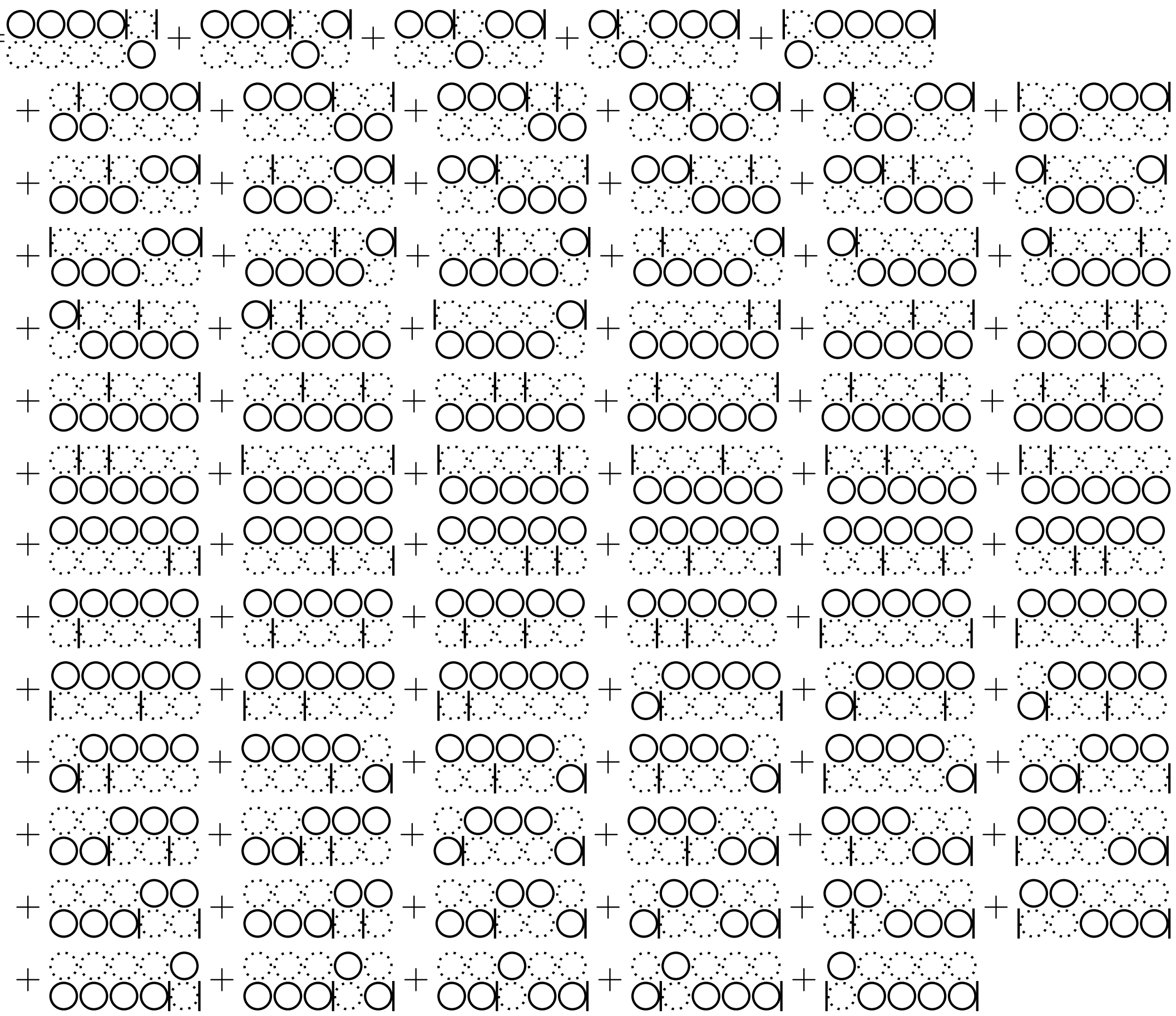

$Q^2_8(5,1) =$ 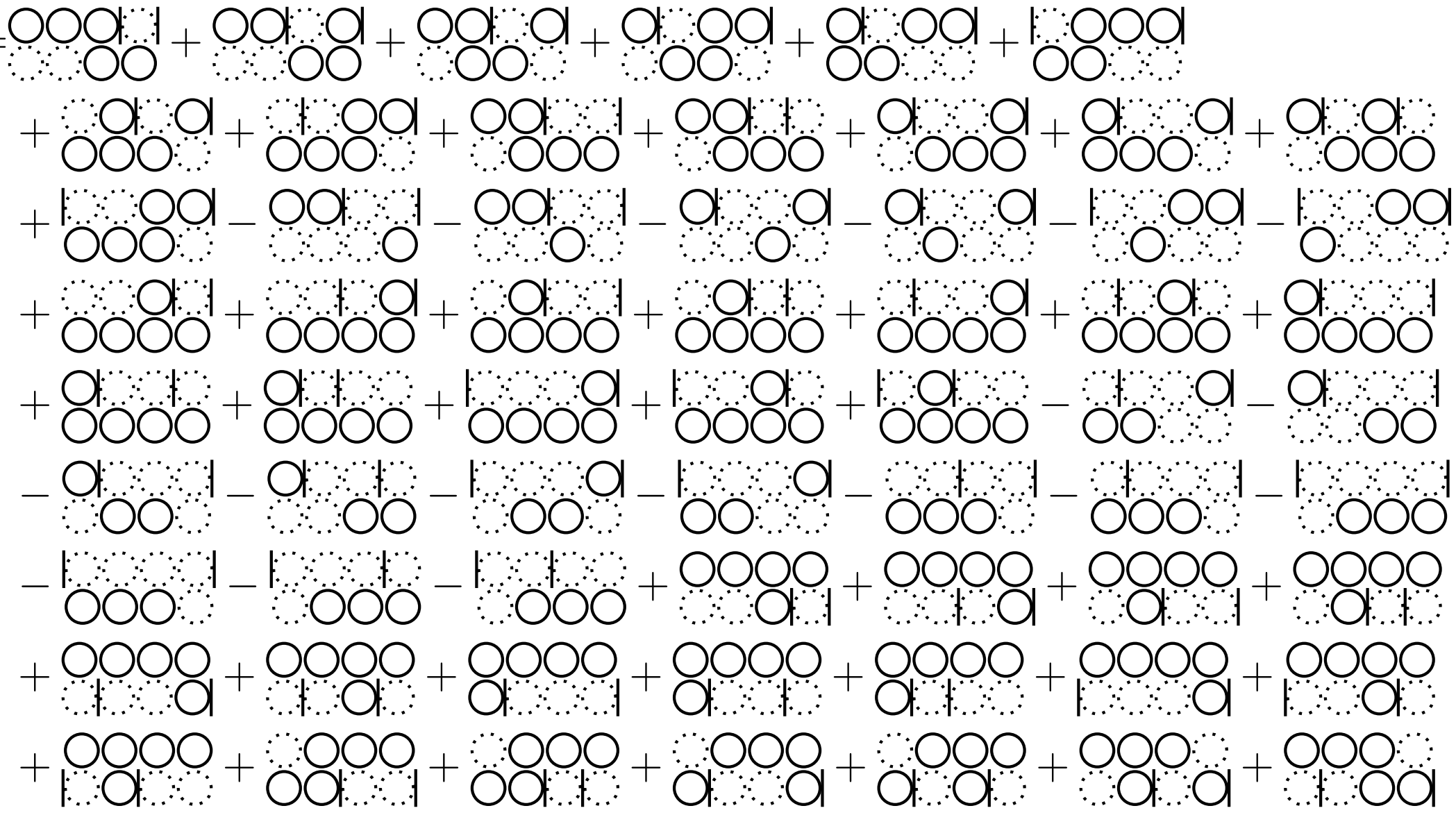

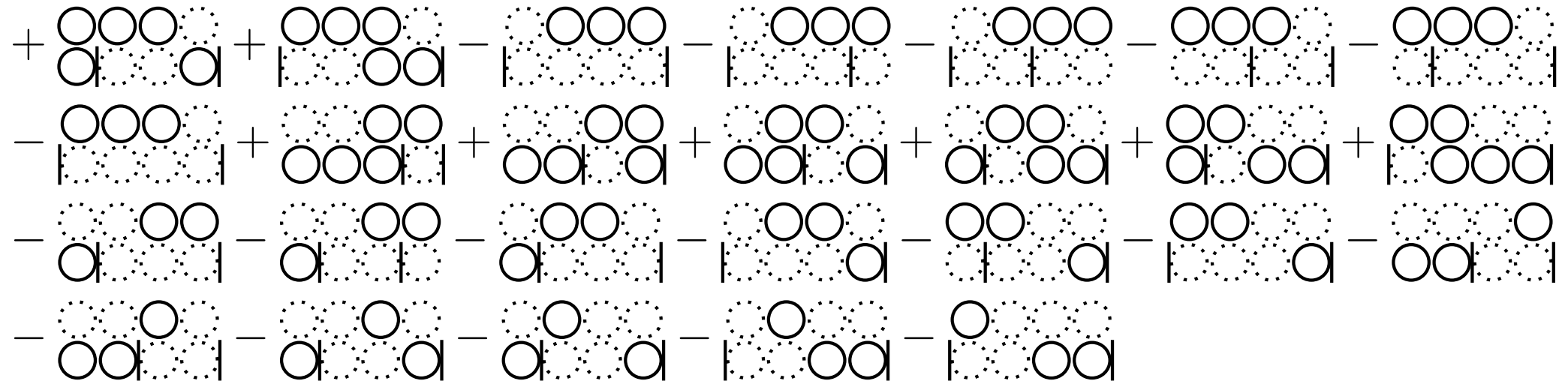

$Q_8^2(4,2) =$ 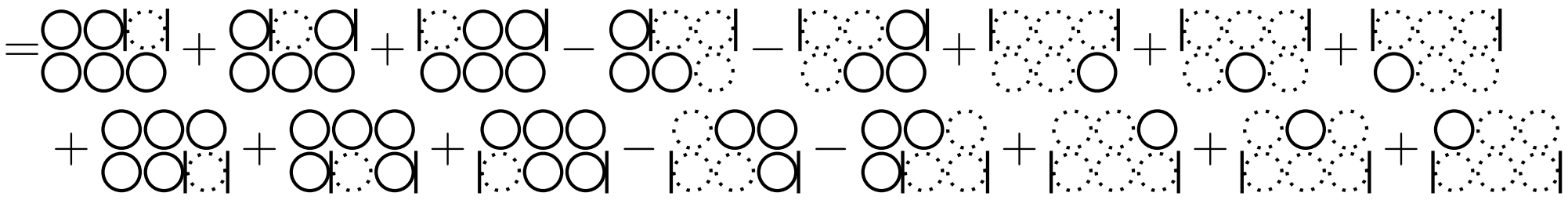

$Q_8^2(4,0) =$ 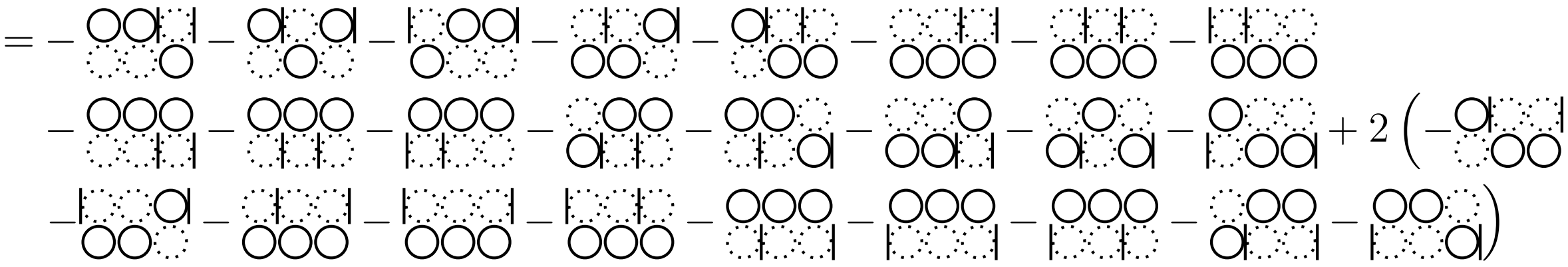

$Q_8^2(3,1) =$ 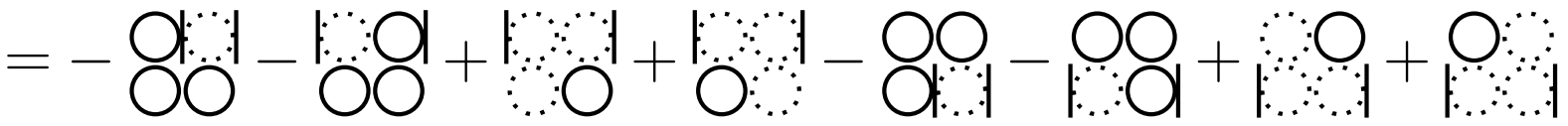

$Q_8^2(2,0) =$ 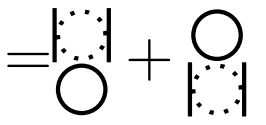

$Q_8^3(5,0) =$ 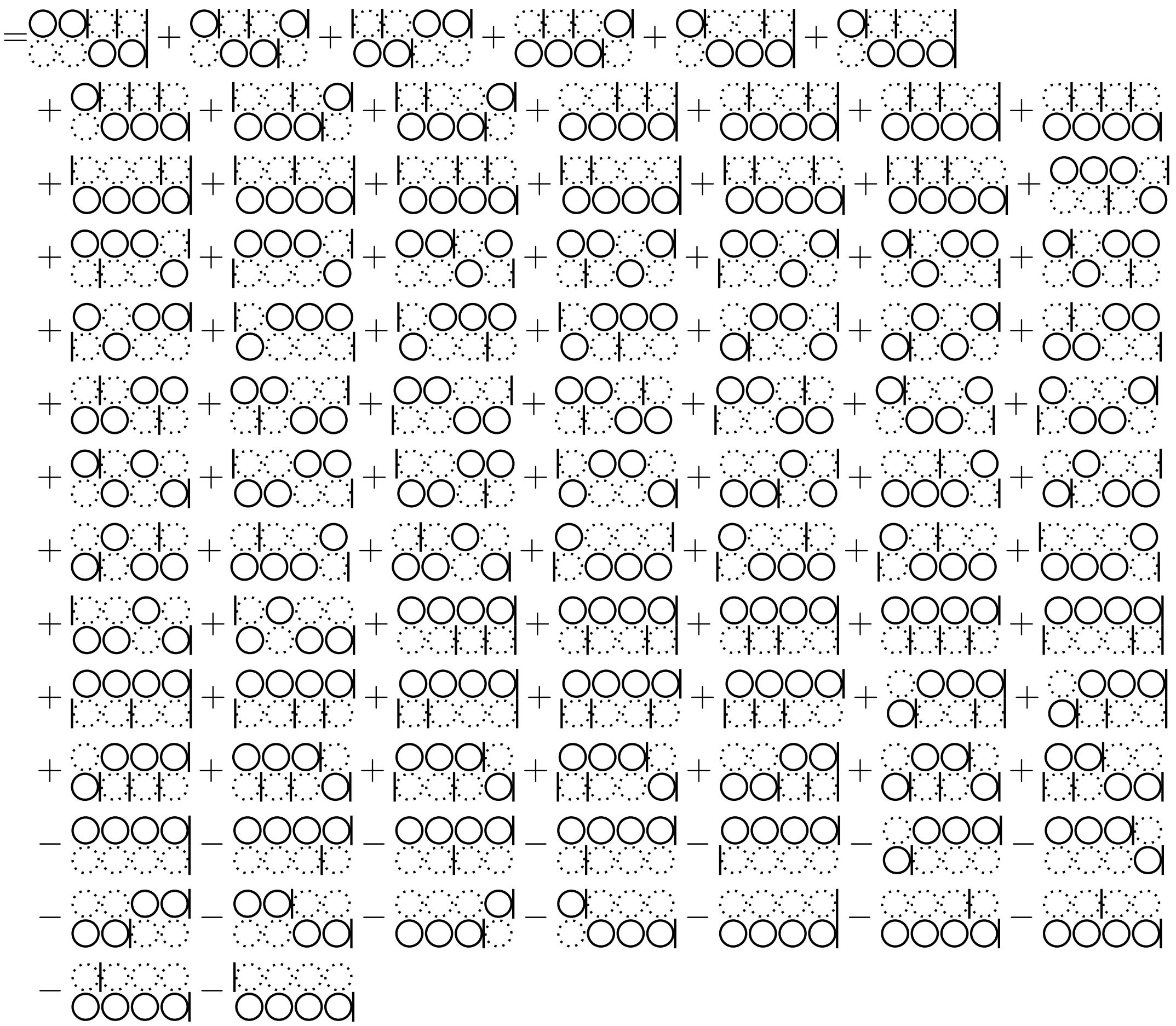

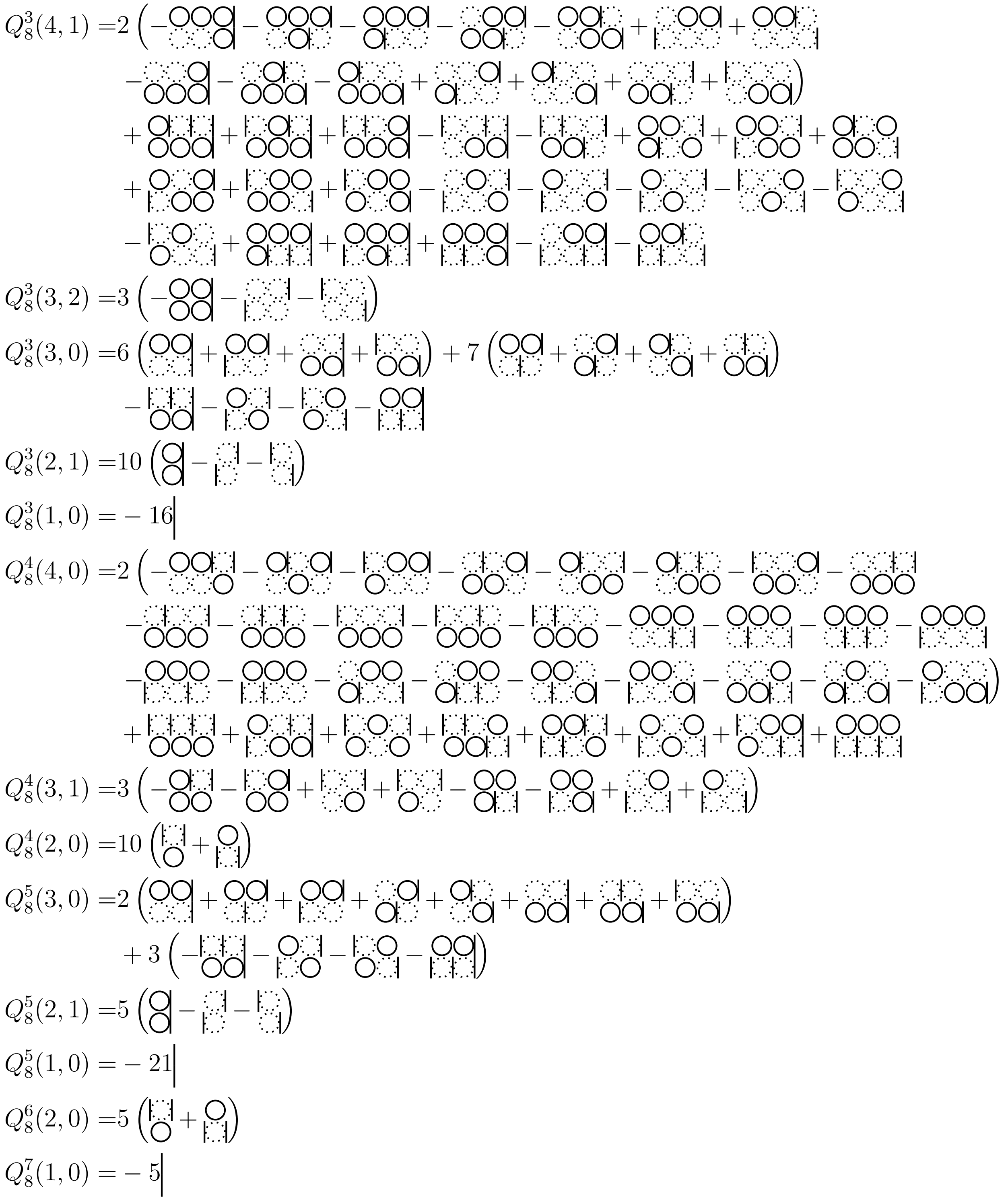
Q83(4,1) =2
Q83(3,2) =3
Q83(3,0) =6
+ 7
Q83(2,1) =10
Q83(1,0) = − 16
Q84(4,0) =2
Q84(3,1) =3
Q84(2,0) =10
Q85(3,0) =2
+ 3
Q85(2,1) =5
Q85(1,0) = − 21
Q86(2,0) =5
Q87(1,0) = − 5

(D.2)

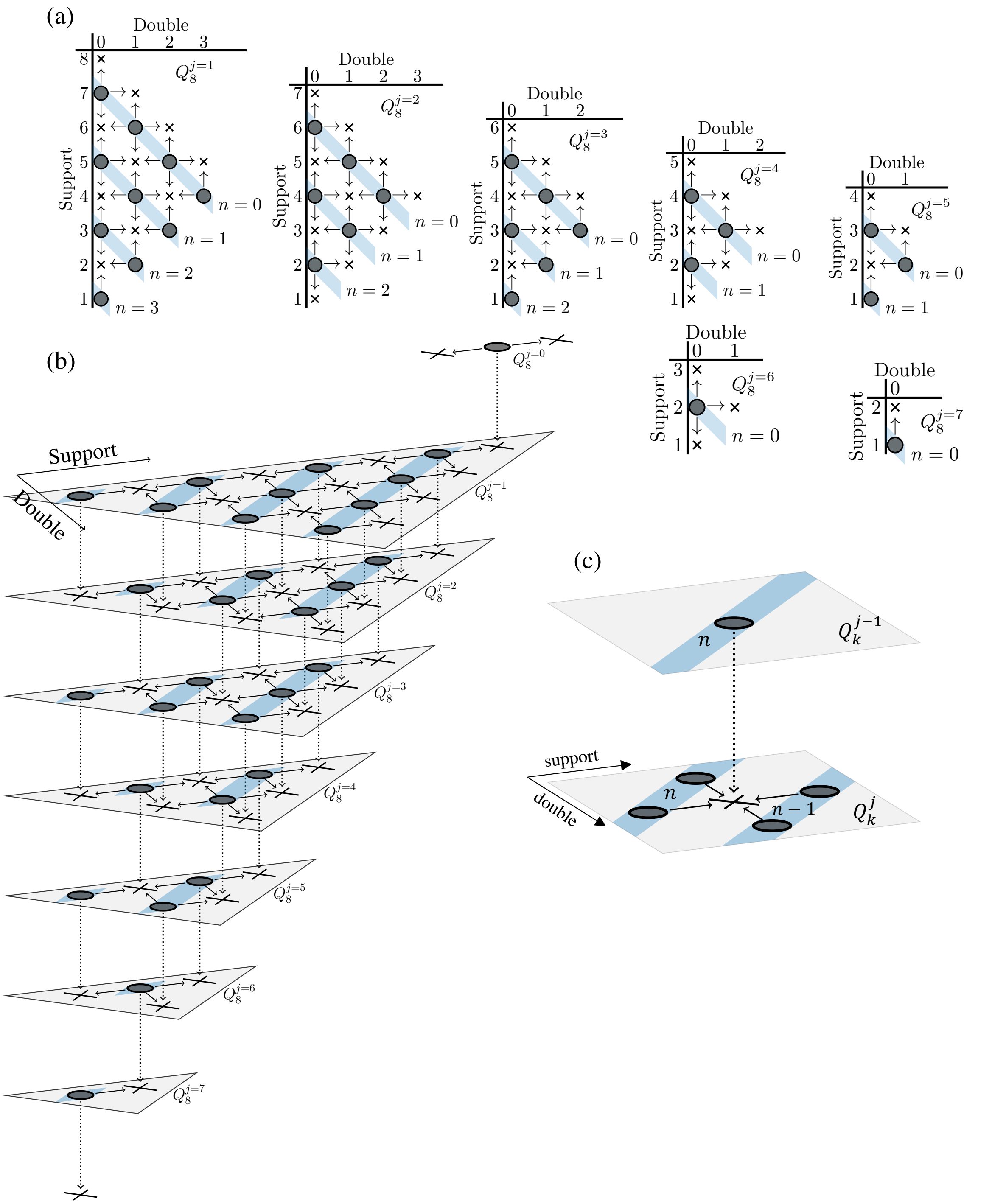

Figure D.2: The structure of $Q_8^j$ for each $j$ (a), all the structure of $Q_8$ (b), and the basic structure of the cancellation of diagrams (c). In (b), each plane represents the structure of $Q_8^j$ in (a), and the axis of support and double are omitted. The solid arrow in planes represents the commutator of diagrams with $H_0$, and the vertical dotted arrow represents the commutator of diagrams with $H_{\text{int}}$. In (c), I only show the circle and arrows related to the cancellation at the crosses.

# Appendix E

# The higher-order charges of $B_k$

In this appendix, I present the explicit expressions for $B_9$, $B_{10}$ and $B_{11}$ as a supplemental material for section 5.5 in Chapter 5.

The explicit expression for $B_9$ is

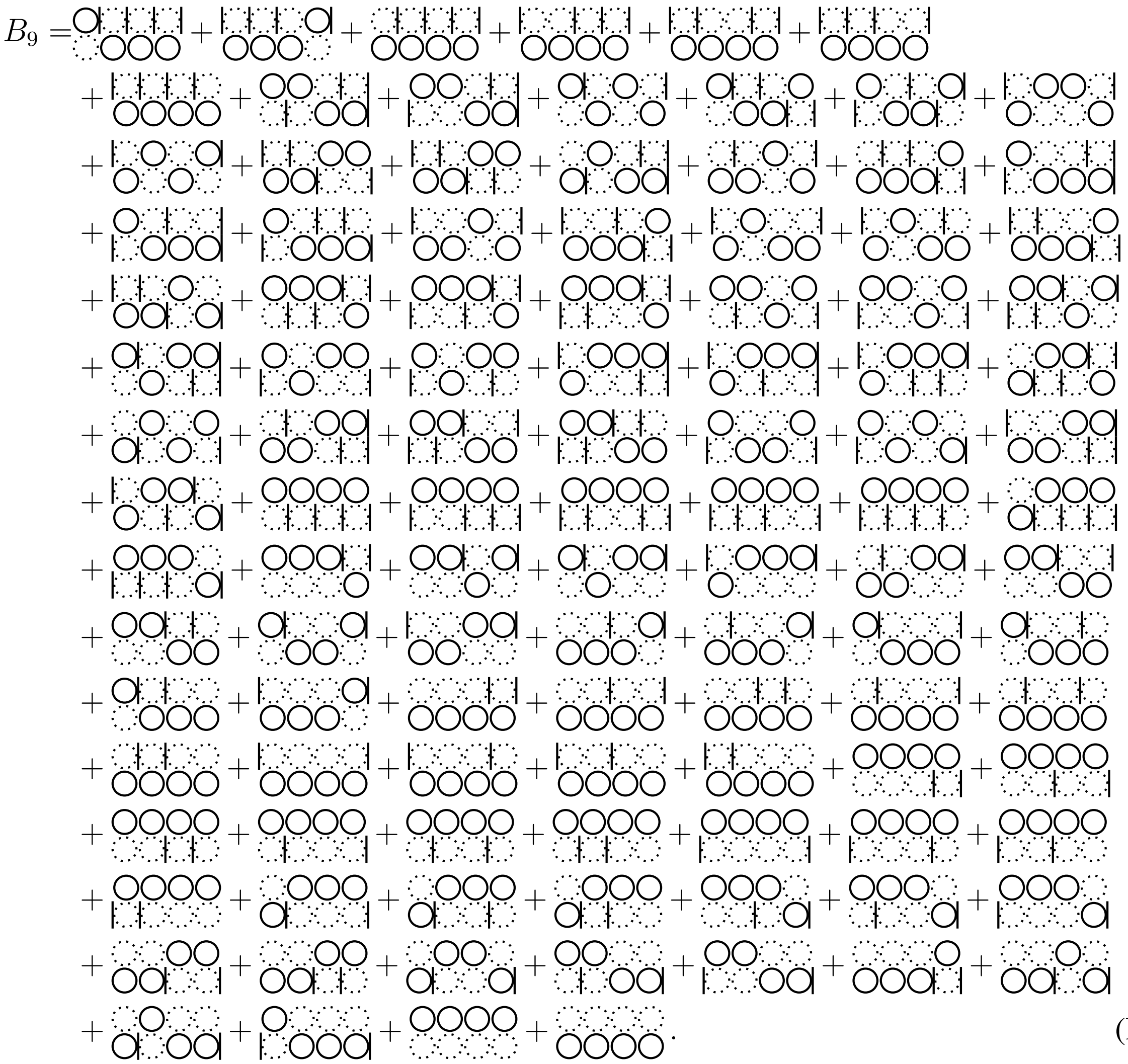

(E.1)

The explicit expression for $B_{10}$ is

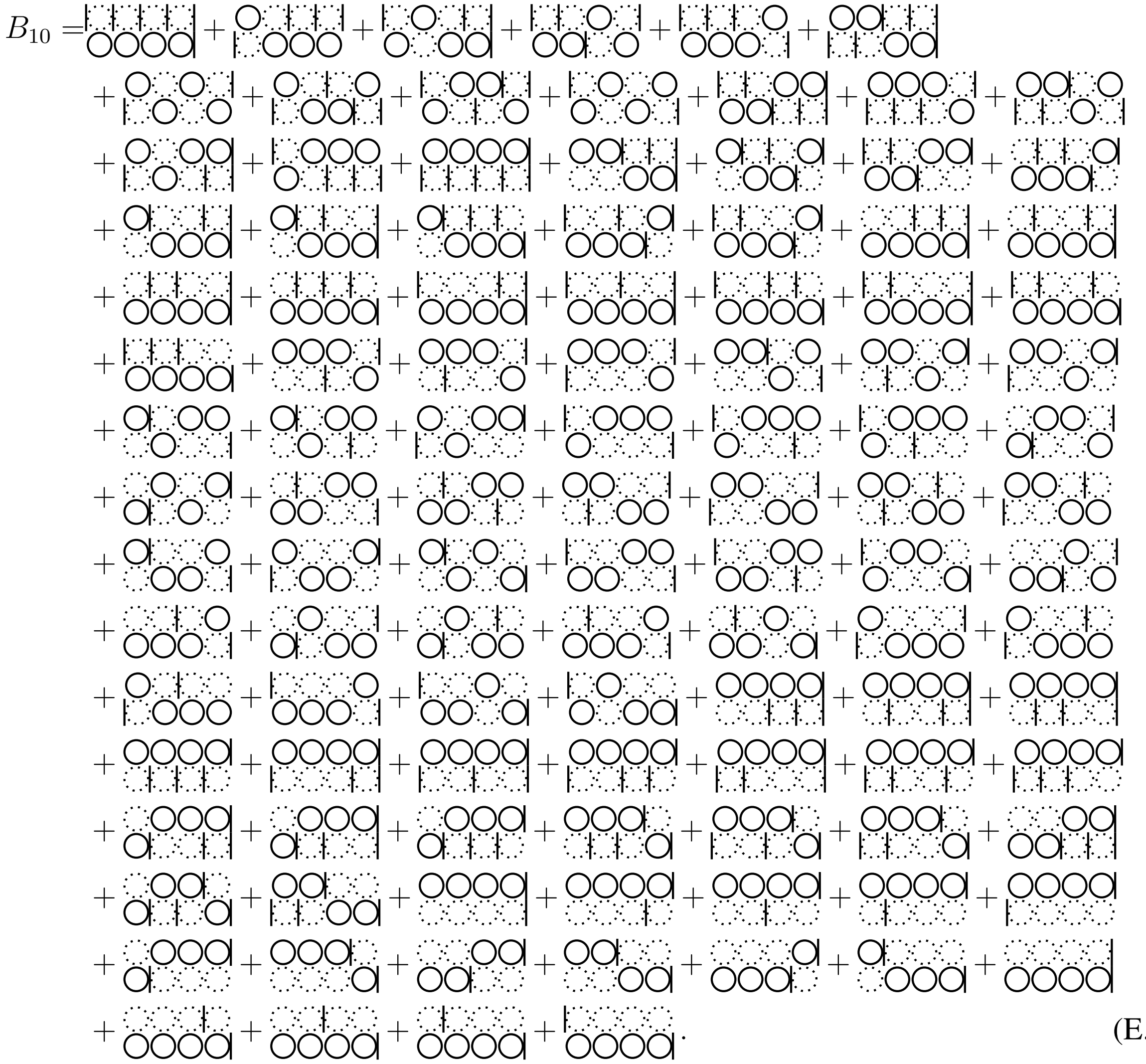

(E.2)

The explicit expression for $B_{11}$ is

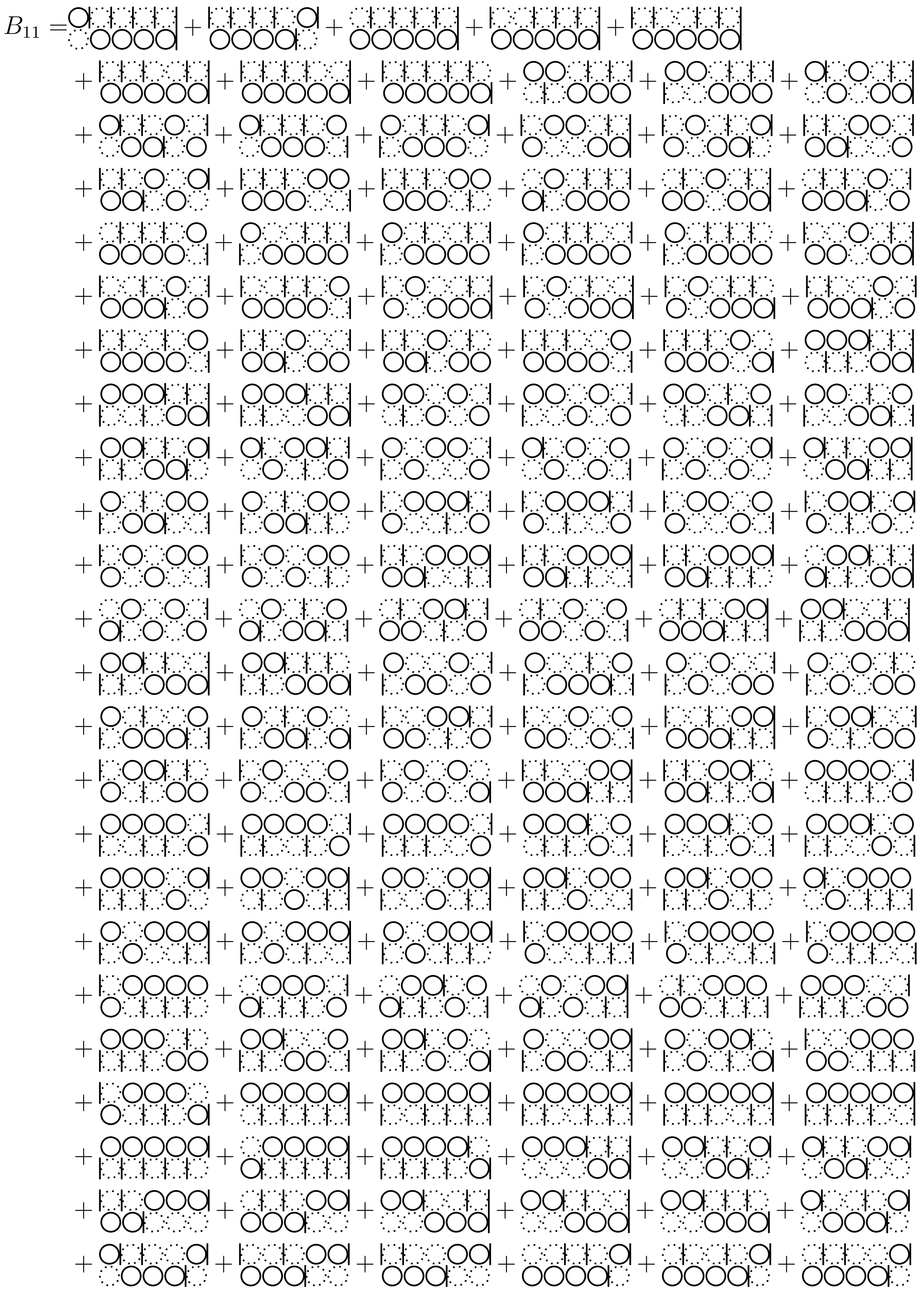

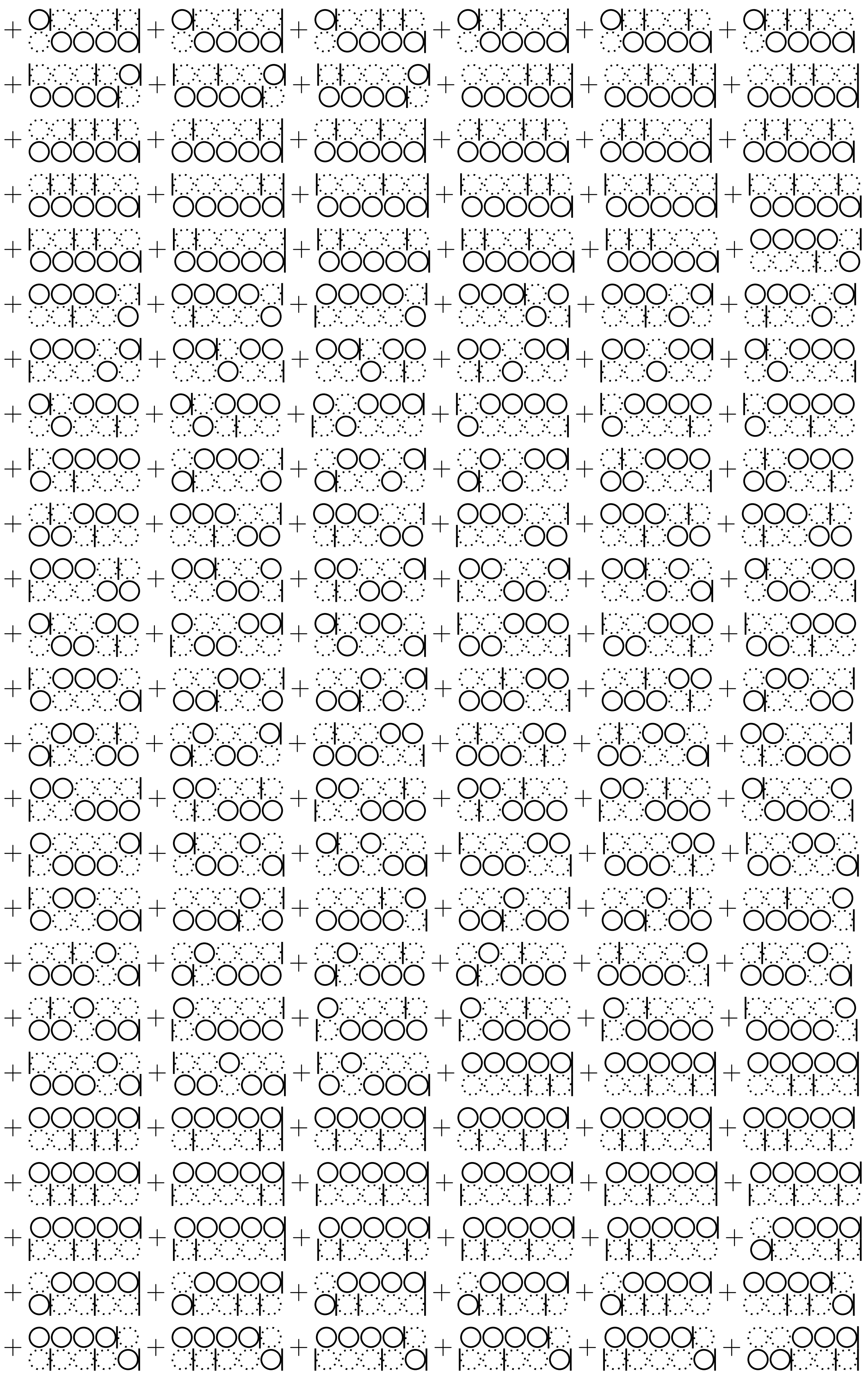

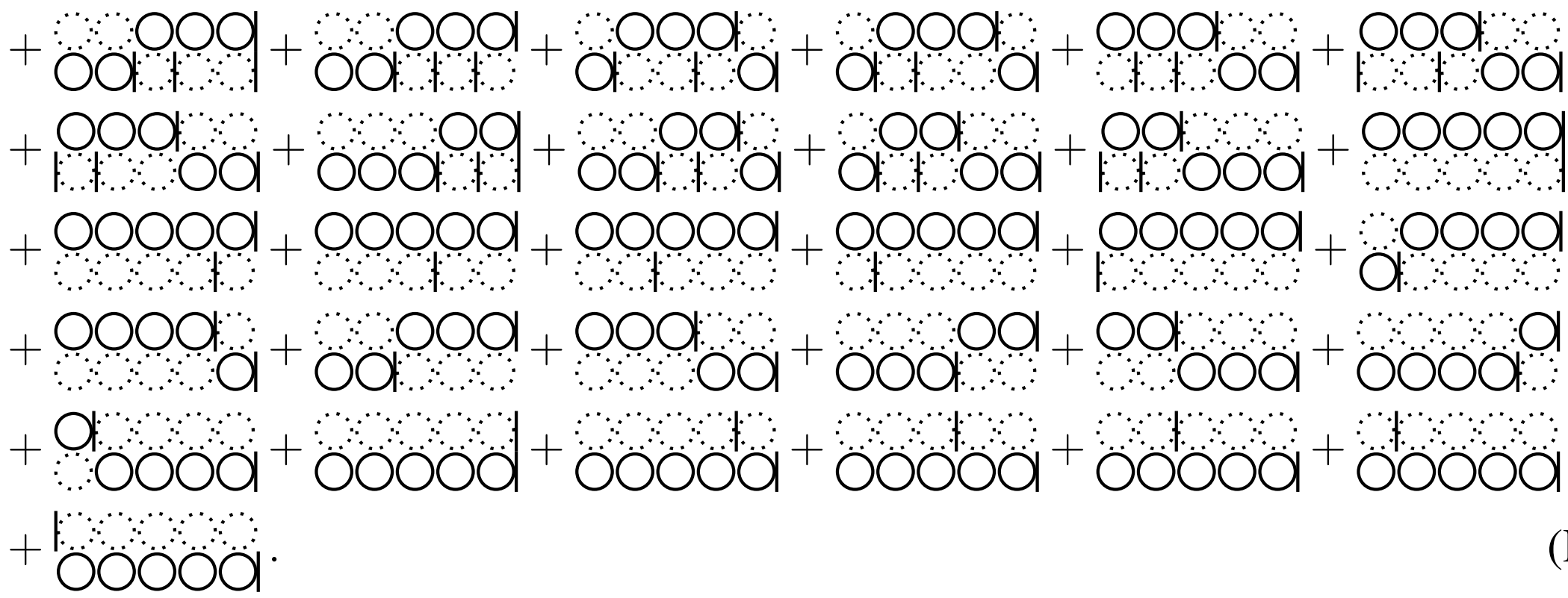

(E.3)

# Appendix F

# Examples of coefficients $C^{j,m}_{n,d}\left(\boldsymbol{\lambda}\right)$

In Appendix F, I present several examples of the coefficients $C^{j,m}_{n,d}\left(\boldsymbol{\lambda}\right)$. The coefficients presented here include enough information to construct $Q_k$ for $k \leq 16$.

I present the table for the value of $C^{j,m}_{n,d}\left(\boldsymbol{\lambda}\right)$ below for $m, n > 0$. Using the symmetry of $C^{j,m}_{n,d}\left(\boldsymbol{\lambda}\right)$ in Eqs. (5.79), (5.80) and (5.81), one can restrict the list $\boldsymbol{\lambda} = \{\lambda_L; \lambda_1, \ldots, \lambda_w; \lambda_R\} = \{\lambda_L; \vec{\lambda}; \lambda_R\}$ to satisfy $0 \leq \lambda_L \leq \lambda_R \leq n$ and $0 \leq \lambda_1 \leq \lambda_2 \leq \cdots \leq \lambda_w \leq n$ without losing generality. Here, I defined $w \equiv j - 1 - 2m$. The columns indicate $(\lambda_L, \lambda_R)$. The rows indicate the value of $d$ and $\{\lambda_1, \lambda_2, \ldots, \lambda_w\} = \vec{\lambda}$ which is represented by $\phi$ in the case of $w = 0$, i.e., the case of $\boldsymbol{\lambda} = \{\lambda_L; \lambda_R\}$. In Figure F.1, I explain how to interpret the table of coefficients.

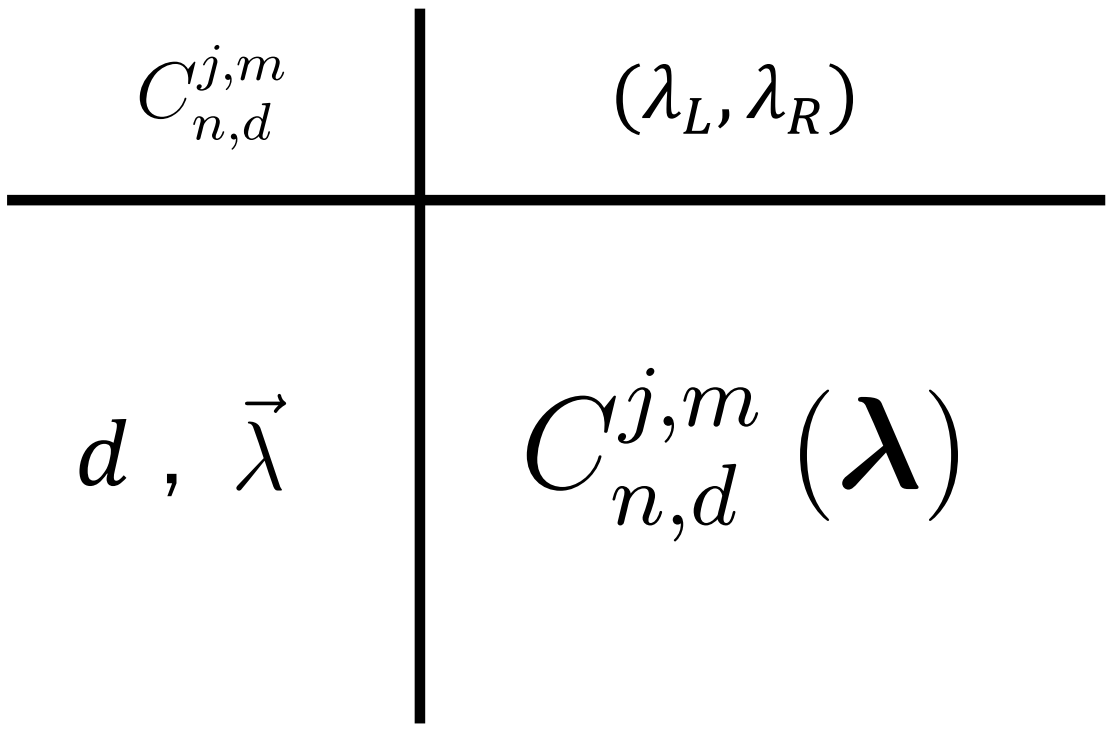

Figure F.1: Schematic picture explaining the Table of Coefficients in Appendix F. The column indicates $(\lambda_L, \lambda_R)$. The row indicates $d$ and $\vec{\lambda}$. In the case of $\boldsymbol{\lambda} = \{\lambda_L; \lambda_R\}$, $\vec{\lambda}$ is represented by $\phi$.

| $C^{j=3,m=1}_{n=1,d}$ | $(0,0)$ | $(0,1)$ | $(1,1)$ |
|---|---|---|---|
| $d=0$ $\phi$ | 5 | 6 | 7 |
| $d=1$ $\phi$ | 10 | 11 | 12 |
| $d=2$ $\phi$ | 15 | 16 | 17 |
| $d=3$ $\phi$ | 20 | 21 | 22 |
| $d=4$ $\phi$ | 25 | 26 | 27 |
| $d=5$ $\phi$ | 30 | 31 | 32 |

| $C^{j=3,m=1}_{n=2,d}$ | $(0,0)$ | $(0,1)$ | $(0,2)$ | $(1,1)$ | $(1,2)$ | $(2,2)$ |
|---|---|---|---|---|---|---|
| $d=0$ $\phi$ | 16 | 20 | 21 | 24 | 25 | 26 |
| $d=1$ $\phi$ | 30 | 34 | 35 | 38 | 39 | 40 |
| $d=2$ $\phi$ | 44 | 48 | 49 | 52 | 53 | 54 |
| $d=3$ $\phi$ | 58 | 62 | 63 | 66 | 67 | 68 |
| $d=4$ $\phi$ | 72 | 76 | 77 | 80 | 81 | 82 |

| $C^{j=3,m=1}_{n=3,d}$ | $(0,0)$ | $(0,1)$ | $(0,2)$ | $(0,3)$ | $(1,1)$ | $(1,2)$ | $(1,3)$ | $(2,2)$ |
|---|---|---|---|---|---|---|---|---|
| $d=0$ $\phi$ | 40 | 50 | 54 | 55 | 60 | 64 | 65 | 68 |
| $d=1$ $\phi$ | 70 | 80 | 84 | 85 | 90 | 94 | 95 | 98 |
| $d=2$ $\phi$ | 100 | 110 | 114 | 115 | 120 | 124 | 125 | 128 |
| $d=3$ $\phi$ | 130 | 140 | 144 | 145 | 150 | 154 | 155 | 158 |

| $C^{j=3,m=1}_{n=3,d}$ | $(2,3)$ | $(3,3)$ |
|---|---|---|
| $d=0$ $\phi$ | 69 | 70 |
| $d=1$ $\phi$ | 99 | 100 |
| $d=2$ $\phi$ | 129 | 130 |
| $d=3$ $\phi$ | 159 | 160 |

| $C^{j=3,m=1}_{n=4,d}$ | $(0,0)$ | $(0,1)$ | $(0,2)$ | $(0,3)$ | $(0,4)$ | $(1,1)$ | $(1,2)$ | $(1,3)$ |
|---|---|---|---|---|---|---|---|---|
| $d=0$ $\phi$ | 85 | 105 | 115 | 119 | 120 | 125 | 135 | 139 |
| $d=1$ $\phi$ | 140 | 160 | 170 | 174 | 175 | 180 | 190 | 194 |
| $d=2$ $\phi$ | 195 | 215 | 225 | 229 | 230 | 235 | 245 | 249 |

| $C^{j=3,m=1}_{n=4,d}$ | $(1,4)$ | $(2,2)$ | $(2,3)$ | $(2,4)$ | $(3,3)$ | $(3,4)$ | $(4,4)$ |
|---|---|---|---|---|---|---|---|
| $d=0$ $\phi$ | 140 | 145 | 149 | 150 | 153 | 154 | 155 |
| $d=1$ $\phi$ | 195 | 200 | 204 | 205 | 208 | 209 | 210 |
| $d=2$ $\phi$ | 250 | 255 | 259 | 260 | 263 | 264 | 265 |

| $C^{j=3,m=1}_{n=5,d}$ | $(0,0)$ | $(0,1)$ | $(0,2)$ | $(0,3)$ | $(0,4)$ | $(0,5)$ | $(1,1)$ | $(1,2)$ |
|---|---|---|---|---|---|---|---|---|
| $d=0$ $\phi$ | 161 | 196 | 216 | 226 | 230 | 231 | 231 | 251 |
| $d=1$ $\phi$ | 252 | 287 | 307 | 317 | 321 | 322 | 322 | 342 |

| $C^{j=3,m=1}_{n=5,d}$ | $(1,3)$ | $(1,4)$ | $(1,5)$ | $(2,2)$ | $(2,3)$ | $(2,4)$ | $(2,5)$ | $(3,3)$ |
|---|---|---|---|---|---|---|---|---|
| $d=0$ $\phi$ | 261 | 265 | 266 | 271 | 281 | 285 | 286 | 291 |
| $d=1$ $\phi$ | 352 | 356 | 357 | 362 | 372 | 376 | 377 | 382 |

| $C^{j=3,m=1}_{n=5,d}$ | $(3,4)$ | $(3,5)$ | $(4,4)$ | $(4,5)$ | $(5,5)$ |
|---|---|---|---|---|---|
| $d=0$ $\phi$ | 295 | 296 | 299 | 300 | 301 |
| $d=1$ $\phi$ | 386 | 387 | 390 | 391 | 392 |

| $C^{j=3,m=1}_{n=6,d}$ | $(0,0)$ | $(0,1)$ | $(0,2)$ | $(0,3)$ | $(0,4)$ | $(0,5)$ | $(0,6)$ | $(1,1)$ |
|---|---|---|---|---|---|---|---|---|
| $d=0$ $\phi$ | 280 | 336 | 371 | 391 | 401 | 405 | 406 | 392 |

| $C^{j=3,m=1}_{n=6,d}$ | $(1,2)$ | $(1,3)$ | $(1,4)$ | $(1,5)$ | $(1,6)$ | $(2,2)$ | $(2,3)$ | $(2,4)$ |
|---|---|---|---|---|---|---|---|---|
| $d=0$ $\phi$ | 427 | 447 | 457 | 461 | 462 | 462 | 482 | 492 |

| $C^{j=3,m=1}_{n=6,d}$ | $(2,5)$ | $(2,6)$ | $(3,3)$ | $(3,4)$ | $(3,5)$ | $(3,6)$ | $(4,4)$ | $(4,5)$ |
|---|---|---|---|---|---|---|---|---|
| $d=0$ $\phi$ | 496 | 497 | 502 | 512 | 516 | 517 | 522 | 526 |

| $C^{j=3,m=1}_{n=6,d}$ | $(4,6)$ | $(5,5)$ | $(5,6)$ | $(6,6)$ |
|---|---|---|---|---|
| $d=0$ $\phi$ | 527 | 530 | 531 | 532 |

| $C_{n=1,d}^{j=4,m=1}$ | | $(0,0)$ | $(0,1)$ | $(1,1)$ |
|---|---|---|---|---|
| $d=0$ | $(0)$ | 10 | 11 | 12 |
| | $(1)$ | 14 | 15 | 16 |
| $d=1$ | $(0)$ | 15 | 16 | 17 |
| | $(1)$ | 20 | 21 | 22 |
| $d=2$ | $(0)$ | 20 | 21 | 22 |
| | $(1)$ | 26 | 27 | 28 |
| $d=3$ | $(0)$ | 25 | 26 | 27 |
| | $(1)$ | 32 | 33 | 34 |
| $d=4$ | $(0)$ | 30 | 31 | 32 |
| | $(1)$ | 38 | 39 | 40 |

| $C_{n=2,d}^{j=4,m=1}$ | | $(0,0)$ | $(0,1)$ | $(0,2)$ | $(1,1)$ | $(1,2)$ | $(2,2)$ |
|---|---|---|---|---|---|---|---|
| | $(0)$ | 30 | 34 | 35 | 38 | 39 | 40 |
| $d=0$ | $(1)$ | 50 | 55 | 56 | 60 | 61 | 62 |
| | $(2)$ | 56 | 61 | 62 | 66 | 67 | 68 |
| | $(0)$ | 44 | 48 | 49 | 52 | 53 | 54 |
| $d=1$ | $(1)$ | 69 | 74 | 75 | 79 | 80 | 81 |
| | $(2)$ | 76 | 81 | 82 | 86 | 87 | 88 |
| | $(0)$ | 58 | 62 | 63 | 66 | 67 | 68 |
| $d=2$ | $(1)$ | 88 | 93 | 94 | 98 | 99 | 100 |
| | $(2)$ | 96 | 101 | 102 | 106 | 107 | 108 |
| | $(0)$ | 72 | 76 | 77 | 80 | 81 | 82 |
| $d=3$ | $(1)$ | 107 | 112 | 113 | 117 | 118 | 119 |
| | $(2)$ | 116 | 121 | 122 | 126 | 127 | 128 |

| $C_{n=3,d}^{j=4,m=1}$ | | $(0,0)$ | $(0,1)$ | $(0,2)$ | $(0,3)$ | $(1,1)$ | $(1,2)$ | $(1,3)$ | $(2,2)$ |
|---|---|---|---|---|---|---|---|---|---|
| $d=0$ | (0) | 70 | 80 | 84 | 85 | 90 | 94 | 95 | 98 |
| | (1) | 128 | 142 | 147 | 148 | 156 | 161 | 162 | 166 |
| | (2) | 158 | 173 | 178 | 179 | 188 | 193 | 194 | 198 |
| | (3) | 166 | 181 | 186 | 187 | 196 | 201 | 202 | 206 |
| $d=1$ | (0) | 100 | 110 | 114 | 115 | 120 | 124 | 125 | 128 |
| | (1) | 172 | 186 | 191 | 192 | 200 | 205 | 206 | 210 |
| | (2) | 207 | 222 | 227 | 228 | 237 | 242 | 243 | 247 |
| | (3) | 216 | 231 | 236 | 237 | 246 | 251 | 252 | 256 |
| $d=2$ | (0) | 130 | 140 | 144 | 145 | 150 | 154 | 155 | 158 |
| | (1) | 216 | 230 | 235 | 236 | 244 | 249 | 250 | 254 |
| | (2) | 256 | 271 | 276 | 277 | 286 | 291 | 292 | 296 |
| | (3) | 266 | 281 | 286 | 287 | 296 | 301 | 302 | 306 |

| $C_{n=3,d}^{j=4,m=1}$ | | $(2,3)$ | $(3,3)$ |
|---|---|---|---|
| $d=0$ | (0) | 99 | 100 |
| | (1) | 167 | 168 |
| | (2) | 199 | 200 |
| | (3) | 207 | 208 |
| $d=1$ | (0) | 129 | 130 |
| | (1) | 211 | 212 |
| | (2) | 248 | 249 |
| | (3) | 257 | 258 |
| $d=2$ | (0) | 159 | 160 |
| | (1) | 255 | 256 |
| | (2) | 297 | 298 |
| | (3) | 307 | 308 |

| $C_{n=4,d}^{j=4,m=1}$ | | $(0,0)$ | $(0,1)$ | $(0,2)$ | $(0,3)$ | $(0,4)$ | $(1,1)$ | $(1,2)$ | $(1,3)$ |
|---|---|---|---|---|---|---|---|---|---|
| | (0) | 140 | 160 | 170 | 174 | 175 | 180 | 190 | 194 |
| | (1) | 270 | 300 | 314 | 319 | 320 | 330 | 344 | 349 |
| $d=0$ | (2) | 356 | 390 | 405 | 410 | 411 | 424 | 439 | 444 |
| | (3) | 396 | 431 | 446 | 451 | 452 | 466 | 481 | 486 |
| | (4) | 406 | 441 | 456 | 461 | 462 | 476 | 491 | 496 |
| | (0) | 195 | 215 | 225 | 229 | 230 | 235 | 245 | 249 |
| | (1) | 355 | 385 | 399 | 404 | 405 | 415 | 429 | 434 |
| $d=1$ | (2) | 455 | 489 | 504 | 509 | 510 | 523 | 538 | 543 |
| | (3) | 500 | 535 | 550 | 555 | 556 | 570 | 585 | 590 |
| | (4) | 511 | 546 | 561 | 566 | 567 | 581 | 596 | 601 |

| $C_{n=4,d}^{j=4,m=1}$ | | $(1,4)$ | $(2,2)$ | $(2,3)$ | $(2,4)$ | $(3,3)$ | $(3,4)$ | $(4,4)$ |
|---|---|---|---|---|---|---|---|---|
| | (0) | 195 | 200 | 204 | 205 | 208 | 209 | 210 |
| | (1) | 350 | 358 | 363 | 364 | 368 | 369 | 370 |
| $d=0$ | (2) | 445 | 454 | 459 | 460 | 464 | 465 | 466 |
| | (3) | 487 | 496 | 501 | 502 | 506 | 507 | 508 |
| | (4) | 497 | 506 | 511 | 512 | 516 | 517 | 518 |
| | (0) | 250 | 255 | 259 | 260 | 263 | 264 | 265 |
| | (1) | 435 | 443 | 448 | 449 | 453 | 454 | 455 |
| $d=1$ | (2) | 544 | 553 | 558 | 559 | 563 | 564 | 565 |
| | (3) | 591 | 600 | 605 | 606 | 610 | 611 | 612 |
| | (4) | 602 | 611 | 616 | 617 | 621 | 622 | 623 |

| $C_{n=5,d}^{j=4,m=1}$ | | $(0,0)$ | $(0,1)$ | $(0,2)$ | $(0,3)$ | $(0,4)$ | $(0,5)$ | $(1,1)$ | $(1,2)$ |
|---|---|---|---|---|---|---|---|---|---|
| | (0) | 252 | 287 | 307 | 317 | 321 | 322 | 322 | 342 |
| | (1) | 502 | 557 | 587 | 601 | 606 | 607 | 612 | 642 |
| $d=0$ | (2) | 692 | 757 | 791 | 806 | 811 | 812 | 822 | 856 |
| | (3) | 806 | 875 | 910 | 925 | 930 | 931 | 944 | 979 |
| | (4) | 856 | 926 | 961 | 976 | 981 | 982 | 996 | 1031 |
| | (5) | 868 | 938 | 973 | 988 | 993 | 994 | 1008 | 1043 |

| $C_{n=5,d}^{j=4,m=1}$ | | $(1,3)$ | $(1,4)$ | $(1,5)$ | $(2,2)$ | $(2,3)$ | $(2,4)$ | $(2,5)$ | $(3,3)$ |
|---|---|---|---|---|---|---|---|---|---|
| $d=0$ | $(0)$ | 352 | 356 | 357 | 362 | 372 | 376 | 377 | 382 |
| | $(1)$ | 656 | 661 | 662 | 672 | 686 | 691 | 692 | 700 |
| | $(2)$ | 871 | 876 | 877 | 890 | 905 | 910 | 911 | 920 |
| | $(3)$ | 994 | 999 | 1000 | 1014 | 1029 | 1034 | 1035 | 1044 |
| | $(4)$ | 1046 | 1051 | 1052 | 1066 | 1081 | 1086 | 1087 | 1096 |
| | $(5)$ | 1058 | 1063 | 1064 | 1078 | 1093 | 1098 | 1099 | 1108 |

| $C_{n=5,d}^{j=4,m=1}$ | | $(3,4)$ | $(3,5)$ | $(4,4)$ | $(4,5)$ | $(5,5)$ |
|---|---|---|---|---|---|---|
| $d=0$ | $(0)$ | 386 | 387 | 390 | 391 | 392 |
| | $(1)$ | 705 | 706 | 710 | 711 | 712 |
| | $(2)$ | 925 | 926 | 930 | 931 | 932 |
| | $(3)$ | 1049 | 1050 | 1054 | 1055 | 1056 |
| | $(4)$ | 1101 | 1102 | 1106 | 1107 | 1108 |
| | $(5)$ | 1113 | 1114 | 1118 | 1119 | 1120 |

| $C_{n=1,d}^{j=5,m=1}$ | | $(0,0)$ | $(0,1)$ | $(1,1)$ |
|---|---|---|---|---|
| $d=0$ | $(0,0)$ | 15 | 16 | 17 |
| | $(0,1)$ | 20 | 21 | 22 |
| | $(1,1)$ | 25 | 26 | 27 |
| $d=1$ | $(0,0)$ | 20 | 21 | 22 |
| | $(0,1)$ | 26 | 27 | 28 |
| | $(1,1)$ | 32 | 33 | 34 |
| $d=2$ | $(0,0)$ | 25 | 26 | 27 |
| | $(0,1)$ | 32 | 33 | 34 |
| | $(1,1)$ | 39 | 40 | 41 |
| $d=3$ | $(0,0)$ | 30 | 31 | 32 |
| | $(0,1)$ | 38 | 39 | 40 |
| | $(1,1)$ | 46 | 47 | 48 |
| $d=4$ | $(0,0)$ | 35 | 36 | 37 |
| | $(0,1)$ | 44 | 45 | 46 |
| | $(1,1)$ | 53 | 54 | 55 |

| $C^{j=5,m=1}_{n=2,d}$ | | $(0,0)$ | $(0,1)$ | $(0,2)$ | $(1,1)$ | $(1,2)$ | $(2,2)$ |
|---|---|---|---|---|---|---|---|
| $d=0$ | $(0,0)$ | 44 | 48 | 49 | 52 | 53 | 54 |
| | $(0,1)$ | 69 | 74 | 75 | 79 | 80 | 81 |
| | $(0,2)$ | 76 | 81 | 82 | 86 | 87 | 88 |
| | $(1,1)$ | 101 | 107 | 108 | 113 | 114 | 115 |
| | $(1,2)$ | 108 | 114 | 115 | 120 | 121 | 122 |
| | $(2,2)$ | 115 | 121 | 122 | 127 | 128 | 129 |
| $d=1$ | $(0,0)$ | 58 | 62 | 63 | 66 | 67 | 68 |
| | $(0,1)$ | 88 | 93 | 94 | 98 | 99 | 100 |
| | $(0,2)$ | 96 | 101 | 102 | 106 | 107 | 108 |
| | $(1,1)$ | 126 | 132 | 133 | 138 | 139 | 140 |
| | $(1,2)$ | 134 | 140 | 141 | 146 | 147 | 148 |
| | $(2,2)$ | 142 | 148 | 149 | 154 | 155 | 156 |
| $d=2$ | $(0,0)$ | 72 | 76 | 77 | 80 | 81 | 82 |
| | $(0,1)$ | 107 | 112 | 113 | 117 | 118 | 119 |
| | $(0,2)$ | 116 | 121 | 122 | 126 | 127 | 128 |
| | $(1,1)$ | 151 | 157 | 158 | 163 | 164 | 165 |
| | $(1,2)$ | 160 | 166 | 167 | 172 | 173 | 174 |
| | $(2,2)$ | 169 | 175 | 176 | 181 | 182 | 183 |
| $d=3$ | $(0,0)$ | 86 | 90 | 91 | 94 | 95 | 96 |
| | $(0,1)$ | 126 | 131 | 132 | 136 | 137 | 138 |
| | $(0,2)$ | 136 | 141 | 142 | 146 | 147 | 148 |
| | $(1,1)$ | 176 | 182 | 183 | 188 | 189 | 190 |
| | $(1,2)$ | 186 | 192 | 193 | 198 | 199 | 200 |
| | $(2,2)$ | 196 | 202 | 203 | 208 | 209 | 210 |

| $C_{n=3,d}^{j=5,m=1}$ | | $(0,0)$ | $(0,1)$ | $(0,2)$ | $(0,3)$ | $(1,1)$ | $(1,2)$ | $(1,3)$ | $(2,2)$ |
|---|---|---|---|---|---|---|---|---|---|
| $d=0$ | $(0,0)$ | 100 | 110 | 114 | 115 | 120 | 124 | 125 | 128 |
| | $(0,1)$ | 172 | 186 | 191 | 192 | 200 | 205 | 206 | 210 |
| | $(0,2)$ | 207 | 222 | 227 | 228 | 237 | 242 | 243 | 247 |
| | $(0,3)$ | 216 | 231 | 236 | 237 | 246 | 251 | 252 | 256 |
| | $(1,1)$ | 279 | 298 | 304 | 305 | 317 | 323 | 324 | 329 |
| | $(1,2)$ | 323 | 343 | 349 | 350 | 363 | 369 | 370 | 375 |
| | $(1,3)$ | 332 | 352 | 358 | 359 | 372 | 378 | 379 | 384 |
| | $(2,2)$ | 367 | 388 | 394 | 395 | 409 | 415 | 416 | 421 |
| | $(2,3)$ | 376 | 397 | 403 | 404 | 418 | 424 | 425 | 430 |
| | $(3,3)$ | 385 | 406 | 412 | 413 | 427 | 433 | 434 | 439 |
| $d=1$ | $(0,0)$ | 130 | 140 | 144 | 145 | 150 | 154 | 155 | 158 |
| | $(0,1)$ | 216 | 230 | 235 | 236 | 244 | 249 | 250 | 254 |
| | $(0,2)$ | 256 | 271 | 276 | 277 | 286 | 291 | 292 | 296 |
| | $(0,3)$ | 266 | 281 | 286 | 287 | 296 | 301 | 302 | 306 |
| | $(1,1)$ | 342 | 361 | 367 | 368 | 380 | 386 | 387 | 392 |
| | $(1,2)$ | 392 | 412 | 418 | 419 | 432 | 438 | 439 | 444 |
| | $(1,3)$ | 402 | 422 | 428 | 429 | 442 | 448 | 449 | 454 |
| | $(2,2)$ | 442 | 463 | 469 | 470 | 484 | 490 | 491 | 496 |
| | $(2,3)$ | 452 | 473 | 479 | 480 | 494 | 500 | 501 | 506 |
| | $(3,3)$ | 462 | 483 | 489 | 490 | 504 | 510 | 511 | 516 |
| $d=2$ | $(0,0)$ | 160 | 170 | 174 | 175 | 180 | 184 | 185 | 188 |
| | $(0,1)$ | 260 | 274 | 279 | 280 | 288 | 293 | 294 | 298 |
| | $(0,2)$ | 305 | 320 | 325 | 326 | 335 | 340 | 341 | 345 |
| | $(0,3)$ | 316 | 331 | 336 | 337 | 346 | 351 | 352 | 356 |
| | $(1,1)$ | 405 | 424 | 430 | 431 | 443 | 449 | 450 | 455 |
| | $(1,2)$ | 461 | 481 | 487 | 488 | 501 | 507 | 508 | 513 |
| | $(1,3)$ | 472 | 492 | 498 | 499 | 512 | 518 | 519 | 524 |
| | $(2,2)$ | 517 | 538 | 544 | 545 | 559 | 565 | 566 | 571 |
| | $(2,3)$ | 528 | 549 | 555 | 556 | 570 | 576 | 577 | 582 |
| | $(3,3)$ | 539 | 560 | 566 | 567 | 581 | 587 | 588 | 593 |

| $C_{n=3,d}^{j=5,m=1}$ | | $(2,3)$ | $(3,3)$ |
|---|---|---|---|
| $d=0$ | $(0,0)$ | 129 | 130 |
| | $(0,1)$ | 211 | 212 |
| | $(0,2)$ | 248 | 249 |
| | $(0,3)$ | 257 | 258 |
| | $(1,1)$ | 330 | 331 |
| | $(1,2)$ | 376 | 377 |
| | $(1,3)$ | 385 | 386 |
| | $(2,2)$ | 422 | 423 |
| | $(2,3)$ | 431 | 432 |
| | $(3,3)$ | 440 | 441 |
| $d=1$ | $(0,0)$ | 159 | 160 |
| | $(0,1)$ | 255 | 256 |
| | $(0,2)$ | 297 | 298 |
| | $(0,3)$ | 307 | 308 |
| | $(1,1)$ | 393 | 394 |
| | $(1,2)$ | 445 | 446 |
| | $(1,3)$ | 455 | 456 |
| | $(2,2)$ | 497 | 498 |
| | $(2,3)$ | 507 | 508 |
| | $(3,3)$ | 517 | 518 |
| $d=2$ | $(0,0)$ | 189 | 190 |
| | $(0,1)$ | 299 | 300 |
| | $(0,2)$ | 346 | 347 |
| | $(0,3)$ | 357 | 358 |
| | $(1,1)$ | 456 | 457 |
| | $(1,2)$ | 514 | 515 |
| | $(1,3)$ | 525 | 526 |
| | $(2,2)$ | 572 | 573 |
| | $(2,3)$ | 583 | 584 |
| | $(3,3)$ | 594 | 595 |

| $C_{n=1,d}^{j=5,m=2}$ | | $(0,0)$ | $(0,1)$ | $(1,1)$ |
|---|---|---|---|---|
| $d=0$ | $\phi$ | 21 | 25 | 29 |
| $d=1$ | $\phi$ | 52 | 57 | 62 |
| $d=2$ | $\phi$ | 92 | 98 | 104 |
| $d=3$ | $\phi$ | 141 | 148 | 155 |
| $d=4$ | $\phi$ | 199 | 207 | 215 |

| $C_{n=2,d}^{j=5,m=2}$ | $(0,0)$ | $(0,1)$ | $(0,2)$ | $(1,1)$ | $(1,2)$ | $(2,2)$ |
|---|---|---|---|---|---|---|
| $d=0$ $\phi$ | 124 | 159 | 165 | 195 | 201 | 207 |
| $d=1$ $\phi$ | 294 | 337 | 344 | 381 | 388 | 395 |
| $d=2$ $\phi$ | 507 | 558 | 566 | 610 | 618 | 626 |
| $d=3$ $\phi$ | 763 | 822 | 831 | 882 | 891 | 900 |

| $C_{n=3,d}^{j=5,m=2}$ | $(0,0)$ | $(0,1)$ | $(0,2)$ | $(0,3)$ | $(1,1)$ | $(1,2)$ | $(1,3)$ | $(2,2)$ |
|---|---|---|---|---|---|---|---|---|
| $d=0$ $\phi$ | 532 | 701 | 752 | 760 | 878 | 930 | 938 | 982 |
| $d=1$ $\phi$ | 1193 | 1398 | 1457 | 1466 | 1610 | 1670 | 1679 | 1730 |
| $d=2$ $\phi$ | 2002 | 2242 | 2309 | 2319 | 2489 | 2557 | 2567 | 2625 |

| $C_{n=3,d}^{j=5,m=2}$ | $(2,3)$ | $(3,3)$ |
|---|---|---|
| $d=0$ $\phi$ | 990 | 998 |
| $d=1$ $\phi$ | 1739 | 1748 |
| $d=2$ $\phi$ | 2635 | 2645 |

| $C_{n=4,d}^{j=5,m=2}$ | $(0,0)$ | $(0,1)$ | $(0,2)$ | $(0,3)$ | $(0,4)$ | $(1,1)$ | $(1,2)$ | $(1,3)$ |
|---|---|---|---|---|---|---|---|---|
| $d=0$ $\phi$ | 1837 | 2434 | 2673 | 2740 | 2750 | 3065 | 3312 | 3380 |
| $d=1$ $\phi$ | 3891 | 4607 | 4882 | 4957 | 4968 | 5351 | 5633 | 5709 |

| $C_{n=4,d}^{j=5,m=2}$ | $(1,4)$ | $(2,2)$ | $(2,3)$ | $(2,4)$ | $(3,3)$ | $(3,4)$ | $(4,4)$ |
|---|---|---|---|---|---|---|---|
| $d=0$ $\phi$ | 3390 | 3560 | 3628 | 3638 | 3696 | 3706 | 3716 |
| $d=1$ $\phi$ | 5720 | 5916 | 5992 | 6003 | 6068 | 6079 | 6090 |

| $C_{n=5,d}^{j=5,m=2}$ | $(0,0)$ | $(0,1)$ | $(0,2)$ | $(0,3)$ | $(0,4)$ | $(0,5)$ | $(1,1)$ | $(1,2)$ |
|---|---|---|---|---|---|---|---|---|
| $d=0$ $\phi$ | 5403 | 7127 | 7948 | 8257 | 8340 | 8352 | 8956 | 9811 |

| $C_{n=5,d}^{j=5,m=2}$ | $(1,3)$ | $(1,4)$ | $(1,5)$ | $(2,2)$ | $(2,3)$ | $(2,4)$ | $(2,5)$ | $(3,3)$ |
|---|---|---|---|---|---|---|---|---|
| $d=0$ $\phi$ | 10128 | 10212 | 10224 | 10674 | 10992 | 11076 | 11088 | 11310 |

| $C_{n=5,d}^{j=5,m=2}$ | $(3,4)$ | $(3,5)$ | $(4,4)$ | $(4,5)$ | $(5,5)$ |
|---|---|---|---|---|---|
| $d=0$ $\phi$ | 11394 | 11406 | 11478 | 11490 | 11502 |

| | $C_{n=1,d}^{j=6,m=1}$ | $(0,0)$ | $(0,1)$ | $(1,1)$ |
|---|---|---|---|---|
| $d=0$ | $(0,0,0)$ | 20 | 21 | 22 |
| | $(0,0,1)$ | 26 | 27 | 28 |
| | $(0,1,1)$ | 32 | 33 | 34 |
| | $(1,1,1)$ | 38 | 39 | 40 |
| $d=1$ | $(0,0,0)$ | 25 | 26 | 27 |
| | $(0,0,1)$ | 32 | 33 | 34 |
| | $(0,1,1)$ | 39 | 40 | 41 |
| | $(1,1,1)$ | 46 | 47 | 48 |
| $d=2$ | $(0,0,0)$ | 30 | 31 | 32 |
| | $(0,0,1)$ | 38 | 39 | 40 |
| | $(0,1,1)$ | 46 | 47 | 48 |
| | $(1,1,1)$ | 54 | 55 | 56 |
| $d=3$ | $(0,0,0)$ | 35 | 36 | 37 |
| | $(0,0,1)$ | 44 | 45 | 46 |
| | $(0,1,1)$ | 53 | 54 | 55 |
| | $(1,1,1)$ | 62 | 63 | 64 |

| | $C^{j=6,m=1}_{n=2,d}$ | $(0,0)$ | $(0,1)$ | $(0,2)$ | $(1,1)$ | $(1,2)$ | $(2,2)$ |
|---|---|---|---|---|---|---|---|
| $d=0$ | $(0,0,0)$ | 58 | 62 | 63 | 66 | 67 | 68 |
| | $(0,0,1)$ | 88 | 93 | 94 | 98 | 99 | 100 |
| | $(0,0,2)$ | 96 | 101 | 102 | 106 | 107 | 108 |
| | $(0,1,1)$ | 126 | 132 | 133 | 138 | 139 | 140 |
| | $(0,1,2)$ | 134 | 140 | 141 | 146 | 147 | 148 |
| | $(0,2,2)$ | 142 | 148 | 149 | 154 | 155 | 156 |
| | $(1,1,1)$ | 172 | 179 | 180 | 186 | 187 | 188 |
| | $(1,1,2)$ | 180 | 187 | 188 | 194 | 195 | 196 |
| | $(1,2,2)$ | 188 | 195 | 196 | 202 | 203 | 204 |
| | $(2,2,2)$ | 196 | 203 | 204 | 210 | 211 | 212 |
| $d=1$ | $(0,0,0)$ | 72 | 76 | 77 | 80 | 81 | 82 |
| | $(0,0,1)$ | 107 | 112 | 113 | 117 | 118 | 119 |
| | $(0,0,2)$ | 116 | 121 | 122 | 126 | 127 | 128 |
| | $(0,1,1)$ | 151 | 157 | 158 | 163 | 164 | 165 |
| | $(0,1,2)$ | 160 | 166 | 167 | 172 | 173 | 174 |
| | $(0,2,2)$ | 169 | 175 | 176 | 181 | 182 | 183 |
| | $(1,1,1)$ | 204 | 211 | 212 | 218 | 219 | 220 |
| | $(1,1,2)$ | 213 | 220 | 221 | 227 | 228 | 229 |
| | $(1,2,2)$ | 222 | 229 | 230 | 236 | 237 | 238 |
| | $(2,2,2)$ | 231 | 238 | 239 | 245 | 246 | 247 |
| $d=2$ | $(0,0,0)$ | 86 | 90 | 91 | 94 | 95 | 96 |
| | $(0,0,1)$ | 126 | 131 | 132 | 136 | 137 | 138 |
| | $(0,0,2)$ | 136 | 141 | 142 | 146 | 147 | 148 |
| | $(0,1,1)$ | 176 | 182 | 183 | 188 | 189 | 190 |
| | $(0,1,2)$ | 186 | 192 | 193 | 198 | 199 | 200 |
| | $(0,2,2)$ | 196 | 202 | 203 | 208 | 209 | 210 |
| | $(1,1,1)$ | 236 | 243 | 244 | 250 | 251 | 252 |
| | $(1,1,2)$ | 246 | 253 | 254 | 260 | 261 | 262 |
| | $(1,2,2)$ | 256 | 263 | 264 | 270 | 271 | 272 |
| | $(2,2,2)$ | 266 | 273 | 274 | 280 | 281 | 282 |

| $C_{n=1,d}^{j=6,m=2}$ | | $(0,0)$ | $(0,1)$ | $(1,1)$ |
|---|---|---|---|---|
| $d=0$ | $(0)$ | 52 | 57 | 62 |
| | $(1)$ | 66 | 71 | 76 |
| $d=1$ | $(0)$ | 92 | 98 | 104 |
| | $(1)$ | 112 | 118 | 124 |
| $d=2$ | $(0)$ | 141 | 148 | 155 |
| | $(1)$ | 168 | 175 | 182 |
| $d=3$ | $(0)$ | 199 | 207 | 215 |
| | $(1)$ | 234 | 242 | 250 |

| $C_{n=2,d}^{j=6,m=2}$ | | $(0,0)$ | $(0,1)$ | $(0,2)$ | $(1,1)$ | $(1,2)$ | $(2,2)$ |
|---|---|---|---|---|---|---|---|
| | $(0)$ | 294 | 337 | 344 | 381 | 388 | 395 |
| $d=0$ | $(1)$ | 435 | 485 | 492 | 536 | 543 | 550 |
| | $(2)$ | 462 | 512 | 519 | 563 | 570 | 577 |
| | $(0)$ | 507 | 558 | 566 | 610 | 618 | 626 |
| $d=1$ | $(1)$ | 706 | 765 | 773 | 825 | 833 | 841 |
| | $(2)$ | 741 | 800 | 808 | 860 | 868 | 876 |
| | $(0)$ | 763 | 822 | 831 | 882 | 891 | 900 |
| $d=2$ | $(1)$ | 1029 | 1097 | 1106 | 1166 | 1175 | 1184 |
| | $(2)$ | 1073 | 1141 | 1150 | 1210 | 1219 | 1228 |

| $C_{n=3,d}^{j=6,m=2}$ | | $(0,0)$ | $(0,1)$ | $(0,2)$ | $(0,3)$ | $(1,1)$ | $(1,2)$ | $(1,3)$ | $(2,2)$ |
|---|---|---|---|---|---|---|---|---|---|
| $d=0$ | $(0)$ | 1193 | 1398 | 1457 | 1466 | 1610 | 1670 | 1679 | 1730 |
| | $(1)$ | 1956 | 2220 | 2288 | 2297 | 2492 | 2561 | 2570 | 2630 |
| | $(2)$ | 2222 | 2495 | 2563 | 2572 | 2776 | 2845 | 2854 | 2914 |
| | $(3)$ | 2266 | 2539 | 2607 | 2616 | 2820 | 2889 | 2898 | 2958 |
| $d=1$ | $(0)$ | 2002 | 2242 | 2309 | 2319 | 2489 | 2557 | 2567 | 2625 |
| | $(1)$ | 3064 | 3371 | 3448 | 3458 | 3686 | 3764 | 3774 | 3842 |
| | $(2)$ | 3406 | 3723 | 3800 | 3810 | 4048 | 4126 | 4136 | 4204 |
| | $(3)$ | 3460 | 3777 | 3854 | 3864 | 4102 | 4180 | 4190 | 4258 |

| $C_{n=3,d}^{j=6,m=2}$ | | $(2,3)$ | $(3,3)$ |
|---|---|---|---|
| $d=0$ | $(0)$ | 1739 | 1748 |
| | $(1)$ | 2639 | 2648 |
| | $(2)$ | 2923 | 2932 |
| | $(3)$ | 2967 | 2976 |
| $d=1$ | $(0)$ | 2635 | 2645 |
| | $(1)$ | 3852 | 3862 |
| | $(2)$ | 4214 | 4224 |
| | $(3)$ | 4268 | 4278 |

| $C_{n=4,d}^{j=6,m=2}$ | | $(0,0)$ | $(0,1)$ | $(0,2)$ | $(0,3)$ | $(0,4)$ | $(1,1)$ | $(1,2)$ | $(1,3)$ |
|---|---|---|---|---|---|---|---|---|---|
| $d=0$ | $(0)$ | 3891 | 4607 | 4882 | 4957 | 4968 | 5351 | 5633 | 5709 |
| | $(1)$ | 6849 | 7840 | 8190 | 8276 | 8287 | 8866 | 9224 | 9311 |
| | $(2)$ | 8253 | 9319 | 9680 | 9766 | 9777 | 10421 | 10790 | 10877 |
| | $(3)$ | 8680 | 9757 | 10118 | 10204 | 10215 | 10870 | 11239 | 11326 |
| | $(4)$ | 8745 | 9822 | 10183 | 10269 | 10280 | 10935 | 11304 | 11391 |

| $C_{n=4,d}^{j=6,m=2}$ | | $(1,4)$ | $(2,2)$ | $(2,3)$ | $(2,4)$ | $(3,3)$ | $(3,4)$ | $(4,4)$ |
|---|---|---|---|---|---|---|---|---|
| $d=0$ | $(0)$ | 5720 | 5916 | 5992 | 6003 | 6068 | 6079 | 6090 |
| | $(1)$ | 9322 | 9583 | 9670 | 9681 | 9757 | 9768 | 9779 |
| | $(2)$ | 10888 | 11160 | 11247 | 11258 | 11334 | 11345 | 11356 |
| | $(3)$ | 11337 | 11609 | 11696 | 11707 | 11783 | 11794 | 11805 |
| | $(4)$ | 11402 | 11674 | 11761 | 11772 | 11848 | 11859 | 11870 |

| $C_{n=1,d}^{j=7,m=2}$ | | $(0,0)$ | $(0,1)$ | $(1,1)$ |
|---|---|---|---|---|
| $d=0$ | $(0,0)$ | 92 | 98 | 104 |
| | $(0,1)$ | 112 | 118 | 124 |
| | $(1,1)$ | 132 | 138 | 144 |
| $d=1$ | $(0,0)$ | 141 | 148 | 155 |
| | $(0,1)$ | 168 | 175 | 182 |
| | $(1,1)$ | 195 | 202 | 209 |
| $d=2$ | $(0,0)$ | 199 | 207 | 215 |
| | $(0,1)$ | 234 | 242 | 250 |
| | $(1,1)$ | 269 | 277 | 285 |
| $d=3$ | $(0,0)$ | 266 | 275 | 284 |
| | $(0,1)$ | 310 | 319 | 328 |
| | $(1,1)$ | 354 | 363 | 372 |

| $C^{j=7,m=2}_{n=2,d}$ | | (0,0) | (0,1) | (0,2) | (1,1) | (1,2) | (2,2) |
|---|---|---|---|---|---|---|---|
| $d=0$ | (0,0) | 507 | 558 | 566 | 610 | 618 | 626 |
| | (0,1) | 706 | 765 | 773 | 825 | 833 | 841 |
| | (0,2) | 741 | 800 | 808 | 860 | 868 | 876 |
| | (1,1) | 940 | 1007 | 1015 | 1075 | 1083 | 1091 |
| | (1,2) | 975 | 1042 | 1050 | 1110 | 1118 | 1126 |
| | (2,2) | 1010 | 1077 | 1085 | 1145 | 1153 | 1161 |
| $d=1$ | (0,0) | 763 | 822 | 831 | 882 | 891 | 900 |
| | (0,1) | 1029 | 1097 | 1106 | 1166 | 1175 | 1184 |
| | (0,2) | 1073 | 1141 | 1150 | 1210 | 1219 | 1228 |
| | (1,1) | 1339 | 1416 | 1425 | 1494 | 1503 | 1512 |
| | (1,2) | 1383 | 1460 | 1469 | 1538 | 1547 | 1556 |
| | (2,2) | 1427 | 1504 | 1513 | 1582 | 1591 | 1600 |
| $d=2$ | (0,0) | 1062 | 1129 | 1139 | 1197 | 1207 | 1217 |
| | (0,1) | 1404 | 1481 | 1491 | 1559 | 1569 | 1579 |
| | (0,2) | 1458 | 1535 | 1545 | 1613 | 1623 | 1633 |
| | (1,1) | 1800 | 1887 | 1897 | 1975 | 1985 | 1995 |
| | (1,2) | 1854 | 1941 | 1951 | 2029 | 2039 | 2049 |
| | (2,2) | 1908 | 1995 | 2005 | 2083 | 2093 | 2103 |

| $C^{j=7,m=3}_{n=1,d}$ | | (0,0) | (0,1) | (1,1) |
|---|---|---|---|---|
| $d=0$ | $\phi$ | 84 | 98 | 112 |
| $d=1$ | $\phi$ | 232 | 252 | 272 |
| $d=2$ | $\phi$ | 453 | 480 | 507 |
| $d=3$ | $\phi$ | 760 | 795 | 830 |

| $C^{j=7,m=3}_{n=2,d}$ | | (0,0) | (0,1) | (0,2) | (1,1) | (1,2) | (2,2) |
|---|---|---|---|---|---|---|---|
| $d=0$ | $\phi$ | 752 | 953 | 980 | 1161 | 1188 | 1215 |
| $d=1$ | $\phi$ | 2010 | 2290 | 2325 | 2578 | 2613 | 2648 |
| $d=2$ | $\phi$ | 3820 | 4191 | 4235 | 4571 | 4615 | 4659 |

| $C^{j=7,m=3}_{n=3,d}$ | | (0,0) | (0,1) | (0,2) | (0,3) | (1,1) | (1,2) | (1,3) | (2,2) |
|---|---|---|---|---|---|---|---|---|---|
| $d=0$ | $\phi$ | 4720 | 6242 | 6613 | 6657 | 7855 | 8235 | 8279 | 8615 |
| $d=1$ | $\phi$ | 12088 | 14167 | 14641 | 14695 | 16340 | 16824 | 16878 | 17308 |

| $C_{n=3,d}^{j=7,m=3}$ | | $(2,3)$ | $(3,3)$ |
|---|---|---|---|
| $d=0$ | $\phi$ | 8659 | 8703 |
| $d=1$ | $\phi$ | 17362 | 17416 |

| $C_{n=4,d}^{j=7,m=3}$ | | $(0,0)$ | $(0,1)$ | $(0,2)$ | $(0,3)$ | $(0,4)$ | $(1,1)$ | $(1,2)$ | $(1,3)$ |
|---|---|---|---|---|---|---|---|---|---|
| $d=0$ | $\phi$ | 23203 | 31302 | 33996 | 34585 | 34650 | 40016 | 42825 | 43425 |

| $C_{n=4,d}^{j=7,m=3}$ | | $(1,4)$ | $(2,2)$ | $(2,3)$ | $(2,4)$ | $(3,3)$ | $(3,4)$ | $(4,4)$ |
|---|---|---|---|---|---|---|---|---|
| $d=0$ | $\phi$ | 43490 | 45645 | 46245 | 46310 | 46845 | 46910 | 46975 |

| $C_{n=1,d}^{j=8,m=2}$ | | $(0,0)$ | $(0,1)$ | $(1,1)$ |
|---|---|---|---|---|
| $d=0$ | $(0,0,0)$ | 141 | 148 | 155 |
| | $(0,0,1)$ | 168 | 175 | 182 |
| | $(0,1,1)$ | 195 | 202 | 209 |
| | $(1,1,1)$ | 222 | 229 | 236 |
| $d=1$ | $(0,0,0)$ | 199 | 207 | 215 |
| | $(0,0,1)$ | 234 | 242 | 250 |
| | $(0,1,1)$ | 269 | 277 | 285 |
| | $(1,1,1)$ | 304 | 312 | 320 |
| $d=2$ | $(0,0,0)$ | 266 | 275 | 284 |
| | $(0,0,1)$ | 310 | 319 | 328 |
| | $(0,1,1)$ | 354 | 363 | 372 |
| | $(1,1,1)$ | 398 | 407 | 416 |

| $C_{n=1,d}^{j=8,m=3}$ | | $(0,0)$ | $(0,1)$ | $(1,1)$ |
|---|---|---|---|---|
| $d=0$ | $(0)$ | 232 | 252 | 272 |
| | $(1)$ | 280 | 300 | 320 |
| $d=1$ | $(0)$ | 453 | 480 | 507 |
| | $(1)$ | 528 | 555 | 582 |
| $d=2$ | $(0)$ | 760 | 795 | 830 |
| | $(1)$ | 870 | 905 | 940 |

| $C^{j=8,m=3}_{n=2,d}$ | | $(0,0)$ | $(0,1)$ | $(0,2)$ | $(1,1)$ | $(1,2)$ | $(2,2)$ |
|---|---|---|---|---|---|---|---|
| | $(0)$ | 2010 | 2290 | 2325 | 2578 | 2613 | 2648 |
| $d=0$ | $(1)$ | 2770 | 3085 | 3120 | 3408 | 3443 | 3478 |
| | $(2)$ | 2880 | 3195 | 3230 | 3518 | 3553 | 3588 |
| | $(0)$ | 3820 | 4191 | 4235 | 4571 | 4615 | 4659 |
| $d=1$ | $(1)$ | 4986 | 5401 | 5445 | 5825 | 5869 | 5913 |
| | $(2)$ | 5140 | 5555 | 5599 | 5979 | 6023 | 6067 |

| $C^{j=8,m=3}_{n=3,d}$ | | $(0,0)$ | $(0,1)$ | $(0,2)$ | $(0,3)$ | $(1,1)$ | $(1,2)$ | $(1,3)$ | $(2,2)$ |
|---|---|---|---|---|---|---|---|---|---|
| | $(0)$ | 12088 | 14167 | 14641 | 14695 | 16340 | 16824 | 16878 | 17308 |
| $d=0$ | $(1)$ | 18358 | 20911 | 21439 | 21493 | 23568 | 24106 | 24160 | 24644 |
| | $(2)$ | 20042 | 22649 | 23177 | 23231 | 25360 | 25898 | 25952 | 26436 |
| | $(3)$ | 20250 | 22857 | 23385 | 23439 | 25568 | 26106 | 26160 | 26644 |

| $C^{j=8,m=3}_{n=3,d}$ | | $(2,3)$ | $(3,3)$ |
|---|---|---|---|
| | $(0)$ | 17362 | 17416 |
| $d=0$ | $(1)$ | 24698 | 24752 |
| | $(2)$ | 26490 | 26544 |
| | $(3)$ | 26698 | 26752 |

| $C^{j=9,m=3}_{n=1,d}$ | | $(0,0)$ | $(0,1)$ | $(1,1)$ |
|---|---|---|---|---|
| | $(0,0)$ | 453 | 480 | 507 |
| $d=0$ | $(0,1)$ | 528 | 555 | 582 |
| | $(1,1)$ | 603 | 630 | 657 |
| | $(0,0)$ | 760 | 795 | 830 |
| $d=1$ | $(0,1)$ | 870 | 905 | 940 |
| | $(1,1)$ | 980 | 1015 | 1050 |
| | $(0,0)$ | 1166 | 1210 | 1254 |
| $d=2$ | $(0,1)$ | 1320 | 1364 | 1408 |
| | $(1,1)$ | 1474 | 1518 | 1562 |

| $C^{j=9,m=4}_{n=1,d}$ | | $(0,0)$ | $(0,1)$ | $(1,1)$ |
|---|---|---|---|---|
| $d=0$ | $\phi$ | 330 | 378 | 426 |
| $d=1$ | $\phi$ | 975 | 1050 | 1125 |
| $d=2$ | $\phi$ | 2035 | 2145 | 2255 |

| $C^{j=9,m=4}_{n=2,d}$ | | (0,0) | (0,1) | (0,2) | (1,1) | (1,2) | (2,2) |
|---|---|---|---|---|---|---|---|
| $d=0$ | $\phi$ | 4035 | 5020 | 5130 | 6040 | 6150 | 6260 |
| $d=1$ | $\phi$ | 11605 | 13101 | 13255 | 14641 | 14795 | 14949 |

| $C^{j=9,m=4}_{n=3,d}$ | | (0,0) | (0,1) | (0,2) | (0,3) | (1,1) | (1,2) | (1,3) | (2,2) |
|---|---|---|---|---|---|---|---|---|---|
| $d=0$ | $\phi$ | 33913 | 44341 | 46487 | 46695 | 55419 | 57619 | 57827 | 59819 |

| $C^{j=9,m=4}_{n=3,d}$ | | (2,3) | (3,3) |
|---|---|---|---|
| $d=0$ | $\phi$ | 60027 | 60235 |

| $C^{j=10,m=4}_{n=1,d}$ | | (0,0) | (0,1) | (1,1) |
|---|---|---|---|---|
| $d=0$ | (0) | 975 | 1050 | 1125 |
| | (1) | 1140 | 1215 | 1290 |
| $d=1$ | (0) | 2035 | 2145 | 2255 |
| | (1) | 2310 | 2420 | 2530 |

| $C^{j=10,m=4}_{n=2,d}$ | | (0,0) | (0,1) | (0,2) | (1,1) | (1,2) | (2,2) |
|---|---|---|---|---|---|---|---|
| $d=0$ | (0) | 11605 | 13101 | 13255 | 14641 | 14795 | 14949 |
| | (1) | 15246 | 16896 | 17050 | 18590 | 18744 | 18898 |
| | (2) | 15675 | 17325 | 17479 | 19019 | 19173 | 19327 |

| $C^{j=11,m=5}_{n=1,d}$ | | (0,0) | (0,1) | (1,1) |
|---|---|---|---|---|
| $d=0$ | $\phi$ | 1287 | 1452 | 1617 |
| $d=1$ | $\phi$ | 3982 | 4257 | 4532 |

| $C^{j=11,m=5}_{n=2,d}$ | | (0,0) | (0,1) | (0,2) | (1,1) | (1,2) | (2,2) |
|---|---|---|---|---|---|---|---|
| $d=0$ | $\phi$ | 20152 | 24618 | 25047 | 29238 | 29667 | 30096 |

| $C^{j=12,m=5}_{n=1,d}$ | | (0,0) | (0,1) | (1,1) |
|---|---|---|---|---|
| $d=0$ | (0) | 3982 | 4257 | 4532 |
| | (1) | 4554 | 4829 | 5104 |

| $C^{j=13,m=6}_{n=1,d}$ | | (0,0) | (0,1) | (1,1) |
|---|---|---|---|---|
| $d=0$ | $\phi$ | 5005 | 5577 | 6149 |

# Acknowledgements

I extend my heartfelt thanks to my supervisor, Prof. Hirokazu Tsunetsugu, for his invaluable advice in enhancing my thesis and for the enriching daily conversations through which I was able to develop my understanding and proficiency in physics.

Special thanks to Dr. Balázs Pozsgay for highlighting the connection between the strong-coupling limit of my local charges and Maassarani's work and for his assistance during my stay in Budapest. I am also grateful for our collaboration on the Temperley-Lieb project, which significantly expanded my understanding of integrable systems and greatly enriched this thesis.

I deeply thank Dr. Lenart Zadnik for his exceptional support during my stay in Trieste and for the insightful conversations we shared at StatPhys23. The memories of the beautiful sea in Trieste served as a source of inspiration and energy during the writing of this doctoral thesis.

My sincere thanks to Dr. Yuan Miao for the inspiring continuous discussions on integrability and his encouragement. I also greatly appreciate our collaboration on the ongoing project.

I convey my deep gratitude to Dr. Yuji Nozawa for the collaborative work on the XYZ project and for the engaging discussions, which continued even after his transition to the industry.

I wish to acknowledge Mr. Kyoichi Yamada for his collaboration on the MPO project and for his astute insights, which have been invaluable to our research.

I am deeply thankful to Dr. Tatsuhiko N. Ikeda, a former research associate of the Tsunetsugu group, for his ongoing encouragement and support throughout my research journey.

My sincere appreciation goes to Dr. Chinzei, Mr. Hideki Ema, Mr. Koji Kawahara, Mr. Yuki Kaneko, Mr. Kazuki Takaji, and Mr. Ichiro Tanaka from the Tsunetsugu group for their support and daily conversations.

I am grateful to Ms. Atsuko Tsuji, the secretary of the Tsunetsugu group, for her constant assistance and support in my research.

I express profound thanks for the financial support received from the Forefront Physics and Mathematics Program to Drive Transformation (FoPM), World-leading Innovative Graduate Study (WINGS) Program, JSR Fellowship, the University of Tokyo, and the Japan Society for the Promotion of Science (JSPS), all of which were crucial to my research endeavors.

Finally, I wish to acknowledge the unwavering and unconditional support from my parents, Hiroki and Yumi, and my sister, Yuka, who have been my pillars of strength throughout this academic journey.